\documentclass[12pt]{article}
\usepackage{jheppub}

\pdfoutput=1

\usepackage{amsmath,bbm,array,amsfonts,graphicx,wrapfig,lscape,float,mathtools,multirow,longtable}
\usepackage[dvipsnames,table]{xcolor}
\usepackage{array}
\usepackage{color}
\usepackage{physics}
\usepackage{subcaption}
\usepackage{tikz,tikz-3dplot}
\usepackage[colorlinks=true]{hyperref}
\newcolumntype{C}[1]{>{\centering\let\newline\\\arraybackslash\hspace{0pt}}m{#1}}

\newcommand{\be}{\begin{equation}}
\newcommand{\ee}{\end{equation}}
\newcommand{\beq}{\begin{equation}}
\newcommand{\beql}[1]{\begin{equation}\label{#1}}
\newcommand{\eeq}{\end{equation}}
\newcommand{\ba}{\begin{array}}
\newcommand{\ea}{\end{array}}
\newcommand{\bea}{\begin{eqnarray}}
\newcommand{\beal}[1]{\begin{eqnarray}\label{#1}}
\newcommand{\eea}{\end{eqnarray}}
\newcommand{\ben}{\begin{enumerate}}
\newcommand{\een}{\end{enumerate}}
\newcommand{\bean}{\begin{eqnarray*}}
\newcommand{\eean}{\end{eqnarray*}}
\newcommand{\eref}[1]{(\ref{#1})}
\newcommand{\sref}[1]{\S\ref{#1}}
\newcommand{\tref}[1]{Table~\ref{#1}}
\newcommand{\nn}{\nonumber}

\newcommand{\fref}[1]{Figure \ref{#1}}
\newcommand{\btab}[1]{\begin{tabular}{#1}}
\newcommand{\etab}{\end{tabular}}

\newcommand{\comment}[1]{}

\newcommand{\IC}{\mathbb{C}}

\newcommand{\qed}{\nobreak \ifvmode \relax \else
      \ifdim\lastskip<1.5em \hskip-\lastskip
      \hskip1.5em plus0em minus0.5em \fi \nobreak
      \vrule height0.75em width0.5em depth0.25em\fi}

\definecolor{darkspringgreen}{rgb}{0.09, 0.45, 0.27}
\definecolor{forestgreen}{rgb}{0.13, 0.55, 0.13}

\title{
Birational Transformations and\\
$2d$ $(0,2)$ Quiver Gauge Theories\\
beyond Toric Fano 3-folds
}

\author[a]{Dongwook Ghim,}
\author[b]{Minsung Kho,}
\author[b,c]{Rak-Kyeong Seong}

\affiliation[a]{
RIKEN Center for Interdisciplinary Theoretical and Mathematical Sciences (iTHEMS), \\
RIKEN, 2-1 Hirosawa, Wako, Saitama 351-0198, Japan
}
	
\affiliation[b]{
Department of Mathematical Sciences, and
\textsuperscript{$c$}Department of Physics,\\
Ulsan National Institute of Science and Technology,\\
50 UNIST-gil, Ulsan 44919, South Korea
}

\emailAdd{dongwook.ghim@riken.jp}
\emailAdd{minsung@unist.ac.kr}
\emailAdd{seong@unist.ac.kr}

\preprint{
\begin{flushright}
UNIST-MTH-25-RS-01 \\
RIKEN-iTHEMS-Report-25
\end{flushright}
}

\abstract{
We show that a family of birational transformations that relate toric Fano 3-folds 
defined by reflexive lattice polytopes can be identified with mass deformations of corresponding $2d$ $(0,2)$ supersymmetric quiver gauge theories.
These theories are
realized by a Type IIA brane configuration known as brane brick models. 
We further show that the same 
family of birational transformations
extends to more general toric Calabi-Yau 4-folds,
including those defined by non-reflexive toric diagrams.
Under these birational transformations,
the mesonic moduli spaces of the associated abelian $2d$ $(0,2)$ supersymmetric gauge theories and brane brick models
share the same number of generators and
the same Hilbert series when refined only under the $U(1)_R$ symmetry. 
Since these transformations
categorize toric Calabi-Yau 4-folds and their corresponding $2d$ $(0,2)$ supersymmetric gauge theories 
into non-trivial equivalence classes, 
we anticipate that our findings will pave the way
for a `Minimal Model Program' for quiver gauge theories corresponding to toric Calabi-Yau manifolds.
}
\begin{document}

\maketitle

\section{Introduction}

Birational geometry is a branch of modern algebraic geometry and is concerned with the classification of algebraic varieties up to birational equivalences. 
It has profoundly influenced our understanding of algebraic curves, surfaces and higher-dimensional varieties, leading to the groundbreaking Minimal Model Program (MMP)
initiated by Mori \cite{mori1979projective, fa4d6420-e404-3d65-bad9-fee06a1f00ea, ab3c5836-bdbd-3fe7-9721-d43dbbc02ff0} and further developed by others \cite{kawamata1985pluricanonical,MR1158625, Kollar_Mori_1998}.
More recently, these advances in classifying algebraic varieties and moduli spaces
have led to significant contributions towards the Minimal Model Program, including the recent advances made by Birkar and others \cite{birkar2010existence}.

\begin{figure}[H]
\begin{center}
\resizebox{0.78\hsize}{!}{
\includegraphics[height=6cm]{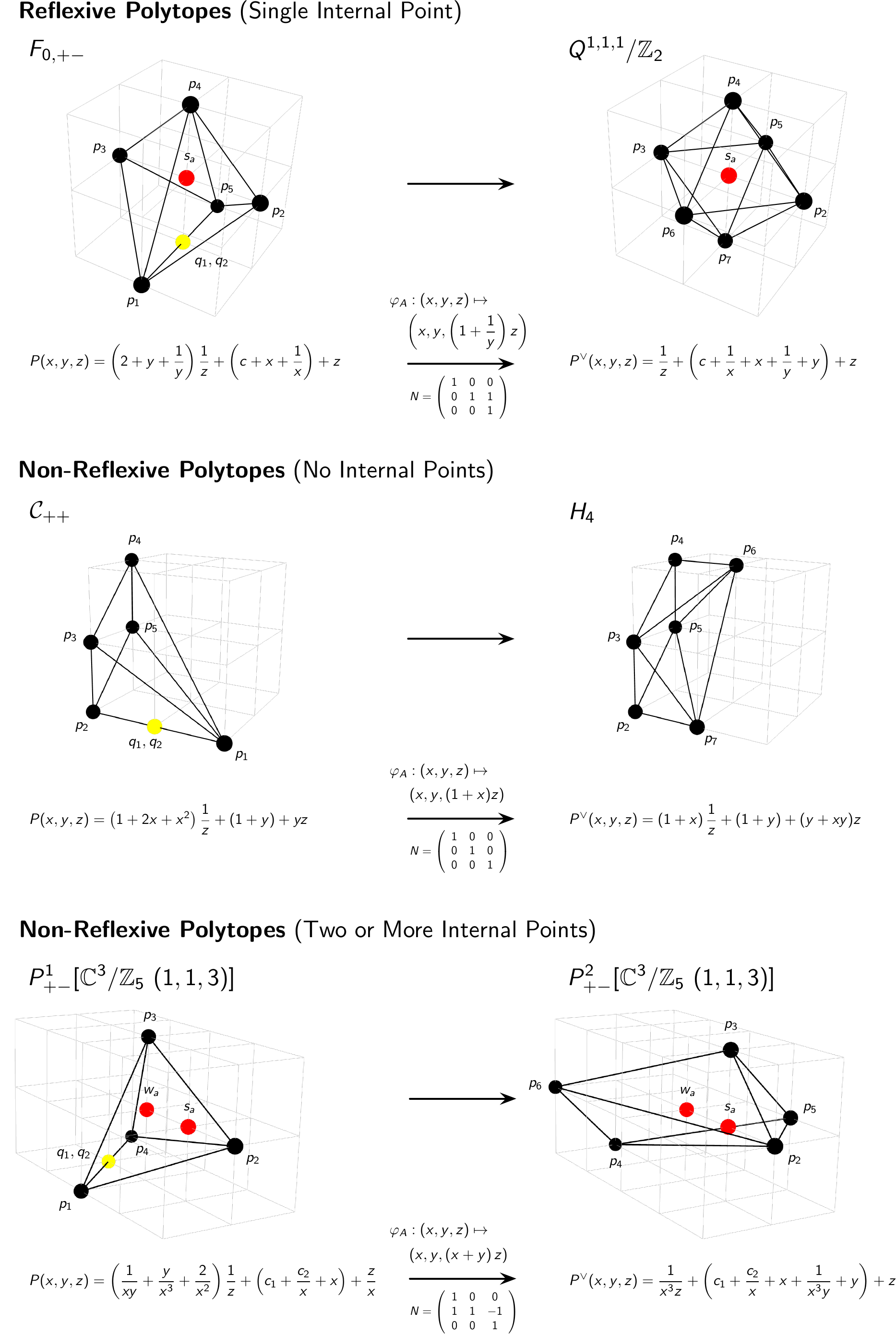} 
}
\caption{
This work considers 3 families of toric Calabi-Yau 4-folds
and their corresponding $2d$ $(0,2)$ supersymmetric gauge theories and brane brick models.
The first family has toric diagrams that are reflexive lattice polytopes in $\mathbb{Z}^3$ with the origin as their unique internal lattice point.
The second and third families have toric diagrams that are not reflexive with no internal lattice points and with $2$ or more internal lattice points, respectively.
In this work, we study birational transformations $\varphi_A$ combined with $GL(3,\mathbb{Z})$ transformations $N$ that act on the Newton polynomials $P(x,y,z)$ of the toric diagrams
and identify pairs of toric Calabi-Yau 4-folds to each other within the same family. 
\label{f_ref_0a}}
 \end{center}
 \end{figure}

Techniques and advances in birational geometry \cite{akhtar2012minkowski, gross2013birational, coates2016quantum, kasprzyk2017minimality, coates2021maximally, Coates2022MirrorSL, corti2023cluster} are increasingly becoming more relevant in string theory, 
particularly in the study of complex manifolds such as Calabi–Yau varieties, 
which are crucial for both string compactification and the vacuum moduli spaces of supersymmetric gauge theories.
In the case of toric Fano 3-folds, 
the complex cone over them forms a family of 
non-compact toric Calabi-Yau 4-folds
that can be probed by D1-branes.
The worldvolume theories on probe D1-branes at toric Calabi-Yau 4-folds 
are $2d$ $(0,2)$ supersymmetric quiver gauge theories,
which admit a Type IIA brane configuration description in string theory known as brane brick models \cite{Franco:2015tna, Franco:2015tya, Franco:2016nwv, Franco:2016qxh}.

The Minimal Model Program in birational geometry aims to simplify 
an algebraic variety by performing a sequence of birational transformations,
ultimately obtaining a minimal representative of the original variety. 
Analogously, 
in the setting of toric Fano 3-folds
-- corresponding to reflexive convex lattice polytopes in $\mathbb{Z}^3$
--
it was shown in \cite{akhtar2012minkowski}
that one can introduce birational transformations that not only 
preserve the period of the toric Fano 3-folds but also 
organize them into distinct birational equivalence classes.
Out of the 4319 reflexive polytopes in $\mathbb{Z}^3$ originally classified by Kreuzer and Skarke \cite{Kreuzer:1995cd, Kreuzer:1998vb, Kreuzer:2000xy},
3025 polytopes fall into 165 non-trivial equivalence classes referred to in \cite{akhtar2012minkowski} as `buckets'. 

Motivated by these findings,
\cite{Ghim:2024asj} initiated a program to investigate the implications of these birational transformations for the
$2d$ $(0,2)$ supersymmetric gauge theories realized in string theory, focusing on the toric Calabi-Yau 4-folds corresponding to toric Fano 3-folds.
In particular, it was demonstrated in \cite{Ghim:2024asj}
that if two 
$2d$ $(0,2)$ supersymmetric gauge theories related to toric Fano 3-folds
can be connected by a mass deformation of the $J$- and $E$-terms \cite{Franco:2023tyf},
then the associated toric Fano 3-folds
can be related by a birational transformation of the form introduced in \cite{akhtar2012minkowski}.
This relationship was explicitly verified in \cite{Ghim:2024asj}
for brane brick models associated with the abelian orbifold $\mathbb{C}^4/\mathbb{Z}_6$ with orbifold action $(1,1,2,2)$ \cite{Davey:2010px, Hanany:2010ne, Franco:2015tna},
which is related by a birational transformation to the toric Calabi-Yau 4-fold known as $P_{+-}^{2} [\text{dP}_0]$ \cite{Ghim:2024asj}.
Both of these toric Calabi-Yau 4-folds are
complex cones over toric Fano 3-folds that belong both to one of the non-trivial `buckets' identified in \cite{akhtar2012minkowski}.

The family of birational transformations for toric Fano 3-folds introduced in \cite{akhtar2012minkowski} 
can be combined with $GL(3,\mathbb{Z})$ transformations of their corresponding toric diagrams in $\mathbb{Z}^3$.
These combined transformations are also referred to as algebraic mutations in \cite{akhtar2012minkowski}, 
and can be interpreted as a combinatorial mutation --
a direct combinatorial transformation of the toric diagram itself.

The family of birational transformations studied in \cite{akhtar2012minkowski} act on the Newton polynomials $P(x,y,z)$ of the convex lattice polytopes in $\mathbb{Z}^3$.
These polytopes serve as toric diagrams of the corresponding toric Calabi-Yau 4-folds. 
The zero locus $\Sigma ~:~ P(x,y,z)=0$
defines a holomorphic surface that
determines the underlying brane brick models \cite{Franco:2015tna, Franco:2015tya, Franco:2016nwv, Franco:2016qxh}, which realize the corresponding $2d$ $(0,2)$ supersymmetric gauge theories. 
Because the Newton polynomial $P(x,y,z)$
simultaneously identifies the toric variety and determines the brane brick model as a Type IIA brane configuration in string theory, 
it is unsurprising that birational transformations acting on $P(x,y,z)$
directly influence the associated $2d$ $(0,2)$ supersymmetric gauge theories and their brane brick models.

In this work, 
we continue the investigation initiated in \cite{Ghim:2024asj}
by studying brane brick models and $2d$ $(0,2)$ supersymmetric gauge theories
corresponding to two toric Fano 3-folds, which are related by a mass deformation of their $J$- and $E$-terms.
We show that the associated toric Fano 3-folds
are then related by a birational transformation.
The toric Fano 3-folds that we study in this work
correspond to the toric Calabi-Yau 4-folds known as $F_{0,+-}$ first studied with its corresponding brane brick model in \cite{Franco:2023tyf}
and the toric Calabi-Yau 4-fold given by the orbifold $Q^{1,1,1}/\mathbb{Z}_2$ whose toric phase was first studied in \cite{Franco:2016nwv, Franco:2016qxh, Franco:2023tyf}.
Our findings further support the evidence presented in \cite{Ghim:2024asj},
that brane brick models and $2d$ $(0,2)$ supersymmetric gauge theories related by a mass deformation
and corresponding to toric Fano 3-folds
are also related by a birational transformation.

Moreover, this work broadens the scope of the investigation initiated in \cite{Ghim:2024asj}
by identifying birational transformations acting on toric varieties defined by non-reflexive convex lattice polytopes in $\mathbb{Z}^3$.
Although these toric varieties do not correspond to toric Fano 3-folds, the complex cone over them still defines a toric Calabi-Yau 4-fold.
We show that algebraic mutations, which combines
these birational transformations with $GL(3,\mathbb{Z})$ transformations,
exist for non-reflexive toric diagrams,  
mapping a non-reflexive toric diagram to another convex lattice polytope in $\mathbb{Z}^3$.
By identifying the corresponding brane brick models and $2d$ $(0,2)$ supersymmetric gauge theories for these toric Calabi-Yau 4-folds,
we illustrate that when the $2d$ $(0,2)$ theories are related by a mass deformation, then 
the corresponding toric Calabi-Yau 4-folds with non-reflexive toric diagrams are related by an algebraic mutation based on a birational transformation.
We present several examples, categorized according to whether the corresponding toric diagrams are given by reflexive polytopes, 
by non-reflexive polytopes with no internal points, or by non-reflexive polytopes with two or more internal points, as illustrated in \fref{f_ref_0a}.

This work thus extends the birational transformations introduced in \cite{akhtar2012minkowski} -- originally for toric Fano 3-folds --
to a broader class of toric varieties defined by non-reflexive convex lattice polytopes.
We conjecture that, in general, whenever two brane brick models and their corresponding $2d$ $(0,2)$ supersymmetric gauge theories are related by a mass deformation, 
the corresponding toric Calabi-Yau 4-folds are related by a birational transformation, as demonstrated in this work. 
We anticipate that these findings will contribute to a \textit{Minimal Model Program for supersymmetric gauge theories realized in string theory}, 
starting with the brane brick models and $2d$ $(0,2)$ supersymmetric gauge theories associated with toric Calabi-Yau 4-folds.
\\

Our work is organized as follows. 
Section \sref{sec:review}
gives a brief introduction to the family of $2d$ $(0,2)$ supersymmetric gauge theories
that are realized as worldvolume theories of a D1-brane probing a toric Calabi-Yau 4-fold.
The section connects the geometry of the toric Calabi-Yau 4-fold given by the holomorphic surface $\Sigma$
with the Type IIA brane configuration giving the brane brick model.
It further explains how mass deformation of the $2d$ $(0,2)$ supersymmetric gauge theories
affects the brane brick model and the corresponding toric Calabi-Yau 4-fold
and summarizes how brane brick models have been studied in the context of toric Fano 3-folds
in \cite{Franco:2022gvl}.
Section \sref{sec:03}
reviews the birational transformations as introduced in \cite{akhtar2012minkowski} for toric Fano 3-folds.
The section reviews how birational transformations combined with $GL(3,\mathbb{Z})$ transformations of the toric diagram 
give a family of transformations called algebraic mutations in \cite{akhtar2012minkowski}.
The section also summarizes how they affect the corresponding $2d$ $(0,2)$ supersymmetric gauge theories and brane brick models. 
Section \sref{sec:examples} then proceeds in 
presenting several examples how birational transformations for toric Fano 3-folds
and more general toric varieties with non-reflexive toric diagrams 
affect the corresponding $2d$ $(0,2)$ supersymmetric gauge theories and brane brick models.
The section here presents examples for brane brick models and their mass deformations corresponding to birational transformations 
for toric Calabi-Yau 4-folds whose toric diagrams have one internal point, no internal points and 2 internal points as summarized in \fref{f_ref_0a}. 
\\

\section{Toric Calabi-Yau 4-folds and $2d$ $(0,2)$ Gauge Theories \label{sec:review} } 

\subsection{$2d$ $(0,2)$ Gauge Theories and Brane Brick Models \label{sec:bbm}}

The worldvolume theories of a D1-brane probing a toric Calabi-Yau 4-fold form a family of $2d$ $(0,2)$ supersymmetric gauge theories. 
The D1-brane at the Calabi-Yau 4-fold is under T-duality 
related to a Type IIA brane configuration  
consisting of D4-branes suspended between a NS5-brane that wraps a holomorphic surface $\Sigma$.
This brane configuration, summarized in \tref{t_brane}, is referred to 
as a \textbf{brane brick model} \cite{Franco:2015tna, Franco:2015tya}.

The holomorphic surface $\Sigma$ is defined by the zero locus of the \textbf{Newton polynomial} $P(x,y,z)$ associated to the toric diagram $\Delta$ of the toric Calabi-Yau 4-fold.
The holomorphic surface is given by, 
\beal{es01a01}\label{newton-poly}
\Sigma ~ : ~  P(x,y,z) = 0 ~,~
\eea
where the Newton polynomial $P(x,y,z)$ is in $x,y,z\in \mathbb{C}^*$.
For a given toric diagram $\Delta$ of a toric Calabi-Yau 4-fold, 
the Newton polynomial $P(x,y,z)$ can be written as 
\beal{es01a01b}
P(x,y,z) = 
\sum_{\mathbf{v} = (v_1,v_2, v_3) \in \Delta}
c_{(v_1,v_2,v_3)} ~x^{v_1} y ^{v_2} z^{v_3} ~,~
\eea
where the sum is over the points $\mathbf{v} = (v_1,v_2, v_3) \in \mathbb{Z}^{3}$ of the toric diagram $\Delta$.
The complex coefficients $c_{(v_1,v_2,v_3)}  \in \mathbb{C}^*$
correspond to the complex structure moduli of the mirror Calabi-Yau 4-fold \cite{Hori:2000kt, Hori:2000ck, Cachazo:2001sg, Feng:2005gw, Franco:2016qxh, Franco:2016tcm}.

In $\Sigma$, the coordinates $x, y, z$ 
can be identified respectively with the $(23)$, $(45)$ and $(67)$ directions in the Type IIA brane configuration in \tref{t_brane}. 
T-duality 
along the $(357)$ directions 
maps the probe D1-brane to the D4-brane.
We take the $(357)$ directions as the arguments of the complex coordinates $x, y, z$ such that the $(357)$ directions are periodically identified to form a 3-torus $T^3$. 
The brane brick model on $T^3$ can be visualized when one projects the holomorphic surface $\Sigma$ onto the angular part of the $x, y, z$ coordinates. This projection is also known as the \textbf{coamoeba projection} \cite{Feng:2005gw, Franco:2016qxh, Franco:2016tcm, Seong:2023njx, Seong:2024wkt} of $\Sigma$ onto $T^3$. 

\begin{table}[h!]
\centering
\begin{tabular}{|c|cc|cccccc|cc|}
\hline
\; & 0 & 1 & 2 & 3 & 4 & 5 & 6 & 7 & 8 & 9 \\
\hline  \hline
D4 & $\times$ & $\times$ & $\cdot$ & $\times$ & $\cdot$ & $\times$ & $\cdot$ & $\times$ & $\cdot$ & $\cdot$
\\
NS5 & $\times$ & $\times$ & \multicolumn{6}{c|}{------ $\Sigma$ ------} & $\cdot$ & $\cdot$
\\
\hline
\end{tabular}
\caption{The Type IIA brane configuration for brane brick models, where the holomorphic surface $\Sigma$ refers to the zero locus of Newton polynomial $P(x,y,z)$ 
for the toric diagram $\Delta$
corresponding to the toric Calabi-Yau 4-fold.}
\label{t_brane}
\end{table}

The D4-brane suspended between a NS5-brane wrapping $\Sigma$ forms a tessellation of $T^3$ consisting of polytopes referred to as \textit{brane bricks}, 
\textit{brick faces} and \textit{brick edges}.
One can introduce a dictionary between these objects forming the tessellation of $T^3$,
the underlying Type IIA brane configuration and the
the corresponding $2d$ $(0, 2)$ supersymmetric gauge theory as follows \cite{Franco:2015tya}:

\begin{figure}[H]
\begin{center}
\resizebox{0.9\hsize}{!}{
\includegraphics[height=6cm]{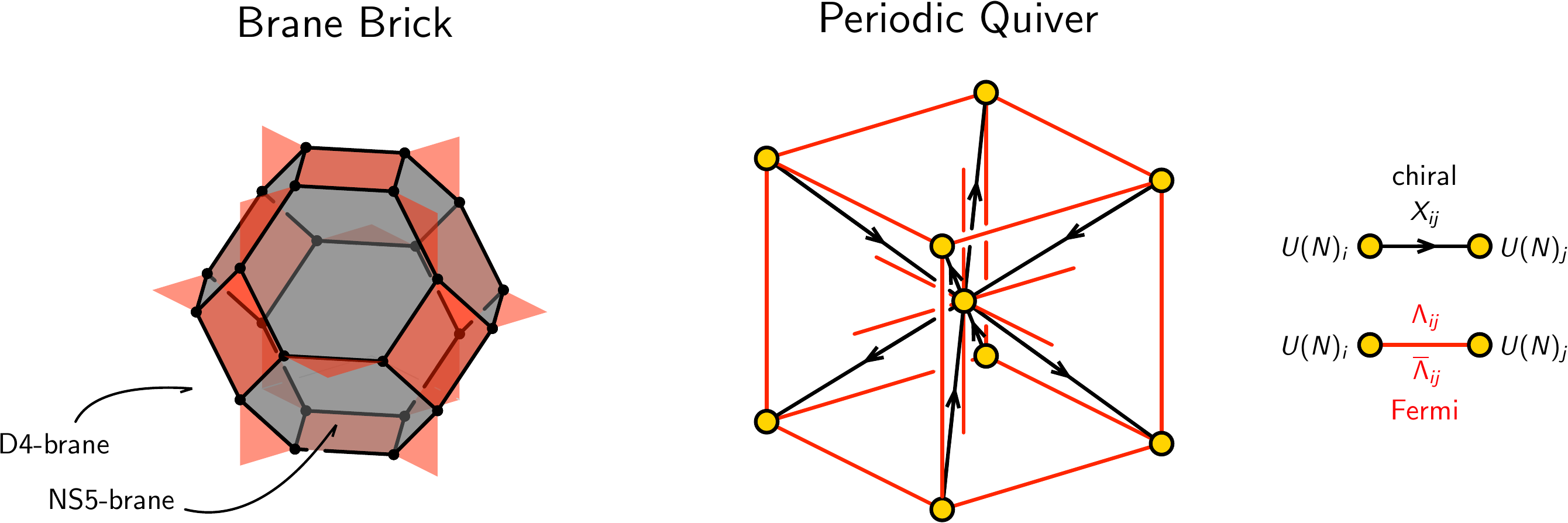} 
}
\caption{
A brane brick taking the form of a truncated octahedron and the corresponding periodic quiver. The interior of the brane brick indicates the location of a D5-brane and the boundary of the brane brick indicates the location of the NS5-brane wrapping $\Sigma$. 
\label{f_in10}}
 \end{center}
 \end{figure}

\begin{itemize}
\item \textbf{Brane Bricks.} As a $3$-dimensional building block of the tessellation of the 3-torus $T^3$, the interior of a brane brick indicates the location of the D4-branes suspended between the NS5-brane, as illustrated in \fref{f_in10}. 
Each brane brick corresponds to a $U(N)_i$ gauge group labelled by $i=1, \dots, G$ in the associated $2d$ $(0, 2)$ supersymmetric gauge theory, 
where $G$ is the total number of $U(N)_i$ gauge groups in the $2d$ $(0,2)$ theory.
For a single D1-brane probing the toric Calabi-Yau 4-fold, we have $N=1$
and an abelian $2d$ $(0,2)$ theory.
\\

\begin{figure}[H]
\begin{center}
\resizebox{0.67\hsize}{!}{
\includegraphics[height=6cm]{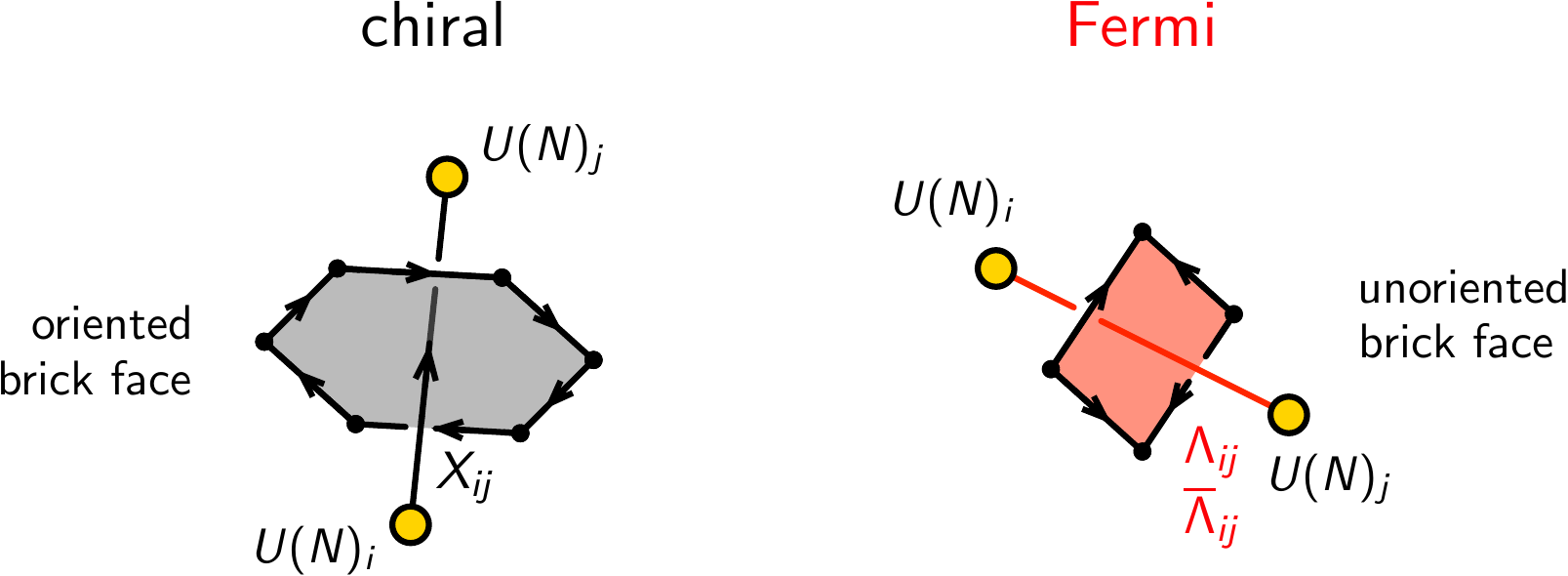} 
}
\caption{
Oriented brick faces corresponding to a bifundamental chiral field $X_{ij}$
and unoriented square brick faces corresponding to a Fermi field and its conjugate, $(\Lambda_{ij}, \overline{\Lambda}_{ij})$.
\label{f_in11}}
 \end{center}
 \end{figure}
    
\item \textbf{Brick Faces.} The boundary of brane bricks consists of polygons, which we refer to as brick faces. These brick faces indicate the location in the 3-torus $T^3$ of the NS5-brane wrapping $\Sigma$.
Forming a boundary between two adjacent brane bricks, the brick faces represent
bifundamental chiral multiplets $X_{ij}$ or Fermi multiplets $\Lambda_{ij}$
in the $2d$ $(0, 2)$ supersymmetric gauge theory.
These fields are under the bifundamental representation of $U(N)_i$ gauge groups associated with the adjacent brane bricks.
As illustrated in \fref{f_in11}, the bifundamental representation on chiral multiplets $X_{ij}$ naturally sets an orientation across the corresponding brick face, defining an orientation around the brick edges that form the boundary of a brick face. 
In comparison, the Fermi multiplets $\Lambda_{ij}$ and their conjugates $\overline{\Lambda}_{ij}$ in conjugate representation refer to the same brick face, and the conjugation of any Fermi multiplet leaves the Lagrangian of the $2d$ $(0, 2)$ supersymmetric gauge theory invariant. 
The brick faces corresponding to the chiral multiplets $X_{ij}$ are even-sided polygons, whereas the Fermi multiplets are associated to square brick faces in the brane brick model.\\

\begin{figure}[H]
\begin{center}
\resizebox{1\hsize}{!}{
\includegraphics[height=6cm]{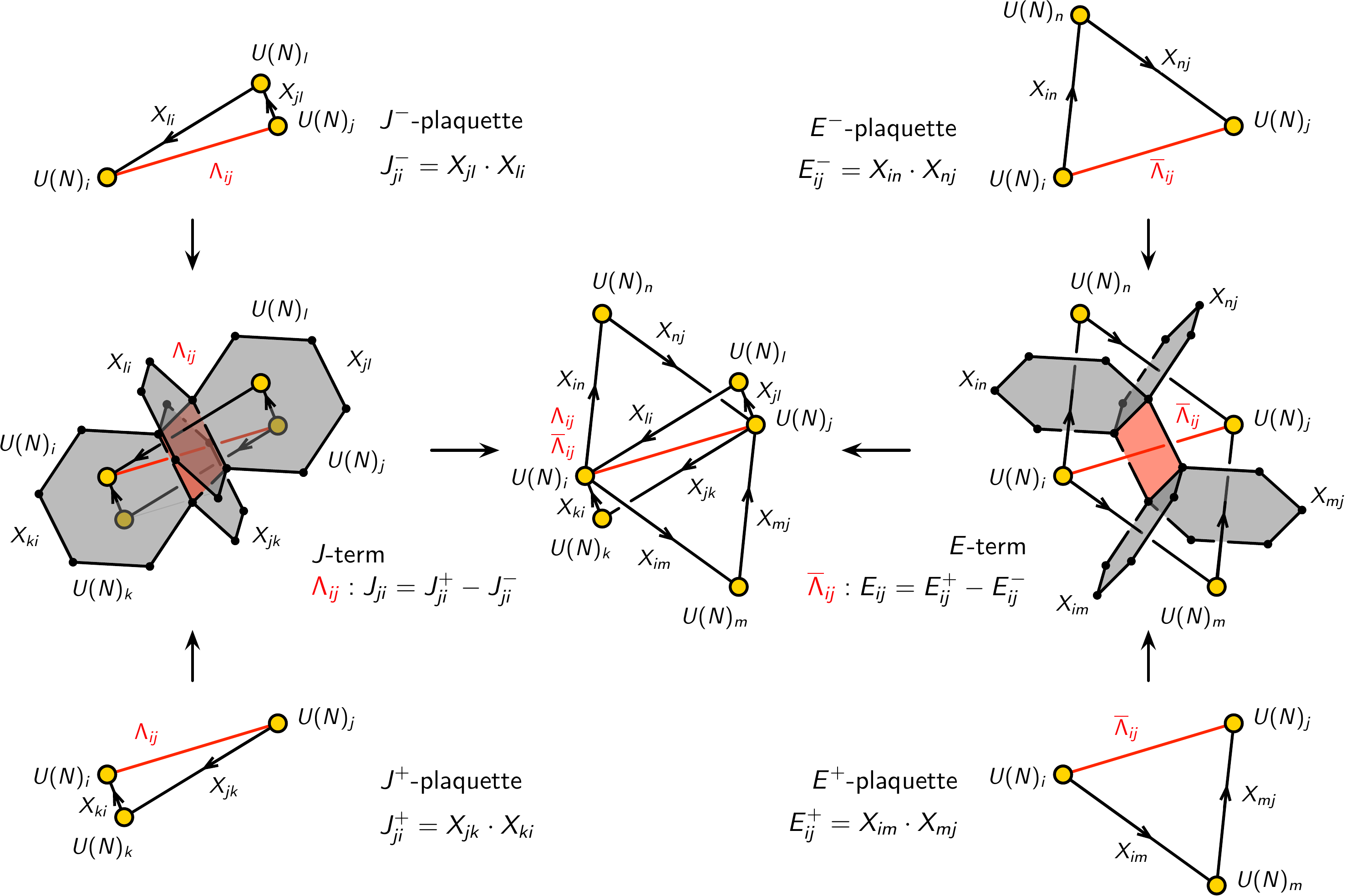} 
}
\caption{
The four plaquettes associated to $(\Lambda_{ij}, \overline{\Lambda}_{ij})$ 
that form a $J$- and $E$-term in the brane brick model.
\label{f_in12}}
 \end{center}
 \end{figure}

\item \textbf{Brick Edges.} 
As mentioned above, the brick faces corresponding to the bifundamental chiral multiplets $X_{ij}$ define an orientation around the brick edges.
This in turn defines a natural cyclic ordering of fields corresponding to the brick faces attached to a given brick edge. 
Amongst these fields is one Fermi multiplet $\Lambda_{ij}$ and a collection of chiral multiplets $X_{ij}$. 
We refer to this collection of fields associated to a brick edge as a \textbf{plaquette} in the brane brick model. 
They form together a gauge invariant combination of chiral and Fermi multiplets. 
Given that Fermi multiplets $\Lambda_{ij}$ and their conjugates are represented by square brick faces, 
each $(\Lambda_{ij}, \overline{\Lambda}_{ij})$ is associated to one of the four following plaquettes,
\beal{es01a02}
\Lambda_{ij} \cdot J^+_{ji} ~,~ \quad \Lambda_{ij} \cdot J^-_{ji} ~,~ \quad 
\overline{\Lambda}_{ij} \cdot E^+_{ij} ~,~ \quad \overline{\Lambda}_{ij} \cdot E^-_{ij} ~,~
\eea
where $J_{ji}^\pm$ and $E_{ij}^\pm$ are monomial products of bifundamental chiral fields. 
As illustrated in \fref{f_in12},
plaquettes at opposite sides of a square Fermi brick face are identified to form either a $J$- or a $E$-term in the $2d$ $(0, 2)$ supersymmetric gauge theory, which take respectively the following form, 
\beal{es01a03}
\ba{rccccc}
\Lambda_{ij} : & J_{ji} &=& J^+_{ji} &-& J^-_{ji} ~,~ \\
\overline{\Lambda}_{ij} : & E_{ij} &=& E^+_{ij} &-& E^-_{ij} ~.~
\ea
\eea
\end{itemize}

We note here that the $J$- and $E$-terms from a brane brick model in $T^3$ are by construction binomial and form a binomial ideal for the case when the gauge groups are all $U(1)$ with $N=1$. 
This binomial ideal is what we see in the following section as part of the definition for the mesonic moduli space $\mathcal{M}^{mes}$ of the $2d$ $(0, 2)$ supersymmetric gauge theory, which for the abelian $N=1$ case becomes the probed toric Calabi-Yau 4-fold. 
Without loss of generality, we assume in the rest of this work that the $2d$ $(0, 2)$ supersymmetric gauge theories realized by brane brick models are all abelian.  
\\

\subsection{Mesonic Moduli Spaces and Hilbert Series} \label{sec:HS}

As discussed in section \sref{sec:bbm}, the
$J$- and $E$-terms of the
$2d$ $(0,2)$ gauge theories given by a brane brick model take the following general form, 
\beal{es01a10}
\ba{rccccc}
\Lambda_{ij} : & J_{ji} &=& J^+_{ji} &-& J^-_{ji} ~,~ \\
\overline{\Lambda}_{ij} : & E_{ij} &=& E^+_{ij} &-& E^-_{ij} ~.~
\ea
\eea
where $J^{\pm}_{ji}$ and $E^{\pm}_{ij}$ are monomials in chiral matter fields $X_{ij}$. 
When the $2d$ $(0,2)$ gauge theory is abelian with $U(1)$ gauge groups, 
the $J$- and $E$-terms in \eref{es01a10} form a binomial ideal $\mathcal{I}_{JE}$
of a toric variety \cite{fulton1993introduction}.
The \textbf{mesonic moduli space $\mathcal{M}^{mes}$} \cite{Franco:2015tna, Franco:2015tya} of the brane brick model 
is then given 
by the following quotient,
\beal{es01a11}
\mathcal{M}^{mes} = \textrm{Spec} 
\left( \IC [ X_{ij} ] / \mathcal{I}^{\textrm{Irr}}_{JE} \right) // U(1)^{G-1} ~,~
\eea
where $\IC [ X_{ij} ] $ is the coordinate ring in chiral fields $X_{ij}$ of the toric variety, and
$\mathcal{I}^{\textrm{Irr}}_{JE}$ is the irreducible component of the ideal formed by the $J$- and $E$-terms. 
The quotient in \eref{es01a11} under $U(1)^{G-1}$
makes sure that the mesonic moduli space $\mathcal{M}^{mes}$
captures the spectrum of gauge invariant operators in chiral fields $X_{ij}$, where $G$ is the number of gauge groups with an overall decoupled $U(1)$.

We can rewrite the mesonic moduli space $\mathcal{M}^{mes}$ 
in terms of GLSM fields $p_a$.
Under the new coordinates, the mesonic moduli space $\mathcal{M}^{mes}$
can be written in terms of the following symplectic quotient \cite{fulton1993introduction, Cox:1993fz},
\beal{es01a12}
\mathcal{M}^{mes} = \textrm{Spec} \left(
\mathbb{C} [ p_a ] // Q_{JE} \right) // Q_D ~,~
\eea
where $a =1 , \dots , c $ labels the GLSM fields $p_a$.
The $J$- and $E$-terms are encoded in the $U(1)$ charge matrix given by $Q_{JE}$, whereas the $D$-terms with the corresponding gauge charges are encoded in the $Q_D$ charge matrix.
These are respectively obtained as follows using the \textbf{forward algorithm} for brane brick models \cite{Feng:2000mi, Franco:2015tna, Franco:2015tya},
\beal{es01a13} \label{Mmes-sym}
Q_{JE} = \textrm{ker} P ~,~  \quad \bar{d} = Q_D \cdot P^{t} ~,~
\eea
where the $P$-matrix encodes the relationship between GLSM fields $p_a$ and bifundamental chiral fields $X_{ij}$ in the brane brick model, and $\overline{d}$ is the reduced incidence matrix of the quiver for the $2d$ $(0,2)$ supersymmetric gauge theory.
We note here that the GLSM fields $p_a$ correspond to brick matchings in the brane brick model, which are combinational objects covering plaquettes in the $J$- and $E$-terms that have been originally introduced in \cite{Franco:2016nwv}. 
We also note that the mesonic moduli space $\mathcal{M}^{mes}$ of abelian brane brick model is 
a toric Calabi-Yau 4-fold.

The \textbf{Hilbert series} \cite{Benvenuti:2006qr, Hanany:2006uc, Butti:2007jv, Feng:2007ur, Hanany:2007zz} of the mesonic moduli space $\mathcal{M}^{mes}$ 
is the generating function of mesonic gauge invariant operators
in the brane brick model. 
In terms of the symplectic quotient in \eqref{es01a11}, the Hilbert series of the mesonic moduli space $\mathcal{M}^{mes}$ can be obtained using the following Molien integral formula,
\beal{es01a15} \label{HS-def}
g(t_a ; \mathcal{M}^{mes} ) = 
\prod_{\mu=1}^{c-4} \oint_{|z_\mu|=1 } \frac{d z_\mu}{2 \pi i z_\mu} \prod_{a=1}^c \frac{1}{1-t_a \prod_{\nu=1}^{c-4} z_\nu ^{(Q_t)_{\nu \alpha}}} ~,~
\eea
where $Q_t = (Q_{JE} ~,~ Q_D ) $ and $c$ is the total number of GLSM fields $p_a$. 
The fugacities $t_a$ in \eqref{HS-def} count the respective degrees in the GLSM fields $p_a$.
Under a change of fugacities,
the Hilbert series can be refined in terms of fugacities that count charges under the global symmetry of the $2d$ $(0,2)$ supersymmetric gauge theory.
Given that the abelian brane brick model has a mesonic moduli space $\mathcal{M}^{mes}$, which is a toric Calabi-Yau 4-fold, 
the overall total rank of the global symmetry is 4 with a $U(1)$ factor corresponding to the $U(1)_R$ symmetry and the remaining symmetry being the mesonic flavor symmetry \cite{Forcella:2008bb, Forcella:2008eh, Franco:2022gvl, Kho:2023dcm}.

Using the 
\textbf{plethystic logarithm} \cite{Feng:2007ur, Hanany:2007zz} of the Hilbert series, 
it is possible to extract algebro-geometric information about the underlying algebraic variety 
associated to the mesonic moduli space $\mathcal{M}^{mes}$ of the brane brick model.
This includes information about the generators and defining relations of the mesonic moduli space $\mathcal{M}^{mes}$.
The plethystic logarithm of Hilbert series is defined as follows,
\beal{es01a18}\label{Plet-def}
PL[g(t_a ;\mathcal{M}^{mes}) ] = \sum_{k=1}^{\infty} \frac{\mu(k)}{k} \text{log}~[g(t_a^k ;\mathcal{M}^{mes})]
\eea
where $g(t_a ;\mathcal{M}^{mes})$ is the refined Hilbert series of the mesonic moduli space $\mathcal{M}^{mes}$ in \eref{es01a15},
and
$\mu(k)$ is the M\"{o}bius function. 
The first positive terms and negative terms in the expansion of the plethystic logarithm correspond to the generators and the defining relations amongst them, respectively. 
All other higher order terms are relations amongst relations, known as
$\textit{syzygies}$ \cite{Benvenuti:2006qr, Gray:2006jb}, that exist for non-complete intersection moduli spaces. 
\\

\subsection{Mass Deformations of Brane Brick Models} \label{sec:deform} 

Recently, it has been shown in \cite{Franco:2023tyf} that brane brick models as $2d$ $(0,2)$ supersymmetric gauge theories can undergo \textbf{mass deformations} by the introduction of supersymmetry preserving mass terms to the $J$- and $E$-terms of the brane brick model. 
In general, these mass terms added to the Lagrangian of the $2d$ $(0,2)$ theories are gauge-invariant terms involving a chiral and a Fermi field. 
They take the following general form, 
\beal{es01a20}
\begin{tabular}{rrclrcl}
$(\Lambda_{ij}, X_{ij} ) \in Q :$ & 
$J'_{ji}$ &=& $J_{ji}$ ~,~ & 
$E'_{ij}$ &=& $\pm m X_{ij} + E_{ij}$ ~,~
\\
$(\overline{\Lambda}_{ij} , X_{ji} ) \in Q : $ &  
$J'_{ji}$ &=& $\pm m X_{ji} + J_{ji}$ ~,~ & 
$\quad E'_{ij}$ &=& $E_{ij}$ ~,~
\end{tabular}
\eea
where $J_{ji}'$ and $E'_{ij}$ are the $J$- and $E$-terms associated to the Fermi $\Lambda_{ij}$ after the mass deformation
and $J_{ji}$ and $E_{ij}$ are the $J$- and $E$-terms before the mass deformation. 
By integrating out the massive chiral-Fermi pairs,
one obtains the following replacements
for
chiral fields in the $J$- and $E$-terms,
\beal{es01a21}
\label{chi-repl}
X_{ij} = \mp \frac{1}{m} (E_{ij}^+ - E_{ij}^-) \quad \textrm{or} \quad X_{ji} = \mp \frac{1}{m} (J^+_{ji} - J^-_{ji} ) ~.~
\eea

The chiral field replacements in \eqref{chi-repl}, in principle, can violate the binomial property of the $J$- and $E$-terms in the brane brick model \cite{Franco:2023tyf}. 
In order to preserve 
the binomial structure of the $J$- and $E$-terms during mass deformation, 
we need to deform the $J$- and $E$-terms further by introducing 
higher-order coupling terms among chiral fields.
These are effectively redefinitions of Fermi-chiral interaction terms that take the following form, 
\beal{es01a22}\label{holo-redef}
\Lambda_{ij}' \cdot X'_{jk} = \Lambda_{ij} \cdot (X_{jk} + \sum_h c_h^{(jk)} X_{jh} X_{hk} ) ~,~
\eea
where $X_{jk}'$ and $\Lambda_{ij}'$ indicate the new fields after the redefinition and $c_h^{(jk)}$ are coefficients specific to the redefinition.

Mass deformation 
in brane brick models can be described in terms of combinatorial objects
known as \textbf{brick matchings} \cite{Franco:2016nwv}.
A brick matching 
is a collection of chiral, Fermi and conjugate Fermi fields in the brane brick model
that cover terms in the $J$- and $E$-terms referred to as plaquettes in the brane brick model uniquely once.
These brick matchings correspond to GLSM fields and points in the toric diagram of the associated toric Calabi-Yau 4-fold.
As a result, they correspond to coordinates in the symplectic quotient description of the mesonic moduli space in \eref{es01a11}.
We refer to \cite{Franco:2016nwv} for more details on the definition of brick matchings.

Under mass deformation, brick matchings in the original brane brick model exhibit properties that allow us to identify them as one of three types \cite{Franco:2023tyf}.
\textit{Massive brick matchings} contain chiral fields in the brane brick model that become massive during mass deformation.
They correspond to extremal points in the toric diagram of the toric Calabi-Yau 4-fold that do not change their relative position during mass deformation.
\textit{Moving brick matchings} do not contain any chiral fields that become massive during mass deformation, but they do correspond to extremal points in the toric diagram that change their relative position during mass deformation.
Finally, there are also \textit{unaffected brick matchings}, whose chiral field content is partially or sometimes completely preserved during mass deformation. 
They can correspond to points in the toric diagram that can be extremal and do not change their relative position. 

We refer the reader to \cite{Franco:2023tyf} for a more comprehensive review on brick matchings and their properties under mass deformation.
In the following work, we connect brick matchings corresponding to points in the toric diagram 
and birational transformations interpreted as polytope mutations that affect the shape of toric diagrams for toric Calabi-Yau 4-folds.
\\

\subsection{Reflexive Polytopes and Toric Fano 3-folds} \label{sec:reflexive} 

\textbf{Reflexive polytopes}
form a special family of convex lattice polytopes $\Delta$ in $\mathbb{Z}^3$ \cite{Batyrev_1982, batyrev1993dual, borisov1993towards,  batyrev1994dual, Batyrev:1994pg, batyrev1999classification, He:2017gam, Franco:2022gvl, Bao:2024nyu}.
They have the defining property that their \textit{dual polytope} given by, 
\beal{es01a49}
\Delta^\circ = 
\{
\mathbf{u} \in \mathbb{Z}^3 
~|~ 
\mathbf{u} \cdot \mathbf{v} \geq -1,~ \forall\mathbf{v} \in \Delta
\}~,~
\eea
is another lattice polytope in $\mathbb{Z}^3$.
In \cite{Kreuzer:1995cd, Kreuzer:1998vb, Kreuzer:2000xy}, Kreuzer and Skarke showed in their classification that there are exactly 4319 of them in $\mathbb{Z}^3$.
Out of the 4319 reflexive polytope, 18 are known to be regular polytopes, where every cone in the fan defined by $\Delta$ has generators that form part of a $\mathbb{Z}$-basis \cite{cox2024toric}.

\begin{figure}[ht!!]
\begin{center}
\resizebox{1\hsize}{!}{
\includegraphics[height=6cm]{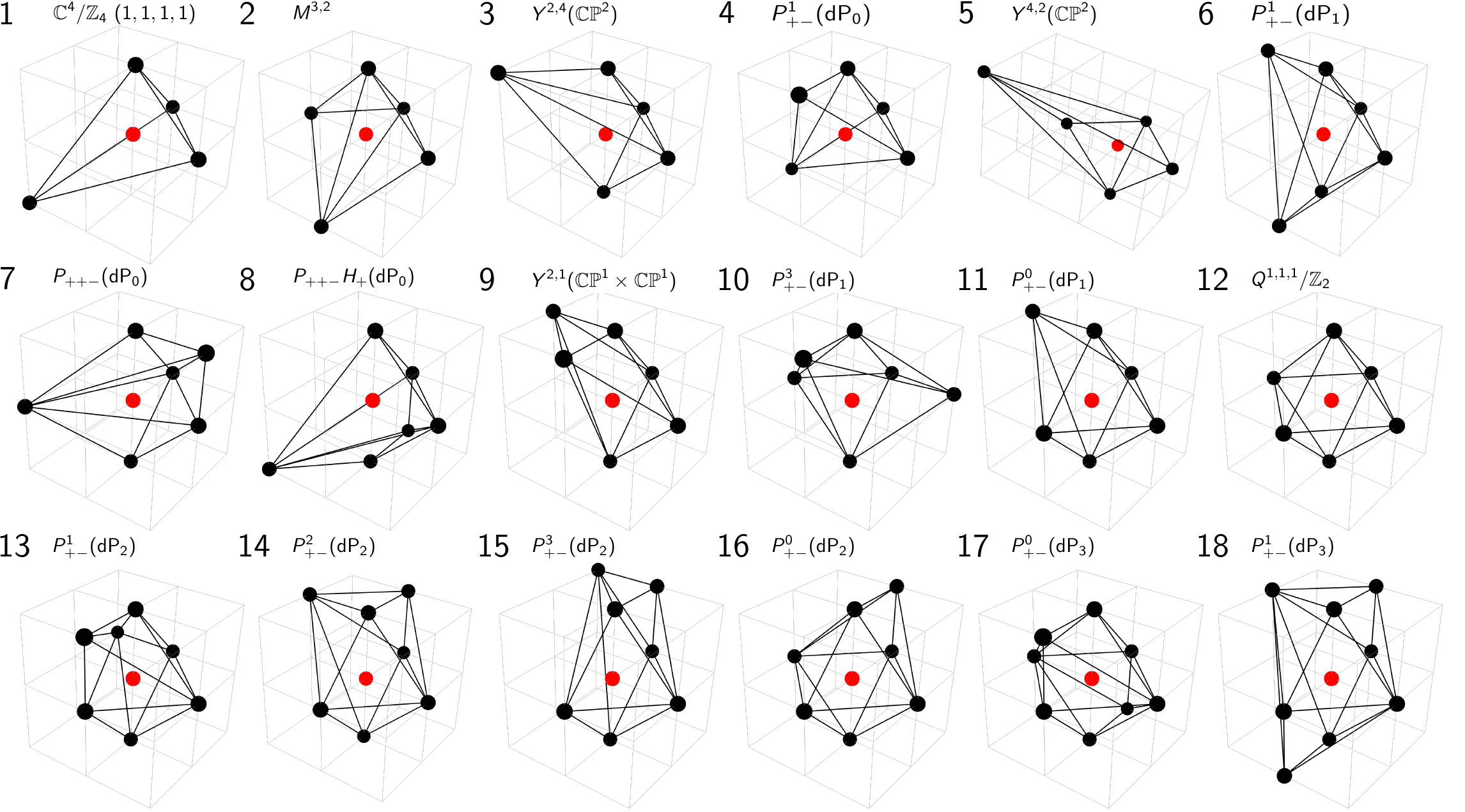} 
}
\caption{
The 18 regular reflexive polytopes in $\mathbb{Z}^3$ corresponding to smooth toric Fano 3-folds.
The corresponding brane brick models have been identified in \cite{Franco:2022gvl}. 
\label{f_in13}}
 \end{center}
 \end{figure}

The toric variety $X(\Delta)$ \cite{fulton1993introduction}
defined over a reflexive polytope $\Delta$ in $\mathbb{Z}^3$
is known as a \textbf{toric Fano $3$-fold} \cite{ewald1988classification, watanabe1982classification, nill2005gorenstein}.
When $\Delta$ is reflexive and regular, the corresponding variety $X(\Delta)$ is also smooth. 
The complex affine cone over $X(\Delta)$ is a non-compact toric Calabi-Yau 4-fold $\mathcal{X}$. 
In \cite{Franco:2022gvl}, in the case of the 18 smooth Fano 3-folds, 
the corresponding brane brick models 
and $2d$ $(0,2)$ supersymmetric gauge theories
were identified explicitly.  
Related work in \cite{He:2017gam, Bao:2024nyu} also studied the associated minimum volumes of the Sasaki-Einstein base of $\mathcal{X}$ as well as corresponding Futaki invariants. 
\fref{f_in13} illustrated the 18 regular reflexive polytopes corresponding to smooth toric Fano 3-folds as studied in \cite{Franco:2022gvl}.
\\

\begin{figure}[ht!!]
\begin{center}
\resizebox{1\hsize}{!}{
\includegraphics[height=6cm]{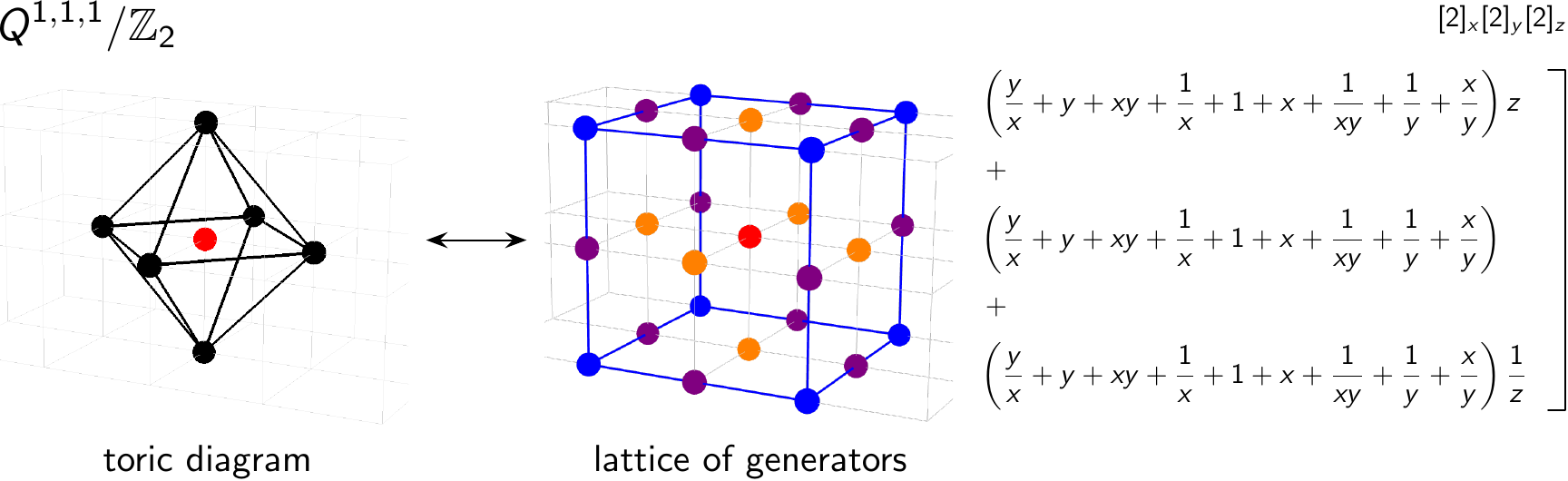} 
}
\caption{
The toric diagram $\Delta$ of the cone over $Q^{1,1,1}/\mathbb{Z}_2$
and its dual polytope $\Delta^\circ$ forming the lattice of generators of the mesonic moduli space $\mathcal{M}^{mes}$ of the corresponding brane brick model. 
The 27 generators of $\mathcal{M}^{mes}$ transform in the $[2]_x [2]_y [2]_z$ irreducible representation of the mesonic flavor symmetry $SU(2)_x \times SU(2)_y \times SU(2)_z$ of the brane brick model for $Q^{1,1,1}/\mathbb{Z}_2$.
\label{f_in14}}
 \end{center}
 \end{figure}

As studied systematically in \cite{Franco:2022gvl}, 
the dual lattice polytope $\Delta^\circ$ of a reflexive polytope $\Delta$ in $\mathbb{Z}^3$
is interpreted as the $\textbf{lattice of generators}$ of the mesonic moduli space $\mathcal{M}^{mes}$ of the corresponding abelian brane brick model as defined in \eref{es01a11}.
As discussed in section \sref{sec:HS}, 
given that the mesonic moduli space $\mathcal{M}^{mes}$ of an abelian brane brick model is a toric Calabi-Yau 4-fold $\mathcal{X}$, 
the overall total rank of the global symmetry of the brane brick model is $4$. 
The mesonic flavor symmetry has a total rank of 3 and the charges under the mesonic flavor symmetry on the generators of $\mathcal{M}^{mes}$
can be scaled such that they are all in $\mathbb{Z}^3$.
Considering these scaled charges on the generators of $\mathcal{M}^{mes}$ as points in $\mathbb{Z}^3$, the convex hull of the set of all points corresponding to generators of $\mathcal{M}^{mes}$ forms the polytope $\Delta^\circ$ whose dual $\Delta$ is the toric diagram of the toric Calabi-Yau 4-fold $\mathcal{X}$.
\fref{f_in14} shows the toric diagram of the cone over $Q^{1,1,1}/\mathbb{Z}_2$, which is a toric Calabi-Yau 4-fold, and its dual polytope forming the lattice of generators of the mesonic moduli space of the corresponding brane brick model. 
\\

\section{Birational Transformations and Brane Brick Models \label{sec:03}}

\subsection{Birational Transformations, Brane Brick Models and Toric Fano 3-folds} \label{sec:back} 

As introduced in section \sref{sec:bbm} as part of the underlying Type IIA brane configuration of a brane brick model, 
given a toric diagram $\Delta$, we can write the corresponding \textbf{Newton polynomial} of the following form, 
\beal{es01a49b1}
P(x,y,z) = 
\sum_{\mathbf{v} = (v_1,v_2, v_3) \in \Delta}
c_{\mathbf{v}}~ x^{v_1} y ^{v_2} z^{v_3} ~,~
\eea
where now we set $c_{\mathbf{v}}=1$ if the corresponding point $\mathbf{v} \in \Delta$ is extremal
and $c_{\mathbf{v}} \in \mathbb{C}^{*}$ otherwise. 

As studied in \cite{akhtar2012minkowski}, 
when $\Delta$ is a reflexive polytope,
one can introduce a family of \textbf{birational transformations}
of the following form,
\beal{es01a49b2}
\varphi_A: (\mathbb{C}^{*})^3 & \rightarrow & (\mathbb{C}^{*})^3
\nn\\
(x,y,z) &\mapsto & (x,y,A(x,y) z) ~,~ 
\eea
where $A(x,y)$ is a Laurent polynomial in $x,y$.
Here, the birational transformations map $P(x,y,z)$
into a new Newton polynomial $P^\vee(x,y,z)$
of the same form as in \eref{es01a49b1}.
The new Newton polynomial $P^\vee(x,y,z)$
corresponds to again a
convex lattice polytope $\Delta^\vee$.
It was shown in \cite{akhtar2012minkowski} that, 
\begin{center}
\textit{
If $\Delta^\vee$ is reflexive and $X(\Delta^\vee)$ corresponds to a toric Fano $3$-fold, then the original polytope $\Delta$ has to be reflexive with $X(\Delta)$ being also a toric Fano $3$-fold.} 
\end{center}
We can summarize the relevant findings of \cite{akhtar2012minkowski} for our work as follows:
\begin{itemize}

\item 
The birational transformation $\varphi_A$ preserves the \textbf{period} \cite{akhtar2012minkowski, givental1996equivariant, beck2008maximal} of $P(x,y,z)$ and the corresponding toric variety $X(\Delta)$ given by,
\beal{es01a49b3}
\pi_{P}(t) = 
\frac{1}{(2\pi i)^3}
\int_{|x|,|y|,|z|=1}
\frac{dx}{x}
\frac{dy}{y}
\frac{dz}{z}~
\frac{1}{
1
- t~ P(x,y,z) 
}
~.~
\eea

\item
The birational transformation $\varphi_A$ preserves the \textbf{Ehrhart polynomial} \cite{ehrhart1977polynomes, stanley1980decompositions} of the dual polytope $\Delta^\circ$ of a reflexive polytope $\Delta$.
For a given reflexive polytope $\Delta$, the Ehrhart polynomial is defined as 
\beal{es01a49b4}
\text{Ehr}_{\Delta} (t) = \sum_{m=0}^{\infty} | m\Delta \cap \mathbb{Z}^3 | t^m ~,~
\eea
where $| m\Delta \cap \mathbb{Z}^3 |$ is the number of points on $\mathbb{Z}^3$
of the $m$-times enlarged lattice polytope $m\Delta$.

\item
The toric varieties $X(\Delta)$
corresponding to Newton polynomials $P(x,y,z)$ related under a birational transformation $\varphi_A$
are \textbf{deformation equivalent}.
They appear as fibers of a flat family over a curve \cite{akhtar2012minkowski}.

\end{itemize}
Out of the 4319 reflexive polytopes in $\mathbb{Z}^3$ classified by Kreuzer and Skarke \cite{Kreuzer:1995cd, Kreuzer:1998vb, Kreuzer:2000xy}, 3025 reflexive polytopes separate into 165 distinct non-trivial equivalence classes called \textbf{buckets} \cite{akhtar2012minkowski} such that any two polytopes in a bucket have the above properties.

In \cite{Ghim:2024asj}, we showed that the $2d$ $(0,2)$ supersymmetric gauge theories that are realized by brane brick models associated to the toric Calabi-Yau 4-folds
$\mathcal{X}$ and $\mathcal{X}^\vee$
are also related under the birational transformation $\varphi_A$ connecting $\Delta$ with $\Delta^\vee$.
Let us summarize the findings in \cite{Ghim:2024asj} as follows:
\begin{itemize}
\item
When 
two $2d$ $(0,2)$ supersymmetric gauge theories and brane brick models 
corresponding to Calabi-Yau 4-folds with reflexive toric diagrams $\Delta$ and $\Delta^\vee$ 
are related by a \textbf{mass deformation} as defined in section \sref{sec:deform},
then the toric Calabi-Yau 4-folds with $\Delta$ and $\Delta^\vee$ 
are related by a birational transformation of the form defined in \eref{es01a49b2}.
We recall that mass deformations
introduce mass terms to the $J$- and $E$-terms of the brane brick model. 
By integrating out the resulting massive chiral-Fermi pairs, 
the defining relations amongst the generators of the mesonic moduli space $\mathcal{M}^{mes}$ defined in \eref{es01a11}
and corresponding to $\Delta$
deform into new defining relations of the new mesonic moduli space corresponding to $\Delta^\vee$.

\item
The \textbf{number of generators} of the mesonic moduli spaces $\mathcal{M}^{mes}$
in the abelian $2d$ $(0,2)$ supersymmetric gauge theories and brane brick models corresponding to $\Delta$ and $\Delta^\vee$
is the same. As discussed in section \sref{sec:HS}, these generators are determined by the plethystic logarithm of the Hilbert series of $\mathcal{M}^{mes}$.
Moreover, since $\Delta$ and $\Delta^\vee$ are reflexive polytopes, 
the \textbf{lattice of generators} given by their dual polytopes $\Delta^{\circ}$ and $(\Delta^\vee)^\circ$ are also reflexive. 
We expect these dual polytopes 
to be related by another birational transformation acting on their respective Newton polynomials. 

\item
The unrefined \textbf{Hilbert series},
expressed in terms of $U(1)_R$ symmetry fugacities,
is the same for the mesonic moduli spaces $\mathcal{M}^{mes}$
of the abelian $2d$ $(0,2)$ supersymmetric gauge theories and brane brick models associated with $\Delta$ and $\Delta^\vee$.
This unrefined Hilbert series is obtained 
from the refined Hilbert series defined in \eref{es01a15}
by restricting to the fugacity that counts only $U(1)_R$ charges. 

\end{itemize}
We note here that the above properties studied in \cite{Ghim:2024asj} directly relate to the properties found in \cite{akhtar2012minkowski}.
The above properties were explicitly studied in \cite{Ghim:2024asj}
for brane brick models corresponding to the abelian orbifold of the form $\mathbb{C}^4/\mathbb{Z}_6$ with orbifold action $(1,1,2,2)$
and $P_{+-}^{2}[\text{dP}_0]$,
which are toric Calabi-Yau 4-folds $\mathcal{X}$
whose toric diagrams $\Delta$, toric varieties $X(\Delta)$ and corresponding Newton polynomials $P(x,y,z)$ are all related to each other by a birational transformation $\varphi_A$.

In this work, we build on the findings of \cite{Ghim:2024asj} by presenting additional examples of toric Calabi-Yau 4-folds 
and the corresponding $2d$ $(0,2)$ supersymmetric gauge theories realized by brane brick models.
In particular, we show that birational transformations of the form in \eref{es01a49b2}
also manifest for non-reflexive lattice polytopes. 
Furthermore, the resulting toric Calabi-Yau 4-folds and their associated 
$2d$ $(0,2)$ supersymmetric gauge theories and brane brick models continue to satisfy the properties outlined above. 
\\

In the following section, we review birational transformations acting on toric diagrams, 
toric varieties and Newton polynomials,
highlighting how they constitute
a family of transformations
called \textit{algebraic mutations} in \cite{akhtar2012minkowski}.
We also review a combinatorial procedure for obtaining 
$\Delta^\vee$ from $\Delta$, known as \textit{combinatorial mutation} in \cite{akhtar2012minkowski}.
\\

\subsection{Birational Transformations: Algebraic and Combinatorial Mutations} \label{sec:mut} 

In the following section,
we introduce the family of birational transformations $\varphi_A$ 
-- originally proposed in \cite{akhtar2012minkowski} --
which form the core of our study.
We also review the combinatorial interpretation of $\varphi_A$ in terms of the corresponding toric diagrams.

\paragraph{Birational Transformations and Algebraic Mutations.}
Under a $GL(3,\mathbb{Z})$ transformation $M$ on $(x,y,z) \in \mathbb{C}^*$, 
the Newton polynomial $P(x,y,z)$ defined in \eref{es01a01b}
can be transformed as follows, 
\beal{es01a50}
M ~:~
P(x,y,z) \mapsto 
P( x^{M_{11}} y^{M_{12}} z^{M_{13}} ,  x^{M_{21}} y^{M_{22}} z^{M_{23}},  x^{M_{31}} y^{M_{32}} z^{M_{33}} )
~.~
\nn\\
\eea
For an appropriate choice for the transformation $M \in GL(3,\mathbb{Z})$, 
the Newton polynomial can be written as the following Laurent polynomial of the form, 
\beal{es01a51}
P (x, y, z) = \sum_{m=a}^b C_m (x, y) z^m
~,~
\eea
where
$a \in \mathbb{Z}_{<0} $ and $b \in \mathbb{Z}_{>0} $, and $C_m (x,y)$ are
Laurent polynomials in $x$ and $y$ for $a \leq m \leq b $.
In this form, 
the \textbf{birational transformation} $\varphi_A$
that we consider in this work acts on $P(x,y,z)$ as follows,
\beal{es01a52}
\varphi_A: (\mathbb{C}^{*})^3 & \rightarrow & (\mathbb{C}^{*})^3
\nn\\
(x,y,z) &\mapsto & (x,y,A(x,y) z) ~,~ 
\eea
where the Laurent polynomial $A(x,y)$ is chosen to satisfy, 
\beal{es01a53}
A(x,y)^{|m|} ~|~ C_{m}(x,y) ~~\text{for every $a\leq m < 0$} ~.~
\eea

In general, we can combine the above $GL(3,\mathbb{Z})$ transformation and the
birational transformation $\varphi_A$ on $P(x,y,z)$
to obtain 
the following composite transformation,
\beal{es01a54}
\varphi_{A; M, N} = N \circ \varphi_A \circ M
~:~
P (x, y, z) \mapsto 
P^{\vee} (x, y, z)
~,~
\eea
where
$A(x,y)$ is chosen such that it satisfies the constraint in \eref{es01a53}, 
$M,N \in GL(3,\mathbb{Z})$
and $N$ can be chosen to be the inverse of $M$. 
As in \cite{akhtar2012minkowski}, we
refer to the composite transformation $\varphi_{A; M, N}$
as an \textbf{algebraic mutation} of the corresponding toric Calabi-Yau 4-fold. 
An algebraic mutation connects Newton polynomials 
$P (x, y, z)$ and $P^{\vee} (x, y, z)$
that are associated respectively to polytopes $\Delta$ and $\Delta^{\vee}$
that do not necessarily correspond to the same toric Calabi-Yau 4-fold. 
\\

\paragraph{Combinatorial Mutations.}

An algebraic mutation $\varphi_{A; M, N}$ of the form defined in \eref{es01a54}
can be described from a combinatorial perspective
based on the
convex lattice polytopes $\Delta$ and $\Delta^{\vee}$ in $\mathbb{Z}^3$
associated to Newton polynomials $P (x, y, z)$ and $P^{\vee} (x, y, z)$, respectively. 
This connection, also referred to as a \textbf{combinatorial mutation}
between convex lattice polytopes, 
has been described in \cite{akhtar2012minkowski} in the context of reflexive polytopes in $\mathbb{Z}^3$
corresponding to Fano 3-folds. 

We specify a primitive vector $\textbf{w} \in \mathbb{Z}^3$
that is used to measure
the \textit{height} $h=\textbf{w} \cdot \textbf{v} $ of a lattice point with coordinate vector $\textbf{v} \in \mathbb{Z}^3$.
This allows us to identify lattice points at height $h$ under $\textbf{w}$ in $\mathbb{Z}^{3}$ given by
$\textrm{H}_{\textbf{w}, h} = \{ \textbf{x} \in \mathbb{Z}^{n} | \textbf{w} \cdot \textbf{x} = h \}$. 
Given a convex lattice polytope $\Delta$, we can identify points in $\Delta$ at a given height $h$ along $\textbf{w}$ as follows, 
\beal{es01a60}
w_{h} ( \Delta ) = \textrm{conv} \left( \textrm{H}_{\textbf{w}, h} \cap \Delta \right) ~,~
\eea
where $\textrm{conv}$ refers to the convex hull of the set of points at height $h$. 
Within the polytope $\Delta$, 
we can define respectively the minimum and maximum height along $\textbf{w}$
as follows,
\beal{es01a61}
h_{min} = \min\{\textbf{w} \cdot \textbf{v}~|~\textbf{v} \in \Delta \} ~,~
h_{max} = \max\{\textbf{w}\cdot \textbf{v}~|~\textbf{v} \in \Delta \} ~.~
\eea
Using the above definitions, we can define the \textit{width} $\text{w}_{\textbf{w}}(\Delta)$ of a convex lattice polytope $\Delta$ along $\textbf{w}$, 
\beal{es01a62}
w_{\textbf{w}}(\Delta)
= h_{max} - h_{min} ~.~
\eea

We also define the \textit{factor} $F \subset \mathbb{Z}^{3} $ of a convex lattice polytope $\Delta$ with respect to $\textbf{w}$
that satisfies $\mathbf{w} \cdot F = 0$
and is located at $h=0$ such that $ F \subset \textrm{H}_{\textbf{w}, 0}$.
The factor $F$ is used to identify a set of lattice polytopes $G_h$ at a height $h$
that satisfy the following constrains, 
\beal{es01a63} \label{layercon}
\textrm{H}_{\textbf{w}, h} \cap V(\Delta)
\subseteq G_{h} + |h| F \subseteq w_h (\Delta) 
~.~
\eea
for every
$h_{min} \leq h < 0 $, where $V (\Delta) $ stands for the extremal points of the toric diagram $\Delta$. 
Note that 
without loss of generality, $G_{h}$ can consist of empty lattice polytopes. 
Furthermore, the addition in $G_{h} + |h| F$ in \eref{es01a63} corresponds to the $\textit{Minkowski sum}$, which is defined as,
\beal{es01a64}
A + B = \{ \textbf{a} + \textbf{b} ~ | ~ \textbf{a} \in A ~,~ \textbf{b} \in B \} 
~.~
\eea

\begin{figure}[H]
\begin{center}
\resizebox{1\hsize}{!}{
\includegraphics[height=6cm]{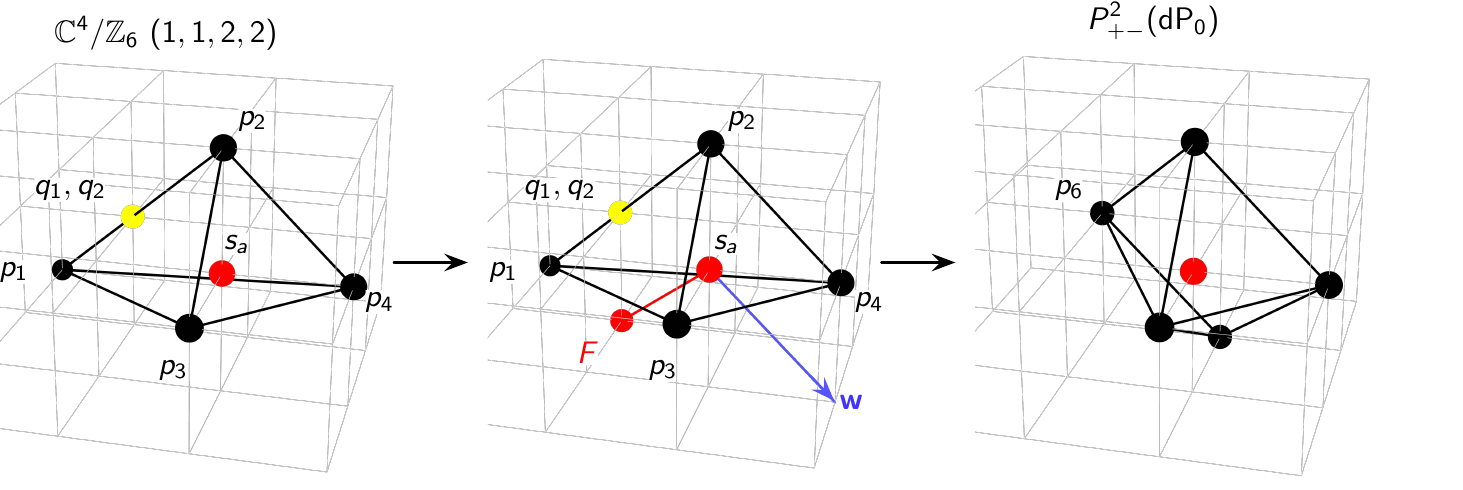} 
}
\caption{
The combinatorial mutation studied in \cite{Ghim:2024asj} 
of the toric diagram corresponding to the abelian orbifold of the form $\mathbb{C}^4/\mathbb{Z}_6$ with orbifold action $(1,1,2,2)$ leading to the toric diagram for the toric Calabi-Yau 4-fold known as $P^2_{+-}[\text{dP}_0]$.
Here, we choose the height vector $\textbf{w}=(0,-1,1)$ and the factor $F=\{(0,0,0),(-1,-1,-1)\}$.
\label{f_in15}}
 \end{center}
 \end{figure}

Using the above definitions, we are now able to define 
the
\textit{combinatorial mutation} $\mu_{\textbf{w}} (\Delta, F ; \{ G_h \}_{h<0} )$ 
of a toric diagram $\Delta$
as follows,
\beal{es01a65}
\mu_{\textbf{w}} (\Delta, F ; \{ G_h \}_{h<0} )
= \text{conv} \left( \bigcup_{h_{min}\leq h<0} G_h \cup \bigcup_{0\leq h \leq h_{max}} \left( w_h (\Delta) + h F \right) \right) ~.~
\eea
We note that
the combinatorial mutation 
is independent of the choice of polytopes $G_h$ satisfying \eref{layercon} once the mutation vector $\mathbf{w}$ and the corresponding factor $F$ is fixed.
As a result, we denote the combinatorial mutation of a toric diagram $\Delta$ along $\mathbf{w}$ as
$\mu_{\mathbf{w}} (\Delta , F )$.
\fref{f_in15} illustrates the combinatorial mutation studied in \cite{Ghim:2024asj} of the toric diagram corresponding to the abelian orbifold of the form $\mathbb{C}^4/\mathbb{Z}_6$ with orbifold action $(1,1,2,2)$ \cite{Davey:2010px, Hanany:2010ne, Franco:2015tna} leading to the toric diagram for the toric Calabi-Yau 4-fold known as $P^2_{+-}[\text{dP}_0]$ \cite{Ghim:2024asj}. 
\\

\subsection{Beyond Reflexive Polytopes and Toric Fano 3-folds} \label{sec:mut} 

The works in \cite{akhtar2012minkowski, Ghim:2024asj}
examined a family of birational transformations $\varphi_A$
acting on the Newton polynomials $P(x,y,z)$ of reflexive polytopes in $\mathbb{Z}^3$ that define toric Fano 3-folds. 
In this work, we show that the same family of birational transformations $\varphi_A$
can be extended to Newton polynomials $P(x,y,z)$ corresponding to general convex lattice polytopes in $\mathbb{Z}^3$,
even when those polytopes are \textit{not} reflexive. 

There are two families of convex lattice polytopes in $\mathbb{Z}^3$ that are not reflexive, as illustrated in \fref{f_ref_0a}:
\begin{itemize}

\item \textbf{Non-Reflexive Polytopes with No Internal Points.}
For these polytopes, we define the corresponding Newton polynomial $P(x,y,z)$ in \eref{es01a49b1} so that the origin $(0,0,0) \in \mathbb{Z}^3$
is one of the extremal points of the polytope. 

\item \textbf{Non-Reflexive Polytopes with Two or More Internal Points.}
For these polytopes, we also define the Newton polynomial $P(x,y,z)$ in \eref{es01a49b1}, but now the origin $(0,0,0) \in \mathbb{Z}^3$ is a fully internal point of the polytope.
\end{itemize}
For both families, we choose coefficients $c_{\textbf{v}}$ in $P(x,y,z)$ for extremal points to be $1$, 
for points along edges of the polytope to be binomial coefficients, 
and for all other internal points of the polytope to be in $\mathbb{C}^*$.
With this choice of coefficients, 
we can identify
birational transformations $\varphi_A$ of the general form defined in \eref{es01a52},
which map the Newton polynomial $P(x,y,z)$ of a non-reflexive convex lattice polytope $\Delta$ in $\mathbb{Z}^3$
to a Newton polynomial $P^\vee(x,y,z)$ of another non-reflexive convex lattice polytope $\Delta^\vee$ in $\mathbb{Z}^3$.
Notably, under this family of birational transformations $\varphi_A$, if $\Delta^\vee$ is not reflexive, then $\Delta$ must also be non-reflexive. 
In general, these birational transformations $\varphi_A$ preserve the total number of fully internal points of the convex lattice polytope.
 
Just as in the reflexive case,
non-reflexive lattice polytopes
$\Delta$ and $\Delta^\vee$ still define toric Calabi-Yau 4-folds 
corresponding to brane brick models that realize $2d$ $(0,2)$ supersymmetric gauge theories. 
We observe that when
these brane brick models 
are related by a mass deformation, 
the corresponding polytopes $\Delta$ and $\Delta^\vee$ are likewise related by a birational transformation $\varphi_A$,
mirroring the situation in the toric Fano 3-fold case.
Furthermore, the abelian $2d$ $(0,2)$ supersymmetric gauge theories
have mesonic moduli spaces $\mathcal{M}^{mes}$ with the same number of generators
and identical Hilbert series when refined soley in terms of the $U(1)_R$ fugacity. 
Although $\Delta$ and $\Delta^\vee$, as non-reflexive lattice polytopes, do \textit{not} have dual lattice polytopes in $\mathbb{Z}^3$, 
we find that the corresponding toric varieties share the same period. 
\\

In the following section, we present several 
examples of birational transformations involving non-reflexive lattice polytopes, 
their corresponding toric Calabi-Yau 4-folds, and the brane brick models realizing $2d$ $(0,2)$ supersymmetric gauge theories.
For non-reflexive polytopes with no internal points, we identify birational transformations 
relating the toric Calabi-Yau 4-fold $\mathcal{C}_{++}$ to $H_4$ in section \sref{sec:c++_deform}, 
and the toric Calabi-Yau 4-fold $\mathcal{C}_{+-}$ to $Q^{1,1,1}$ in section \sref{sec:c+-_deform}.
Subsequently, in section \sref{sec:hpmc3z5_deform}, we illustrate a birational transformation relating two non-reflexive polytopes 
corresponding to toric Calabi-Yau 4-folds $P^{1}_{+-}[\mathbb{C}^3/\mathbb{Z}_5~(1,1,3)]$ and $P^{2}_{+-}[\mathbb{C}^3/\mathbb{Z}_5~(1,1,3)]$.
\\

\section{Examples \label{sec:examples}}

\subsection{Reflexive Case: $F_{0,+-}$ and $Q^{1,1,1}/\mathbb{Z}_2 $ \label{sec:f0+-_deform} }

\begin{figure}[H]
\begin{center}
\resizebox{0.85\hsize}{!}{
\includegraphics[height=6cm]{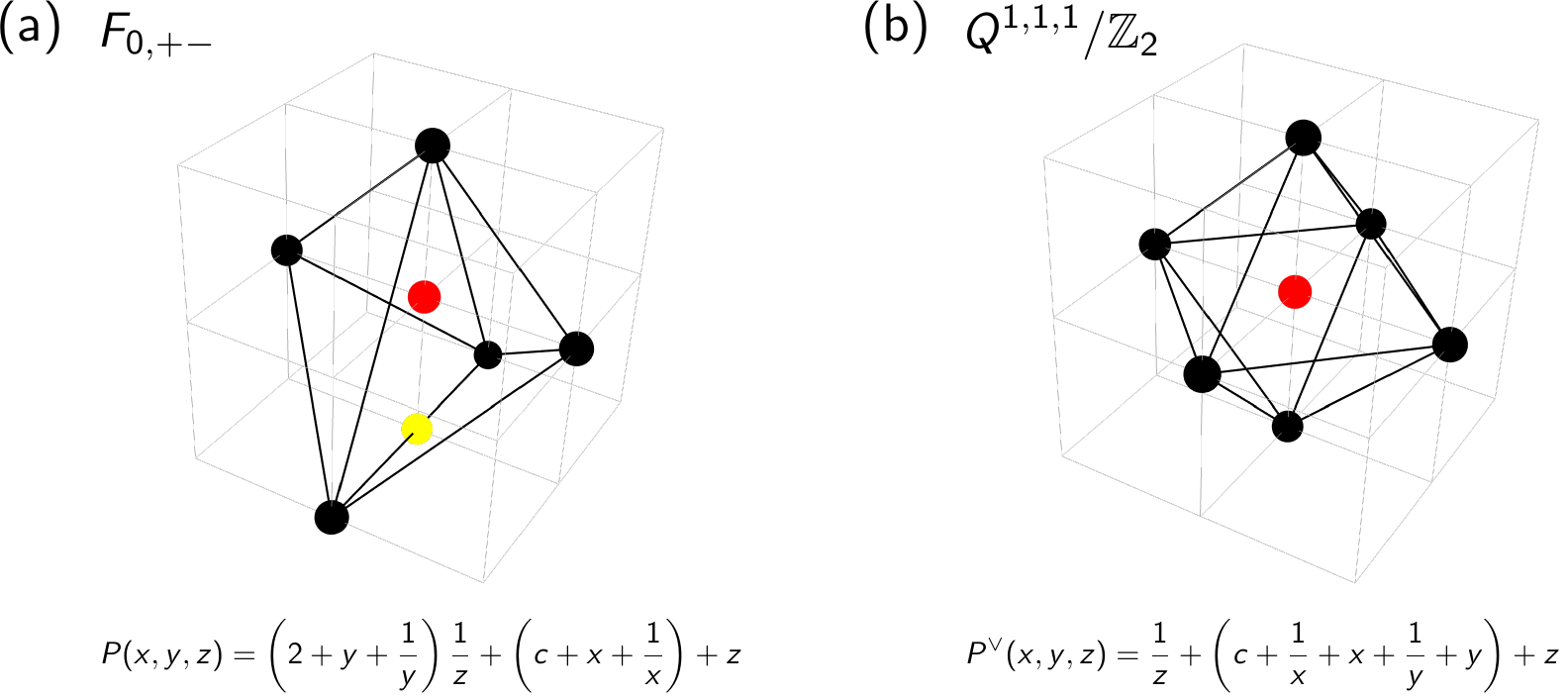} 
}
\caption{
The toric diagrams for the toric Calabi-Yau 4-folds known as (a) $F_{0,+-}$ and (b) $Q^{1,1,1}/\mathbb{Z}_2$,
and the corresponding Newton polynomials. 
\label{f_ref_ex1}}
 \end{center}
 \end{figure}

The toric diagram $\Delta$ of the $F_{0,+-}$ model, which is shown in \fref{f_ref_ex1}(a), is a reflexive polytope. 
The points of the toric diagram $\Delta$ for $F_{0,+-}$ are given by,
\beal{es02a01} \label{f0+-_point}
&&
\Delta = \{  ( 0, -1, -1 ), (1,0,0) , (-1, 0, 0) ,
( 0, 0, 1 ),  (0,1,-1) ,  (0,0, -1), (0,0,0) 
\} ~.~
\eea
\\

\paragraph{Algebraic Mutation.}
The Newton polynomial corresponding to $F_{0,+-}$ 
can be written using the point coordinates in \eref{es02a01} as follows,
\beal{es02a02}
P ( x, y, z) = \left( 
2 + y+ \frac{1}{y}
\right) \frac{1}{z} 
+ \left(c + x+ \frac{1}{x}
\right) + z ~,~
\eea
where we have chosen the coefficients for the extremal points to be $1$, 
for the point on the edge to be $2$,
and for the internal point at the origin to be $c \in \mathbb{C}^*$.
Under these choices of coefficients, we can introduce a birational transform $\varphi_A$ of the following form,
\beal{es02a03}
\varphi_A ~:~
(x,y,z) \mapsto 
\left(
x,y, \left(1+ \frac{1}{y}\right)z
\right)
~,~ 
\text{where $A(x,y) = 1+\frac{1}{y}$}
~.~
\eea
Under the above birational transformation,
we obtain a new Laurent polynomial $P^\prime (x,y,z)$,
which takes the following form,
\beal{es02a03b}
P^\prime (x,y,z)
=
(1+y) z^{-1}
+
\left(
c +\frac{1}{x}
+ x 
\right)
+
\left(
1 + \frac{1}{y} 
\right)
z
~.~
\eea
The above Laurent polynomial $P^\prime (x,y,z)$
 can be transformed under 
a $GL (3, \mathbb{Z})$ transformation given by
\beal{es02a04}
N =
    \left(
    \begin{array}{ccc}
    1 & 0 & 0 \\
    0 & 1 & 1 \\
    0 & 0 & 1
    \end{array}
    \right) ~,~
\eea
to become the following form
\beal{es02a05}
 P^{\vee} (x, y, z) =  
 \frac{1}{z}
 +
 \left(
 c+ \frac{1}{x} + x
+ \frac{1}{y}  + y 
 \right)
 + z
 ~.~
\eea
The above is a 
Newton polynomial corresponding to the toric diagram of $Q^{1,1,1}/\mathbb{Z}_2 $,
which is shown in \fref{f_ref_ex1}(b).
Here, we note that like in the original toric diagram $\Delta$ for $F_{0,+-}$ in \fref{f_ref_ex1}(a), 
the internal point at the origin in the toric diagram $\Delta^\vee$ for $Q^{1,1,1}/\mathbb{Z}_2 $
has a coefficient in $ P^{\vee} (x, y, z)$ given by $c \in \mathbb{C}^*$.
In other words, the birational transformation and the resulting algebraic mutation remains unaffected for any choice of coefficient $c \in \mathbb{C}^*$ in the Newton polynomial
for the internal point at the origin of the toric diagram. 
\\

\begin{figure}[H]
\begin{center}
\resizebox{0.85\hsize}{!}{
\includegraphics[height=6cm]{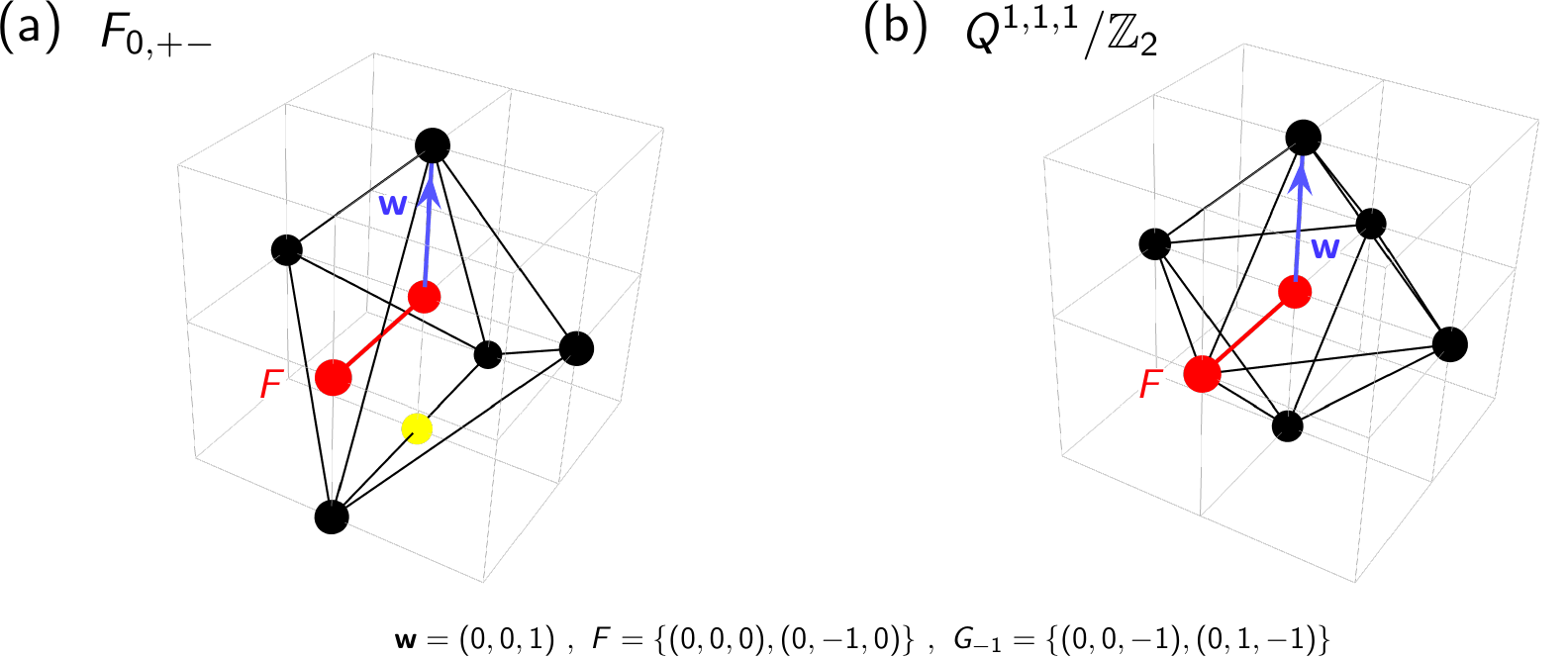} 
}
\caption{
The toric diagrams for the toric Calabi-Yau 4-folds known as (a) $F_{0,+-}$ and (b) $Q^{1,1,1}/\mathbb{Z}_2$
with height measurement vector $\textbf{w}=(0,0,1)$ and factor $F$ used for combinatorial mutation.
\label{f_ref_ex1b}}
 \end{center}
 \end{figure}

\paragraph{Combinatorial Mutation.}
Together with the choice on the height measurement vector $\mathbf{w} = ( 0, 0, 1)$, we fix
the factor $F$ and identify the polytopes $G_h$ as follows, 
\beal{es02a06}
F=  \{ ( 0, 0, 0) ~,~ ( 0, -1, 0) \} ~,~
G_{-1} =  \{ (0, 0, -1) ~,~ (0, 1, -1) \}  ~.~
\eea
\fref{f_ref_ex1b}(a) shows the factor $F$ and height vector $\mathbf{w}$ with the toric diagrams for $F_{0,+-}$. 
Following \eref{es01a65}, we obtain the following polytope using combinatorial mutation, 
\beal{es02a07}
\begin{tabular}{rcl}
$\underline{ h=-1 : }$  &  $(0 , 0 , -1 ) ~,~ ( 0 , 1 , -1)$ & $\in G_{-1}$ ~,~
\\
$\underline{ h=  0  : }$ & $( 0, 0 , 0 )~,~ (1 , 0 , 0) ~,~ (-1, 0, 0 )$ & $\in w_0 (\Delta)$ ~,~
\\
$\underline{ h=+1 : }$ & $(0 ,0 ,  1 ) ~,~ (0 , -1 , 1 )$ & $\in w_1 (\Delta) + F$ ~,~
\end{tabular}
\eea
which we identify as the toric diagram of $Q^{1,1,1}/ \mathbb{Z}_2 $ as illustrated in \fref{f_ref_ex1b}(b).
\\

\begin{figure}[H]
\begin{center}
\resizebox{0.85\hsize}{!}{
\includegraphics[height=6cm]{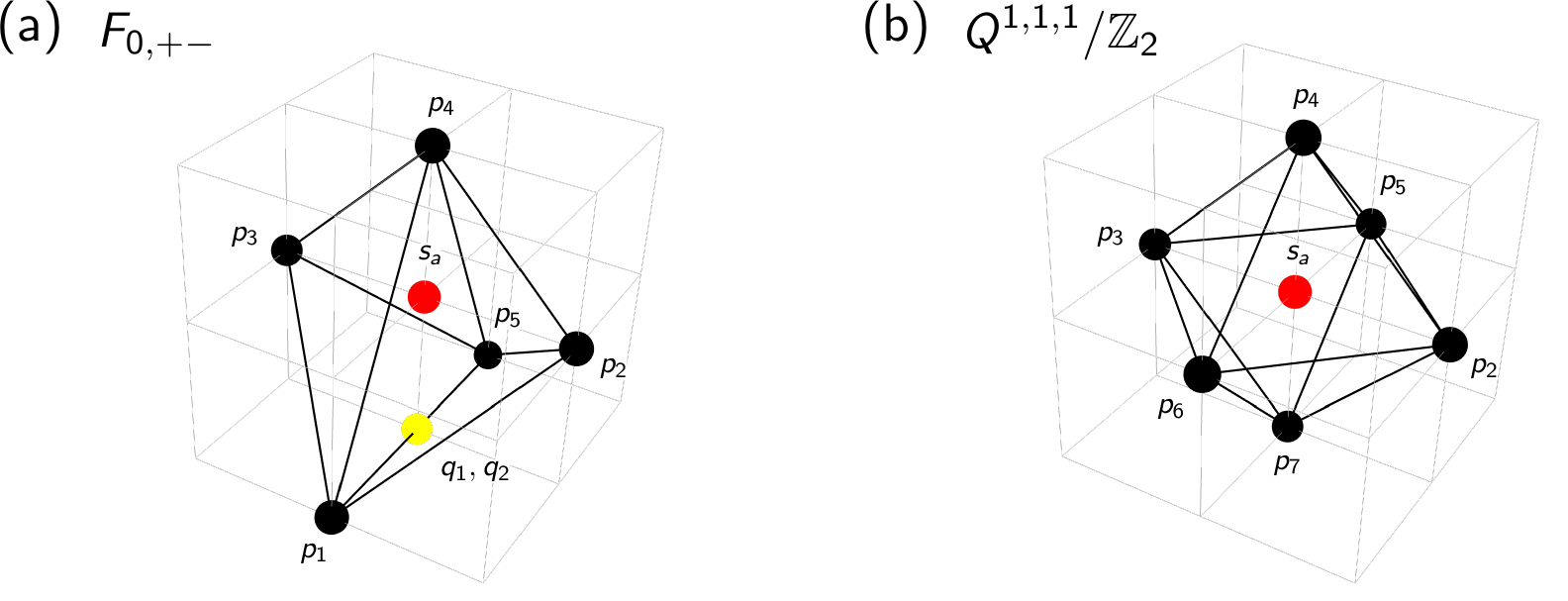} 
}
\caption{
The toric diagrams for the toric Calabi-Yau 4-folds known as (a) $F_{0,+-}$ and (b) $Q^{1,1,1}/\mathbb{Z}_2$
with points labelled by the GLSM fields in the corresponding brane brick models.
\label{f_ref_ex1c}}
 \end{center}
 \end{figure}

\begin{figure}[H]
\begin{center}
\resizebox{0.8\hsize}{!}{
\includegraphics[height=6cm]{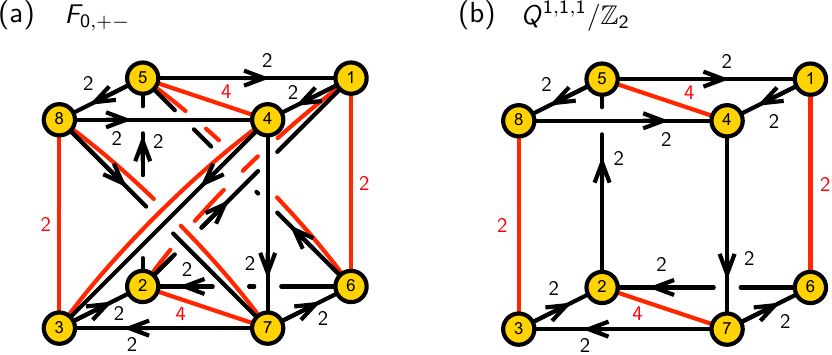} 
}
\caption{
The quiver diagrams for the brane brick models corresponding to (a) $F_{0,+-}$ and (b) $Q^{1,1,1}/\mathbb{Z}_2$.
\label{f_ref_ex1d}}
 \end{center}
 \end{figure}

\paragraph{Brane Brick Model for $F_{0,+-}$.}
The quiver for the $F_{0,+-}$ brane brick model is shown in \fref{f_ref_ex1d}(a) \cite{Franco:2023tyf}. 
The corresponding
$J$- and $E$-terms take the following form, 
\beal{es10a21}
\resizebox{0.6\textwidth}{!}{$
\begin{array}{rrrclcrcl}
& & & J  & & &  & E   \\
\Lambda_{21} : & \ & Y_{14} \cdot Y_{43} \cdot X_{32}  &-&  X_{14} \cdot Y_{43} \cdot Y_{32}   & \ \ \ \ &  U_{25} \cdot P_{51}  &-&  V_{25} \cdot Q_{51}  \\
\Lambda_{43} : & \ & X_{32} \cdot Y_{21} \cdot Y_{14}  &-&  Y_{32} \cdot Y_{21} \cdot X_{14}   & \ \ \ \ &  V_{47} \cdot  Q_{73}  &-&  U_{47} \cdot P_{73}  \\
\Lambda_{65} : & \ & Y_{58} \cdot Y_{87} \cdot X_{76}  &-&  X_{58} \cdot Y_{87} \cdot Y_{76}   & \ \ \ \ & Q_{62} \cdot  V_{25}  &-&  P_{62} \cdot U_{25}   \\
 \Lambda_{87} : & \ & X_{76} \cdot Y_{65} \cdot Y_{58}  &-&  Y_{76} \cdot Y_{65} \cdot X_{58}   & \ \ \ \ &  P_{84} \cdot U_{47}  &-&  Q_{84} \cdot  V_{47}   \\
 \Lambda^1_{54} : & \ & V_{47} \cdot X_{76} \cdot Y_{65}  &-&  Y_{43} \cdot X_{32} \cdot V_{25}   & \ \ \ \ &  Y_{58} \cdot  Q_{84}  &-&  Q_{51} \cdot Y_{14}  \\
 \Lambda^2_{54} : & \ & V_{47} \cdot Y_{76} \cdot Y_{65}  &-&  Y_{43} \cdot Y_{32} \cdot V_{25}   & \ \ \ \ &  Q_{51} \cdot X_{14}  &-&  X_{58} \cdot  Q_{84}   \\
 \Lambda^3_{54} : & \ & U_{47} \cdot Y_{76} \cdot Y_{65}  &-&   Y_{43} \cdot Y_{32} \cdot U_{25}   & \ \ \ \ &  X_{58} \cdot  P_{84}  &-&  P_{51} \cdot X_{14}    \\
 \Lambda^4_{54} : & \ & U_{47} \cdot X_{76} \cdot Y_{65}  &-&   Y_{43} \cdot X_{32} \cdot U_{25}   & \ \ \ \ &  P_{51} \cdot Y_{14}  &-&  Y_{58} \cdot  P_{84}   \\
 \Lambda^1_{72} : & \ & Y_{21} \cdot Y_{14} \cdot V_{47}  &-&   V_{25} \cdot Y_{58} \cdot Y_{87}   & \ \ \ \ &  X_{76} \cdot  Q_{62}  &-&  Q_{73} \cdot X_{32}  \\
 \Lambda^2_{72} : & \ & Y_{21} \cdot Y_{14} \cdot U_{47}  &-&   U_{25} \cdot Y_{58} \cdot Y_{87}   & \ \ \ \ &  P_{73} \cdot X_{32}  &-&  X_{76} \cdot  P_{62}   \\
 \Lambda^3_{72} : & \ & Y_{21} \cdot X_{14} \cdot U_{47}  &-&   U_{25} \cdot X_{58} \cdot Y_{87}   & \ \ \ \ &  Y_{76} \cdot  P_{62}   &-&  P_{73} \cdot Y_{32}   \\
 \Lambda^4_{72} : & \ & Y_{21} \cdot X_{14} \cdot V_{47}  &-&   V_{25} \cdot X_{58} \cdot Y_{87}   & \ \ \ \ &  Q_{73} \cdot Y_{32}  &-&  Y_{76} \cdot  Q_{62}   \\
 \Lambda^1_{61} : & \ &  Y_{14} \cdot V_{47} \cdot X_{76} &-&   X_{14} \cdot V_{47} \cdot Y_{76}   & \ \ \ \ & Y_{65} \cdot  Q_{51}  &-&   Q_{62} \cdot Y_{21}   \\
 \Lambda^2_{61} : & \ & Y_{14} \cdot U_{47} \cdot X_{76} &-&   X_{14} \cdot U_{47} \cdot Y_{76}   & \ \ \ \ &  P_{62} \cdot Y_{21}  &-&  Y_{65} \cdot  P_{51}   \\
\Lambda^1_{83} : & \ &  Y_{32} \cdot V_{25} \cdot X_{58} &-&   X_{32} \cdot V_{25} \cdot Y_{58}   & \ \ \ \ &  Y_{87} \cdot  Q_{73}  &-&  Q_{84} \cdot Y_{43}   \\
 \Lambda^2_{83} : & \ &  Y_{32} \cdot U_{25} \cdot X_{58} &-&   X_{32} \cdot U_{25} \cdot Y_{58}   & \ \ \ \ &  P_{84} \cdot Y_{43}  &-&  Y_{87} \cdot  P_{73}   \\
\end{array}
$}
~.~
\eea
Using the forward algorithm \cite{Feng:2000mi, Franco:2015tna, Franco:2015tya},
we obtain from the $J$- and $E$-terms the 
corresponding $P$-matrix,
\beal{es02a11}
P =
\resizebox{0.6\textwidth}{!}{$
\left(
\ba{c|ccccc|cc|cccccccccc|cccccc}
 & p_1 & p_2 & p_3 & p_4 & p_5 & q_1 & q_2 & s_1 & s_2 & s_3 & s_4 & s_5 & s_6 & s_7 & s_8 & s_9 & s_{10} & o_1 & o_2 & o_3 & o_4 & o_5 & o_6 \\
\hline
 P_{51} & 1 & 0 & 0 & 0 & 0 & 0 & 1 & 0 & 0 & 0 & 0 & 0 & 0 & 0 & 0 & 1 & 1 & 0 & 0 & 1 & 1 & 1 & 1 \\ 
 P_{62} & 1 & 0 & 0 & 0 & 0 & 0 & 1 & 0 & 0 & 0 & 0 & 0 & 0 & 1 & 1 & 0 & 0 & 0 & 1 & 0 & 1 & 1 & 1 \\ 
 P_{73} & 1 & 0 & 0 & 0 & 0 & 0 & 1 & 0 & 0 & 0 & 0 & 0 & 1 & 0 & 1 & 0 & 0 & 1 & 1 & 1 & 0 & 0 & 1 \\ 
 P_{84} & 1 & 0 & 0 & 0 & 0 & 0 & 1 & 0 & 0 & 0 & 0 & 1 & 0 & 0 & 0 & 0 & 1 & 1 & 1 & 1 & 0 & 1 & 0 \\ 
 Q_{51} & 0 & 0 & 0 & 0 & 1 & 0 & 1 & 0 & 0 & 0 & 0 & 0 & 0 & 0 & 0 & 1 & 1 & 0 & 0 & 1 & 1 & 1 & 1 \\ 
 Q_{62} & 0 & 0 & 0 & 0 & 1 & 0 & 1 & 0 & 0 & 0 & 0 & 0 & 0 & 1 & 1 & 0 & 0 & 0 & 1 & 0 & 1 & 1 & 1 \\ 
 Q_{73} & 0 & 0 & 0 & 0 & 1 & 0 & 1 & 0 & 0 & 0 & 0 & 0 & 1 & 0 & 1 & 0 & 0 & 1 & 1 & 1 & 0 & 0 & 1 \\ 
 Q_{84} & 0 & 0 & 0 & 0 & 1 & 0 & 1 & 0 & 0 & 0 & 0 & 1 & 0 & 0 & 0 & 0 & 1 & 1 & 1 & 1 & 0 & 1 & 0 \\ 
 U_{25} & 0 & 0 & 0 & 0 & 1 & 1 & 0 & 0 & 0 & 0 & 1 & 0 & 0 & 0 & 0 & 0 & 0 & 1 & 1 & 1 & 0 & 0 & 0 \\ 
 U_{47} & 0 & 0 & 0 & 0 & 1 & 1 & 0 & 0 & 0 & 1 & 0 & 0 & 0 & 0 & 0 & 0 & 0 & 0 & 0 & 0 & 1 & 1 & 1 \\ 
 V_{25} & 1 & 0 & 0 & 0 & 0 & 1 & 0 & 0 & 0 & 0 & 1 & 0 & 0 & 0 & 0 & 0 & 0 & 1 & 1 & 1 & 0 & 0 & 0 \\ 
 V_{47} & 1 & 0 & 0 & 0 & 0 & 1 & 0 & 0 & 0 & 1 & 0 & 0 & 0 & 0 & 0 & 0 & 0 & 0 & 0 & 0 & 1 & 1 & 1 \\ 
 X_{14} & 0 & 1 & 0 & 0 & 0 & 0 & 0 & 0 & 1 & 0 & 0 & 1 & 0 & 0 & 0 & 0 & 0 & 1 & 1 & 0 & 0 & 0 & 0 \\ 
 X_{32} & 0 & 1 & 0 & 0 & 0 & 0 & 0 & 1 & 0 & 0 & 0 & 0 & 0 & 1 & 0 & 0 & 0 & 0 & 0 & 0 & 1 & 1 & 0 \\ 
 X_{58} & 0 & 1 & 0 & 0 & 0 & 0 & 0 & 0 & 1 & 0 & 0 & 0 & 0 & 0 & 0 & 1 & 0 & 0 & 0 & 0 & 1 & 0 & 1 \\ 
 X_{76} & 0 & 1 & 0 & 0 & 0 & 0 & 0 & 1 & 0 & 0 & 0 & 0 & 1 & 0 & 0 & 0 & 0 & 1 & 0 & 1 & 0 & 0 & 0 \\ 
 Y_{14} & 0 & 0 & 1 & 0 & 0 & 0 & 0 & 0 & 1 & 0 & 0 & 1 & 0 & 0 & 0 & 0 & 0 & 1 & 1 & 0 & 0 & 0 & 0 \\ 
 Y_{21} & 0 & 0 & 0 & 1 & 0 & 0 & 0 & 0 & 0 & 0 & 1 & 0 & 0 & 0 & 0 & 1 & 1 & 0 & 0 & 1 & 0 & 0 & 0 \\ 
 Y_{32} & 0 & 0 & 1 & 0 & 0 & 0 & 0 & 1 & 0 & 0 & 0 & 0 & 0 & 1 & 0 & 0 & 0 & 0 & 0 & 0 & 1 & 1 & 0 \\ 
 Y_{43} & 0 & 0 & 0 & 1 & 0 & 0 & 0 & 0 & 0 & 1 & 0 & 0 & 1 & 0 & 1 & 0 & 0 & 0 & 0 & 0 & 0 & 0 & 1 \\ 
 Y_{58} & 0 & 0 & 1 & 0 & 0 & 0 & 0 & 0 & 1 & 0 & 0 & 0 & 0 & 0 & 0 & 1 & 0 & 0 & 0 & 0 & 1 & 0 & 1 \\ 
 Y_{65} & 0 & 0 & 0 & 1 & 0 & 0 & 0 & 0 & 0 & 0 & 1 & 0 & 0 & 1 & 1 & 0 & 0 & 0 & 1 & 0 & 0 & 0 & 0 \\ 
 Y_{76} & 0 & 0 & 1 & 0 & 0 & 0 & 0 & 1 & 0 & 0 & 0 & 0 & 1 & 0 & 0 & 0 & 0 & 1 & 0 & 1 & 0 & 0 & 0 \\ 
 Y_{87} & 0 & 0 & 0 & 1 & 0 & 0 & 0 & 0 & 0 & 1 & 0 & 1 & 0 & 0 & 0 & 0 & 1 & 0 & 0 & 0 & 0 & 1 & 0
\ea
\right)
$}~,~
\eea
where we have $5$ GLSM fields corresponding to the 5 extremal points of the toric diagram in \fref{f_ref_ex1c}(a). 
The $U(1)$ charges on GLSM fields corresponding to the $J$- and $E$-terms as well as the $D$-terms are given respectively as follows, 
\beal{es2a12}
Q_{JE} =
\resizebox{0.6\textwidth}{!}{$
\left(
\ba{ccccc|cc|cccccccccc|cccccc}
 p_1 & p_2 & p_3 & p_4 & p_5 & q_1 & q_2 & s_1 & s_2 & s_3 & s_4 & s_5 & s_6 & s_7 & s_8 & s_9 & s_{10} & o_1 & o_2 & o_3 & o_4 & o_5 & o_6 \\
\hline
 1 & 0 & 0 & 0 & 1 & 0 & 0 & 0 & 0 & 0 & 0 & 0 & 0 & 0 & 1 & 0 & 1 & 1 & -1 & -1 & 1 & -1
   & -1 \\
 0 & 1 & 1 & 0 & 0 & 0 & 0 & 0 & 0 & 0 & 0 & 0 & 0 & 0 & -1 & 0 & -1 & -2 & 1 & 1 & -2 &
   1 & 1 \\
 0 & 0 & 0 & 1 & 0 & 0 & 0 & 0 & 0 & 0 & -1 & 0 & 0 & 0 & -1 & 0 & -1 & -1 & 1 & 1 & 0 &
   0 & 0 \\
 0 & 0 & 0 & 0 & 0 & 1 & 0 & 0 & 0 & 0 & -1 & 0 & 0 & 0 & 1 & 0 & 1 & 0 & 0 & 0 & 1 & -1
   & -1 \\
 0 & 0 & 0 & 0 & 0 & 0 & 1 & 0 & 0 & 0 & 1 & 0 & 0 & 0 & 0 & 0 & 0 & 1 & -1 & -1 & 0 & 0
   & 0 \\
 0 & 0 & 0 & 0 & 0 & 0 & 0 & 1 & 0 & 0 & 0 & 0 & 0 & 0 & -1 & 0 & 0 & -1 & 1 & 0 & -1 & 0
   & 1 \\
 0 & 0 & 0 & 0 & 0 & 0 & 0 & 0 & 1 & 0 & 0 & 0 & 0 & 0 & 0 & 0 & -1 & -1 & 0 & 1 & -1 & 1
   & 0 \\
 0 & 0 & 0 & 0 & 0 & 0 & 0 & 0 & 0 & 1 & -1 & 0 & 0 & 0 & 0 & 0 & 0 & -1 & 1 & 1 & 1 & -1
   & -1 \\
 0 & 0 & 0 & 0 & 0 & 0 & 0 & 0 & 0 & 0 & 0 & 1 & 0 & 0 & 0 & 0 & -1 & -1 & 0 & 1 & 0 & 0
   & 0 \\
 0 & 0 & 0 & 0 & 0 & 0 & 0 & 0 & 0 & 0 & 0 & 0 & 1 & 0 & -1 & 0 & 0 & -1 & 1 & 0 & 0 & 0
   & 0 \\
 0 & 0 & 0 & 0 & 0 & 0 & 0 & 0 & 0 & 0 & 0 & 0 & 0 & 1 & -1 & 0 & 0 & 0 & 0 & 0 & -1 & 0
   & 1 \\
 0 & 0 & 0 & 0 & 0 & 0 & 0 & 0 & 0 & 0 & 0 & 0 & 0 & 0 & 0 & 1 & -1 & 0 & 0 & 0 & -1 & 1
   & 0 \\
\ea
\right)
$}~,~
\eea
and
\beal{es2a13}
Q_D =
\resizebox{0.6\textwidth}{!}{$
\left(
\ba{ccccc|cc|cccccccccc|cccccc}
 p_1 & p_2 & p_3 & p_4 & p_5 & q_1 & q_2 & s_1 & s_2 & s_3 & s_4 & s_5 & s_6 & s_7 & s_8 & s_9 & s_{10} & o_1 & o_2 & o_3 & o_4 & o_5 & o_6 \\
\hline
 0 & 0 & 0 & 0 & 0 & 0 & 0 & 0 & 1 & 0 & 0 & 0 & 0 & 0 & 0 & -1 & 0 & 0 & 0 & 0 & 0 & 0 & 0 \\ 
 0 & 0 & 0 & 0 & 0 & 0 & 0 & 0 & 0 & 0 & 1 & 0 & 0 & -1 & 0 & 0 & 0 & 0 & 0 & 0 & 0 & 0 & 0 \\ 
 0 & 0 & 0 & 0 & 0 & 0 & 0 & 1 & 0 & 0 & 0 & 0 & -1 & 0 & 0 & 0 & 0 & 0 & 0 & 0 & 0 & 0 & 0 \\ 
 0 & 0 & 0 & 0 & 0 & 0 & 0 & 0 & 0 & 1 & 0 & -1 & 0 & 0 & 0 & 0 & 0 & 0 & 0 & 0 & 0 & 0 & 0 \\ 
 0 & 0 & 0 & 0 & 0 & 0 & 0 & 0 & 0 & 0 & -1 & 0 & 0 & 0 & 0 & 1 & 0 & 0 & 0 & 0 & 0 & 0 & 0 \\ 
 0 & 0 & 0 & 0 & 0 & 0 & 0 & -1 & 0 & 0 & 0 & 0 & 0 & 1 & 0 & 0 & 0 & 0 & 0 & 0 & 0 & 0 & 0 \\ 
 0 & 0 & 0 & 0 & 0 & 0 & 0 & 0 & 0 & -1 & 0 & 0 & 1 & 0 & 0 & 0 & 0 & 0 & 0 & 0 & 0 & 0 & 0
\ea
\right)
$}
~.~
\eea
The corresponding toric diagram in \fref{f_ref_ex1c}(a) is then given by, 
\beal{es02a14}
G_t =
\resizebox{0.6\textwidth}{!}{$
\left(
\ba{ccccc|cc|cccccccccc|cccccc}
 p_1 & p_2 & p_3 & p_4 & p_5 & q_1 & q_2 & s_1 & s_2 & s_3 & s_4 & s_5 & s_6 & s_7 & s_8 & s_9 & s_{10} & o_1 & o_2 & o_3 & o_4 & o_5 & o_6 \\
\hline
 0 & 1 & -1 & 0 & 0 & 0 & 0 & 0 & 0 & 0 & 0 & 0 & 0 & 0 & 0 & 0 & 0 & 0 & 0 & 0 & 0 & 0 & 0 \\ 
 -1 & 0 & 0 & 0 & 1 & 0 & 0 & 0 & 0 & 0 & 0 & 0 & 0 & 0 & 0 & 0 & 0 & 0 & 0 & 0 & 0 & 0 & 0 \\ 
 -1 & 0 & 0 & 1 & -1 & -1 & -1 & 0 & 0 & 0 & 0 & 0 & 0 & 0 & 0 & 0 & 0 & -1 & -1 & -1 & -1 & -1 & -1 \\ 
 \hline
 1 & 1 & 1 & 1 & 1 & 1 & 1 & 1 & 1 & 1 & 1 & 1 & 1 & 1 & 1 & 1 & 1 & 2 & 2 & 2 & 2 & 2 & 2
\ea
\right)
$}
~.~
\eea
We note here that the forward algorithm results in an over-parameterization of the mesonic moduli space $\mathcal{M}^{mes}_{F_{0,+-}}$ in terms of extra GLSM fields $o_1, \dots, o_6$ \cite{Franco:2015tna} that do not correspond to any points in the toric diagram in \fref{f_ref_ex1c}(a).

From the $Q_{JE}$ and $Q_D$ charge matrices, we can identify that the global symmetry of the brane brick model is enhanced to the following form, 
\beal{es02a15}
SU(2)_x \times SU(2)_y \times U(1)_f \times U(1)_R ~,~
\eea
where $SU(2)_x \times SU(2)_y \times U(1)_f$ is the mesonic flavor symmetry. 
The global symmetry charges on the extremal GLSM fields are summarized in \tref{tab_10}.

\begin{table}[ht!]
\centering
\begin{tabular}{|c|c|c|c|c|l|}
\hline
\; & $SU(2)_x$ & $SU(2)_y$ & $U(1)_f$ & $U(1)_R$ & fugacity \\
\hline
$p_1$ & $ +1 $ & $ 0 $ & $0 $ & $ r_1$ &  $t_1 = x \overline{t}_1 $ \\
$p_2$ & $ 0 $ & $ +1 $ & $+1 $ & $ r_2 $ &  $t_2 = yf \overline{t}_2 $ \\
$p_3$ & $ 0 $ & $ -1 $ & $+1 $ & $ r_2 $ &  $t_3 = y^{-1}f \overline{t}_2 $ \\
$p_4$ & $ 0 $ & $ 0 $ & $-2 $ & $ 2 r_1 $ &  $t_4 = f^{-2} \overline{t}_1^2$ \\
$p_5$ & $ -1 $ & $ 0 $  & $0 $ & $ r_1 $ &  $t_5 = x^{-1} \overline{t}_1 $ \\
\hline
\end{tabular}
\caption{Charges under the global symmetry of the $F_{0,+-} $ model on the extremal GLSM fields $p_a$.
Here, $U(1)_R$ charges $r_1$ and $r_2$ are chosen such that 
the $J$- and $E$-terms coupled to Fermi fields have an overall $U(1)_R$ charge of $2$
with $4r_1 + 2 r_2 =2$. 
 \label{tab_10}}
\end{table}

The Hilbert series of the mesonic moduli space $\mathcal{M}^{mes}_{F_{0,+-}}$ of the $F_{0,+-}$ model can be obtained using the Molien integral formula in \eref{es01a15}, and takes the following form, 
\beal{es02a16}
&&
g(t_a,y_q,y_s,y_o;\mathcal{M}^{mes}_{F_{0,+-}})= \frac{P(t_a,y_q,y_s,y_o;\mathcal{M}^{mes}_{F_{0,+-}})}{(1-y_s y_o t_2^2 t_4^2 ) (1-y_s y_o t_3^2 t_4^2 ) 
(1-y_q^2 y_s y_o^3 t_1^4 t_2^2 ) }
\nn\\
&&\hspace{1cm}\times
\frac{1}{
(1-y_q^2 y_s y_o^3 t_1^4 t_3^2 ) (1-y_q^2 y_s y_o^3 t_2^2 t_5^4 ) (1-y_q^2 y_s y_o^3 t_3^2 t_5^4 )} 
~,~
\eea
where $t_a$ is the fugacity corresponding to extremal GLSM fields $p_a$.
We also have the fugacities $y_q$, $y_s$ and $y_o$ which correspond to products of non-extremal GLSM fields $q_1 q_2$, $s_1 \dots s_{10}$, and $o_1 \dots o_{6}$, respectively. 
The numerator $P(t_a,y_q,y_s,y_o;\mathcal{M}^{mes}_{F_{0,+-}})$ of the Hilbert series in \eref{es02a16} is presented in appendix \sref{app_num_01}. 

Under the following fugacity map, 
\beal{es02a17b}
\overline{t}_1 = t_1^{1/2} t_5^{1/2}~,~
\overline{t}_2 = \frac{t_2^{1/2} t_3^{1/2} t_4^{1/2}}{t_1^{1/2} t_5^{1/2}}~,~ 
x= \frac{t_1^{1/2}}{t_5^{1/2}}~,~ 
y=\frac{t_2^{1/2}}{t_3^{1/2}} ~,~ 
f =\frac{t_1^{1/2} t_5^{1/2}}{t_4^{1/2}} ~,~
\eea
where fugacities corresponding to non-extremal GLSM fields are set to $y_q=y_s=y_o=1$, 
we can express the Hilbert series 
of the mesonic moduli space $\mathcal{M}^{mes}_{F_{0,+-}}$
in terms of characters of irreducible representations of $SU(2)_x \times SU(2)_y$ as follows, 
\beal{es02a17} 
g(\overline{t}_1,\overline{t}_2,x,y,f;\mathcal{M}^{mes}_{F_{0,+-}}) 
&=& 
\sum_{n_1,n_2=0}^{\infty} [4n_1;2n_1+2n_2] f^{2n_1-2n_2} 
 \overline{t}_1^{4n_1+4n_2} \overline{t}_2^{2n_1+2n_2} \nn\\
&&
+
\sum_{n_1,n_2=0}^{\infty} [4n_1+2;2n_1+2n_2+2] f^{2n_1-2n_2}  \overline{t}_1^{4n_1+4n_2+4} \overline{t}_2^{2n_1+2n_2+2}
~,~\nn\\
\eea
where the fugacity $x$ corresponds to $SU(2)_x$, 
$y$ corresponds to $SU(2)_y$, $f$ corresponds to the $U(1)_f$ factor in the global symmetry, 
and $\overline{t}_1$ and $\overline{t}_2$ is chosen to correspond to $U(1)_R$ charges $r_1$ and $r_2$ defined in \tref{tab_10}.
$[n;m]=[n]_x [m]_y$ are the characters of irreducible representation of $SU(2)_x \times SU(2)_y$ 
with highest weight $(n),(m)$.
The corresponding highest weight generating function \cite{Hanany:2014dia} is given by
\beal{es41a006}
h(\overline{t}_1,\overline{t}_2,\mu,\nu,f;\mathcal{M}^{mes}_{F_{0,+-}})=\frac{1+\mu^2 \nu^2 \overline{t}_1^4 \overline{t}_2^2}{(1-\mu^4 \nu^2 f^2 \overline{t}_1^4 \overline{t}_2^2)(1-\nu^2 f^{-2} \overline{t}_1^4 \overline{t}_2^2)} 
~,~
\eea
where $\mu^m \nu^n$ counts characters of the form $[m]_x [n]_y$.
\\

\begin{table}[H]
\centering
\resizebox{0.77\textwidth}{!}{
\begin{tabular}{|c|c|c|c|c|c|c|c|}
\hline
PL term&generators  & GLSM fields & $SU(2)_{x}$ & $SU(2)_{y}$ & $U(1)_{f}$ & fugacity\\
\hline\hline
\multirow{5}{*}{$+[2]_y f^{-2} \overline{t}_1^4 \overline{t}_2^2$}  &$X_{76} Y_{65} X_{58} Y_{87}= X_{14} Y_{43} X_{32} Y_{21} $ & $p_2^2 p_4^2 s o$ & $0$ & $+2$ & $-2$ & $y^2 f^{-2} \overline{t}_1^4 \overline{t}_2^2$\\
&$$ & $$ & $$ &$$&$$ & $$ \\
&$X_{76} Y_{65} Y_{58} Y_{87}= X_{58} Y_{87} Y_{76} Y_{65} = X_{32} Y_{21} Y_{14} Y_{43}= Y_{21} X_{14} Y_{43} Y_{32} $ & $p_2 p_3 p_4^2 s o$ & $0$ & $0$ & $-2$ & $f^{-2} \overline{t}_1^4 \overline{t}_2^2$\\
&$$ & $$ & $$ &$$&$$ & $$ \\
&$Y_{58} Y_{87} Y_{76} Y_{65}= Y_{43} Y_{32} Y_{21} Y_{14} $ & $p_3^2 p_4^2 s o$ & $0$ & $-2$ & $-2$ & $y^{-2} f^{-2} \overline{t}_1^4 \overline{t}_2^2$\\
\hline
\multirow{39}{*}{$+[2]_x [2]_y \overline{t}_1^4 \overline{t}_2^2$}  & \begin{tabular}{l}$
 V_{47} X_{76} Y_{65} X_{58} P_{84} = 
 V_{47} X_{76} P_{62} Y_{21} X_{14} = 
 V_{47} P_{73} X_{32} Y_{21} X_{14} = 
 Y_{87} X_{76} P_{62} V_{25} X_{58} $ \\ $ =  
 Y_{87} P_{73} X_{32} V_{25} X_{58} = 
 P_{84} Y_{43} X_{32} V_{25} X_{58} = 
 P_{51} X_{14} V_{47} X_{76} Y_{65} = 
 P_{51} X_{14} Y_{43} X_{32} V_{25} $ \end{tabular}  
 
 & $p_1^2 p_2^2 p_4 q s o^2$ & $+2$ & $+2$ & $0$ & $x^2 y^2 \overline{t}_1^4 \overline{t}_2^2$\\
 & $$ & $$ & $$ &$$&$$ & $$ \\
& \begin{tabular}{l}$
 P_{84} V_{47} X_{76} Y_{65} Y_{58}  = 
 P_{62} Y_{21}  Y_{14} V_{47} X_{76}  = 
 P_{84} V_{47} Y_{76} Y_{65} X_{58} = 
 P_{73} X_{32}  Y_{21} Y_{14} V_{47} $ \\ $ =  
 P_{62} Y_{21}  X_{14} V_{47} Y_{76}= 
 P_{73} Y_{32}  Y_{21} X_{14} V_{47} = 
 P_{62} V_{25} Y_{58} Y_{87} X_{76} =
  P_{62} V_{25} X_{58} Y_{87} Y_{76} $ \\ $ =  
  P_{73} Y_{32} V_{25} X_{58} Y_{87}= 
  P_{84} Y_{43} Y_{32} V_{25} X_{58} = 
  P_{73} X_{32} V_{25} Y_{58} Y_{87}= 
  P_{84} Y_{43} X_{32} V_{25} Y_{58}$ \\ $ =  
  P_{51} Y_{14} V_{47} X_{76} Y_{65}= 
  P_{51} X_{14} V_{47} Y_{76} Y_{65} = 
  P_{51} Y_{14} Y_{43}  X_{32} V_{25} = 
  P_{51} X_{14} Y_{43} Y_{32} V_{25} $ \end{tabular} 

& $p_1^2 p_2 p_3 p_4 q s o^2$ & $+2$ & $0$ & $0$ & $x^2 \overline{t}_1^4 \overline{t}_2^2$\\
&$$ & $$ & $$ &$$&$$ & $$ \\
& \begin{tabular}{l}$
 P_{84} V_{47} Y_{76} Y_{65} Y_{58} =
 P_{62} Y_{21} Y_{14} V_{47} Y_{76}=
 P_{73} Y_{32} Y_{21} Y_{14} V_{47} = 
 P_{62} V_{25} Y_{58} Y_{87} Y_{76} $ \\ $ = 
 P_{73} Y_{32} V_{25} Y_{58} Y_{87}= 
 P_{84} Y_{43} Y_{32} V_{25} Y_{58}= 
 P_{51} Y_{14} V_{47} Y_{76} Y_{65} = 
 P_{51} Y_{14} Y_{43} Y_{32} V_{25} $ \end{tabular}  

& $p_1^2 p_3^2 p_4 q s o^2$ & $+2$ & $-2$ & $0$ & $x^2 y^{-2} \overline{t}_1^4 \overline{t}_2^2$\\
&$$ & $$ & $$ &$$&$$ & $$ \\
& \begin{tabular}{l}$P_{62} U_{25} X_{58} Y_{87} X_{76} = 
 P_{73} X_{32} U_{25} X_{58} Y_{87}= 
 P_{84} Y_{43} X_{32} U_{25} X_{58} = 
 Q_{84} V_{47} X_{76} Y_{65} X_{58} $ \\ $= 
 Q_{51} X_{14} V_{47} X_{76} Y_{65}= 
 Q_{62} Y_{21} X_{14} V_{47} X_{76} = 
 Q_{73} X_{32} Y_{21} X_{14} V_{47} = 
 Q_{62} V_{25} X_{58} Y_{87} X_{76} $ \\ $=
 Q_{73} X_{32} V_{25} X_{58} Y_{87}= 
 Q_{84} Y_{43} X_{32} V_{25} X_{58} = 
 Q_{51} X_{14} Y_{43} X_{32} V_{25} = 
 P_{84} U_{47} X_{76} Y_{65} X_{58} $ \\ $= 
 P_{62} Y_{21} X_{14} U_{47} X_{76} = 
 P_{73} X_{32} Y_{21} X_{14} U_{47} = 
 P_{51} X_{14} Y_{43} X_{32} U_{25} = 
 P_{51} X_{14} U_{47} X_{76} Y_{65}$ \end{tabular}   

& $p_1 p_2^2 p_4 p_5 q s o^2$ & $0$ & $+2$ & $0$ & $y^2 \overline{t}_1^4 \overline{t}_2^2$\\
&$$ & $$ & $$ &$$&$$ & $$ \\
& \begin{tabular}{l}$
 P_{62} U_{25} Y_{58} Y_{87} X_{76} = 
 P_{62} U_{25} X_{58} Y_{87} Y_{76} = 
 P_{73} Y_{32}  U_{25} X_{58} Y_{87}= 
 P_{84} Y_{43}  Y_{32} U_{25} X_{58}  $ \\ $=
 P_{73} X_{32}  U_{25} Y_{58} Y_{87}= 
 P_{84} Y_{43}  X_{32} U_{25} Y_{58}= 
 Q_{84} V_{47} X_{76} Y_{65} Y_{58} = 
 Q_{51} Y_{14} V_{47} X_{76} Y_{65}$ \\ $ =
 Q_{62} Y_{21} Y_{14} V_{47} X_{76} = 
 Q_{84} V_{47} Y_{76} Y_{65} X_{58} = 
 Q_{73} X_{32} Y_{21} Y_{14} V_{47} = 
 Q_{51} X_{14} V_{47} Y_{76} Y_{65} $ \\ $= 
 Q_{62} Y_{21} X_{14} V_{47} Y_{76}= 
 Q_{73} Y_{32} Y_{21} X_{14} V_{47} = 
 Q_{62} V_{25} Y_{58} Y_{87} X_{76} = 
 Q_{62} V_{25} X_{58} Y_{87} Y_{76} $ \\ $= 
 Q_{73} Y_{32} V_{25} X_{58} Y_{87}= 
 Q_{84} Y_{43} Y_{32} V_{25} X_{58}  = 
Q_{73} X_{32} V_{25} Y_{58} Y_{87}= 
Q_{84} Y_{43}  X_{32} V_{25} Y_{58}$ \\ $= 
Q_{51} Y_{14} Y_{43} X_{32} V_{25}  = 
Q_{51} X_{14} Y_{43} Y_{32} V_{25} =
 P_{84} U_{47} X_{76} Y_{65} Y_{58} = 
 P_{62} Y_{21} Y_{14} U_{47} X_{76} $ \\ $= 
 P_{84} U_{47} Y_{76} Y_{65} X_{58} = 
 P_{73} X_{32} Y_{21} Y_{14} U_{47} = 
 P_{62} Y_{21} X_{14} U_{47} Y_{76}= 
 P_{73} Y_{32} Y_{21} X_{14} U_{47} $ \\ $= 
 P_{51} Y_{14} Y_{43} X_{32} U_{25} = 
 P_{51} X_{14} Y_{43} Y_{32} U_{25} = 
 P_{51} Y_{14} U_{47} X_{76} Y_{65}= 
 P_{51} X_{14} U_{47} Y_{76} Y_{65} $ \end{tabular}  

& $p_1 p_2 p_3 p_4 p_5 q s o^2$ & $0$ & $0$ & $0$ & $\overline{t}_1^4 \overline{t}_2^2$\\
&$$ & $$ & $$ &$$&$$ & $$ \\
& \begin{tabular}{l}$
 P_{62} U_{25} Y_{58} Y_{87} Y_{76} = 
 P_{73} Y_{32} U_{25} Y_{58} Y_{87}= 
 P_{84} Y_{43} Y_{32} U_{25} Y_{58}= 
 Q_{84} V_{47} Y_{76} Y_{65} Y_{58} $ \\ $= 
 Q_{51} Y_{14} V_{47} Y_{76} Y_{65} = 
 Q_{62} Y_{21} Y_{14} V_{47} Y_{76}= 
 Q_{73} Y_{32} Y_{21}  Y_{14} V_{47} = 
 Q_{62} V_{25} Y_{58} Y_{87} Y_{76} $ \\ $= 
 Q_{73} Y_{32} V_{25} Y_{58} Y_{87}= 
 Q_{84} Y_{43} Y_{32} V_{25} Y_{58}= 
 Q_{51} Y_{14} Y_{43} Y_{32} V_{25} = 
 P_{84} U_{47} Y_{76} Y_{65} Y_{58} $ \\ $= 
 P_{62} Y_{21} Y_{14} U_{47} Y_{76}= 
 P_{73} Y_{32} Y_{21} Y_{14} U_{47} = 
 P_{51} Y_{14} Y_{43} Y_{32} U_{25} = 
 P_{51} Y_{14} U_{47} Y_{76} Y_{65} $ \end{tabular} 

& $p_1 p_3^2 p_4 p_5 q s o^2$ & $0$ & $-2$ & $0$ & $y^{-2} \overline{t}_1^4 \overline{t}_2^2$\\
&$$ & $$ & $$ &$$&$$ & $$ \\
& \begin{tabular}{l}$
 Q_{62} U_{25} X_{58} Y_{87} X_{76} =  
 Q_{73} X_{32} U_{25} X_{58} Y_{87}= 
 Q_{84} Y_{43} X_{32} U_{25} X_{58} = 
 Q_{51} X_{14} Y_{43} X_{32} U_{25} $ \\ $= 
 Q_{84} U_{47} X_{76} Y_{65} X_{58} =
  Q_{51} X_{14} U_{47} X_{76} Y_{65}= 
  Q_{62} Y_{21} X_{14} U_{47} X_{76} = 
  Q_{73} X_{32} Y_{21} X_{14} U_{47} $ \end{tabular}  

& $p_2^2 p_4 p_5^2q s o^2$ & $-2$ & $+2$ & $0$ & $x^{-2} y^2\overline{t}_1^4 \overline{t}_2^2$\\
&$$ & $$ & $$ &$$&$$ & $$ \\
& \begin{tabular}{l}$
 Q_{62} U_{25} Y_{58} Y_{87} X_{76} = 
 Q_{62} U_{25} X_{58} Y_{87} Y_{76} = 
 Q_{73} Y_{32} U_{25} X_{58} Y_{87}= 
 Q_{84} Y_{43} Y_{32} U_{25} X_{58} $ \\ $= 
 Q_{73} X_{32} U_{25} Y_{58} Y_{87}= 
 Q_{84} Y_{43} X_{32} U_{25} Y_{58}= 
 Q_{51} Y_{14} Y_{43} X_{32} U_{25} = 
 Q_{51} X_{14} Y_{43} Y_{32} U_{25} $ \\ $= 
 Q_{84} U_{47} X_{76} Y_{65} Y_{58} = 
 Q_{51} Y_{14} U_{47} X_{76} Y_{65}= 
 Q_{62} Y_{21} Y_{14} U_{47} X_{76} = 
 Q_{84} U_{47} Y_{76} Y_{65} X_{58} $ \\ $=
 Q_{73} X_{32} Y_{21} Y_{14} U_{47} = 
 Q_{51} X_{14} U_{47} Y_{76} Y_{65} = 
 Q_{62} Y_{21} X_{14} U_{47} Y_{76}= 
 Q_{73} Y_{32} Y_{21} X_{14} U_{47} $ \end{tabular}  

& $p_2 p_3 p_4 p_5^2 q s o^2$ & $-2$ & $0$ & $0$ & $x^{-2} \overline{t}_1^4 \overline{t}_2^2$\\
&$$ & $$ & $$ &$$&$$ & $$ \\
&\begin{tabular}{l}$
Q_{62} U_{25} Y_{58} Y_{87} Y_{76} = 
Q_{73} Y_{32} U_{25} Y_{58} Y_{87}= 
Q_{84} Y_{43} Y_{32} U_{25} Y_{58}= 
Q_{51} Y_{14} Y_{43} Y_{32} U_{25} $ \\ $= 
Q_{84} U_{47} Y_{76} Y_{65} Y_{58} = 
Q_{51} Y_{14} U_{47} Y_{76} Y_{65} = 
Q_{62} Y_{21} Y_{14} U_{47} Y_{76}= 
Q_{73} Y_{32} Y_{21} Y_{14} U_{47} $ \end{tabular} 

& $p_3^2 p_4 p_5^2 q s o^2$ & $-2$ & $-2$ & $0$ & $x^{-2} y^{-2} \overline{t}_1^4 \overline{t}_2^2$\\
\hline
\end{tabular}}
\caption{
Generators of the mesonic moduli space $\mathcal{M}^{mes}_{F_{0,+-}}$ of the $F_{0, +-} $ brane brick model in terms of chiral fields 
and GLSM fields with the corresponding mesonic flavor symmetry charges. 
Here, $q$, $s$ and $o$ denote products of GLSM fields $q_1 q_2$, $s_1 \dots s_{10}$ and $o_1 \dots o_6$, respectively. \textbf{(Part 1)}
\label{tab_105_1}
}
\end{table}

\begin{table}[H]
\centering
\resizebox{0.87\textwidth}{!}{
\begin{tabular}{|c|c|c|c|c|c|c|c|}
\hline
PL term& generators  & GLSM fields & $SU(2)_{x}$ & $SU(2)_{y}$ & $U(1)_{f}$ & fugacity\\
\hline\hline
\multirow{78}{*}{$+[4]_x [2]_y f^2 \overline{t}_1^4 \overline{t}_2^2$} & $P_{62} V_{25} X_{58} P_{84} V_{47} X_{76}= P_{73} X_{32} V_{25} 
X_{58} P_{84} V_{47} = V_{25} P_{51} X_{14} V_{47} X_{76} P_{62} = P_{51} X_{14}  V_{47} P_{73} X_{32} V_{25} $

& $p_1^4 p_2^2 q^2 s o^3$ & $+4$ & $+2$ & $+2$ & $x^4 y^2 f^2 \overline{t}_1^4 \overline{t}_2^2$\\
&$$ & $$ & $$ &$$&$$ & $$\\
 &\begin{tabular}{l}$
 P_{62} V_{25} Y_{58} P_{84} V_{47} X_{76} = 
 P_{62} V_{25} X_{58} P_{84} V_{47} Y_{76}=
  P_{73} Y_{32} V_{25} X_{58} P_{84} V_{47}  = 
  P_{73} X_{32} V_{25} Y_{58} P_{84} V_{47} $ \\ $=
  P_{51} Y_{14} V_{47} X_{76} P_{62} V_{25}  = 
  P_{51} Y_{14} V_{47} P_{73} X_{32} V_{25}  = 
  P_{51} X_{14} V_{47} Y_{76} P_{62} V_{25} = 
  P_{51} X_{14} V_{47} P_{73} Y_{32} V_{25} $ \end{tabular} 

& $p_1^4 p_2 p_3 q^2 s o^3$ & $+4$ & $0$ & $+2$ & $x^4 f^2 \overline{t}_1^4 \overline{t}_2^2$\\
&$$ & $$ & $$ &$$&$$ & $$ \\
& \begin{tabular}{l}$
 P_{62} V_{25} Y_{58} P_{84} V_{47} Y_{76}= 
 P_{73} Y_{32} V_{25} Y_{58} P_{84} V_{47}= 
 P_{51} Y_{14} V_{47} Y_{76} P_{62} V_{25} = 
 P_{51} Y_{14} V_{47} P_{73} Y_{32} V_{25} $ \end{tabular}  

& $p_1^4 p_3^2 q^2 s o^3$ & $+4$ & $-2$ & $+2$ & $x^4 y^{-2} f^2 \overline{t}_1^4 \overline{t}_2^2$\\
&$$ & $$ & $$ &$$&$$ & $$ \\
& \begin{tabular}{l}$
 P_{62} U_{25} X_{58} P_{84} V_{47} X_{76}= 
 P_{73} X_{32} U_{25} X_{58} P_{84} V_{47} = 
 P_{84} V_{47} X_{76} Q_{62} V_{25} X_{58} = 
P_{62} V_{25} X_{58} Q_{84} V_{47} X_{76}$ \\ $= 
P_{84} V_{47} Q_{73} X_{32} V_{25} X_{58}= 
P_{73} X_{32} V_{25} X_{58} Q_{84} V_{47} = 
P_{62} V_{25} Q_{51} X_{14} V_{47} X_{76}= 
P_{73} X_{32} V_{25} Q_{51} X_{14} V_{47} $ \\ $= 
P_{62} V_{25} X_{58} P_{84} U_{47} X_{76}= 
P_{73} X_{32} V_{25} X_{58} P_{84} U_{47} = 
P_{51} X_{14} V_{47} X_{76} P_{62} U_{25} = 
P_{51} X_{14} V_{47} P_{73} X_{32} U_{25} $ \\ $= 
P_{51} X_{14} V_{47} X_{76} Q_{62} V_{25} = 
P_{51} X_{14} V_{47} Q_{73} X_{32} V_{25} = 
P_{51} X_{14} U_{47} X_{76} P_{62} V_{25} = 
P_{51} X_{14} U_{47} P_{73} X_{32} V_{25} $ \end{tabular}  

& $p_1^3 p_2^2 p_5 q^2 s o^3$ & $+2$ & $+2$ & $+2$ & $x^2 y^2 f^2 \overline{t}_1^4 \overline{t}_2^2$\\
&$$ & $$ & $$ &$$&$$ & $$\\
& \begin{tabular}{l}$
 P_{62} U_{25} Y_{58} P_{84} V_{47} X_{76}  = 
 P_{62} U_{25} X_{58} P_{84} V_{47} Y_{76} = 
 P_{73} Y_{32} U_{25} X_{58} P_{84}V_{47}  = 
P_{73} X_{32} U_{25} Y_{58} P_{84} V_{47} $ \\ $= 
P_{84} V_{47} X_{76} Q_{62} V_{25} Y_{58}= 
P_{62} V_{25} Y_{58} Q_{84} V_{47} X_{76} = 
P_{62} V_{25} Q_{51} Y_{14} V_{47} X_{76} = 
P_{84} V_{47} Y_{76} Q_{62} V_{25} X_{58} $ \\ $= 
P_{62} V_{25} X_{58} Q_{84} V_{47} Y_{76}= 
P_{84} V_{47} Q_{73} Y_{32} V_{25} X_{58} = 
P_{73} Y_{32} V_{25} X_{58} Q_{84} V_{47} = 
P_{84} V_{47} Q_{73} X_{32} V_{25} Y_{58}$ \\ $= 
P_{73} X_{32} V_{25} Y_{58} Q_{84} V_{47} = 
P_{73} X_{32} V_{25} Q_{51} Y_{14} V_{47} = 
P_{62} V_{25} Q_{51} X_{14} V_{47} Y_{76}= 
P_{73} Y_{32} V_{25} Q_{51} X_{14} V_{47} $ \\ $= 
P_{62} V_{25} Y_{58} P_{84} U_{47} X_{76} = 
P_{62} V_{25} X_{58} P_{84} U_{47} Y_{76}= 
P_{73} Y_{32} V_{25} X_{58} P_{84} U_{47} = 
P_{73} X_{32} V_{25} Y_{58} P_{84} U_{47} $ \\ $= 
P_{51} Y_{14} V_{47} X_{76} P_{62} U_{25} = 
P_{51} Y_{14} V_{47} P_{73} X_{32} U_{25} = 
P_{51} X_{14} V_{47} Y_{76} P_{62} U_{25} = 
P_{51} X_{14} V_{47} P_{73} Y_{32} U_{25} $ \\ $= 
P_{51} Y_{14} V_{47} X_{76} Q_{62} V_{25} = 
P_{51} Y_{14} V_{47} Q_{73} X_{32} V_{25} = 
P_{51} X_{14} V_{47} Y_{76} Q_{62} V_{25} = 
P_{51} X_{14} V_{47} Q_{73} Y_{32} V_{25} $ \\ $= 
P_{51} Y_{14} U_{47} X_{76} P_{62} V_{25}  = 
P_{51} Y_{14} U_{47} P_{73} X_{32} V_{25} = 
P_{51} X_{14} U_{47} Y_{76} P_{62} V_{25} = 
P_{51} X_{14} U_{47} P_{73} Y_{32} V_{25} $ \end{tabular}  

& $p_1^3 p_2 p_3 p_5 q^2 s o^3$ & $+2$ & $0$ & $+2$& $x^2 f^2 \overline{t}_1^4 \overline{t}_2^2$\\
&$$ & $$ & $$ &$$&$$ & $$ \\
 &\begin{tabular}{l}$
 P_{62} U_{25} Y_{58} P_{84} V_{47} Y_{76}= 
 P_{73} Y_{32} U_{25} Y_{58} P_{84} V_{47} = 
 P_{84} V_{47} Y_{76} Q_{62} V_{25} Y_{58} =
 P_{62} V_{25} Y_{58} Q_{84}  V_{47} Y_{76}$ \\ $= 
 P_{84} V_{47} Q_{73} Y_{32} V_{25} Y_{58}= 
 P_{73} Y_{32} V_{25} Y_{58} Q_{84} V_{47} = 
P_{62} V_{25} Q_{51} Y_{14} V_{47} Y_{76}= 
P_{73} Y_{32}V_{25} Q_{51} Y_{14} V_{47} $ \\ $= 
P_{62} V_{25} Y_{58} P_{84} U_{47} Y_{76}= 
P_{73} Y_{32} V_{25} Y_{58} P_{84} U_{47} = 
P_{51} Y_{14} V_{47} Y_{76} P_{62} U_{25} = 
P_{51} Y_{14} V_{47} P_{73} Y_{32} U_{25} $ \\ $= 
P_{51} Y_{14} V_{47} Y_{76} Q_{62} V_{25} = 
P_{51} Y_{14} V_{47} Q_{73} Y_{32} V_{25} = 
P_{51} Y_{14} U_{47} Y_{76} P_{62} V_{25} = 
P_{51} Y_{14} U_{47} P_{73} Y_{32} V_{25} $ \end{tabular}  

& $p_1^3 p_3^2 p_5 q^2 s o^3$ & $+2$ & $-2$ & $+2$ & $x^2 y^{-2} f^2\overline{t}_1^4 \overline{t}_2^2$\\
&$$ & $$ & $$ &$$&$$ & $$\\
& \begin{tabular}{l}$
 P_{84} V_{47} X_{76} Q_{62} U_{25} X_{58} = 
 P_{62} U_{25} X_{58} Q_{84} V_{47} X_{76}= 
 P_{84} V_{47} Q_{73} X_{32} U_{25} X_{58}= 
P_{73} X_{32} U_{25} X_{58} Q_{84} V_{47} $ \\ $= 
P_{62} U_{25} Q_{51} X_{14} V_{47} X_{76}= 
P_{73} X_{32} U_{25} Q_{51} X_{14} V_{47} = 
Q_{62} V_{25} X_{58} Q_{84} V_{47}  X_{76}= 
Q_{73} X_{32} V_{25} X_{58} Q_{84} V_{47} $ \\ $= 
Q_{51} X_{14} V_{47} X_{76} Q_{62} V_{25} = 
Q_{51} X_{14} V_{47} Q_{73} X_{32} V_{25} = 
P_{62} U_{25} X_{58} P_{84} U_{47} X_{76}= 
P_{73} X_{32} U_{25} X_{58} P_{84} U_{47} $ \\ $= 
P_{84} U_{47} X_{76} Q_{62} V_{25} X_{58} = 
P_{62} V_{25} X_{58} Q_{84} U_{47} X_{76}= 
P_{84} U_{47} Q_{73} X_{32} V_{25} X_{58}= 
P_{73} X_{32} V_{25} X_{58} Q_{84} U_{47} $ \\ $= 
P_{62} V_{25} Q_{51} X_{14} U_{47} X_{76}= 
P_{73} X_{32} V_{25} Q_{51} X_{14} U_{47} = 
P_{51} X_{14} V_{47} X_{76} Q_{62} U_{25} = 
P_{51}  X_{14} V_{47} Q_{73} X_{32} U_{25} $ \\ $= 
P_{51} X_{14} U_{47} X_{76} P_{62} U_{25} = 
P_{51} X_{14} U_{47} P_{73} X_{32} U_{25} = 
P_{51} X_{14} U_{47} X_{76} Q_{62}  V_{25}  = 
P_{51} X_{14} U_{47} Q_{73} X_{32}  V_{25}  $ \end{tabular}  

& $p_1^2 p_2^2 p_5^2 q^2 s o^3$ & $0$ & $+2$ & $+2$ & $y^2 f^2 \overline{t}_1^4 \overline{t}_2^2$\\
&$$ & $$ & $$ &$$&$$ & $$\\
& \begin{tabular}{l}$
 P_{84} V_{47} X_{76} Q_{62} U_{25} Y_{58}= 
 P_{84} U_{47} Q_{73} Y_{32} V_{25} X_{58}= 
 P_{84} V_{47} Y_{76} Q_{62} U_{25} X_{58} =
 P_{84} V_{47} Q_{73} Y_{32} U_{25} X_{58} $ \\ $= 
 P_{84} V_{47} Q_{73} X_{32} U_{25}  Y_{58}=
 P_{84} U_{47} X_{76} Q_{62}  V_{25} Y_{58}= 
 P_{84} U_{47} Y_{76} Q_{62}  V_{25} X_{58} =
 P_{84} U_{47} Q_{73} X_{32} V_{25} Y_{58}$ \\ $= 

P_{73} Y_{32} U_{25} X_{58} Q_{84} V_{47} = 
P_{73} X_{32} U_{25}  Y_{58} Q_{84} V_{47} = 
P_{73} X_{32} U_{25} Q_{51} Y_{14} V_{47} = 
P_{73} Y_{32} U_{25} Q_{51} X_{14} V_{47}  $ \\ $= 
Q_{73} Y_{32}  V_{25} Q_{84} V_{47} X_{58} = 
Q_{73} X_{32}  V_{25} Q_{84} V_{47} Y_{58}= 
P_{73} Y_{32} U_{25} P_{84} U_{47} X_{58} = 
P_{73} X_{32} U_{25} P_{84} U_{47} Y_{58}$ \\ $= 
P_{73} Y_{32} V_{25} X_{58} Q_{84} U_{47}  = 
P_{73} X_{32} V_{25} Y_{58} Q_{84} U_{47} = 
P_{73} X_{32} V_{25} Q_{51} Y_{14} U_{47} = 
P_{73} Y_{32} V_{25} Q_{51} X_{14} U_{47} $ \\ $= 
 P_{62}  U_{25} Q_{51} Y_{14} V_{47} X_{76} =
 P_{62}  U_{25} X_{58} Q_{84} V_{47} Y_{76}= 
 P_{62}  U_{25} Y_{58} Q_{84} V_{47} X_{76} = 
P_{62} U_{25}   Q_{51} X_{14} V_{47} Y_{76}$ \\ $= 
Q_{62}  V_{25} Y_{58} Q_{84} V_{47} X_{76} = 
Q_{62}  V_{25} X_{58} Q_{84} V_{47} Y_{76}= 
P_{62} U_{25}  Y_{58}  P_{84} U_{47} X_{76} = 
P_{62} U_{25}   X_{58} P_{84} U_{47} Y_{76}$ \\ $= 
P_{62} V_{25}  Y_{58} Q_{84} U_{47} X_{76} = 
P_{62} V_{25}  Q_{51} Y_{14} U_{47} X_{76} = 
P_{62} V_{25}  X_{58} Q_{84} U_{47} Y_{76}= 
P_{62} V_{25}  Q_{51} X_{14} U_{47} Y_{76}$ \\ $= 
Q_{51} Y_{14} V_{47} X_{76} Q_{62} V_{25} = 
Q_{51} Y_{14} V_{47} Q_{73} X_{32} V_{25} = 
Q_{51} X_{14} V_{47} Y_{76} Q_{62} V_{25} = 
Q_{51} X_{14} V_{47} Q_{73}Y_{32} V_{25} $ \\ $= 
P_{51} Y_{14} V_{47} X_{76} Q_{62} U_{25}  = 
P_{51} Y_{14} V_{47} Q_{73}  X_{32} U_{25} = 
P_{51} X_{14} V_{47} Y_{76} Q_{62} U_{25}  = 
P_{51} X_{14} V_{47} Q_{73} Y_{32} U_{25}  $ \\ $= 
P_{51} Y_{14} U_{47} X_{76} P_{62} U_{25}  = 
P_{51} Y_{14} U_{47} P_{73} X_{32} U_{25} = 
P_{51} X_{14} U_{47} Y_{76} P_{62} U_{25} = 
P_{51} X_{14} U_{47} P_{73} Y_{32} U_{25} $ \\ $= 
P_{51} Y_{14} U_{47} X_{76} Q_{62} V_{25} = 
P_{51} Y_{14} U_{47} Q_{73} X_{32} V_{25} = 
P_{51} X_{14} U_{47} Y_{76} Q_{62} V_{25} = 
P_{51} X_{14} U_{47} Q_{73} Y_{32} V_{25} $ \end{tabular}  
& $p_1^2 p_2 p_3 p_5^2 q^2 s o^3$ & $0$ & $0$ & $+2$ & $f^2 \overline{t}_1^4 \overline{t}_2^2$\\
&$$ & $$ & $$ &$$&$$ & $$\\
 &\begin{tabular}{l}$
 P_{84} V_{47} Y_{76} Q_{62} U_{25} Y_{58} = 
 P_{62} U_{25} Y_{58} Q_{84} V_{47}  Y_{76}= 
 P_{84} V_{47}  Q_{73} Y_{32} U_{25} Y_{58}= 
P_{73} Y_{32} U_{25} Y_{58} Q_{84} V_{47} $ \\ $= 
P_{62} U_{25} Q_{51} Y_{14} V_{47} Y_{76}= 
P_{73} Y_{32} U_{25} Q_{51} Y_{14} V_{47}= 
Q_{62} V_{25} Y_{58} Q_{84} V_{47} Y_{76}= 
Q_{73} Y_{32} V_{25} Y_{58} Q_{84} V_{47} $ \\ $= 
Q_{51} Y_{14} V_{47} Y_{76} Q_{62} V_{25} = 
Q_{51} Y_{14} V_{47} Q_{73} Y_{32} V_{25} = 
P_{62} U_{25} Y_{58} P_{84} U_{47} Y_{76}= 
P_{73} Y_{32} U_{25} Y_{58}  P_{84} U_{47} $ \\ $= 
P_{84} U_{47} Y_{76} Q_{62} V_{25} Y_{58} = 
P_{62} V_{25} Y_{58} Q_{84} U_{47} Y_{76}= 
P_{84} U_{47} Q_{73} Y_{32} V_{25} Y_{58}= 
P_{73} Y_{32} V_{25} Y_{58} Q_{84} U_{47} $ \\ $= 
P_{62} V_{25} Q_{51} Y_{14} U_{47} Y_{76}= 
P_{73} Y_{32} V_{25} Q_{51} Y_{14} U_{47}  = 
P_{51} Y_{14} V_{47} Y_{76} Q_{62} U_{25}   = 
P_{51} Y_{14} V_{47} Q_{73} Y_{32} U_{25}  $ \\ $= 
P_{51} Y_{14} U_{47} Y_{76} P_{62} U_{25} = 
P_{51} Y_{14} U_{47} P_{73} Y_{32} U_{25} = 
P_{51} Y_{14} U_{47} Y_{76} Q_{62} V_{25}  = 
P_{51} Y_{14} U_{47} Q_{73} Y_{32} V_{25}  $ \end{tabular}  

& $p_1^2 p_3^2 p_5^2 q^2 s o^3$ & $0$ & $-2$ & $+2$ & $y^{-2} f^2\overline{t}_1^4 \overline{t}_2^2$\\
&$$ & $$ & $$ &$$&$$ & $$\\
& \begin{tabular}{l}$
 Q_{62} U_{25} X_{58} Q_{84} V_{47}  X_{76}= 
 Q_{73} X_{32} U_{25} X_{58} Q_{84} V_{47}  = 
 Q_{51}  X_{14} V_{47} X_{76} Q_{62} U_{25} = 
Q_{51} X_{14}  V_{47}  Q_{73} X_{32} U_{25} $ \\ $= 
P_{84} U_{47} X_{76} Q_{62} U_{25} X_{58} = 
P_{62} U_{25} X_{58} Q_{84} U_{47} X_{76}= 
P_{84} U_{47} Q_{73} X_{32} U_{25} X_{58}= 
P_{73} X_{32} U_{25} X_{58} Q_{84} U_{47} $ \\ $= 
P_{62} U_{25} Q_{51} X_{14} U_{47} X_{76}= 
P_{73} X_{32} U_{25} Q_{51} X_{14} U_{47}  = 
Q_{62} V_{25} X_{58} Q_{84} U_{47} X_{76}= 
Q_{73} X_{32} V_{25} X_{58} Q_{84} U_{47} $ \\ $= 
Q_{51} X_{14} U_{47} X_{76} Q_{62} V_{25}  = 
Q_{51} X_{14} U_{47} Q_{73} X_{32} V_{25} = 
P_{51} X_{14} U_{47} X_{76} Q_{62} U_{25} = 
P_{51} X_{14} U_{47} Q_{73} X_{32} U_{25} $ \end{tabular}  

& $p_1 p_2^2 p_5^3 q^2 s o^3$ & $-2$ & $+2$ & $+2$ & $x^{-2} y^2 f^2\overline{t}_1^4 \overline{t}_2^2$\\
&$$ & $$ & $$ &$$&$$ & $$\\
 &\begin{tabular}{l}$
 Q_{62} U_{25} Y_{58} Q_{84} V_{47} X_{76} = 
 Q_{62} U_{25} Q_{51} Y_{14} V_{47} X_{76}  = 
 Q_{62}  U_{25} X_{58} Q_{84}V_{47} Y_{76}= 
Q_{73} Y_{32} U_{25} X_{58} Q_{84} V_{47} $ \\ $= 
Q_{73} X_{32} U_{25} Y_{58} Q_{84} V_{47} = 
Q_{51} Y_{14} V_{47} Q_{73} X_{32} U_{25} = 
Q_{51} X_{14} V_{47} Y_{76} Q_{62} U_{25} = 
Q_{51} X_{14} V_{47} Q_{73} Y_{32} U_{25} $ \\ $= 
P_{84} U_{47} X_{76} Q_{62} U_{25} Y_{58}= 
P_{62} U_{25} Y_{58} Q_{84} U_{47} X_{76} = 
P_{62} U_{25} Q_{51} Y_{14} U_{47} X_{76} = 
P_{84} U_{47}  Y_{76} Q_{62} U_{25} X_{58}$\\$=  
P_{62} U_{25} X_{58} Q_{84} U_{47} Y_{76}= 
P_{84} U_{47} Q_{73} Y_{32} U_{25} X_{58} = 
P_{73} Y_{32} U_{25} X_{58} Q_{84} U_{47} = 
P_{84} U_{47} Q_{73} X_{32} U_{25} Y_{58}$\\$= 
P_{73} X_{32} U_{25} Y_{58} Q_{84} U_{47} = 
P_{73}  X_{32} U_{25} Q_{51} Y_{14} U_{47} = 
P_{62}  U_{25} Q_{51} X_{14} U_{47} Y_{76}= 
P_{73} Y_{32} U_{25} Q_{51} X_{14} U_{47} $\\$= 
Q_{62} V_{25} Y_{58} Q_{84} U_{47} X_{76} = 
Q_{51} Y_{14} U_{47} X_{76} Q_{62} V_{25}  = 
Q_{62} V_{25} X_{58} Q_{84} U_{47} Y_{76}= 
Q_{73} Y_{32} V_{25} X_{58} Q_{84} U_{47} $\\$= 
Q_{73} X_{32} V_{25} Y_{58} Q_{84} U_{47} = 
Q_{51}Y_{14}  U_{47} Q_{73}  X_{32} V_{25} = 
Q_{51} X_{14} U_{47} Y_{76} Q_{62} V_{25} = 
Q_{51} X_{14} U_{47} Q_{73} Y_{32} V_{25} $\\$= 
P_{51} Y_{14} U_{47} X_{76} Q_{62} U_{25}  = 
P_{51} Y_{14} U_{47} Q_{73} X_{32} U_{25}  = 
P_{51} X_{14} U_{47} Y_{76} Q_{62} U_{25} = 
P_{51} X_{14} U_{47} Q_{73} Y_{32} U_{25} $ \end{tabular}  

& $p_1 p_2 p_3 p_5^3 q^2 s o^3$ & $-2$ & $0$ & $+2$ & $x^{-2} f^2 \overline{t}_1^4 \overline{t}_2^2$\\
&$$ & $$ & $$ &$$&$$ & $$ \\
& \begin{tabular}{l}$
 Q_{62} U_{25} Y_{58} Q_{84} V_{47} Y_{76}= 
 Q_{73} Y_{32} U_{25} Y_{58} Q_{84} V_{47}  = 
 Q_{51}  Y_{14} V_{47}Y_{76} Q_{62} U_{25} = 
Q_{51} Y_{14} V_{47}  Q_{73} Y_{32} U_{25} $\\$=  
P_{84} U_{47} Y_{76} Q_{62} U_{25} Y_{58} = 
P_{62} U_{25} Y_{58} Q_{84} U_{47} Y_{76}= 
P_{84} U_{47} Q_{73} Y_{32} U_{25} Y_{58}= 
P_{73} Y_{32} U_{25} Y_{58} Q_{84} U_{47} $\\$= 
P_{62} U_{25} Q_{51} Y_{14} U_{47} Y_{76}= 
P_{73} Y_{32} U_{25} Q_{51} Y_{14}  U_{47} = 
Q_{62} V_{25} Y_{58} Q_{84} U_{47} Y_{76}= 
Q_{73} Y_{32} V_{25} Y_{58} Q_{84} U_{47} $\\$=  
Q_{51} Y_{14} U_{47} Y_{76} Q_{62} V_{25}  = 
Q_{51} Y_{14} U_{47} Q_{73} Y_{32} V_{25}  = 
P_{51} Y_{14} U_{47} Y_{76} Q_{62} U_{25} = 
P_{51} Y_{14} U_{47} Q_{73} Y_{32} U_{25} $ \end{tabular}  

& $p_1 p_3^2 p_5^3 q^2 s o^3$ & $-2$ & $-2$ & $+2$ & $x^{-2} y^{-2} f^2 \overline{t}_1^4 \overline{t}_2^2$\\
&$$ & $$ & $$ &$$&$$ & $$ \\
& \begin{tabular}{l}$
 Q_{62} U_{25} X_{58} Q_{84} U_{47} X_{76}= 
 Q_{73} X_{32} U_{25} X_{58} Q_{84} U_{47} = 
 Q_{51} X_{14} U_{47} X_{76} Q_{62} U_{25} = 
Q_{51} X_{14} U_{47}  Q_{73} X_{32} U_{25} $ \end{tabular}  

& $p_2^2 p_5^4 q^2 s o^3$ & $-4$ & $+2$ & $+2$ & $x^{-4} y^2 f^2\overline{t}_1^4 \overline{t}_2^2$\\
&$$ & $$ & $$ &$$&$$ & $$ \\
& \begin{tabular}{l}$
 Q_{62} U_{25} Y_{58} Q_{84} U_{47} X_{76} = 
 Q_{51} Y_{14} U_{47} X_{76} Q_{62} U_{25} = 
 Q_{62} U_{25} X_{58} Q_{84} U_{47} Y_{76}= 
Q_{73} Y_{32} U_{25} X_{58} Q_{84} U_{47}  $\\$= 
Q_{73} X_{32} U_{25} Y_{58} Q_{84} U_{47} = 
Q_{51} Y_{14} U_{47} Q_{73} X_{32} U_{25}  = 
Q_{51} X_{14} U_{47} Y_{76} Q_{62} U_{25} = 
Q_{51} X_{14} U_{47} Q_{73} Y_{32} U_{25} $ \end{tabular}  

& $p_2 p_3 p_5^4 q^2 s o^3$ & $-4$ & $0$ & $+2$ & $x^{-4} f^2 \overline{t}_1^4 \overline{t}_2^2$\\
&$$ & $$ & $$ &$$&$$ & $$ \\
&$Q_{62} U_{25} Y_{58} Q_{84} U_{47} Y_{76}= 
Q_{73} Y_{32} U_{25} Y_{58} Q_{84} U_{47} = 
Q_{51} Y_{14} U_{47} Y_{76} Q_{62} U_{25} = 
Q_{51} Y_{14} U_{47} Q_{73} Y_{32} U_{25} $ 

& $p_3^2 p_5^4 q^2 s o^3$ & $-4$ & $-2$ & $+2$ & $x^{-4} y^{-2} f^2 \overline{t}_1^4 \overline{t}_2^2$\\
\hline
\end{tabular}}
\caption{
Generators of the mesonic moduli space $\mathcal{M}^{mes}_{F_{0,+-}}$ of the $F_{0, +-} $ brane brick model in terms of chiral fields 
and GLSM fields with the corresponding mesonic flavor symmetry charges. 
Here, $q$, $s$ and $o$ denote products of GLSM fields $q_1 q_2$, $s_1 \dots s_{10}$ and $o_1 \dots o_6$, respectively. \textbf{(Part 2)}
\label{tab_105_2}
}
\end{table} 

The plethystic logarithm of the refined Hilbert series in \eref{es02a17} is as follows,
\beal{es02a19}\label{es41a007}
&&PL[g(\overline{t}_1,\overline{t}_2,x,y,f;\mathcal{M}^{mes}_{F_{0,+-}}) ]
=
([4]_x [2]_y f^2+[2]_x [2]_y+[2]_y f^{-2}) \overline{t}_1^4 \overline{t}_2^2
\nn\\
&&
\hspace{1cm}
-(1+f^{-4}+f^4+[2]_x f^{-2}+[2]_x f^2+[2]_x [2]_y+[2]_x [2]_y f^{-2}+[2]_x [2]_y f^2\nn\\&&\hspace{1cm}
+[2]_x [2]_y f^4+[2]_x [4]_y f^2+2 [4]_x+[4]_x f^2 +[4]_x f^4+[4]_y+[4]_y f^4+[4]_x [2]_y\nn\\&&\hspace{1cm} +[4]_x [2]_y f^2+[4]_x [4]_y+[4]_x [4]_y f^2+[4]_x [4]_y f^4+[6]_x f^2+[6]_x [2]_y f^2\nn\\&&\hspace{1cm}+[6]_x [2]_y f^4+[8]_x f^4) \overline{t}_1^8 \overline{t}_2^4 +\dots
\dots 
~.~
\eea
We can see from the infinite expansion of the plethystic logarithm that the mesonic moduli space $\mathcal{M}^{mes}_{F_{0,+-}}$ is not a complete intersection.
The first positive terms of the plethystic logarithm correspond to the generators of the 
mesonic moduli space $\mathcal{M}^{mes}_{F_{0,+-}}$.
The generators of the mesonic moduli space $\mathcal{M}^{mes}_{F_{0,+-}}$ are summarized in \tref{tab_105_1} and \tref{tab_105_2} 
in terms of the corresponding chiral fields as well as GLSM fields in the brane brick model and their corresponding mesonic flavor symmetry charges. 
\\

\paragraph{Mass Deformation and the Brane Brick Model for $Q^{1,1,1}/ \mathbb{Z}_2$.}
As studied in \cite{Franco:2023tyf},
we can introduce mass terms to the $E$-terms of the $F_{0,+-}$ brane brick model in \eref{es10a21} as follows, 
\beal{es20a20}
\resizebox{0.65\textwidth}{!}{$
\begin{array}{rrrclcrcl}
& & & J  & & &  E & \textcolor{blue}{+}  & \textcolor{blue}{\Delta E} \\
\Lambda_{21} : & \ \ \ & Y_{14} \cdot Y_{43} \cdot X_{32}  &-&  X_{14} \cdot Y_{43} \cdot Y_{32}   & \ \ \ \ &  \textcolor{blue}{-Y_{21}} + U_{25} \cdot P_{51}  &-&  V_{25} \cdot Q_{51}  \\
\Lambda_{43} : & \ \ \ & X_{32} \cdot Y_{21} \cdot Y_{14}  &-&  Y_{32} \cdot Y_{21} \cdot X_{14}   & \ \ \ \ &  \textcolor{blue}{+Y_{43}} + V_{47} \cdot  Q_{73}  &-&  U_{47} \cdot P_{73}   \\
\Lambda_{65} : & \ \ \ & Y_{58} \cdot Y_{87} \cdot X_{76}  &-&  X_{58} \cdot Y_{87} \cdot Y_{76}   & \ \ \ \ &  \textcolor{blue}{+Y_{65}} + Q_{62} \cdot  V_{25}  &-&  P_{62} \cdot U_{25}  \\
 \Lambda_{87} : & \ \ \ & X_{76} \cdot Y_{65} \cdot Y_{58}  &-&  Y_{76} \cdot Y_{65} \cdot X_{58}   & \ \ \ \ &  \textcolor{blue}{-Y_{87}} + P_{84} \cdot U_{47}  &-&  Q_{84} \cdot  V_{47}   \\
\end{array}
$}
 ~.~
\eea
We note here that the original $E$-terms in \eref{es20a20} 
consist of chiral fields corresponding to extremal brick matchings $p_1$ and $p_5$, 
whereas the mass terms introduced in \eref{es20a20} consist of chiral fields corresponding to extremal brick matching $p_4$.
Based on the $U(1)_R$ charges on extremal brick matchings summarized in \tref{tab_10},
we note that the mass terms introduced in \eref{es20a20} satisfy the overall $U(1)_R$ charge constraint set by the original $E$-terms as follows, 
\beal{es20a20b}
r[p_1] + r[p_5] = r[p_4] = 2r_1 ~.~
\eea

By integrating out the mass terms, we obtain the $J$- and $E$-terms of a toric phase corresponding to $Q^{1,1,1}/ \mathbb{Z}_2$ \cite{Franco:2016nwv, Franco:2016qxh, Franco:2023tyf}, whose quiver diagram is shown in \fref{f_ref_ex1d}(b).
The $J$- and $E$-terms take the following form, 
\beal{es20a21}
\resizebox{0.8\textwidth}{!}{$
\begin{array}{rrrclcrcl}
& & & J  & & &  & E  \\
 \Lambda^1_{54} : & \ \ \ & V_{47} \cdot X_{76} \cdot P_{62} \cdot U_{25}  &-&  U_{47} \cdot X_{76} \cdot P_{62} \cdot V_{25}   & \ \ \ \ &  Q_{51} \cdot Y_{14}  &-&  Y_{58} \cdot  Q_{84}   \\
 \Lambda^2_{54} : & \ \ \ & V_{47} \cdot Y_{76} \cdot P_{62} \cdot U_{25}  &-&  U_{47} \cdot Y_{76} \cdot P_{62} \cdot V_{25}   & \ \ \ \ &  Q_{51} \cdot X_{14}  &-&  X_{58} \cdot  Q_{84}   \\
 \Lambda^3_{54} : & \ \ \ & V_{47} \cdot Y_{76} \cdot Q_{62} \cdot U_{25}  &-&  U_{47} \cdot Y_{76} \cdot Q_{62} \cdot V_{25}   & \ \ \ \ &  P_{51} \cdot X_{14}  &-&  X_{58} \cdot  P_{84}   \\
 \Lambda^4_{54} : & \ \ \ & V_{47} \cdot X_{76} \cdot Q_{62} \cdot U_{25}  &-&  U_{47} \cdot X_{76} \cdot Q_{62} \cdot V_{25}   & \ \ \ \ &  P_{51} \cdot Y_{14}  &-&  Y_{58} \cdot  P_{84}   \\
 \Lambda^1_{72} : & \ \ \ & V_{25} \cdot Y_{58} \cdot P_{84} \cdot U_{47}  &-&  U_{25} \cdot Y_{58} \cdot P_{84} \cdot V_{47}   & \ \ \ \ &  Q_{73} \cdot X_{32}  &-&  X_{76} \cdot  Q_{62}   \\
 \Lambda^2_{72} : & \ \ \ & V_{25} \cdot Y_{58} \cdot Q_{84} \cdot U_{47}  &-&  U_{25} \cdot Y_{58} \cdot Q_{84} \cdot V_{47}   & \ \ \ \ &  P_{73} \cdot X_{32}  &-&  X_{76} \cdot  P_{62}   \\
 \Lambda^3_{72} : & \ \ \ &  V_{25} \cdot X_{58} \cdot Q_{84} \cdot U_{47}  &-&  U_{25} \cdot X_{58} \cdot Q_{84} \cdot V_{47}   & \ \ \ \ &  P_{73} \cdot Y_{32}  &-&  Y_{76} \cdot  P_{62}   \\
 \Lambda^4_{72} : & \ \ \ & V_{25} \cdot X_{58} \cdot P_{84} \cdot U_{47}  &-&  U_{25} \cdot X_{58} \cdot P_{84} \cdot V_{47}   & \ \ \ \ &  Q_{73} \cdot Y_{32}  &-&  Y_{76} \cdot  Q_{62}   \\
 \Lambda^1_{61} : & \ \ \ &  Y_{14} \cdot V_{47} \cdot X_{76} &-&   X_{14} \cdot V_{47} \cdot Y_{76}   & \ \ \ \ &  Q_{62} \cdot U_{25} \cdot P_{51}  &-&  P_{62} \cdot U_{25} \cdot  Q_{51}   \\
 \Lambda^2_{61} : & \ \ \ & Y_{14} \cdot U_{47} \cdot X_{76} &-&   X_{14} \cdot U_{47} \cdot Y_{76}   & \ \ \ \ &  Q_{62} \cdot V_{25} \cdot P_{51}  &-&  P_{62} \cdot V_{25} \cdot  Q_{51}   \\
\Lambda^1_{83} : & \ \ \ &  Y_{32} \cdot V_{25} \cdot X_{58} &-&   X_{32} \cdot V_{25} \cdot Y_{58}   & \ \ \ \ &  Q_{84} \cdot U_{47} \cdot P_{73}  &-&  P_{84} \cdot U_{47} \cdot  Q_{73}   \\
 \Lambda^2_{83} : & \ \ \ &  Y_{32} \cdot U_{25} \cdot X_{58} &-&   X_{32} \cdot U_{25} \cdot Y_{58}   & \ \ \ \ &  Q_{84} \cdot V_{47} \cdot P_{73}  &-&  P_{84} \cdot V_{47} \cdot  Q_{73}   \\
\end{array}
$}
~.~
\eea

Using the forward algorithm \cite{Feng:2000mi, Franco:2015tna, Franco:2015tya}, we obtain the $P$-matrix for the $Q^{1,1,1}/ \mathbb{Z}_2$ model, 
\beal{es02a22}
P =
\resizebox{0.47\textwidth}{!}{$
\left(
\ba{c|cccccc|cccccccccc}
 & p_2 & p_3 & p_4 & p_5 & p_6 & p_7 & s_1 & s_2 & s_3 & s_4 & s_5 & s_6 & s_7 & s_8 & s_9 & s_{10} \\
\hline
 P_{51} & 0 & 0 & 0 & 0 & 1 & 0 & 0 & 0 & 0 & 0 & 0 & 0 & 0 & 0 & 1 & 1 \\ 
 P_{62} & 0 & 0 & 0 & 0 & 1 & 0 & 0 & 0 & 0 & 0 & 0 & 0 & 1 & 1 & 0 & 0 \\ 
 P_{73} & 0 & 0 & 0 & 0 & 1 & 0 & 0 & 0 & 0 & 0 & 0 & 1 & 0 & 1 & 0 & 0 \\ 
 P_{84} & 0 & 0 & 0 & 0 & 1 & 0 & 0 & 0 & 0 & 0 & 1 & 0 & 0 & 0 & 0 & 1 \\ 
 Q_{51} & 0 & 0 & 0 & 1 & 0 & 0 & 0 & 0 & 0 & 0 & 0 & 0 & 0 & 0 & 1 & 1 \\ 
 Q_{62} & 0 & 0 & 0 & 1 & 0 & 0 & 0 & 0 & 0 & 0 & 0 & 0 & 1 & 1 & 0 & 0 \\ 
 Q_{73} & 0 & 0 & 0 & 1 & 0 & 0 & 0 & 0 & 0 & 0 & 0 & 1 & 0 & 1 & 0 & 0 \\ 
 Q_{84} & 0 & 0 & 0 & 1 & 0 & 0 & 0 & 0 & 0 & 0 & 1 & 0 & 0 & 0 & 0 & 1 \\ 
 U_{25} & 0 & 0 & 1 & 0 & 0 & 0 & 0 & 0 & 0 & 1 & 0 & 0 & 0 & 0 & 0 & 0 \\ 
 U_{47} & 0 & 0 & 1 & 0 & 0 & 0 & 0 & 0 & 1 & 0 & 0 & 0 & 0 & 0 & 0 & 0 \\ 
 V_{25} & 0 & 0 & 0 & 0 & 0 & 1 & 0 & 0 & 0 & 1 & 0 & 0 & 0 & 0 & 0 & 0 \\ 
 V_{47} & 0 & 0 & 0 & 0 & 0 & 1 & 0 & 0 & 1 & 0 & 0 & 0 & 0 & 0 & 0 & 0 \\ 
 X_{14} & 1 & 0 & 0 & 0 & 0 & 0 & 0 & 1 & 0 & 0 & 1 & 0 & 0 & 0 & 0 & 0 \\ 
 X_{32} & 1 & 0 & 0 & 0 & 0 & 0 & 1 & 0 & 0 & 0 & 0 & 0 & 1 & 0 & 0 & 0 \\ 
 X_{58} & 1 & 0 & 0 & 0 & 0 & 0 & 0 & 1 & 0 & 0 & 0 & 0 & 0 & 0 & 1 & 0 \\ 
 X_{76} & 1 & 0 & 0 & 0 & 0 & 0 & 1 & 0 & 0 & 0 & 0 & 1 & 0 & 0 & 0 & 0 \\ 
 Y_{14} & 0 & 1 & 0 & 0 & 0 & 0 & 0 & 1 & 0 & 0 & 1 & 0 & 0 & 0 & 0 & 0 \\ 
 Y_{32} & 0 & 1 & 0 & 0 & 0 & 0 & 1 & 0 & 0 & 0 & 0 & 0 & 1 & 0 & 0 & 0 \\ 
 Y_{58} & 0 & 1 & 0 & 0 & 0 & 0 & 0 & 1 & 0 & 0 & 0 & 0 & 0 & 0 & 1 & 0 \\ 
 Y_{76} & 0 & 1 & 0 & 0 & 0 & 0 & 1 & 0 & 0 & 0 & 0 & 1 & 0 & 0 & 0 & 0
\ea
\right)
$}
~.~
\eea
The $U(1)$ charges under the $J$- and $E$-terms in \eref{es20a21}
as well as the $D$-terms are given by the following charge matrices, 
\beal{es02a23}
Q_{JE} =
\resizebox{0.5\textwidth}{!}{$
\left(
\ba{cccccc|cccccccccc}
 p_2 & p_3 & p_4 & p_5 & p_6 & p_7 & s_1 & s_2 & s_3 & s_4 & s_5 & s_6 & s_7 & s_8 & s_9 & s_{10} \\
\hline
0 & 0 & 0 & 1 & 1 & 0 & 1 & 1 & 0 & 0 & -1 & -1 & -1 & 0 & -1 & 0 \\ 
0 & 0 & 0 & 0 & 0 & 0 & 0 & 1 & 0 & 0 & -1 & 0 & 0 & 0 & -1 & 1 \\ 
0 & 0 & 0 & 0 & 0 & 0 & 1 & 0 & 0 & 0 & 0 & -1 & -1 & 1 & 0 & 0 \\ 
0 & 0 & 1 & 0 & 0 & 1 & 0 & 0 & -1 & -1 & 0 & 0 & 0 & 0 & 0 & 0 \\ 
1 & 1 & 0 & 0 & 0 & 0 & -1 & -1 & 0 & 0 & 0 & 0 & 0 & 0 & 0 & 0 
\ea
\right)
$}
~,~
\eea
and
\beal{es02a24}
Q_D =
\resizebox{0.5\textwidth}{!}{$
\left(
\ba{cccccc|cccccccccc}
 p_2 & p_3 & p_4 & p_5 & p_6 & p_7 & s_1 & s_2 & s_3 & s_4 & s_5 & s_6 & s_7 & s_8 & s_9 & s_{10} \\
\hline
0 & 0 & 0 & 0 & 0 & 0 & 0 & 1 & 0 & 0 & 0 & 0 & 0 & 0 & -1 & 0 \\ 
 0 & 0 & 0 & 0 & 0 & 0 & 0 & 0 & 0 & 1 & 0 & 0 & -1 & 0 & 0 & 0 \\ 
 0 & 0 & 0 & 0 & 0 & 0 & 1 & 0 & 0 & 0 & 0 & -1 & 0 & 0 & 0 & 0 \\ 
 0 & 0 & 0 & 0 & 0 & 0 & 0 & 0 & 1 & 0 & -1 & 0 & 0 & 0 & 0 & 0 \\ 
 0 & 0 & 0 & 0 & 0 & 0 & 0 & 0 & 0 & -1 & 0 & 0 & 0 & 0 & 1 & 0 \\ 
 0 & 0 & 0 & 0 & 0 & 0 & -1 & 0 & 0 & 0 & 0 & 0 & 1 & 0 & 0 & 0 \\ 
 0 & 0 & 0 & 0 & 0 & 0 & 0 & 0 & -1 & 0 & 0 & 1 & 0 & 0 & 0 & 0
\ea
\right)
$}
~.~
\eea
The resulting toric diagram for the $Q^{1,1,1}/\mathbb{Z}_2$ model
is given by the following $G_t$-matrix, 
\beal{es02a25}
G_t =
\resizebox{0.45\textwidth}{!}{$
\left(
\ba{cccccc|cccccccccc}
 p_2 & p_3 & p_4 & p_5 & p_6 & p_7 & s_1 & s_2 & s_3 & s_4 & s_5 & s_6 & s_7 & s_8 & s_9 & s_{10} \\
\hline
 1 & -1 & 0 & 0 & 0 & 0 & 0 & 0 & 0 & 0 & 0 & 0 & 0 & 0 & 0 & 0 \\
 0 & 0 & 0 & 1 & -1 & 0 & 0 & 0 & 0 & 0 & 0 & 0 & 0 & 0 & 0 & 0 \\
 0 & 0 & 1 & 0 & 0 & -1 & 0 & 0 & 0 & 0 & 0 & 0 & 0 & 0 & 0 & 0 \\
  \hline
 1 & 1 & 1 & 1 & 1 & 1 & 1 & 1 & 1 & 1 & 1 & 1 & 1 & 1 & 1 & 1
\ea
\right)
$}
~,~
\eea
where we can see that the $Q^{1,1,1}/\mathbb{Z}_2$ model has $6$ extremal GLSM fields corresponding to the $6$ extremal points in the toric diagram shown in \fref{f_ref_ex1c}(b).

Based on the $Q_{JE}$ and $Q_D$ matrices, we observe that the $Q^{1,1,1}/\mathbb{Z}_2$ model has an enhanced
global symmetry of the form, 
\beal{es02a26}
SU(2)_x \times SU(2)_y \times SU(2)_z \times U(1)_R ~,~
\eea
where $SU(2)_x \times SU(2)_y \times SU(2)_z$ is the enhanced mesonic flavor symmetry. 
\tref{tab_12} summarizes how the extremal GLSM fields $p_a$ are charged under the global symmetry of the $Q^{1,1,1}/\mathbb{Z}_2$ model.

\begin{table}[ht!]
\centering
\begin{tabular}{|c|c|c|c|c|l|}
\hline
\; & $SU(2)_x$ & $SU(2)_y$ & $ SU(2)_z$ & $U(1)_R$ & fugacity \\
\hline
$p_2$ & $ +1 $ & $ 0 $ & $ 0 $ & $ r_2$ & $t_2 = x \overline{t}_2 $ \\
$p_3$ & $ -1 $ & $ 0 $ & $ 0 $ & $ r_2 $ & $t_3 = x^{-1} \overline{t}_2 $ \\
$p_4$ & $ 0 $ & $ +1 $ & $ 0 $ & $ r_1 $ & $t_4 = y \overline{t}_1 $ \\
$p_5$ & $ 0 $ & $ 0 $ & $ +1 $ & $ r_1 $ & $t_5 = z \overline{t}_1 $ \\
$p_6$ & $ 0 $ & $ 0 $ & $ -1 $ & $ r_1 $ & $t_6 = z^{-1} \overline{t}_1 $ \\
$p_7$ & $ 0 $ & $ -1 $ & $ 0 $ & $ r_1 $ & $t_7 = y^{-1} \overline{t}_1 $ \\
\hline
\end{tabular}
\caption{Charges under the global symmetry of the $Q^{1,1,1}/\mathbb{Z}_2$ model on the extremal GLSM fields $p_a$.
Here, $U(1)_R$ charges $r_1$ and $r_2$ are chosen such that 
the $J$- and $E$-terms coupled to Fermi fields have an overall $U(1)_R$ charge of $2$
with $4r_1+2r_2 =2$.
\label{tab_12}}
\end{table}

\begin{table}[H]
\centering
\resizebox{0.78\textwidth}{!}{
\begin{tabular}{|c|c|c|c|c|c|c|c|}
\hline
PL term&generator  & GLSM fields & $SU(2)_{x}$ & $SU(2)_{y}$ & $SU(2)_{z}$  & fugacity\\
\hline\hline
\multirow{90}{*}{$+[2]_x [2]_y [2]_z \overline{t}_1^4 \overline{t}_2^2$}&$X_{76} Q_{62} U_{25} X_{58} Q_{84} U_{47}=Q_{51} X_{14} U_{47} X_{76} Q_{62} U_{25}=Q_{73} X_{32} U_{25} X_{58} Q_{84} U_{47}=Q_{51} X_{14} U_{47} Q_{73} X_{32} U_{25}$ 

& $p_2^2 p_4^2 p_5^2 s$ & $+2$ &$+2$&$+2$ & $x^2 y^2 z^2 \overline{t}_1^4 \overline{t}_2^2$\\

&$$ & $$ & $$ &$$&$$ & $$ \\

& \begin{tabular}{l}$
 U_{47} X_{76} Q_{62} U_{25} X_{58} P_{84}=U_{47} X_{76} P_{62} U_{25}  X_{58} Q_{84} = U_{47} X_{76} P_{62} U_{25} Q_{51} X_{14} 
= U_{47} X_{76} Q_{62} U_{25} P_{51}  X_{14} $ \\ $ =   U_{47} Q_{73} X_{32} U_{25} X_{58} P_{84} = U_{47} Q_{73} X_{32} U_{25} P_{51} X_{14} 
= U_{47} P_{73} X_{32} U_{25} X_{58} Q_{84} = U_{47} P_{73} X_{32} U_{25} Q_{51} X_{14} $\\ \end{tabular} 

& $p_2^2 p_4^2 p_5 p_6 s$ & $+2$ &$+2$&$0$ & $x^2 y^2 \overline{t}_1^4 \overline{t}_2^2$\\
&$$ & $$ & $$ &$$&$$ & $$ \\
&$U_{47} X_{76} P_{62} U_{25} X_{58} P_{84} = U_{47} X_{76} P_{62} U_{25} P_{51}  X_{14} = 
  U_{47} P_{73} X_{32} U_{25} X_{58} P_{84} = U_{47} P_{73} X_{32} U_{25} P_{51}  X_{14} $ 
 
& $p_2^2 p_4^2 p_6^2 s$ & $+2$ &$+2$&$-2$ & $x^2 y^2 z^{-2} \overline{t}_1^4 \overline{t}_2^2$\\
& $$ & $$ & $$ &$$&$$ & $$ \\
& \begin{tabular}{l}$
 U_{47} X_{76} Q_{62} V_{25} X_{58} Q_{84}  = V_{47} X_{76} Q_{62} U_{25}  X_{58} Q_{84} = 
 U_{47} X_{76} Q_{62}  V_{25} Q_{51}  X_{14} = V_{47} X_{76} Q_{62} U_{25} Q_{51}  X_{14} $ \\ $ =  
 U_{47} Q_{73} X_{32} V_{25}  X_{58} Q_{84}  = V_{47} Q_{73} X_{32} U_{25} X_{58} Q_{84} = 
 U_{47} Q_{73} X_{32} V_{25} Q_{51} X_{14} = V_{47}  Q_{73}  X_{32} U_{25} Q_{51}  X_{14} $ \end{tabular} 
 
& $p_2^2 p_4 p_5^2 p_7 s$ & $+2$ &$0$&$+2$ & $x^2 z^2 \overline{t}_1^4 \overline{t}_2^2$\\
&$$ & $$ & $$ &$$&$$ & $$ \\
& \begin{tabular}{l}$
U_{47} X_{76} Q_{62} V_{25} X_{58} P_{84} =V_{47} X_{76} Q_{62} U_{25}  X_{58} P_{84} = 
  U_{47} X_{76} P_{62} V_{25} X_{58} Q_{84} =V_{47} X_{76} P_{62} U_{25} X_{58} Q_{84} $ \\ $ = 
  U_{47} X_{76} P_{62} V_{25} Q_{51} X_{14} =  V_{47} X_{76} P_{62} U_{25} Q_{51} X_{14}= 
  U_{47} X_{76} Q_{62} V_{25}  P_{51} X_{14} = V_{47} X_{76} Q_{62} U_{25} P_{51} X_{14}$ \\ $ = 
  U_{47} Q_{73} X_{32} V_{25}  X_{58} P_{84} = V_{47} Q_{73} X_{32} U_{25} X_{58} P_{84} = 
  U_{47} Q_{73} X_{32} V_{25} P_{51} X_{14}  = V_{47} Q_{73} X_{32} U_{25} P_{51} X_{14} $ \\ $ =
  U_{47} P_{73} X_{32} V_{25} X_{58} Q_{84}  = V_{47} P_{73} X_{32} U_{25} X_{58} Q_{84} = 
  U_{47} P_{73} X_{32} V_{25} Q_{51} X_{14}  = V_{47} P_{73} X_{32} U_{25} Q_{51} X_{14} $ \end{tabular} 
 
& $p_2^2 p_4 p_5 p_6 p_7 s$ & $+2$ &$0$&$0$ & $x^2 \overline{t}_1^4 \overline{t}_2^2$\\
&$$ & $$ & $$ &$$&$$ & $$ \\
 &\begin{tabular}{l}$U_{47} X_{76} P_{62} V_{25} X_{58} P_{84} = V_{47} X_{76} P_{62} U_{25} X_{58} P_{84} =
 U_{47} X_{76} P_{62} V_{25} P_{51} X_{14} = V_{47} X_{76} P_{62} U_{25} P_{51} X_{14} $ \\ $ =  
 U_{47} P_{73} X_{32} V_{25} X_{58} P_{84} = V_{47} P_{73} X_{32} U_{25} X_{58} P_{84} = 
 U_{47} P_{73} X_{32} V_{25} P_{51} X_{14} = V_{47} P_{73} X_{32} U_{25} P_{51} X_{14} $ \end{tabular}   

& $p_2^2 p_4 p_6^2 p_7 s$ & $+2$ &$0$&$-2$ & $x^2 z^{-2} \overline{t}_1^4 \overline{t}_2^2$\\
&$$ & $$ & $$ &$$&$$ & $$ \\
&$V_{47} X_{76} Q_{62} V_{25} X_{58} Q_{84} =V_{47} X_{76} Q_{62} V_{25} Q_{51} X_{14} =
 V_{47} Q_{73} X_{32} V_{25} X_{58} Q_{84} =V_{47}  Q_{73} X_{32} V_{25} Q_{51} X_{14} $ 

& $p_2^2 p_5^2 p_7^2 s$ & $+2$ &$-2$&$+2$ & $x^2 y^{-2} z^2 \overline{t}_1^4 \overline{t}_2^2$\\
&$$ & $$ & $$ &$$&$$ & $$ \\
 &\begin{tabular}{l}$
V_{47}  X_{76} Q_{62} V_{25} X_{58} P_{84} =V_{47} X_{76} P_{62} V_{25} X_{58} Q_{84} = 
V_{47}  X_{76} P_{62} V_{25} Q_{51} X_{14} =V_{47} X_{76} Q_{62} V_{25} P_{51} X_{14} $ \\ $ = 
V_{47}  Q_{73} X_{32} V_{25} X_{58} P_{84} =V_{47} Q_{73} X_{32} V_{25} P_{51} X_{14} = 
V_{47}  P_{73} X_{32} V_{25} X_{58} Q_{84} =V_{47} P_{73} X_{32} V_{25} Q_{51} X_{14} $ \end{tabular}  

& $p_2^2 p_5 p_6 p_7^2 s$ & $+2$ &$-2$&$0$ & $x^2 y^{-2} \overline{t}_1^4 \overline{t}_2^2$\\
&$$ & $$ & $$ &$$&$$ & $$ \\
&$V_{47} X_{76} P_{62} V_{25} X_{58} P_{84} =V_{47} X_{76} P_{62} V_{25} P_{51} X_{14} = 
 V_{47} P_{73} X_{32} V_{25}  X_{58} P_{84} =V_{47} P_{73} X_{32} V_{25} P_{51} X_{14} $ 
 
& $p_2^2 p_6^2 p_7^2 s$ & $+2$ &$-2$&$-2$ & $x^2 y^{-2} z^{-2} \overline{t}_1^4 \overline{t}_2^2$\\
& \begin{tabular}{l}$
U_{47} Q_{73} Y_{32} U_{25} X_{58} P_{84} = 
U_{47} Q_{73} X_{32} U_{25} Y_{58} P_{84} = 
U_{47} X_{76} Q_{62} U_{25} Y_{58} P_{84} = 
U_{47} Y_{76} Q_{62} U_{25} X_{58} P_{84} $ \\ $ = 
U_{47} X_{76} P_{62} U_{25} Q_{51} Y_{14} = 
U_{47} P_{73} X_{32} U_{25} Q_{51} Y_{14}= 
U_{47} Y_{76} P_{62} Q_{51} U_{25} X_{14} = 
U_{47} P_{73} Y_{32} U_{25} Q_{51} X_{14} $ \\ $ =
U_{47} Q_{73} X_{32} U_{25} P_{51} Y_{14}= 
U_{47} X_{76} Q_{62} U_{25} P_{51} Y_{14}= 
U_{47} Q_{73} Y_{32} U_{25} P_{51} X_{14} = 
U_{47} Y_{76} Q_{62} U_{25} P_{51} X_{14} $ \\ $ =
U_{47} X_{76} P_{62} U_{25} Y_{58} Q_{84} = 
U_{47} Y_{76} P_{62} U_{25} X_{58} Q_{84} = 
U_{47} P_{73} Y_{32} U_{25} X_{58} Q_{84} = 
U_{47} P_{73} X_{32} U_{25} Y_{58} Q_{84} $ \end{tabular}  

&$p_2 p_3 p_4^2 p_5^2 s$ & $0$ &$+2$&$+2$ & $y^2 z^2 \overline{t}_1^4 \overline{t}_2^2$\\
&$$ & $$ & $$ &$$&$$ & $$ \\
 &\begin{tabular}{l}$
U_{47} Q_{73} Y_{32} U_{25} X_{58} P_{84} = 
U_{47} Q_{73} X_{32} U_{25} Y_{58} P_{84} = 
U_{47} X_{76} Q_{62} U_{25} Y_{58} P_{84} = 
U_{47} Y_{76} Q_{62} U_{25} X_{58} P_{84} $ \\ $ =
U_{47} X_{76} P_{62} U_{25} Q_{51} Y_{14}= 
U_{47} P_{73} X_{32} U_{25} Q_{51} Y_{14}= 
U_{47} Y_{76} P_{62} U_{25} Q_{51} X_{14} =
U_{47} P_{73} Y_{32} U_{25} Q_{51} X_{14} $ \\ $ = 
U_{47} Q_{73} X_{32} U_{25} P_{51} Y_{14}= 
U_{47} X_{76} Q_{62} U_{25} P_{51} Y_{14}= 
U_{47} Q_{73} Y_{32} U_{25} P_{51} X_{14} = 
U_{47} Y_{76} Q_{62} U_{25} P_{51} X_{14} $ \\ $ = 
U_{47} X_{76} P_{62} U_{25} Y_{58} Q_{84} = 
U_{47} Y_{76} P_{62} U_{25} X_{58} Q_{84} = 
U_{47} P_{73} Y_{32} U_{25} X_{58} Q_{84} = 
U_{47} P_{73} X_{32} U_{25} Y_{58} Q_{84} $ \end{tabular} 

& $p_2 p_3 p_4^2 p_5 p_6 s$ & $0$ &$+2$&$0$ & $y^2 \overline{t}_1^4 \overline{t}_2^2$\\
&$$ & $$ & $$ &$$&$$ & $$ \\
& \begin{tabular}{l}$
U_{47} X_{76} P_{62} U_{25} Y_{58} P_{84} = 
U_{47} Y_{76} P_{62} U_{25} X_{58} P_{84}  = 
U_{47} P_{73} Y_{32} U_{25} X_{58} P_{84} = 
U_{47} P_{73} X_{32} U_{25} Y_{58} P_{84} $ \\ $ =  
U_{47} X_{76} P_{62} U_{25} P_{51} Y_{14}= 
U_{47} P_{73} X_{32} U_{25} P_{51} Y_{14}= 
U_{47} Y_{76} P_{62} U_{25} P_{51} X_{14} =
U_{47} P_{73} Y_{32} U_{25} P_{51} X_{14} 
$ \end{tabular} 

& $p_2 p_3 p_4^2 p_6^2 s$ & $0$ &$+2$&$-2$ & $y^2 z^{-2} \overline{t}_1^4 \overline{t}_2^2$\\
&$$ & $$ & $$ &$$&$$ & $$ \\
& \begin{tabular}{l}$
U_{47} Q_{73} X_{32} V_{25} Q_{51} Y_{14}= 
U_{47} X_{76} Q_{62} V_{25} Q_{51} Y_{14}= 
U_{47} Q_{73} Y_{32} V_{25} Q_{51} X_{14} = 
U_{47} Y_{76} Q_{62} V_{25} Q_{51} X_{14} $ \\ $ =
U_{47} Q_{73} Y_{32} V_{25} X_{58} Q_{84} = 
U_{47} Q_{73} X_{32} V_{25} Y_{58} Q_{84} = 
U_{47} X_{76} Q_{62} V_{25} Y_{58} Q_{84} = 
U_{47} Y_{76} Q_{62} V_{25} X_{58} Q_{84} $ \\ $ = 
V_{47} Q_{73} X_{32} U_{25} Q_{51} Y_{14}= 
V_{47} X_{76} Q_{62} U_{25} Q_{51} Y_{14}= 
V_{47} Q_{73} Y_{32} U_{25} Q_{51} X_{14} = 
V_{47} Y_{76} Q_{62} U_{25} Q_{51} X_{14} $ \\ $ = 
V_{47} Q_{73} Y_{32} U_{25} X_{58} Q_{84}  = 
V_{47} Q_{73} X_{32} U_{25} Y_{58} Q_{84} = 
V_{47} X_{76} Q_{62} U_{25} Y_{58} Q_{84} = 
V_{47} Y_{76} Q_{62} U_{25} X_{58} Q_{84} $  \end{tabular} 

& $p_2 p_3 p_4 p_5^2 p_7 s$ & $0$ &$0$&$+2$ & $z^2 \overline{t}_1^4 \overline{t}_2^2$\\
&$$ & $$ & $$ &$$&$$ & $$ \\
& \begin{tabular}{l}$
U_{47} Q_{73} Y_{32} V_{25} X_{58} P_{84} = 
U_{47} Q_{73} X_{32} V_{25} Y_{58} P_{84} = 
U_{47} X_{76} Q_{62} V_{25} Y_{58} P_{84} = 
U_{47} Y_{76} Q_{62} V_{25} X_{58} P_{84} $ \\ $ = 
U_{47} X_{76} P_{62} V_{25} Q_{51} Y_{14}= 
U_{47} P_{73} X_{32} V_{25} Q_{51} Y_{14}= 
U_{47} Y_{76} P_{62} V_{25} Q_{51} X_{14} = 
U_{47} P_{73} Y_{32} V_{25} Q_{51} X_{14} $ \\ $ =
U_{47} Q_{73} X_{32} V_{25} P_{51} Y_{14}= 
U_{47} X_{76} Q_{62} V_{25} P_{51} Y_{14}= 
U_{47} Q_{73} Y_{32} V_{25} P_{51} X_{14} = 
U_{47} Y_{76} Q_{62} V_{25} P_{51} X_{14} $ \\ $ =
U_{47} X_{76} P_{62} V_{25} Y_{58} Q_{84} = 
U_{47} Y_{76} P_{62} V_{25} X_{58} Q_{84} = 
U_{47} P_{73} Y_{32} V_{25} X_{58} Q_{84} = 
U_{47} P_{73} X_{32} V_{25} Y_{58} Q_{84} $ \\ $ =
V_{47} Q_{73} Y_{32} U_{25} X_{58} P_{84} = 
V_{47} Q_{73} X_{32} U_{25} Y_{58} P_{84} = 
V_{47} X_{76} Q_{62} U_{25} Y_{58} P_{84} = 
V_{47} Y_{76} Q_{62} U_{25} X_{58} P_{84} $ \\ $ =
V_{47} X_{76} P_{62} U_{25} Q_{51} Y_{14}= 
V_{47} P_{73} X_{32} U_{25} Q_{51} Y_{14}= 
V_{47} Y_{76} P_{62} U_{25} Q_{51} X_{14} = 
V_{47} P_{73} Y_{32} U_{25} Q_{51} X_{14} $ \\ $ = 
V_{47} Q_{73} X_{32} U_{25} P_{51} Y_{14}= 
V_{47} X_{76} Q_{62} U_{25} P_{51} Y_{14}= 
V_{47} Q_{73} Y_{32} U_{25} P_{51} X_{14} = 
V_{47} Y_{76} Q_{62} U_{25} P_{51} X_{14} $ \\ $ =
V_{47} X_{76} P_{62} U_{25} Y_{58} Q_{84} =
V_{47} Y_{76} P_{62} U_{25} X_{58} Q_{84} = 
V_{47} P_{73} Y_{32} U_{25} X_{58} Q_{84} = 
V_{47} P_{73} X_{32} U_{25} Y_{58} Q_{84} $  \end{tabular} 

& $p_2 p_3 p_4 p_5 p_6 p_7 s$ & $0$ &$0$&$0$ & $\overline{t}_1^4 \overline{t}_2^2$\\
&$$ & $$ & $$ &$$&$$ & $$\\
& \begin{tabular}{l}$
U_{47} X_{76} P_{62} V_{25} Y_{58} P_{84} = 
U_{47} Y_{76} P_{62} V_{25} X_{58} P_{84} = 
U_{47} P_{73} Y_{32} V_{25} X_{58} P_{84} = 
U_{47} P_{73} X_{32} V_{25} Y_{58} P_{84} $ \\ $ = 
U_{47} X_{76} P_{62} V_{25} P_{51} Y_{14}= 
U_{47} P_{73} X_{32} V_{25} P_{51} Y_{14}= 
U_{47} Y_{76} P_{62} V_{25} P_{51} X_{14} = 
U_{47} P_{73} Y_{32} V_{25} P_{51} X_{14} $ \\ $ = 
V_{47} X_{76} P_{62} U_{25} Y_{58} P_{84} = 
V_{47} Y_{76} P_{62} U_{25} X_{58} P_{84} = 
V_{47} P_{73} Y_{32} U_{25} X_{58} P_{84} = 
V_{47} P_{73} X_{32} U_{25} Y_{58} P_{84} $ \\ $ = 
V_{47} X_{76} P_{62} U_{25} P_{51} Y_{14}= 
V_{47} P_{73} X_{32} U_{25} P_{51} Y_{14}= 
V_{47} Y_{76} P_{62} U_{25} P_{51} X_{14} = 
V_{47} P_{73} Y_{32} U_{25} P_{51} X_{14} 
$  \end{tabular}

& $p_2 p_3 p_4 p_6^2 p_7 s $ & $0$ &$0$&$-2$ & $z^{-2} \overline{t}_1^4 \overline{t}_2^2$\\
&$$ & $$ & $$ &$$&$$ & $$ \\
& \begin{tabular}{l}$
V_{47} Q_{73} X_{32} V_{25} Q_{51} Y_{14}= 
V_{47} X_{76} Q_{62} V_{25} Q_{51} Y_{14}= 
V_{47} Q_{73} Y_{32} V_{25} Q_{51} X_{14} = 
V_{47} Y_{76} Q_{62} V_{25} Q_{51} X_{14} $ \\ $ =
V_{47} Q_{73} Y_{32} V_{25} X_{58} Q_{84} = 
V_{47} Q_{73} X_{32} V_{25} Y_{58} Q_{84} = 
V_{47} X_{76}  Q_{62} V_{25} Y_{58} Q_{84} = 
V_{47} Y_{76}  Q_{62} V_{25} X_{58} Q_{84} 
$  \end{tabular}

& $p_2 p_3 p_5^2 p_7^2 s$ & $0$ &$-2$&$+2$ & $y^{-2} z^2 \overline{t}_1^4 \overline{t}_2^2$\\
&$$ & $$ & $$ &$$&$$ & $$ \\
& \begin{tabular}{l}$
V_{47} Q_{73} V_{25} X_{58} Y_{32} P_{84} = 
V_{47} Q_{73} X_{32} V_{25} Y_{58} P_{84} = 
V_{47} X_{76} Q_{62} V_{25} Y_{58} P_{84} = 
V_{47} Y_{76} Q_{62} V_{25} X_{58} P_{84} $ \\ $ =  
V_{47} X_{76} P_{62} V_{25} Q_{51} Y_{14} = 
V_{47} P_{73} X_{32} V_{25} Q_{51} Y_{14} = 
V_{47} Y_{76} P_{62} V_{25} Q_{51} X_{14} = 
V_{47} P_{73} Y_{32} V_{25} Q_{51} X_{14} $ \\ $ =  
V_{47} Q_{73} X_{32} V_{25} P_{51} Y_{14} = 
V_{47} X_{76} Q_{62} V_{25} P_{51} Y_{14} = 
V_{47} Q_{73} Y_{32} V_{25} P_{51} X_{14} = 
V_{47} Y_{76} Q_{62} V_{25} P_{51} X_{14} $ \\ $ = 
V_{47} X_{76} P_{62} V_{25} Y_{58} Q_{84} = 
V_{47} Y_{76} P_{62} V_{25} X_{58} Q_{84} = 
V_{47} P_{73} Y_{32} V_{25} X_{58} Q_{84} = 
V_{47} P_{73} X_{32} V_{25} Y_{58} Q_{84} 
$  \end{tabular}

& $p_2 p_3 p_5 p_6 p_7^2 s$ & $0$ &$-2$&$0$ & $y^{-2} \overline{t}_1^4 \overline{t}_2^2$\\
&$$ & $$ & $$ &$$&$$ & $$ \\
& \begin{tabular}{l}$
V_{47} X_{76} P_{62} V_{25} Y_{58} P_{84} = 
V_{47} Y_{76} P_{62} V_{25} X_{58} P_{84}  = 
V_{47} P_{73} Y_{32} V_{25} X_{58} P_{84} = 
V_{47} P_{73} X_{32} V_{25} Y_{58} P_{84}  $ \\ $ = 
V_{47} X_{76} P_{62} V_{25} P_{51} Y_{14}= 
V_{47} P_{73} X_{32} V_{25} P_{51} Y_{14}= 
V_{47} Y_{76} P_{62} V_{25} P_{51} X_{14} =
V_{47} P_{73} Y_{32} V_{25} P_{51} X_{14} 
$  \end{tabular}

& $p_2 p_3 p_6^2 p_7^2 s$ & $0$ &$-2$&$-2$ & $y^{-2} z^{-2} \overline{t}_1^4 \overline{t}_2^2$\\
&$
 U_{47} Q_{73} Y_{32} U_{25} Q_{51} Y_{14} =
 U_{47} Y_{76} Q_{62} U_{25} Q_{51} Y_{14} = 
 U_{47} Q_{73} Y_{32} U_{25} Y_{58} Q_{84} = 
 U_{47} Y_{76} Q_{62} U_{25} Y_{58} Q_{84} $ 
& $p_3^2 p_4^2 p_5^2 s$ & $-2$ &$+2$&$+2$ & $x^{-2}y^2 z^2 \overline{t}_1^4 \overline{t}_2^2$\\
&$$ & $$ & $$ &$$&$$ & $$ \\
& \begin{tabular}{l}$
U_{47} Q_{73} Y_{32} U_{25} Y_{58} P_{84} = 
U_{47} Y_{76} Q_{62} U_{25} Y_{58} P_{84} = 
U_{47} Y_{76} P_{62} U_{25} Q_{51} Y_{14} = 
U_{47} P_{73} Y_{32} U_{25} Q_{51} Y_{14} $ \\ $ = 
U_{47} Q_{73} Y_{32} U_{25} P_{51} Y_{14} = 
U_{47} Y_{76} Q_{62} U_{25} P_{51} Y_{14} = 
U_{47} Y_{76} P_{62} U_{25} Y_{58} Q_{84} = 
U_{47} P_{73} Y_{32} U_{25} Y_{58} Q_{84} $ \end{tabular}  
& $p_3^2 p_4^2 p_5 p_6 s$ & $-2$ &$+2$&$0$ & $x^{-2}y^2 \overline{t}_1^4 \overline{t}_2^2$\\
&$$ & $$ & $$ &$$&$$ & $$ \\
&$
U_{47} Y_{76} P_{62} U_{25} Y_{58} P_{84} = 
U_{47} P_{73} Y_{32} U_{25} Y_{58} P_{84} = 
U_{47} Y_{76} P_{62} U_{25} P_{51} Y_{14} = 
U_{47} P_{73} Y_{32} U_{25} P_{51} Y_{14} $ 
& $p_3^2 p_4^2 p_6^2 s$ & $-2$ &$+2$&$-2$ & $x^{-2}y^2z^{-2} \overline{t}_1^4 \overline{t}_2^2$\\
&$$ & $$ & $$ &$$&$$ & $$ \\
& \begin{tabular}{l}$
U_{47} Q_{73} Y_{32} V_{25} Q_{51} Y_{14} = 
U_{47} Y_{76} Q_{62} V_{25} Q_{51} Y_{14} = 
U_{47} Q_{73} Y_{32} V_{25} Y_{58} Q_{84} = 
U_{47} Y_{76} Q_{62} V_{25} Y_{58} Q_{84} $ \\ $ =
V_{47} Q_{73} Y_{32} U_{25} Q_{51} Y_{14} = 
V_{47} Y_{76} Q_{62} U_{25} Q_{51} Y_{14} = 
V_{47} Q_{73} Y_{32} U_{25} Y_{58} Q_{84} = 
V_{47} Y_{76} Q_{62} U_{25} Y_{58} Q_{84} $ \end{tabular} 
& $p_3^2 p_4 p_5^2 p_7 s$ & $-2$ &$0$&$+2$ & $x^{-2}z^2 \overline{t}_1^4 \overline{t}_2^2$\\
&$$ & $$ & $$ &$$&$$ & $$ \\
 &\begin{tabular}{l}$
U_{47} Q_{73} Y_{32} V_{25} Y_{58} P_{84} = 
U_{47} Y_{76} Q_{62} V_{25} Y_{58} P_{84} = 
U_{47} Y_{76} P_{62} V_{25} Q_{51} Y_{14} = 
U_{47} P_{73} Y_{32} V_{25} Q_{51} Y_{14} $ \\ $ = 
U_{47} Q_{73} Y_{32} V_{25} P_{51} Y_{14} = 
U_{47} Y_{76} Q_{62} V_{25} P_{51} Y_{14} = 
U_{47} Y_{76} P_{62} V_{25} Y_{58} Q_{84} = 
U_{47} P_{73} Y_{32} V_{25} Y_{58} Q_{84} $ \\ $ = 
V_{47} Q_{73} Y_{32} U_{25} Y_{58} P_{84} = 
V_{47} Y_{76} Q_{62} U_{25} Y_{58} P_{84} = 
V_{47} Y_{76} P_{62} U_{25} Q_{51} Y_{14} = 
V_{47} P_{73} Y_{32} U_{25} Q_{51} Y_{14} $ \\ $ =
V_{47} Q_{73} Y_{32} U_{25} P_{51} Y_{14} = 
V_{47} Y_{76} Q_{62} U_{25} P_{51} Y_{14} = 
V_{47} Y_{76} P_{62} U_{25} Y_{58} Q_{84} = 
V_{47} P_{73} Y_{32} U_{25} Y_{58} Q_{84} $ \end{tabular} 
& $p_3^2 p_4 p_5 p_6 p_7 s$ & $-2$ &$0$&$0$ & $x^{-2} \overline{t}_1^4 \overline{t}_2^2$\\
&$$ & $$ & $$ &$$&$$ & $$ \\
 &\begin{tabular}{l}$
U_{47} Y_{76} P_{62} V_{25} Y_{58} P_{84} = 
U_{47} P_{73} Y_{32} V_{25} Y_{58} P_{84} = 
U_{47} Y_{76} P_{62} V_{25} P_{51} Y_{14} = 
U_{47} P_{73} Y_{32} V_{25} P_{51} Y_{14} $ \\ $ = 
V_{47} Y_{76} P_{62} U_{25} Y_{58} P_{84} = 
V_{47} P_{73} Y_{32} U_{25} Y_{58} P_{84} = 
V_{47} Y_{76} P_{62} U_{25} P_{51} Y_{14} =
V_{47} P_{73} Y_{32} U_{25} P_{51} Y_{14} $ \end{tabular} 
& $p_3^2 p_4 p_6^2 p_7 s$ & $-2$ &$0$&$-2$ & $x^{-2}z^{-2} \overline{t}_1^4 \overline{t}_2^2$\\
&$$ & $$ & $$ &$$&$$ & $$ \\
&$
V_{47} Q_{73} Y_{32} V_{25} Q_{51} Y_{14} = 
V_{47} Y_{76} Q_{62} V_{25} Q_{51} Y_{14} = 
V_{47} Q_{73} Y_{32} V_{25} Y_{58} Q_{84} = 
V_{47} Y_{76} Q_{62} V_{25} Y_{58} Q_{84} $ 
& $p_3^2 p_5^2 p_7^2 s$ & $-2$ &$-2$&$+2$ & $x^{-2} y^{-2} z^2 \overline{t}_1^4 \overline{t}_2^2$\\
&$$ & $$ & $$ &$$&$$ & $$ \\
 &\begin{tabular}{l}$
V_{47} Q_{73} Y_{32} V_{25} Y_{58} P_{84} = 
V_{47} Y_{76} Q_{62} V_{25} Y_{58} P_{84} = 
V_{47} Y_{76} P_{62} V_{25} Q_{51} Y_{14} = 
V_{47} P_{73} Y_{32} V_{25} Q_{51} Y_{14} $ \\ $ = 
V_{47} Q_{73} Y_{32} V_{25} P_{51} Y_{14} = 
V_{47} Y_{76} Q_{62} V_{25} P_{51} Y_{14} = 
V_{47} Y_{76} P_{62} V_{25} Y_{58} Q_{84} = 
V_{47} P_{73} Y_{32} V_{25} Y_{58} Q_{84} $ \end{tabular} &
 $p_3^2 p_5 p_6 p_7^2 s$ & $-2$ &$-2$&$0$ & $x^{-2}y^{-2} \overline{t}_1^4 \overline{t}_2^2$\\
 &$$ & $$ & $$ &$$&$$ & $$ \\
&$
V_{47} Y_{76} P_{62} V_{25} Y_{58} P_{84} = 
V_{47} P_{73} Y_{32} V_{25} Y_{58} P_{84} = 
V_{47} Y_{76} P_{62} V_{25} P_{51} Y_{14} = 
V_{47} P_{73} Y_{32} V_{25} P_{51} Y_{14} $ 
& $p_3^2 p_6^2 p_7^2 s$ & $-2$ &$-2$&$-2$ & $x^{-2}y^{-2}z^{-2} \overline{t}_1^4 \overline{t}_2^2$\\
\hline
\end{tabular}}
\caption{
Generators of the mesonic moduli space $\mathcal{M}^{mes}_{Q^{1,1,1}/\mathbb{Z}_2}$ of the $Q^{1,1,1}/\mathbb{Z}_2$ model in terms of chiral fields and GLSM fields with the corresponding mesonic flavor symmetry charges. 
Here,  $s$ denotes the product of GLSM fields $s_1 \dots s_{10}$.
\label{tab_110}
}
\end{table}

The Hilbert series of the mesonic moduli space $\mathcal{M}^{mes}_{Q^{1,1,1}/\mathbb{Z}_2}$ 
for the $Q^{1,1,1}/\mathbb{Z}_2$ model 
is obtained using the Molien integral formula in \eref{es01a15}, and takes the following form,
\beal{es02a27} \label{es41a008}
&&g(t_a,y_s;\mathcal{M}^{mes}_{Q^{1,1,1}/ \mathbb{Z}_2})=\frac{P(t_a,y_s;\mathcal{M}^{mes}_{Q^{1,1,1}/ \mathbb{Z}_2})}{(1-y_s t_2^2 t_4^2 t_5^2) (1-y_s t_3^2 t_4^2 t_5^2) (1-y_s 
t_2^2 t_4^2 t_6^2)(1-y_s t_3^2 t_4^2 t_6^2)}\nn
\\
&&
\hspace{1cm}\times\frac{1}{ (1-y_s t_2^2 
t_5^2 t_7^2) (1-y_s t_3^2 t_5^2 t_7^2) (1-y_s t_2^2 t_6^2 
t_7^2) (1-y_s t_3^2 t_6^2 t_7^2)} ~,~
\eea
where the fugacities $t_a$ correspond to extremal GLSM fields $p_a$, and
 $y_s$ corresponds to the product of GLSM fields $s_1,\dots , s_{10}$. 
The full numerator $P(t_a,y_s;\mathcal{M}^{mes}_{Q^{1,1,1}/ \mathbb{Z}_2})$ in the Hilbert series \eref{es41a008} 
is presented in appendix \sref{app_num_02}.

We can rewrite the Hilbert series in \eref{es02a27}
in terms of characters of irreducible representations of the non-abelian part of the global symmetry, $SU(2)_x \times SU(2)_y \times SU(2)_z$.
We introduce the following fugacity map,
\beal{es02a28} \label{es41a014}
\overline{t}_1= t_4^{1/4} t_5^{1/4} t_6^{1/4} t_7^{1/4}~,~ 
\overline{t}_2= t_2^{1/2} t_3^{1/2}~,~ 
x= \frac{t_2^{1/2}}{t_3^{1/2}}~,~ 
y=\frac{t_4^{1/2}}{t_7^{1/2}}~,~
z =\frac{t_5^{1/2}}{t_6^{1/2}}~,~
\eea 
where the fugacities $x$, $y$ and $z$
correspond to the factors $SU(2)_x$, $SU(2)_y$ and $SU(2)_z$, respectively.
Using the fugacity map in \eref{es02a28} with $y_s=1$, we obtain the following character expansion of the Hilbert series in \eref{es02a27}, 
\beal{es02a29}
g(\overline{t}_1,\overline{t}_2,x,y,z;\mathcal{M}^{mes}_{Q^{1,1,1}/\mathbb{Z}_2})=\sum_{n=0}^{\infty} [2n;2n;2n] \overline{t}_1^{4n} \overline{t}_2^{2n} 
~,~
\eea
where $[n;m;k]=[n]_x [m]_y [k]_z$ are characters of irreducible representations of $SU(2)_x \times SU(2)_y \times SU(2)_z$ with highest weight given by $(m),(n),(k)$. 
The highest weight generating function corresponding to the Hilbert series in \eref{es02a29} takes the following form,
\beal{es02a29b}
h(\overline{t}_1,\overline{t}_2,\mu,\nu,\kappa;\mathcal{M}^{mes}_{Q^{1,1,1}/\mathbb{Z}_2})=\frac{1}{(1-\mu^2 \nu^2 \kappa^2 \overline{t}_1^4 \overline{t}_2^2)} 
~,~
\eea
where $\mu^m \nu^n \kappa ^l$ corresponds to  $[m]_x [n]_y [l]_z$. 

The plethystic logarithm of the Hilbert series in \eref{es02a29} is given by 
\beal{es02a30}
&&PL[g(\overline{t}_1,\overline{t}_2,x,y,z;\mathcal{M}^{mes}_{Q^{1,1,1}/\mathbb{Z}_2})]
=
[2]_x[2]_y[2]_z \overline{t}_1^4 \overline{t}_2^2
-(1+[4]_x+[4]_y+[4]_z+[2]_x[2]_y\nn\\&&\hspace{0.5cm}+[2]_x[2]_z+[2]_y[2]_z+[4]_x[4]_y+[4]_x[4]_z+[4]_y[4]_z+[4]_x[2]_y[2]_z+[2]_x[4]_y[2]_z\nn\\&&\hspace{0.5cm}+[2]_x[2]_y[4]_z)\overline{t}_1^8 \overline{t}_2^4+\dots ~,~
\eea
where the infinite expansion of the plethystic logarithm indicates that the mesonic moduli space $\mathcal{M}^{mes}_{Q^{1,1,1}/\mathbb{Z}_2}$ is not a complete intersection. 
We can extract the generators of the mesonic moduli space $\mathcal{M}^{mes}_{Q^{1,1,1}/\mathbb{Z}_2}$ from the first positive terms in the plethystic logarithm. 
The generators of the mesonic moduli space $\mathcal{M}^{mes}_{Q^{1,1,1}/\mathbb{Z}_2}$ in terms of chiral fields and GLSM fields with their mesonic flavor symmetry charges are 
summarized in \tref{tab_110}.
\\

\paragraph{Comparison between $F_{0,+-}$ and $Q^{1,1,1}/\mathbb{Z}_2$.}
The generators of the mesonic moduli space for $\mathcal{M}^{mes}_{F_{0,+-}}$ transform under the following irreducible representations 
of the mesonic flavor symmetry $SU(2)_x \times SU(2)_y \times U(1)_f$ based on \eref{es02a19}, 
\beal{es02a40}
[2]_y f^{-2} ~,~ [2]_x[2]_y ~,~ [4]_x[2]_y f^2  ~,~
\eea
which give in total 27 generators. 
In comparison, the generators of the 
mesonic moduli space $\mathcal{M}^{mes}_{Q^{1,1,1}/\mathbb{Z}_2}$ 
transform under the following irreducible representations of the mesonic flavor symmetry 
$SU(2)_x \times SU(2)_y \times SU(2)_z$ based on \eref{es02a30},
\beal{es02a41}
[2]_x[2]_y[2]_z ~,~
\eea
which give also 27 generators. We can see here that the total number of generators of the mesonic moduli space stays invariant 
between brane brick models
$F_{0,+-}$ and $Q^{1,1,1}/\mathbb{Z}_2$,
where the toric Calabi-Yau 4-folds are related under a birational transformation
as defined in \eref{es02a03}.

\begin{figure}[H]
\begin{center}
\resizebox{0.85\hsize}{!}{
\includegraphics[height=6cm]{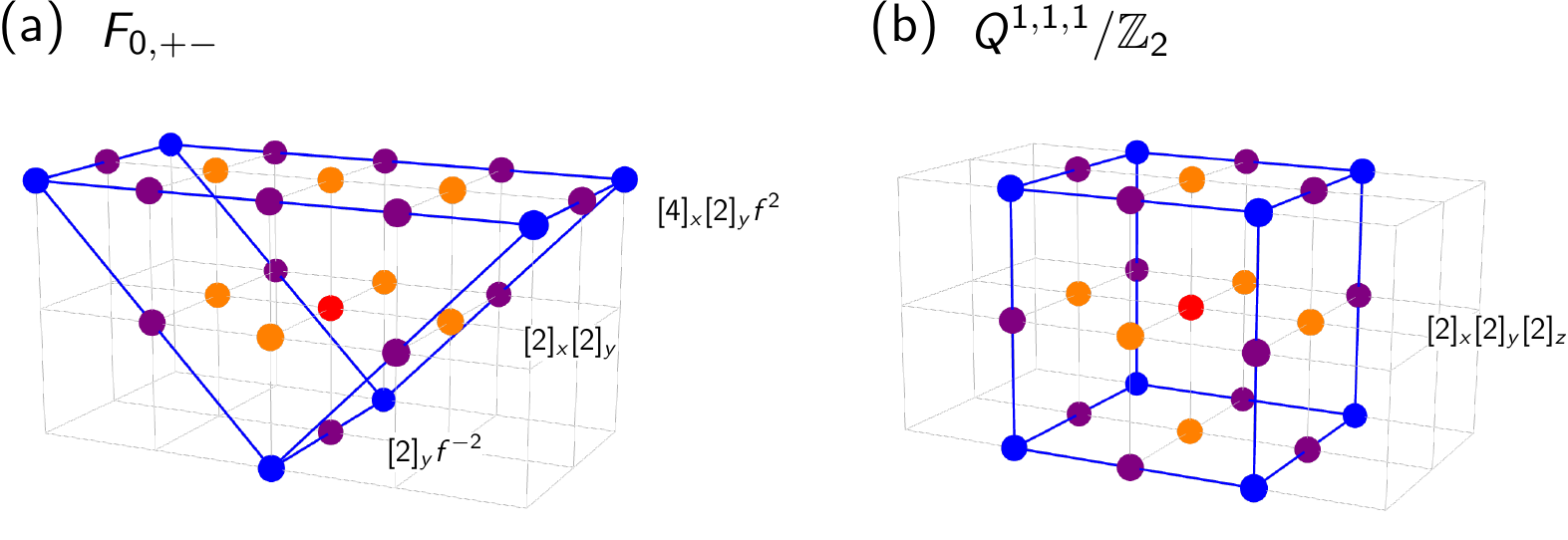} 
}
\caption{
The lattice of generators for the brane brick models corresponding to (a) $F_{0,+-}$ and (b) $Q^{1,1,1}/\mathbb{Z}_2$.
These lattice of generators form dual reflexive polytopes of the corresponding toric diagrams for the mesonic moduli spaces. 
We note here that the total number of generators of the mesonic moduli spaces is preserved under the birational transformation and the corresponding algebraic mutation. 
\label{f_ref_ex1e}}
 \end{center}
 \end{figure}

Given that the toric Calabi-Yau 4-folds
$F_{0,+-}$ and $Q^{1,1,1}/\mathbb{Z}_2$
have reflexive polytopes as toric diagrams
as shown in \fref{f_ref_ex1c},
we expect the dual polytopes to be also reflexive in $\mathbb{Z}^3$.
Using the following fugacity map, 
\beal{es02a41}
\widetilde{x} = \frac{t_1^{1/4}}{t_5^{1/4}}~,~ 
\widetilde{y} =\frac{t_2^{1/4}}{t_3^{1/4}} ~,~ 
\widetilde{f} =\frac{t_1^{1/4} t_5^{1/4}}{t_4^{1/4}} ~,~
\eea
the mesonic flavor charges under 
$SU(2)_x \times SU(2)_y \times U(1)_f$
on the generators of the mesonic moduli space $\mathcal{M}^{mes}_{F_{0,+-}}$ become $\mathbb{Z}$-valued. 
Similarly, we can introduce the following fugacity map, 
\beal{es02a42}
\widetilde{x} = \frac{t_2^{1/4}}{t_3^{1/4}}~,~ 
\widetilde{y} =\frac{t_4^{1/4}}{t_7^{1/4}}~,~
\widetilde{z}  =\frac{t_5^{1/4}}{t_6^{1/4}}~,~
\eea
which rescales the mesonic flavor charges 
under $SU(2)_x \times SU(2)_y \times SU(2)_z$
on the generators of the mesonic moduli space $\mathcal{M}^{mes}_{Q^{1,1,1}/\mathbb{Z}_2}$ such that they become $\mathbb{Z}$-valued. 
Using the $\mathbb{Z}$-values mesonic flavor charges on generators as coordinates on associated points, we can draw the lattice of generators on $\mathbb{Z}^3$
for both the $\mathcal{M}^{mes}_{F_{0,+-}}$ and $\mathcal{M}^{mes}_{Q^{1,1,1}/\mathbb{Z}_2}$ brane brick models as illustrated in \fref{f_ref_ex1e}.
We note here that the generator lattices form convex lattice polytopes in $\mathbb{Z}^3$,
which are the dual reflexive polytopes for the toric diagrams of $F_{0,+-}$ and $Q^{1,1,1}/\mathbb{Z}_2$ in \fref{f_ref_ex1c}.

We note here that the lattice of generators for the $F_{0,+-}$ and $Q^{1,1,1}/\mathbb{Z}_2$ brane brick models in \fref{f_ref_ex1e}
have the same number of points corresponding to the 27 generators of the mesonic moduli spaces. 
Moreover, given that the lattice of generators are the dual reflexive polytopes 
of the reflexive toric diagrams for $F_{0,+-}$ and $Q^{1,1,1}/\mathbb{Z}_2$ in \fref{f_ref_ex1c},
we note that the lattice of generators themselves are associated to Newton polynomials 
that are identified to each other by a birational transformation of them form 
\eref{es02a03}
as introduced for the dual toric diagrams for $F_{0,+-}$ and $Q^{1,1,1}/\mathbb{Z}_2$. 
\\

Taking a closer look at refined Hilbert series in \eref{es02a17} for $\mathcal{M}^{mes}_{F_{0,+-}}$, 
we unrefine the Hilbert series by setting the mesonic flavor fugacities $x=y=f=1$.
In terms of the remaining fugacities $\overline{t}_1$ and $\overline{t}_2$
which are associated to $U(1)_R$ charges $r_1$ and $r_2$, respectively, as defined in \tref{tab_10},
the resulting unrefined Hilbert series takes the following form,
\beal{es02a44}
g(\overline{t}_1,\overline{t}_2;\mathcal{M}^{mes}_{F_{0,+-}})
= 
\frac{(1+\overline{t}_1^4 \overline{t}_2^2) (1+22 \overline{t}_1^4 \overline{t}_2^2+\overline{t}_1^8 \overline{t}_2^4)}{(1-\overline{t}_1^4 \overline{t}_2^2)^4}
~.~
\eea
Similarly, we unrefine the Hilbert series in \eref{es02a29} for $\mathcal{M}^{mes}_{Q^{1,1,1}/\mathbb{Z}_2}$
by setting the mesonic flavor fugacities $x=y=z=1$.
This allows us to express the Hilbert series in terms of just the $U(1)_R$ charge fugacities $\overline{t}_1$ and $\overline{t}_2$ corresponding to $r_1$ and $r_2$, respectively, as summarized in \tref{tab_12}. 
The resulting unrefined Hilbert series take the following form, 
\beal{es02a46}
g(\tilde{t}_1,\tilde{t}_2;\mathcal{M}^{mes}_{Q^{1,1,1}/\mathbb{Z}_2})
=
\frac{(1+\overline{t}_1^4 \overline{t}_2^2) (1+22 \overline{t}_1^4 \overline{t}_2^2+\overline{t}_1^8 \overline{t}_2^4)}{(1-\overline{t}_1^4 \overline{t}_2^2)^4}
~.~
\eea

By comparing the unrefined Hilbert series in \eref{es02a44} and \eref{es02a46}, 
we clearly observe that the 
unrefined Hilbert series
refined only under fugacities corresponding to $U(1)_R$ charges
are identical.
From this observation, we conclude that the
Hilbert series refined only under $U(1)_R$ fugacities
stays invariant
for brane brick models
that are connected by a mass deformation which is associated to a birational transformation of 
the corresponding toric Calabi-Yau 4-folds, in this case $F_{0,+-}$ and $Q^{1,1,1}/\mathbb{Z}_2$.
\\

\subsection{Non-reflexive Case: $\mathcal{C}_{++}$ and $H_4$ \label{sec:c++_deform}}

\begin{figure}[H]
\begin{center}
\resizebox{0.85\hsize}{!}{
\includegraphics[height=6cm]{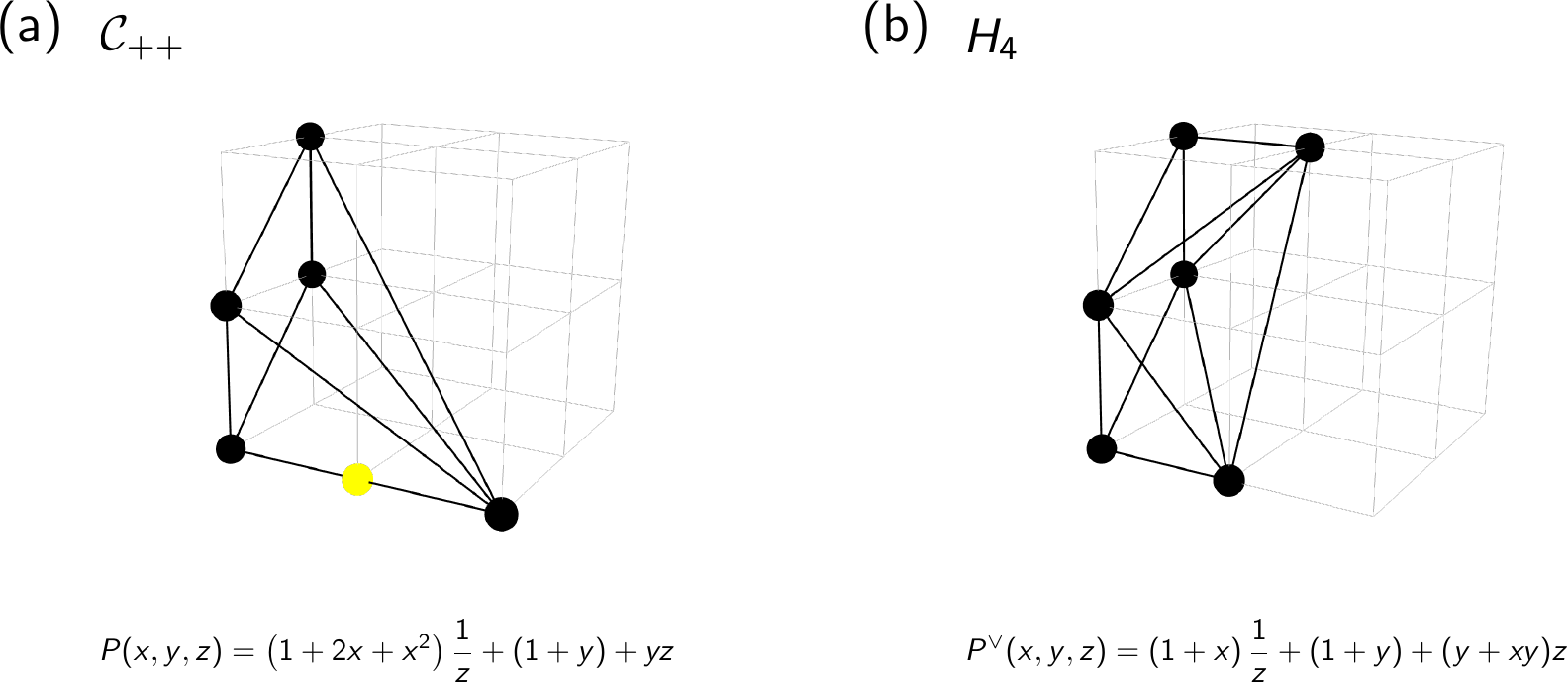} 
}
\caption{
The toric diagrams for the toric Calabi-Yau 4-folds known as (a) $\mathcal{C}_{++}$ and (b) $H_4$, 
and the corresponding Newton polynomials.
\label{f_ref_ex2}}
 \end{center}
 \end{figure}

The toric diagram $\Delta$ for the $\mathcal{C}_{++}$ model
is shown in \fref{f_ref_ex2}(a).
We note here that the toric diagram for $\mathcal{C}_{++}$  is not reflexive.
The coordinates of its points are given as follows,
\beal{es03a01}
\Delta =  \{ (2, 0, -1), ( 0, 0, -1 ), ( 0, 0, 0 ),  ( 0, 1 , 1), ( 0, 1, 0 ), (1 , 0,  -1) \} ~.~
\eea
\\

\paragraph{Algebraic Mutation.}
The Newton polynomial for the $\mathcal{C}_{++}$ model 
takes the following form based on the point coordinates in \eref{es03a01}, 
\beal{es03a02}
P (x, y, z) = \left( 1+ 2x+ x^2 \right) \frac{1}{z} + \left( 1 + y \right) + y z ~,~
\eea
where we have chosen the coefficients for the extremal points to be $1$,
and
for the point on the edge to be $2$.
Under these choices of coefficients, 
we can introduce the following birational transformation, 
\beal{es03a03}
\varphi_A ~:~
(x,y,z)
\mapsto
\left(
x, y, (1+x)z
\right)
~,~
\text{where $A(x,y) = 1+x$}
~.~
\eea
Under the above birational transformation, we obtain the following Newton polynomial,
\beal{es03a04}\label{newt-h4}
P^\vee (x, y, z) = \left(1+x \right) \frac{1}{z} + (1+ y) + (y + xy ) z 
~,~
\eea
which corresponds to the toric diagram of the $H_4$ brane brick model first studied in \cite{Franco:2017cjj}.
The toric diagram for the $H_4$ model is illustrated in \fref{f_ref_ex2}(b).
\\

\begin{figure}[H]
\begin{center}
\resizebox{0.85\hsize}{!}{
\includegraphics[height=6cm]{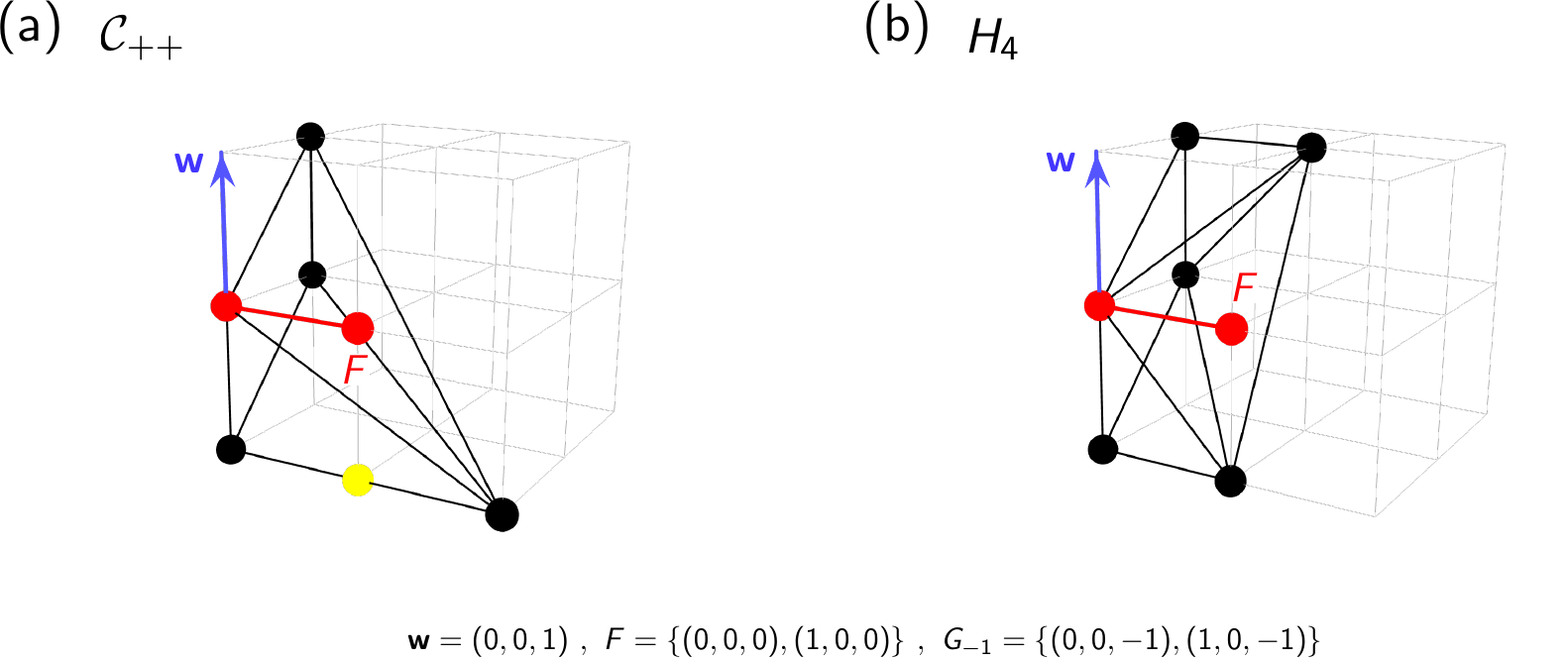} 
}
\caption{
The toric diagrams for the toric Calabi-Yau 4-folds known as (a) $\mathcal{C}_{++}$ and (b) $H_4$, 
with height measurement vector $\textbf{w}=(0,0,1)$ and factor $F$ used for combinatorial mutation.
\label{f_ref_ex2b}}
 \end{center}
 \end{figure}

\paragraph{Combinatorial Mutation.}
For combinatorial mutation, we
choose the height vector as $ \mathbf{w} = ( 0 , 0, 1 ) $. 
Accordingly, the following shows the choice for the factor $F$
and the corresponding polytopes $G_h$, 
\beal{es03a05}
    F = \{ (0 , 0 , 0 ) ~,~ ( 1  , 0 , 0 )  \}  ~,~
    G_{-1} = \{ (0 , 0 ,  -1) ~,~ (1 , 0 , -1 ) \}  ~,~
\eea
which allows us to combinatorial mutate the toric diagram for $\mathcal{C}_{++}$
in \fref{f_ref_ex2b}(a).
The resulting convex polytope in $\mathbb{Z}^3$ is given as follows,
\beal{es03a06}
\begin{tabular}{rcl}
$\underline{ h=-1 : }$ & $(0 , 0 , -1 ) ~,~ ( 1 , 0 , -1) $ & $\in G_{-1}$ ~,~\\
$\underline{ h=  0  : }$ & $( 0, 0 , 0 ) ~,~ (0 , 1 , 0) $ & $\in w_0 (\Delta)$~,~ \\
$\underline{ h=+1 : }$ & $ (0 ,1 ,  1 ) ~,~ (1 , 1 , 1 ) $ & $\in w_1 (\Delta) + F$~,~ \\
\end{tabular}
\eea
which we identify as the toric diagram of $H_4$ as illustrated in \fref{f_ref_ex2b}(b). 
\\

\begin{figure}[H]
\begin{center}
\resizebox{0.85\hsize}{!}{
\includegraphics[height=6cm]{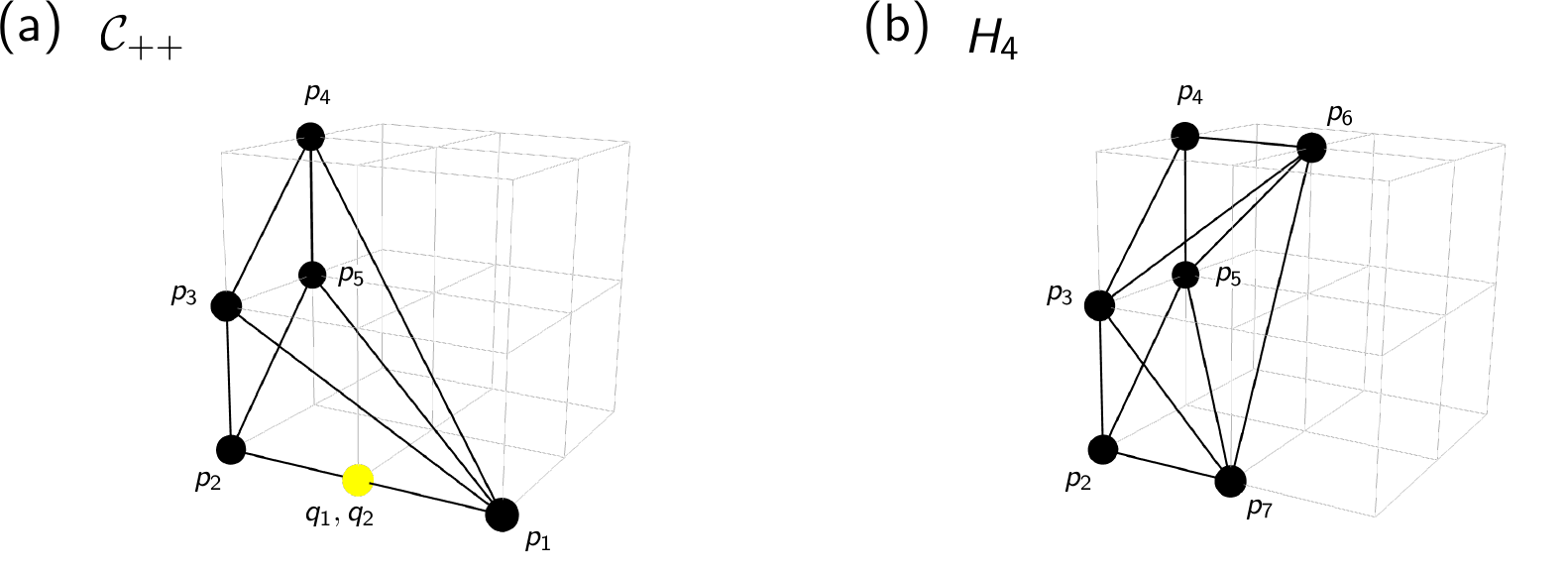} 
}
\caption{
The toric diagrams for the toric Calabi-Yau 4-folds known as (a) $\mathcal{C}_{++}$ and (b) $H_4$, 
with points labelled by the GLSM fields in the corresponding brane brick models.
\label{f_ref_ex2c}}
 \end{center}
 \end{figure}

\begin{figure}[H]
\begin{center}
\resizebox{0.8\hsize}{!}{
\includegraphics[height=6cm]{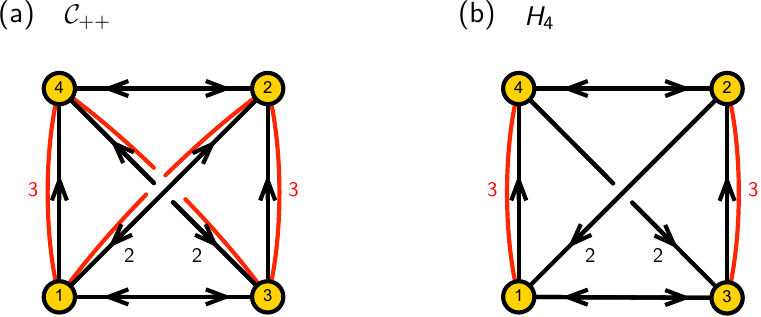} 
}
\caption{
The quiver diagrams for the brane brick models corresponding to  (a) $\mathcal{C}_{++}$ and (b) $H_4$.
\label{f_ref_ex2d}}
 \end{center}
 \end{figure}

\paragraph{Brane Brick Model for $\mathcal{C}_{++}$.}
\fref{f_ref_ex2d}(a) shows the quiver diagram for the $\mathcal{C}_{++}$ brane brick model.
The corresponding $J$- and $E$-terms \cite{Franco:2016fxm} take the following form, 
\beal{es03a10}
\resizebox{0.65\textwidth}{!}{$
\begin{array}{rrrclcrcl}
& & & J  & & &  & E  \\
 \Lambda_{12} : & \ \ \  & X_{21} \cdot Y_{12} \cdot Y_{21} &-&  Y_{21} \cdot Y_{12} \cdot X_{21} &  \ \ \ \  & Z_{13} \cdot X_{32} &-& X_{14} \cdot Z_{42} \\ 
  \Lambda_{34} : & \ \ \  & X_{43} \cdot Y_{34} \cdot Y_{43} &-& Y_{43} \cdot Y_{34} \cdot X_{43} &  \ \ \ \  &  Z_{31}  \cdot X_{14} &-& X_{32} \cdot Z_{24} \\
 \Lambda_{14} : & \ \ \  & Y_{43} \cdot X_{32} \cdot X_{21}&-& X_{43} \cdot X_{32} \cdot Y_{21} &  \ \ \ \  & Z_{13} \cdot Y_{34} &-&Y_{12} \cdot Z_{24} \\
 \Lambda_{32} : & \ \ \  & Y_{21} \cdot X_{14} \cdot X_{43} &-& X_{21} \cdot X_{14} \cdot Y_{43}  &  \ \ \ \  & Z_{31} \cdot Y_{12}&-&Y_{34} \cdot Z_{42} \\
 \Lambda_{23}^{1} : & \ \ \  & Y_{34} \cdot Y_{43} \cdot X_{32} &-& X_{32} \cdot Y_{21} \cdot Y_{12} &  \ \ \ \  & Z_{24} \cdot X_{43}&-& X_{21} \cdot Z_{13} \\ 
 \Lambda_{23}^{2} : & \ \ \  & X_{32} \cdot X_{21} \cdot Y_{12} &-& Y_{34} \cdot X_{43} \cdot X_{32} &  \ \ \ \  & Z_{24} \cdot Y_{43} &-& Y_{21} \cdot Z_{13} \\
 \Lambda_{41}^{1} : & \ \ \  & Y_{12} \cdot Y_{21} \cdot X_{14}&-& X_{14} \cdot Y_{43} \cdot Y_{34} &  \ \ \ \  & Z_{42} \cdot X_{21} &-& X_{43} \cdot Z_{31} \\
 \Lambda_{41}^{2} : & \ \ \  & X_{14} \cdot X_{43} \cdot Y_{34} &-& Y_{12} \cdot X_{21} \cdot X_{14} &  \ \ \ \  & Z_{42} \cdot Y_{21} &-& Y_{43} \cdot Z_{31} 
 \end{array}
$}
~.~
\eea
We obtain from the $J$- and $E$-terms the 
corresponding $P$-matrix using the forward algorithm \cite{Feng:2000mi, Franco:2015tna, Franco:2015tya},
\beal{es03a11}
P =
\resizebox{0.25\textwidth}{!}{$
\left(
\ba{c|ccccc|cc|cc}
 & p_1 & p_2 & p_3 & p_4 & p_5 & q_1 & q_2 & o_1 & o_2 \\
\hline
Z_{13} & 1 & 0 & 0 & 0 & 0 & 0 & 1 & 0 & 1 \\ 
 Z_{24} & 1 & 0 & 0 & 0 & 0 & 0 & 1 & 1 & 0 \\ 
 Z_{31} & 1 & 0 & 0 & 0 & 0 & 1 & 0 & 1 & 0 \\ 
 Z_{42} & 1 & 0 & 0 & 0 & 0 & 1 & 0 & 0 & 1 \\ 
 X_{14} & 0 & 1 & 0 & 0 & 0 & 0 & 1 & 0 & 0 \\ 
 X_{32} & 0 & 1 & 0 & 0 & 0 & 1 & 0 & 0 & 0 \\ 
 X_{21} & 0 & 0 & 0 & 0 & 1 & 0 & 0 & 1 & 0 \\ 
 X_{43} & 0 & 0 & 0 & 0 & 1 & 0 & 0 & 0 & 1 \\ 
 Y_{12} & 0 & 0 & 0 & 1 & 0 & 0 & 0 & 0 & 1 \\ 
 Y_{21} & 0 & 0 & 1 & 0 & 0 & 0 & 0 & 1 & 0 \\ 
 Y_{34} & 0 & 0 & 0 & 1 & 0 & 0 & 0 & 1 & 0 \\ 
 Y_{43} & 0 & 0 & 1 & 0 & 0 & 0 & 0 & 0 & 1
\ea
\right)
$}
.~
\eea
where we observe that there are $5$ GLSM fields corresponding to the 5 extremal points of the toric diagram in \fref{f_ref_ex2}(a). 
Under the GLSM fields, the $J$- and $E$-terms as well as the $D$-terms are given 
in terms $U(1)$ charges summarized in the following charge matrices,
\beal{es03a12}
Q_{JE} =
\resizebox{0.23\textwidth}{!}{$
\left(
\ba{ccccc|cc|cc}
 p_1 & p_2 & p_3 & p_4 & p_5 & q_1 & q_2 & o_1 & o_2 \\
\hline
 1 & 0 & 1 & 1 & 1 & 0 & 0 & -1 & -1 \\ 
 0 & -1 & 1 & 1 & 1 & 1 & 1 & -1 & -1
\ea
\right)
$}
~,~
Q_D =
\resizebox{0.23\textwidth}{!}{$
\left(
\ba{ccccc|cc|cc}
 p_1 & p_2 & p_3 & p_4 & p_5 & q_1 & q_2 & o_1 & o_2 \\
\hline
 0 & 1 & -1 & 0 & -1 & -1 & 0 & 0 & 1 \\ 
 0 & 0 & 0 & -1 & 0 & -1 & 0 & 1 & 0 \\ 
 0 & 0 & 0 & 1 & 0 & 1 & 0 & 0 & -1
\ea
\right)
$}
~.~
\eea
The toric diagram in \fref{f_ref_ex2c}(a) is then obtained as follows, 
\beal{es03a14}
G_t =
\resizebox{0.23\textwidth}{!}{$
\left(
\ba{ccccc|cc|cc}
 p_1 & p_2 & p_3 & p_4 & p_5 & q_1 & q_2 & o_1 & o_2 \\
\hline
 2 & 0 & 0 & 0 & 0 & 1 & 1 & 1 & 1 \\ 
 0 & 0 & 0 & 1 & 1 & 0 & 0 & 1 & 1 \\ 
 -1 & -1 & 0 & 1 & 0 & -1 & -1 & 0 & 0 \\ 
  \hline
 1 & 1 & 1 & 1 & 1 & 1 & 1 & 2 & 2
\ea
\right)
$}
~,~
\eea
where we note that we have $2$ extra GLSM fields $o_1, o_2$ \cite{Franco:2015tna} that over-parameterize the mesonic moduli space 
$\mathcal{M}^{mes}_{\mathcal{C}_{++}}$
and 
do not correspond to any points in the toric diagram in \fref{f_ref_ex2c}(a).

We observe from the $Q_{JE}$ and $Q_D$ matrices that the global symmetry of the brane brick model is enhanced to the following form, 
\beal{es03a15}
SU(2)_x \times U(1)_{f_1} \times U(1)_{f_2} \times U(1)_R ~,~
\eea
where $SU(2)_x \times U(1)_{f_1} \times U(1)_{f_2}$ is the mesonic flavor symmetry. 
The charges on the extremal GLSM fields due to the global symmetry are summarized in \tref{tab_20}.

\begin{table}[ht!] 
\centering
\begin{tabular}{|c|c|c|c|c|l|}
\hline
\; & $SU(2)_x$ & $U(1)_{f_1}$ & $U(1)_{f_2}$ & $U(1)_R$ & fugacity \\
\hline
$p_1$ & $ 0 $ & $ +2 $ & $ 0 $ & $r_1 $ & $t_1 = f_1^{2} \overline{t}_1 $ \\
$p_2$ & $ 0 $ & $ 0 $ & $ +1 $ & $r_2 $ & $t_2 =  f_2 \overline{t}_2 $ \\
$p_3$ & $ +1 $ & $ -1 $ & $ 0 $ & $r_3 $ & $t_3 = x f_1^{-1} \overline{t}_3 $ \\
$p_4$ & $ 0 $ & $ 0 $ & $ -1 $ & $r_1+ r_2 $ & $t_4 = f_2^{-1} \overline{t}_1 \overline{t}_2 $ \\
$p_5$ & $ -1 $ & $ -1 $ & $ 0 $ & $r_3 $ & $t_5 = x^{-1} f_1^{-1}\overline{t}_3 $ \\
\hline
\end{tabular}
\caption{Charges under the global symmetry of the $\mathcal{C}_{++} $ model on the extremal GLSM fields $p_a$.
Here, $U(1)_R$ charges $r_1$, $r_2$ and $r_3$ are chosen such that 
the $J$- and $E$-terms coupled to Fermi fields have an overall $U(1)_R$ charge of $2$
with $2r_1 + 2 r_2+2 r_3 =2$. \label{tab_20}}
\end{table}

The Hilbert series of the mesonic moduli space $\mathcal{M}^{mes}_{\mathcal{C}_{++}}$ can be obtained using the Molien integral formula in \eref{es01a15}.
It takes the following form,
\beal{es03a16}
g(t_a,y_q,y_o;\mathcal{M}^{mes}_{\mathcal{C}_{++}})= \frac{P(t_a,y_q,y_o;\mathcal{M}^{mes}_{\mathcal{C}_{++}})}{(1-y_q y_o t_1^2) (1-y_q y_o t_2^2 t_3^2) (1-y_o t_3 t_4) 
(1-y_o t_4 t_5) (1-y_q y_o t_2^2 t_5^2)}
~,~
\nn\\
\eea
where $t_a$ are the fugacities corresponding to extremal GLSM fields $p_a$.
Furthermore, 
the fugacities $y_q$, $y_o$ correspond to 
products of non-extremal GLSM fields $q_1 q_2$ and $o_1 o_{2}$, respectively. 
The full numerator $P(t_a,y_q,y_o;\mathcal{M}^{mes}_{\mathcal{C}_{++}})$ in \eref{es03a16} 
is presented in appendix \sref{app_num_03}.

We introduce the following fugacity map, 
\beal{es03a17}
\overline{t}_1= t_1^{1/2} ~,~ 
\overline{t}_2=  \frac{t_2^{1/2} t_4^{1/2}}{t_1^{1/4}}~,~
\overline{t}_3=  t_1^{1/4} t_3^{1/2} t_5^{1/2}~,~
x = \frac{t_3^{1/2}}{t_5^{1/2}}~,~
f_1 = t_1^{1/4}~,~
f_2 = \frac{t_1^{1/4} t_2^{1/2}}{t_4^{1/2}}
~,~
\nn\\
\eea
where
fugacities
$\overline{t}_1, \overline{t}_2, \overline{t}_3$
correspond respectively to $U(1)_R$ charges $r_1, r_2, r_3$ defined in \tref{tab_20},
fugacity $x$ corresponds to the $SU(2)_x$ factor,
and fugacities $f_1, f_2$ correspond respectively to the $U(1)_{f_1}$ and  $U(1)_{f_2}$ factors in the mesonic flavor symmetry.
By imposing the above fugacity map and by further setting $y_q=y_o=1$,
we can rewrite the Hilbert series in \eref{es03a16}, 
in terms of characters of irreducible representations of $SU(2)_x$ as follows, 
\beal{es03a18}
&&
g(\overline{t}_1,\overline{t}_2,\overline{t}_3,x,f_1,f_2;\mathcal{M}^{mes}_{\mathcal{C}_{++}}) = 
\nn\\
&&
\hspace{0.3cm}
\sum_{n_1,n_2,n_3=0}^{\infty}
[2n_1+n_2]_x f_1^{-2n_1-n_2+4n_3}f_2^{2n_1-n_2} \overline{t}_1^{n_2+2n_3} \overline{t}_2^{2n_1+n_2} \overline{t}_3^{2n_1+n_2}
\nn\\
&&\hspace{0.5cm}+\sum_{n_1,n_2,n_3=0}^{\infty}
[2n_1+n_2+1]_x f_1^{-2n_1-n_2+4n_3+1}f_2^{2n_1-n_2+1} \overline{t}_1^{n_2+2n_3+1} \overline{t}_2^{2n_1+n_2+1} \overline{t}_3^{2n_1+n_2+1}
~,~
\nn\\
\eea
where $[n]_x$ is the character of an irreducible representation of $SU(2)_x$ with highest weight given by $(n)$. 
The highest weight generating function corresponding to the Hilbert series in \eref{es03a18} takes the following form,
\beal{es03a19}
h(\overline{t}_1,\overline{t}_2,\overline{t}_3,f_1,f_2,\mu;\mathcal{M}^{mes}_{\mathcal{C}_{++}})=\frac{1+\mu f_1 f_2 \overline{t}_1 \overline{t}_2 \overline{t}_3}{(1-\mu^2 f_1^{-2} f_2^{2} \overline{t}_2^2 \overline{t}_3^2) (1-\mu f_1^{-1} f_2^{-1} \overline{t}_1 \overline{t}_2 \overline{t}_3)(1-f_1^4 \overline{t}_1^2)} 
~,~
\nn\\
\eea
where $\mu^m$ counts characters of the form $[m]_x$.

The plethystic logarithm of the Hilbert series in \eref{es03a18} is given by,
\beal{es03a20}
&&
PL[g(\overline{t}_1,\overline{t}_2,\overline{t}_3,x,f_1,f_2;\mathcal{M}^{mes}_{\mathcal{C}_{++}})]=
f_1^4 \overline{t}_1^2+[1]_x f_1^{-1} f_2^{-1} \overline{t}_1 \overline{t}_2 \overline{t}_3 
+[1]_x f_1 f_2 \overline{t}_1 \overline{t}_2 \overline{t}_3
\nn\\
&& 
\hspace{1.5cm}
+[2]_x f_1^{-2} f_2^2 \overline{t}_2^2 \overline{t}_3^2
-(\overline{t}_1^2 \overline{t}_2^2 \overline{t}_3^2
+[1]_x f_1^{-3} f_2 \overline{t}_1 \overline{t}_2^3 \overline{t}_3^3
+[2]_x f_1^2 f_2^2\overline{t}_1^2 \overline{t}_2^2 \overline{t}_3^2
\nn\\
&&
\hspace{1.5cm}
+[1]_x f_1^{-1} f_2^3\overline{t}_1 \overline{t}_2^3 \overline{t}_3^3
+f_1^{-4} f_2^4 \overline{t}_2^4 \overline{t}_3^4)+\dots
~.~
\eea
Here, the infinite expansion of the plethystic logarithm indicates that the mesonic moduli space $\mathcal{M}^{mes}_{\mathcal{C}_{++}}$ is not a complete intersection.
The generators of $\mathcal{M}^{mes}_{\mathcal{C}_{++}}$
are given by the first positive terms in the plethystic logarithm
and are summarized with their mesonic flavor symmetry charges in \tref{tab_75}.

\begin{table}[ht!]
\centering
\resizebox{0.98\textwidth}{!}{
\begin{tabular}{|c|c|c|c|c|c|c|c|}
\hline
PL term& generators  & GLSM fields & $SU(2)_{x}$ & $U(1)_{f_1}$ & $U(1)_{f_2}$  & fugacity\\
\hline\hline
\multirow{2}{*}{$+[1]_x f_1 f_2 \overline{t}_1 \overline{t}_2 \overline{t}_3$}&$Y_{34} Y_{43}= Y_{12} Y_{21}$ & $p_3 p_4 q o$ & $+1$ & $+1$ & $+1$ & $x f_1 f_2  \overline{t}_1 \overline{t}_2 \overline{t}_3$\\
&$X_{21} Y_{12}= X_{43} Y_{34}$ & $p_4 p_5 q o$ & $-1$ & $+1$ & $+1$ & $x^{-1} f_1 f_2 \overline{t}_1 \overline{t}_2 \overline{t}_3$\\
\hline
\multirow{1}{*}{$+f_1^4 \overline{t}_1^2$}&$Z_{24} Z_{42}= Z_{13} Z_{31}$ & $p_1^2 q o$ & $0$ & $+4$ & $0$ & $f_1^4 \overline{t}_1^2$\\
\hline
\multirow{2}{*}{$+[1]_x f_1^{-1} f_2^{-1}  \overline{t}_1 \overline{t}_2 \overline{t}_3$}&$X_{14} Y_{43} Z_{31}= X_{32} Y_{43} Z_{24}= X_{14} Y_{21} Z_{42}=X_{32} Y_{21} Z_{13}$ & $p_1 p_2 p_3 o$ & $+1$ & $-1$ & $-1$ & $x f_1^{-1} f_2^{-1}  \overline{t}_1 \overline{t}_2 \overline{t}_3$\\
&$X_{14} X_{21} Z_{42}=X_{14} X_{43} Z_{31}=X_{32} X_{43} Z_{24}=X_{21}X_{32}Z_{13}$ & $p_1 p_2 p_5 o$ & $-1$ & $-1$ & $-1$ & $x^{-1} f_1^{-1} f_2^{-1}   \overline{t}_1 \overline{t}_2 \overline{t}_3$\\
\hline
\multirow{3}{*}{$+[2]_x f_1^{-2} f_2^2  \overline{t}_1^2 \overline{t}_2^2 \overline{t}_3^2$}&$X_{14} Y_{43} X_{32} Y_{21} $ & $p_2^2 p_3^2 q o$ & $+2$ & $-2$ & $+2$ & $x^2 f_1^{-2} f_2^2 \overline{t}_1^2 \overline{t}_2^2 \overline{t}_3^2$\\
&$X_{14} X_{21} X_{32} Y_{43}=X_{14} X_{32} X_{43} Y_{21}$ & $p_2^2 p_3 p_5 q o$ & $0$ & $-2$ & $+2$ & $f_1^{-2} f_2^2 \overline{t}_1^2 \overline{t}_2^2 \overline{t}_3^2$\\
&$X_{14} X_{21} X_{32} X_{43}$ & $p_2^2 p_5^2 q o$ & $-2$ & $-2$ & $+2$ & $x^{-2} f_1^{-2} f_2^2 \overline{t}_1^2 \overline{t}_2^2 \overline{t}_3^2$\\
\hline
\end{tabular}}
\caption{
Generators of the mesonic moduli space $\mathcal{M}^{mes}_{\mathcal{C}_{++}}$
of the $\mathcal{C}_{++}$ brane brick model 
in terms of chiral fields and GLSM fields with the corresponding mesonic flavor symmetry charges. 
Here, $q$ and $o$ denote products of GLSM fields $q_1 q_2$ and $o_1 o_2$, respectively. 
\label{tab_75}
}.
\end{table}

\paragraph{Mass Deformation and the Brane Brick Model for $H_4$.}
As studied in \cite{Franco:2023tyf}, 
we can introduce mass terms to the $E$-terms in \eref{es03a10} of the $\mathcal{C}_{++}$ brane brick model.
These take the following form, 
\beal{es03a30}
\resizebox{0.7\textwidth}{!}{$
\begin{array}{rrrclcrcl}
& & & J   & & & E & \textcolor{blue}{+}  & \textcolor{blue}{\Delta E} \\
 \Lambda_{12} : & \ \ \  & X_{21} \cdot Y_{12} \cdot Y_{21}  &-&   Y_{21} \cdot Y_{12} \cdot X_{21}  & \ \ \ \  & \textcolor{blue}{ - Y_{12}}+ Z_{13} \cdot X_{32}  &-&  X_{14} \cdot Z_{42} \\ 
 \Lambda_{34} : & \ \ \  & X_{43} \cdot Y_{34} \cdot Y_{43}  &-&  Y_{43} \cdot Y_{34} \cdot X_{43}  & \ \ \ \  &  \textcolor{blue}{+Y_{34}}+Z_{31} \cdot X_{14}  &-&  X_{32} \cdot Z_{24}  \\
 \end{array} 
 $}
 ~.~
\eea
We note here that the original $E$-terms in \eref{es03a10}
consist of chiral fields associated to the extremal brick matchings $p_1$ and $p_2$.
The mass terms introduced in \eref{es03a30} consist of chiral fields
associated to extremal brick matching $p_4$.
When we consider the $U(1)_R$ charges on these extremal brick matchings summarized in \tref{tab_20}, 
we observe that the mass terms have a consistent $U(1)_R$ charge $r[p_4]$
corresponding to the $U(1)_R$ charge $r[p_1]+r[p_2]$ of the original $E$-terms,
\beal{es03a30b}
r[p_1] + r[p_2] = r[p_4] = r_1 + r_2 ~.~
\eea

By integrating out the mass terms in \eref{es03a30}, we obtain the $J$- and $E$-terms of a toric phase corresponding to the $H_4$ brane brick model \cite{Franco:2017cjj}.
The $J$- and $E$-terms take the following form, 
\beal{es03a31}
\resizebox{0.8\textwidth}{!}{$
\begin{array}{rrrclcrcl}
& & & J  & & &  & E  \\
\Lambda_{14} : & \ \ \  & X_{43} \cdot X_{32} \cdot Y_{21}  &-&  Y_{43} \cdot X_{32} \cdot X_{21}   & \ \ \ \  & Z_{13} \cdot Z_{31}\cdot X_{14}  &-&  X_{14} \cdot Z_{42}\cdot Z_{24}   \\
 \Lambda_{32} : & \ \ \  & X_{21} \cdot X_{14} \cdot Y_{43}  &-&  Y_{21} \cdot X_{14} \cdot X_{43}   & \ \ \ \  & Z_{31} \cdot Z_{13} \cdot X_{32} &-& X_{32}\cdot Z_{24} \cdot Z_{42}   \\ 
 \Lambda_{23}^{1} : & \ \ \  & Z_{31} \cdot X_{14} \cdot Y_{43} \cdot X_{32}  &-&  X_{32} \cdot Y_{21}\cdot X_{14} \cdot Z_{42}   & \ \ \ \  & Z_{24} \cdot X_{43}  &-&  X_{21} \cdot Z_{13}   \\ 
 \Lambda_{23}^{2} : & \ \ \  & Z_{31} \cdot X_{14} \cdot X_{43} \cdot X_{32}  &-&  X_{32} \cdot X_{21} \cdot X_{14} \cdot Z_{42}    & \ \ \ \  & Z_{24} \cdot Y_{43}  &-&  Y_{21} \cdot Z_{13}   \\ 
 \Lambda_{41}^{1} : & \ \ \  & X_{14} \cdot Y_{43} \cdot X_{32} \cdot Z_{24}  &-&  Z_{13} \cdot X_{32} \cdot Y_{21} \cdot X_{14}   & \ \ \ \  & Z_{42} \cdot X_{21}  &-&  X_{43} \cdot Z_{31}   \\
 \Lambda_{41}^{2} : & \ \ \  & X_{14} \cdot X_{43} \cdot X_{32} \cdot Z_{24}  &-&  Z_{13} \cdot X_{32} \cdot X_{21} \cdot X_{14}   & \ \ \ \  & Z_{42} \cdot Y_{21}  &-&  Y_{43} \cdot Z_{31}   \\
 \end{array} 
$}
~.~
\eea

The $P$-matrix for the $H_4$ model takes the following form,
\beal{es03a41}
P =
\resizebox{0.25\textwidth}{!}{$
\left(
\ba{c|cccccc|cc}
 & p_2 & p_3 & p_4 & p_5 & p_6 & p_7 & o_1 & o_2 \\
\hline
 Z_{13} & 0 & 0 & 0 & 0 & 0 & 1 & 0 & 1 \\ 
 Z_{24} & 0 & 0 & 0 & 0 & 0 & 1 & 1 & 0 \\ 
 Z_{31} & 0 & 0 & 0 & 0 & 1 & 0 & 1 & 0 \\ 
 Z_{42} & 0 & 0 & 0 & 0 & 1 & 0 & 0 & 1 \\ 
 X_{14} & 0 & 0 & 1 & 0 & 0 & 0 & 0 & 0 \\ 
 X_{32} & 1 & 0 & 0 & 0 & 0 & 0 & 0 & 0 \\ 
 X_{21} & 0 & 0 & 0 & 1 & 0 & 0 & 1 & 0 \\ 
 X_{43} & 0 & 0 & 0 & 1 & 0 & 0 & 0 & 1 \\ 
 Y_{21} & 0 & 1 & 0 & 0 & 0 & 0 & 1 & 0 \\ 
 Y_{43} & 0 & 1 & 0 & 0 & 0 & 0 & 0 & 1
\ea
\right)
$}
~,~
\eea
and the $U(1)$ charges under the $J$- and $E$-terms in \eref{es03a31}
as well as the $D$-terms are given by the following charge matrices, 
\beal{es03a42}
Q_{JE} =
\resizebox{0.22\textwidth}{!}{$
\left(
\ba{cccccc|cc}
 p_2 & p_3 & p_4 & p_5 & p_6 & p_7 & o_1 & o_2 \\
\hline
0 & 1 & 0 & 1 & 1 & 1 & -1 & -1 
\ea
\right)
$}
~,~
Q_{D} =
\resizebox{0.22\textwidth}{!}{$
\left(
\ba{cccccc|cc}
 p_2 & p_3 & p_4 & p_5 & p_6 & p_7 & o_1 & o_2 \\
\hline
 0 & 0 & 1 & 0 & 0 & 1 & -1 & 0 \\ 
 -1 & 0 & 0 & 0 & -1 & 0 & 1 & 0 \\ 
 1 & 0 & 0 & 0 & 1 & 0 & 0 & -1
\ea
\right)
$}
.~
\eea
Using these charge matrices, we obtain the toric diagram for the $H_4$ brane brick model
given by, 
\beal{es03a43}
G_t =
\resizebox{0.22\textwidth}{!}{$
\left(
\ba{cccccc|cc}
 p_2 & p_3 & p_4 & p_5 & p_6 & p_7 & o_1 & o_2 \\
\hline
 0 & 0 & 0 & 0 & 1 & 1 & 1 & 1 \\ 
 0 & 0 & 1 & 1 & 1 & 0 & 1 & 1 \\ 
 -1 & 0 & 1 & 0 & 1 & -1 & 0 & 0 \\ 
  \hline
 1 & 1 & 1 & 1 & 1 & 1 & 2 & 2
\ea
\right)
$}
.~
\eea
where we see that the $H_4$ model has $6$ extremal GLSM fields corresponding to the $6$ extremal points in the toric diagram shown in \fref{f_ref_ex2c}(b).
We also note that there are two extra GLSM fields $o_1, o_2$ that over-parameterize the mesonic moduli space $\mathcal{M}^{mes}_{H_4}$
and do not have corresponding points in the toric diagram for $H_4$. 

Based on the $Q_{JE}$ and $Q_D$ matrices, we can see that the $H_4$ model has an enhanced
global symmetry of the form, 
\beal{es03a46}
SU(2)_x \times U(1)_{f_1} \times U(1)_{f_2} \times U(1)_R ~,~
\eea
where $SU(2)_x \times U(1)_{f_1} \times U(1)_{f_2}$ is the enhanced mesonic flavor symmetry of the global symmetry. 
\tref{tab_22} summarizes how the extremal GLSM fields $p_a$ are charged under the global symmetry of the $H_4$ model.

\begin{table}[ht!]
\centering
\begin{tabular}{|c|c|c|c|c|l|}
\hline
\; & $SU(2)_x$ & $U(1)_{f_1}$ & $U(1)_{f_2}$ & $U(1)_R$ & fugacity \\
\hline
$p_2$ & $ 0 $ & $ +2 $ & $ 0 $ & $ r_2 $ & $t_2 = f_1^2 \overline{t}_2 $ \\
$p_3$ & $ +1 $ & $ 0 $ & $ 0 $ & $ r_3 $ & $t_3 = x \overline{t}_3 $ \\
$p_4$ & $ 0 $ & $ -1 $ & $ 0 $ & $ r_2 $ & $t_4 = f_1^{-1} \overline{t}_2 $ \\
$p_5$ & $ -1 $ & $ 0 $ & $ 0 $ & $  r_3 $ & $t_5 = x^{-1} \overline{t}_3 $ \\
$p_6$ & $ 0 $ & $ -1 $ & $ +1 $ & $ r_1 $ & $t_6 = f_1^{-1} f_2 \overline{t}_1 $ \\
$p_7$ & $ 0 $ & $ 0 $ & $ -1 $ & $ r_1 $ & $t_7 = f_2^{-1} \overline{t}_1$ \\
\hline
\end{tabular}
\caption{Charges under the global symmetry of the $H_4$ model on the extremal GLSM fields $p_a$.
Here, $U(1)_R$ charges $r_1$, $r_2$ and $r_3$ are chosen such that 
the $J$- and $E$-terms coupled to Fermi fields have an overall $U(1)_R$ charge of $2$
with $2r_1+2r_2+2r_3 =2$. \label{tab_22}}
\end{table}

The Hilbert series of the mesonic moduli space $\mathcal{M}^{mes}_{H_4}$ takes the following form,
\beal{es03a47}
&&
g(t_a,y_o;\mathcal{M}^{mes}_{H_4})= \frac{P(t_a,y_o;\mathcal{M}^{mes}_{H_4})}{(1-y_o t_2 t_3^2 t_4) (1-y_o t_2 t_4 t_5^2) (1-y_o t_3 t_4 
t_6) (1-y_o t_4 t_5 t_6) }\nn\\&&\hspace{4cm}\times
\frac{1}{(1-y_o t_2 
t_5 t_7) (1-y_o t_6 t_7)(1-y_o t_2 t_3 t_7)}
~,~
\eea
where $t_a$ are the fugacities corresponding to extremal GLSM fields $p_a$, whereas the
fugacity $y_o$ counts the product of extra GLSM fields $o_1 o_2$. 
The numerator $P(t_a,y_o;\mathcal{M}^{mes}_{H_4})$ for the Hilbert series \eref{es03a47} is presented in appendix \sref{app_num_04}.

We can rewrite the Hilbert series in \eref{es03a47}
in terms of characters of irreducible representations of $SU(2)_x$ in the mesonic flavor symmetry.
In order to do so, we introduce the following fugacity map,
\beal{es03a48}
&
\overline{t}_1=\frac{t_2^{1/6} t_6^{1/2} t_7^{1/2}}{t_4^{1/6}}~,~ 
\overline{t}_2=(t_2 t_4^2)^{1/3} ~,~ 
\overline{t}_3=(t_3 t_5)^{1/2} ~,~  
&
\nn\\
&
x= \frac{t_3^{1/2}}{t_5^{1/2}}~,~ 
f_1= \frac{t_2^{1/3}}{t_4^{1/3}}~,~ 
f_2=  \frac{t_2^{1/6} t_6^{1/2}}{t_4^{1/6} t_7^{1/2}}
~,~
&
\eea
where the fugacity $x$ corresponds to $SU(2)_x$,
and the fugacities $f_1$ and $f_2$ correspond to $U(1)_{f_1}$ and $U(1)_{f_1}$ in the mesonic flavor symmetry, respectively. 
Using the fugacity map in \eref{es03a48} with $y_o=1$, we obtain the following character expansion of the Hilbert series in \eref{es03a47}, 
\beal{es03a49}
&&
g(\overline{t}_1,\overline{t}_2,\overline{t}_3,x,f_1,f_2;\mathcal{M}^{mes}_{H_4}) = 
 \sum_{n_1,n_2,n_3=0}^{\infty} [n_1+2n_2]_x f_1^{2n_1+n2-n3}f_2^{-n_1}
\overline{t}_1^{n_1+2n_3} \overline{t}_2^{n_1+2n_2} \overline{t}_3^{n_1+2n_2} \nn\\
&&
\hspace{0.5cm}
+\sum_{n_1,n_2,n_3=0}^{\infty} [1+n_1+2n_2]_x 
f_1^{-2-2n_1+n_2-n_3}f_2^{1+n_1} \overline{t}_1^{n_1+2n_3+1} \overline{t}_2^{n_1+2n_2+1} \overline{t}_3^{n_1+2n_2+1}  
~,~
\eea
where $[n]_x$ is the character of the irreducible representation of $SU(2)_x $ with highest weight given by $(n)$. 
The corresponding highest weight generating function takes the following form,
\beal{es03a50}
&&
h(\overline{t}_1,\overline{t}_2,\overline{t}_3,\mu,f_1,f_2;\mathcal{M}^{mes}_{H_4})
=
\nn\\
&&
\hspace{1.5cm}
\frac{1- \mu ^2 \overline{t}_1^2 \overline{t}_2^2 \overline{t}_3^2}{(1-f_1^{-1} \overline{t}_1^2)(1-\mu f_1^2 f_2^{-1} \overline{t}_1 \overline{t}_2 \overline{t}_3) (1-\mu f_1^{-2} f_2  \overline{t}_1 \overline{t}_2 \overline{t}_3)(1-\mu^2 f_1 \overline{t}_2^2 \overline{t}_3^2)} 
~,~
\eea
where $\mu^m$ counts characters of the form $[m]_x$. 

The plethystic logarithm of the Hilbert series \eref{es03a49} is given as follows,
\beal{es03a51}
&&
PL[g(\overline{t}_1,\overline{t}_2,\overline{t}_3,x,f_1,f_2; \mathcal{M}^{mes}_{H_4} )] = 
f_1^{-1} \overline{t}_1^2+([1]_x f_1^2 f_2^{-1} + [1]_x f_1^{-2} f_2 )\overline{t}_1 \overline{t}_2 \overline{t}_3
\nn\\
&&
\hspace{1.5cm}
+[2]_x f_1 \overline{t}_2^2 \overline{t}_3^2 
-[2]_x \overline{t}_1^2 \overline{t}_2^2 \overline{t}_3^2
-([1]_x f_1^{-1} f_2 + [1]_x f_1^3 f_2^{-1} )\overline{t}_1 \overline{t}_2^3 \overline{t}_3^3 -f_1^2 \overline{t}_2^4 \overline{t}_3^4 +\dots 
~.~ 
\nn\\
\eea
We note here that the mesonic moduli space $\mathcal{M}^{mes}_{H_4}$ is not a complete intersection.
We can extract the generators for $\mathcal{M}^{mes}_{H_4}$ from the first positive terms in the plethystic logarithm. 
All generators for $\mathcal{M}^{mes}_{H_4}$ are summarized with their mesonic flavor symmetry charges in \tref{tab_85}.
\\

\begin{table}[ht]
\centering
\resizebox{0.95\textwidth}{!}{
\begin{tabular}{|c|c|c|c|c|c|c|c|}
\hline
PL term&generators  & GLSM fields & $SU(2)_{x}$ & $U(1)_{f_1}$ & $U(1)_{f_2}$  & fugacity\\
\hline
\hline
\multirow{1}{*}{$+f_1^{-1} \overline{t}_1^2$}&$Z_{13} Z_{31}=Z_{24} Z_{42}$ & $p_6 p_7 o$ & $0$ & $-1$ & $0$ & $f_1^{-1} \overline{t}_1^2$\\
\hline
\multirow{2}{*}{$+[1]_x f_1^{-2} f_2 \overline{t}_1 \overline{t}_2 \overline{t}_3$}&$X_{14} Z_{42} Y_{21}=X_{14} Y_{43} Z_{31}$ & $p_3 p_4 p_6 o$ & $+1$ & $-2$ & $+1$ & $x f_1^{-2} f_2 \overline{t}_1 \overline{t}_2 \overline{t}_3$\\
&$X_{14} X_{43} Z_{31}=X_{14} Z_{42} X_{21}$ & $p_4 p_5 p_6 o$ & $-1$ & $-2$ & $+1$ & $x^{-1} f_1^{-2} f_2 \overline{t}_1 \overline{t}_2 \overline{t}_3$\\
\hline
\multirow{2}{*}{$+[1]_x f_1^2 f_2^{-1} \overline{t}_1 \overline{t}_2 \overline{t}_3$}&$X_{32} Y_{21} Z_{13}=X_{32} Z_{24} Y_{43}$ & $p_2 p_3 p_7 o$ & $+1$ & $+2$ & $-1$ & $x f_1^2 f_2^{-1} \overline{t}_1 \overline{t}_2 \overline{t}_3$\\
&$X_{32} Z_{24} X_{43}=X_{32} X_{21} Z_{13}$ & $p_2 p_5 p_7 o$ & $-1$ & $+2$ & $-1$ & $x^{-1} f_1^2 f_2^{-1} \overline{t}_1 \overline{t}_2 \overline{t}_3$\\
\hline
\multirow{3}{*}{$+[2]_x f_1 \overline{t}_2^2 \overline{t}_3^2$}&$X_{14} Y_{43} X_{32} Y_{21} $ & $p_2 p_3^2 p_4 o$ & $+2$ & $0$ & $+1$ & $x^2 f_1 \overline{t}_2^2 \overline{t}_3^2$\\
&$X_{14} X_{43} X_{32} Y_{21}=X_{14} Y_{43} X_{32} X_{21}$ & $p_2 p_3 p_4 p_5 o$ & $0$ & $0$ & $+1$ & $f_1 \overline{t}_2^2 \overline{t}_3^2$\\
&$X_{14} X_{43} X_{32} X_{21}$& $p_2 p_4 p_5^2 o$ & $-2$ & $0$ & $+1$ & $x^{-2} f_1 \overline{t}_2^2 \overline{t}_3^2$\\
\hline
\end{tabular}}
\caption{
Generators of the mesonic moduli space $\mathcal{M}^{mes}_{H_4}$ of the $H_4$ brane brick model in terms of chiral fields and GLSM fields with the corresponding mesonic flavor symmetry charges. Here $o$ denotes the product of extra GLSM fields $o_1 o_2$.
\label{tab_85}
}
\end{table}

\paragraph{Comparison between $\mathcal{C}_{++}$ and $H_4$.}
Based on our observation in \eref{es03a20}, 
we note that the mesonic moduli space $\mathcal{M}^{mes}_{\mathcal{C}_{++}}$
has generators that transform under the following irreducible representations of the mesonic flavor symmetry $SU(2)_x \times U(1)_{f_1} \times U(1)_{f_2}$, 
\beal{es03a55}
f_1^4 ~,~
[1]_x f_1^{-1} f_2^{-1} ~,~
[1]_x f_1 f_2 ~,~
[2]_x f_1^{-2} f_2^{-2} ~,~
\eea
which give us in total 8 generators.
In comparison, based on the plethystic logarithm in \eref{es03a51}, 
the mesonic moduli space $\mathcal{M}^{mes}_{H_4}$ has generators that transform under the following irreducible representations of the mesonic flavor symmetry $SU(2)_x \times U(1)_{f_1} \times U(1)_{f_2}$,
\beal{es03a56}
f_1^{-1} ~,~
[1]_x f_1^2 f_2^{-1} ~,~
[1]_x f_1^{-1} f_2^2 ~,~
[2]_x f_1 ~,~
\eea
which gives us in total 8 generators. 
We see here that the total number of generators
of the mesonic moduli spaces
stays invariant between brane brick models $\mathcal{C}_{++}$ and $H_4$.
Accordingly, we confirm that when the toric Calabi-Yau 4-folds are related under a birational transformation, as in this case shown in \eref{es03a03},
then the corresponding brane brick models have mesonic moduli spaces with the same total number of generators. 

Let us also have a closer look at the refined Hilbert series in \eref{es03a18} for $\mathcal{M}^{mes}_{\mathcal{C}_{++}}$.
By setting the mesonic flavor symmetry fugacities to $f_1=f_2=x=1$, we can unrefine the Hilbert series such that it is only in terms of the $U(1)_R$ symmetry fugacities $\overline{t}_1,\overline{t}_2,\overline{t}_3$ corresponding to $U(1)_R$ charges $r_1,r_2, r_3$ in \tref{tab_20}.
The resulting unrefined Hilbert series then takes the form,
\beal{es03a60}
g(\overline{t}_1,\overline{t}_2,\overline{t}_3;\mathcal{M}^{mes}_{\mathcal{C}_{++}})= \frac{1+\overline{t}_2^2 \overline{t}_3^2-4 \overline{t}_1^2 \overline{t}_2^2 \overline{t}_3^2-4 \overline{t}_1 \overline{t}_2^3 \overline{t}_3^3+4 \overline{t}_1^3 \overline{t}_2^3 \overline{t}_3^3+4 \overline{t}_1^2 \overline{t}_2^4
   \overline{t}_3^4-\overline{t}_1^4 \overline{t}_2^4 \overline{t}_3^4-\overline{t}_1^4 \overline{t}_2^6 \overline{t}_3^6}{(1-\overline{t}_1^2) (1-\overline{t}_1 \overline{t}_2 \overline{t}_3)^4 (1-\overline{t}_2^2 \overline{t}_3^2)^2}~.~
\nn\\
\eea
A similar refinement can be done on the Hilbert series in \eref{es03a49}
for $\mathcal{M}^{mes}_{H_4}$ by setting the mesonic flavor symmetry fugacities to $f_1=f_2=x=1$.
In terms of the remaining 
$U(1)_R$ symmetry fugacities $\overline{t}_1,\overline{t}_2,\overline{t}_3$ corresponding to $U(1)_R$ charges $r_1,r_2, r_3$ in \tref{tab_22},
the unrefined Hilbert series takes the following form,
\beal{es03a61} 
g(\overline{t}_a,\overline{t}_b,\overline{t}_c; \mathcal{M}^{\text{mes}}_{H_4}) = \frac{1+\overline{t}_2^2 \overline{t}_3^2-4 \overline{t}_1^2 \overline{t}_2^2 \overline{t}_3^2-4 \overline{t}_1 \overline{t}_2^3 \overline{t}_3^3+4 \overline{t}_1^3 \overline{t}_2^3 \overline{t}_3^3+4 \overline{t}_1^2 \overline{t}_2^4
   \overline{t}_3^4-\overline{t}_1^4 \overline{t}_2^4 \overline{t}_3^4-\overline{t}_1^4 \overline{t}_2^6 \overline{t}_3^6}{(1-\overline{t}_1^2) (1-\overline{t}_1 \overline{t}_2 \overline{t}_3)^4 (1-\overline{t}_2^2 \overline{t}_3^2)^2}~.~
\nn\\
\eea

By comparing the unrefined Hilbert series in \eref{es03a60} and \eref{es03a61}, 
we observe that for brane brick models $\mathcal{C}_{++}$ and $H_4$
the Hilbert series refined only under fugacities counting $U(1)_R$ charges
are identical. 
In fact, this observation further confirms that unrefined Hilbert series only under $U(1)_R$ symmetry fugacities
remain invariant if the brane brick models correspond to toric Calabi-Yau 4-folds that are related under a birational transformation, in this case the one in \eref{es03a03} relating $\mathcal{C}_{++}$ with $H_4$.
This observation also shows that birational transformations keep unrefined Hilbert series invariant 
beyond toric Fano 3-folds and their corresponding brane brick models.
\\

\subsection{Non-reflexive Case: $\mathcal{C}_{+-} $ and $Q^{1,1,1}$
\label{sec:c+-_deform} }

\begin{figure}[H]
\begin{center}
\resizebox{0.85\hsize}{!}{
\includegraphics[height=6cm]{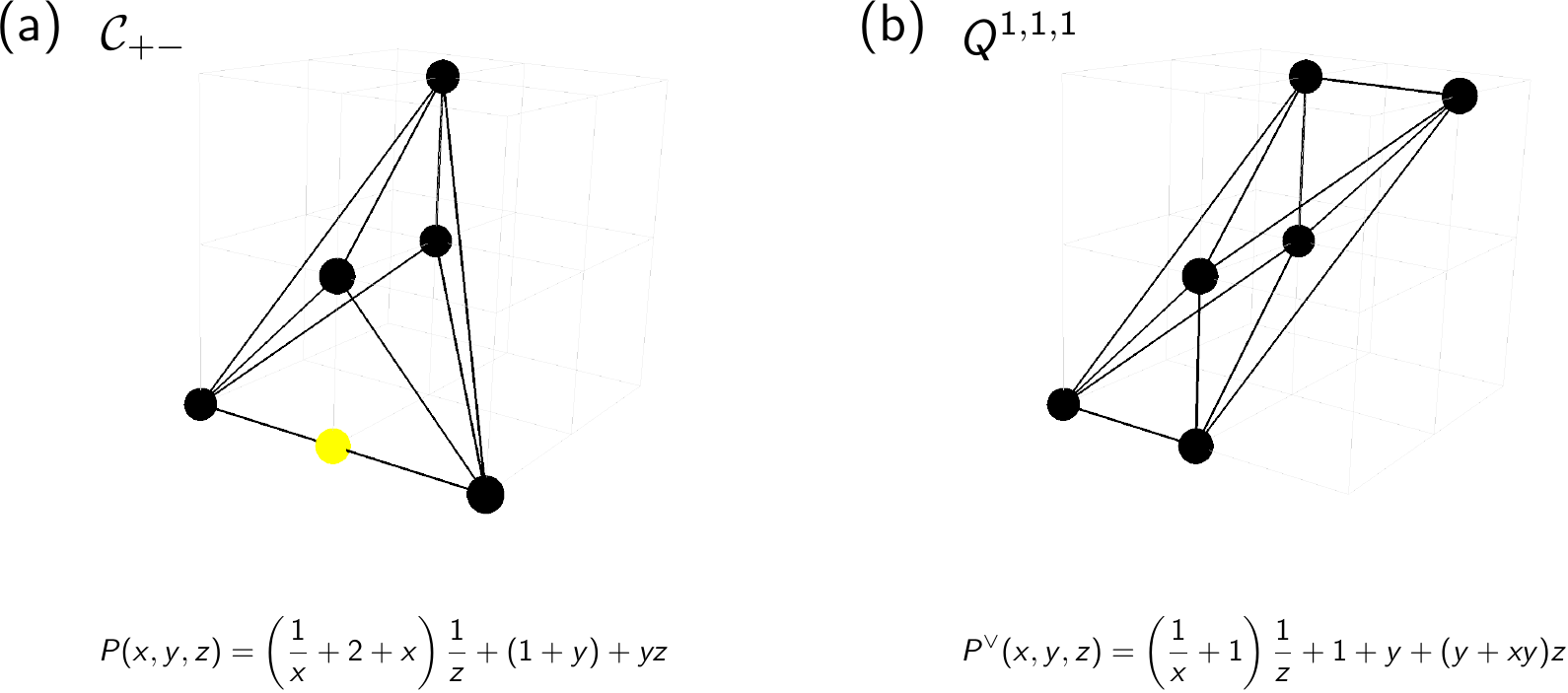} 
}
\caption{
The toric diagrams for the toric Calabi-Yau 4-folds known as (a) $\mathcal{C}_{+-}$ and (b) $Q^{1,1,1}$,
and the corresponding Newton polynomials. 
\label{f_ref_ex3}}
 \end{center}
 \end{figure}

The toric diagram $\Delta$ for the $\mathcal{C}_{+-}$ brane brick model is not reflexive and is shown in \fref{f_ref_ex3}(a).
The coordinates of points in the toric diagram are given by,
\beal{es05a01}
\Delta = \{ (1,0,-1), (-1,0,-1), (0,0,0), (0,1,1), (0,1,0), (0,0,-1) \} ~.~
\eea
\\

\paragraph{Algebraic Mutation.}
The Newton polynomial for the $\mathcal{C}_{+-}$ model takes the following form in terms based on the point coordinates shown in \eref{es05a01},
\beal{es05a02} 
P (x, y, z) = \left( \frac{1}{x}+ 2+ x \right) \frac{1}{z} + (1 + y) + y z ~,~
\eea
where we have chosen the coefficients for extremal points to be $1$, and for the point on the edge to be $2$.
Under these choices of coefficients, we have the following birational transformation,
\beal{es05a03}
\varphi_A ~:~
(x,y,z)
\mapsto
\left(
x, y, (1+x)z
\right)
~,~
\text{where $A(x,y) = 1+x$}
~,~
\eea
which maps \eref{es05a02} to a new Newton polynomial of the following form,
\beal{es05a04}
P^\vee (x,y,z) = \left( \frac{1}{x} + 1 \right) \frac{1}{z} + 1 + y + (y + x y) z ~,~
\eea
whose corresponding toric diagram is shown in \fref{f_ref_ex3}(b).
We identify the toric diagram to belong to the cone over $Q^{1,1,1}$, whose corresponding brane brick model was first studied in \cite{Franco:2015tna}.
\\

\begin{figure}[H]
\begin{center}
\resizebox{0.85\hsize}{!}{
\includegraphics[height=6cm]{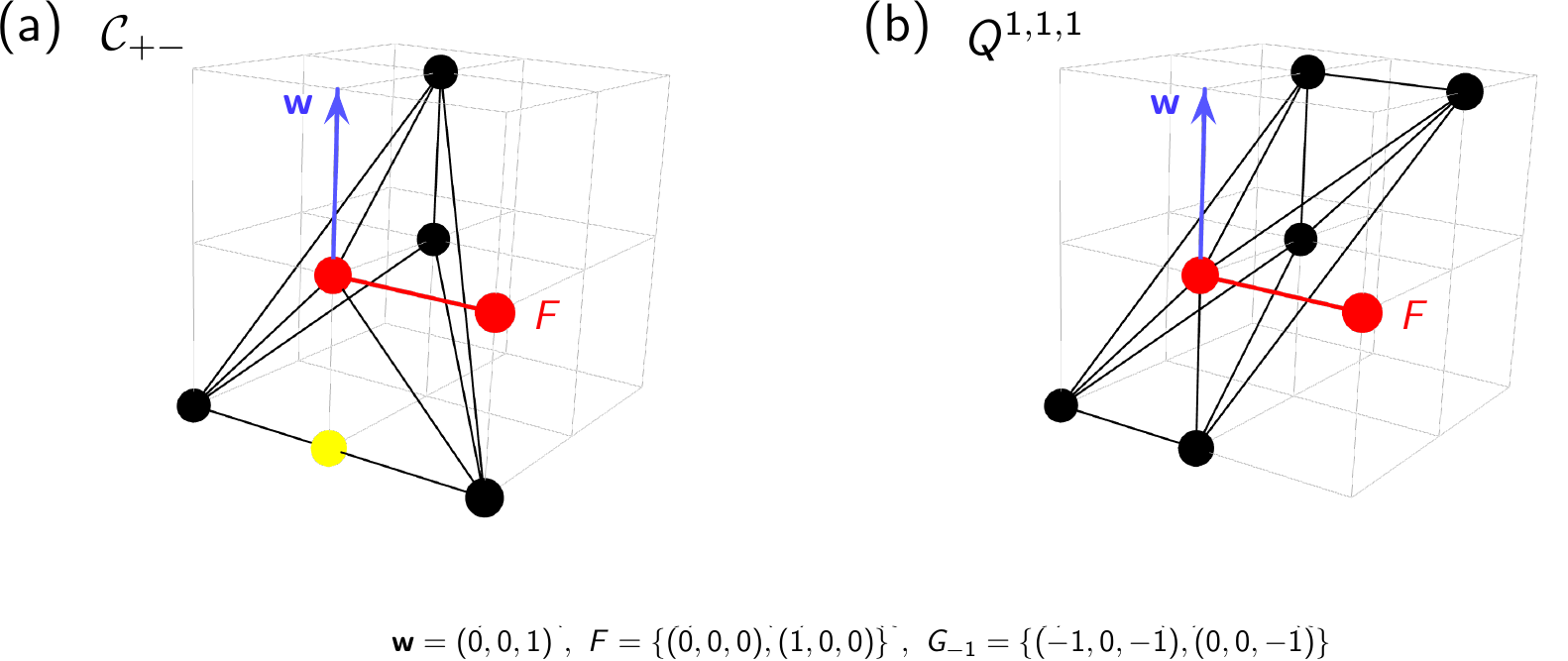} 
}
\caption{
The toric diagrams for the toric Calabi-Yau 4-folds known as (a) $\mathcal{C}_{+-}$ and (b) $Q^{1,1,1}$,
with height measurement vector $\textbf{w}=(0,0,1)$ and factor $F$ used for combinatorial mutation.
\label{f_ref_ex3b}}
 \end{center}
 \end{figure}

\paragraph{Combinatorial Mutation.}
We choose the height vector as $\mathbf{w} = (0, 0, 1) $
for combinatorial mutation of the $\mathcal{C}_{+-}$ model with the toric diagram given in \fref{f_ref_ex3b}(a).
The following shows the choice for the factor $F$ and the corresponding polytopes $G_h$,
\beal{es05a05}
F = \{ (0,0,0) ~,~ (1,0,0) \} ~,~
G_{-1} = \{ (-1,0,-1) ~,~ (0,0,-1) \} ~,~
\eea
which leads to a combinatorial mutation of the toric diagram for $\mathcal{C}_{+-}$.
The resulting convex polytope in $\mathbb{Z}^3$ is given as follows, 
\beal{es05a06}
\begin{tabular}{rcl}
$\underline{ h=-1:}$ & $(-1,0,-1) ~,~ (0,0,-1)$ & $\in G_{-1}$ ~,~ \\
$\underline{ h= 0 : }$ & $(0,0,0) ~,~ (0,1,0)$ & $\in w_0 (\Delta)$ ~,~ \\
$\underline{ h=+1:}$ & $(0,1,1) ~,~ (1,1,1)$ &  $\in w_1 (\Delta) + F$~,~ 
\end{tabular}
\eea
which we identify as the toric diagram of $Q^{1,1,1}$
as shown in \fref{f_ref_ex3b}(b).
\\

\begin{figure}[H]
\begin{center}
\resizebox{0.85\hsize}{!}{
\includegraphics[height=6cm]{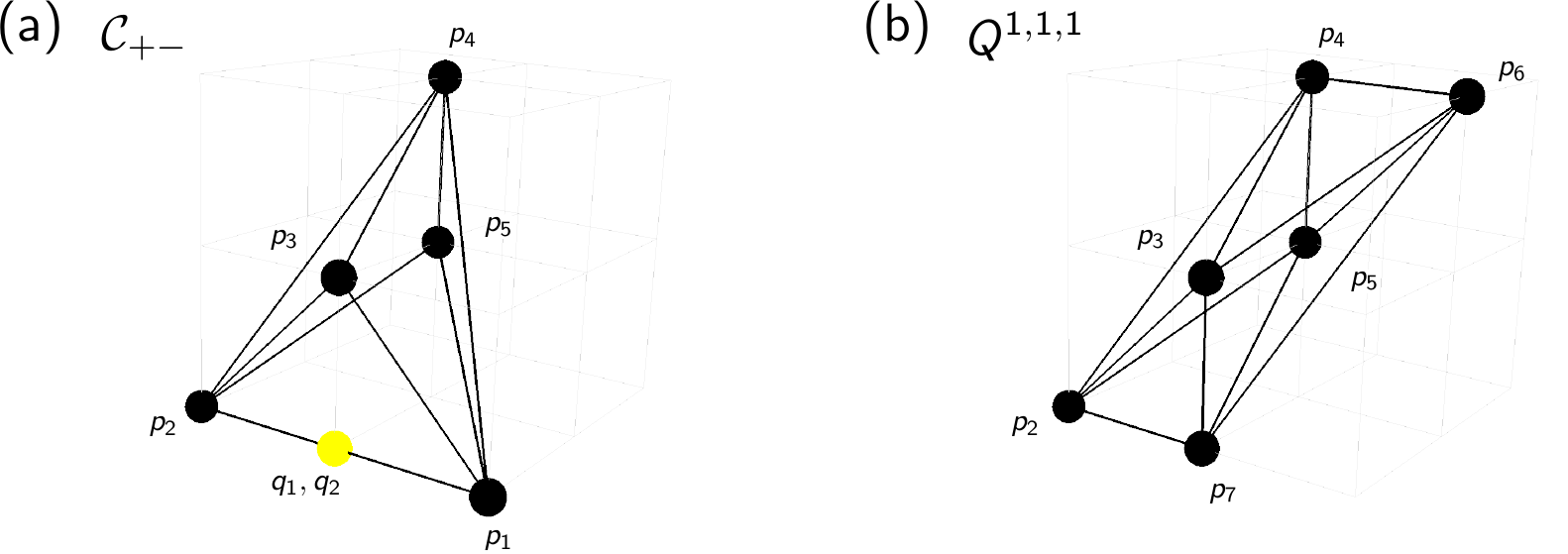} 
}
\caption{
The toric diagrams for the toric Calabi-Yau 4-folds known as (a) $\mathcal{C}_{+-}$ and (b) $Q^{1,1,1}$,
with points labelled by the GLSM fields in the corresponding brane brick models.
\label{f_ref_ex3c}}
 \end{center}
 \end{figure}

\begin{figure}[H]
\begin{center}
\resizebox{0.8\hsize}{!}{
\includegraphics[height=6cm]{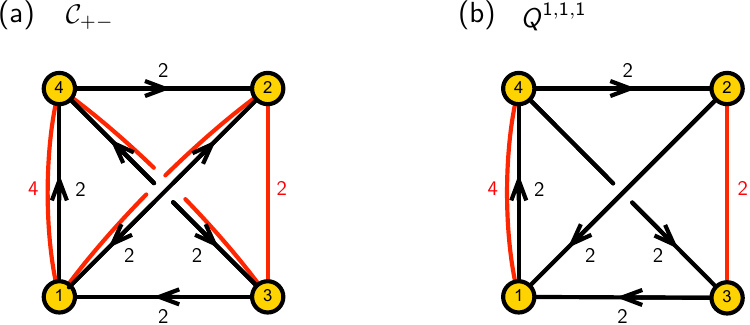} 
}
\caption{
The quiver diagrams for the brane brick models corresponding to (a) $\mathcal{C}_{+-}$ and (b) $Q^{1,1,1}$.
\label{f_ref_ex3d}}
 \end{center}
 \end{figure}

\paragraph{Brane Brick Model for $\mathcal{C}_{+-}$.}
The quiver diagram for the $\mathcal{C}_{+-}$
brane brick model is shown in \fref{f_ref_ex3d}(a)
and the corresponding 
$J$-and $E$-terms take the following form,
\beal{es05a10}
\resizebox{0.67\textwidth}{!}{$
\begin{array}{rrrclcrcl}
& & & J  & & &  & E   \\
  \Lambda_{12} : & \ & X_{21} \cdot Y_{12} \cdot Y_{21} &-& Y_{21} \cdot Y_{12} \cdot X_{21}  & \ \ \ \  & V_{14} \cdot Q_{42} &-& X_{14} \cdot P_{42}  \\ 
  \Lambda_{34} : & \ \ \  & X_{43} \cdot Y_{34} \cdot Y_{43} &-& Y_{43} \cdot Y_{34} \cdot X_{43} & \ \ \ \  & P_{31} \cdot X_{14} &-& Q_{31} \cdot V_{14}  \\ 
 \Lambda_{41}^{1} : & \ \ \  & Y_{12} \cdot Y_{21} \cdot X_{14} &-& X_{14} \cdot Y_{43} \cdot Y_{34} &  \ \ \ \  & P_{42} \cdot X_{21} &-& X_{43} \cdot P_{31}  \\ 
 \Lambda_{41}^{2} : & \ \ \  & X_{14} \cdot X_{43} \cdot Y_{34} &-& Y_{12} \cdot X_{21} \cdot X_{14} & \ \ \ \  & P_{42} \cdot Y_{21} &-& Y_{43} \cdot P_{31}  \\ 
 \Lambda_{41}^{3} : & \ \ \  & Y_{12} \cdot Y_{21} \cdot V_{14} &-& V_{14} \cdot Y_{43} \cdot Y_{34} & \ \ \ \  & Q_{42} \cdot X_{21} &-& X_{43} \cdot Q_{31} \\ 
 \Lambda_{41}^{4} : & \ \ \  & V_{14} \cdot X_{43} \cdot Y_{34} &-& Y_{12} \cdot X_{21} \cdot V_{14} & \ \ \ \  & Q_{42} \cdot Y_{21} &-& Y_{43} \cdot Q_{31}  \\ 
 \Lambda_{32}^{1} : & \ \ \  & Y_{21} \cdot X_{14} \cdot X_{43} &-& X_{21} \cdot X_{14} \cdot Y_{43} & \ \ \ \  & P_{31} \cdot Y_{12} &-& Y_{34} \cdot P_{42}  \\ 
 \Lambda_{32}^{2} : & \ \ \  & Y_{21} \cdot V_{14} \cdot X_{43} &-& X_{21} \cdot V_{14} \cdot Y_{43} & \ \ \ \  & Q_{31} \cdot Y_{12} &-& Y_{34} \cdot Q_{42} 
 \end{array} 
$}
~.~
\eea
Using the forward algorithm \cite{Feng:2000mi, Franco:2015tna, Franco:2015tya},
we obtain from the $J$- and $E$-terms the 
corresponding $P$-matrix,
\beal{es05a11}
P =
\resizebox{0.25\textwidth}{!}{$
\left(
\ba{c|ccccc|cc|cc}
 & p_1 & p_2 & p_3 & p_4 & p_5 & q_1 & q_2 & o_1 & o_2 \\
\hline
  P_{31} & 0 & 1 & 0 & 0 & 0 & 0 & 1 & 0 & 1 \\ 
  P_{42} & 0 & 1 & 0 & 0 & 0 & 0 & 1 & 1 & 0 \\ 
  Q_{31} & 1 & 0 & 0 & 0 & 0 & 0 & 1 & 0 & 1 \\ 
  Q_{42} & 1 & 0 & 0 & 0 & 0 & 0 & 1 & 1 & 0 \\ 
  V_{14} & 0 & 1 & 0 & 0 & 0 & 1 & 0 & 0 & 0 \\ 
  X_{14} & 1 & 0 & 0 & 0 & 0 & 1 & 0 & 0 & 0 \\ 
  X_{21} & 0 & 0 & 0 & 0 & 1 & 0 & 0 & 0 & 1 \\ 
  X_{43} & 0 & 0 & 0 & 0 & 1 & 0 & 0 & 1 & 0 \\ 
  Y_{12} & 0 & 0 & 0 & 1 & 0 & 0 & 0 & 1 & 0 \\ 
  Y_{21} & 0 & 0 & 1 & 0 & 0 & 0 & 0 & 0 & 1 \\ 
  Y_{34} & 0 & 0 & 0 & 1 & 0 & 0 & 0 & 0 & 1 \\ 
  Y_{43} & 0 & 0 & 1 & 0 & 0 & 0 & 0 & 1 & 0
\ea
\right)
$}
~,~
\eea
where we have $5$ GLSM fields corresponding to the 5 extremal points of the toric diagram in \fref{f_ref_ex3c}(a). 
The $U(1)$ charges on GLSM fields due to the $J$- and $E$-terms as well as the $D$-terms 
are summarized in the following charge matrices, 
\beal{es05a12}
Q_{JE} =
\resizebox{0.23\textwidth}{!}{$
\left(
\ba{ccccc|cc|cc}
 p_1 & p_2 & p_3 & p_4 & p_5 & q_1 & q_2 & o_1 & o_2 \\
\hline
  0 & 0 & 1 & 1 & 1 & 0 & 1 & -1 & -1 \\ 
  1 & 1 & 1 & 1 & 1 & -1 & 0 & -1 & -1
\ea
\right)
$}
~,~ 
Q_{D} =
\resizebox{0.23\textwidth}{!}{$
\left(
\ba{ccccc|cc|cc}
 p_1 & p_2 & p_3 & p_4 & p_5 & q_1 & q_2 & o_1 & o_2 \\
\hline
  0 & 0 & 0 & 1 & 0 & 1 & 0 & 0 & -1 \\ 
  0 & 0 & 1 & 0 & 1 & 0 & 0 & -1 & 0 \\ 
  0 & 0 & -1 & 0 & -1 & 0 & 0 & 0 & 1
\ea
\right)
$}
.~
\eea
The resulting toric diagram in \fref{f_ref_ex3c}(a) 
of the $\mathcal{C}_{+-}$ model is given by,
\beal{es05a13}
G_t =
\resizebox{0.23\textwidth}{!}{$
\left(
\ba{ccccc|cc|cc}
 p_1 & p_2 & p_3 & p_4 & p_5 & q_1 & q_2 & o_1 & o_2 \\
\hline
  1 & -1 & 0 & 0 & 0 & 0 & 0 & 0 & 0 \\ 
  0 & 0 & 0 & 1 & 1 & 0 & 0 & 1 & 1 \\ 
  -1 & -1 & 0 & 1 & 0 & -1 & -1 & 0 & 0 \\ 
   \hline
  1 & 1 & 1 & 1 & 1 & 1 & 1 & 2 & 2
\ea
\right)
$}
~,~
\eea
where we note that we have 2 extra GLSM fields $o_1, o_2$ \cite{Franco:2015tna},
which are due to an over-parameterization of the mesonic moduli space $\mathcal{M}^{mes}_{\mathcal{C}_{+-}}$
and do not correspond to any points in the toric diagram in \fref{f_ref_ex3c}(a).

From the $Q_{JE}$ and $Q_D$ charge matrices, 
we observe that the global symmetry of the brane brick model for $\mathcal{C}_{+-}$
is enhanced to the following form,
\beal{es05a15}
SU(2)_x \times SU(2)_{y} \times U(1)_{f} \times U(1)_R ~,~
\eea
where $SU(2)_x \times SU(2)_{y} \times U(1)_{f}$ is the mesonic flavor symmetry. 
The charges on the extremal GLSM fields due to the global symmetry are summarized in \tref{tab_30}.

\begin{table}[ht!]
\centering
\begin{tabular}{|c|c|c|c|c|l|}
\hline
\; & $SU(2)_x$ & $SU(2)_y$ & $U(1)_f$ & $U(1)_R$ & fugacity \\
\hline
$p_1$ & $ +1 $ & $ 0 $ & $ +1 $ & $ r_1 $ & $t_1 = x f \overline{t}_1$ \\
$p_2$ & $ -1 $ & $ 0 $ & $ +1 $ & $ r_1$ & $t_2 = x^{-1} f \overline{t}_1$ \\
$p_3$ & $ 0 $ & $ +1 $ & $ 0 $ & $ r_2 $ & $t_3 = y \overline{t}_2$ \\
$p_4$ & $ 0 $ & $ 0 $ & $ -2 $ & $ 2 r_1 $ & $t_4 = f^{-2} \overline{t}_1^2 $ \\
$p_5$ & $ 0 $ & $ -1 $ & $ 0 $ & $r_2 $ & $t_5 = y^{-1} \overline{t}_2 $ \\
\hline
\end{tabular}
\caption{Charges under the global symmetry of the $\mathcal{C}_{+-}$ model on the extremal GLSM fields $p_a$.
Here, $U(1)_R$ charges $r_1$ and $r_2$ are chosen such that 
the $J$- and $E$-terms coupled to Fermi fields have an overall $U(1)_R$ charge of $2$
with $4r_1+2r_2=2$. \label{tab_30}}
\end{table}

The Hilbert series of the mesonic moduli space $\mathcal{M}^{mes}_{\mathcal{C}_{+-}}$ takes the following form,
\beal{es05a14}
&&
g(t_a,y_q,y_o;\mathcal{M}^{mes}_{\mathcal{C}_{+-}})
= 
\frac{P(t_a,y_q,y_o;\mathcal{M}^{mes}_{\mathcal{C}_{+-}})}{(1-y_q y_o t_1^2 t_3) (1-y_q y_o t_2^2 t_3) (1-y_o t_3 t_4) 
(1-y_q y_o t_1^2 t_5) }
\nn\\
&&
\hspace{4cm}
\times
\frac{1}{(1-y_q y_o t_2^2 t_5) (1-y_o t_4 
t_5) }
~,~
\eea
where $t_a$ are the fugacities associated to the extremal GLSM fields $p_a $ 
and $y_q$,$y_o$ are fugacities that count the products of GLSM fields $q_1 q_2$ and $o_1 o_2$, respectively.
The numerator $P(t_a,y_q,y_s,y_o;\mathcal{M}^{mes}_{\mathcal{C}_{+-}})$ of the Hilbert series in \eref{es05a14} is presented in appendix \sref{app_num_05}. 

Let us consider the following fugacity map,
\beal{es05a15}
\overline{t}_1= t_1^{1/4} t_2^{1/4} t_4^{1/4} ~,~
\overline{t}_2= t_3^{1/2} t_5^{1/2}~,~
x= \frac{t_1^{1/2}}{t_2^{1/2}}~,~ 
y= \frac{t_3^{1/2}}{t_5^{1/2}}~,~ 
f= \frac{t_1^{1/4} t_2^{1/4}}{t_4^{1/4}}
~,~
\eea
where fugacities $\overline{t}_1,\overline{t}_2$
correspond respectively to $U(1)_R$ charges $r_1,r_2$ defined in \tref{tab_30},
fugacities $x,y$ correspond respectively to mesonic flavor symmetry factors $SU(2)_x$ and $SU(2)_y$,
and fugacity $f$ corresponds to mesonic flavor symmetry factor $U(1)_f$.
Using the above fugacity maps
and setting $y_q=y_o=1$, 
we can express the Hilbert series in \eref{es05a14}
in terms of characters of irreducible representations of $SU(2)_x \times SU(2)_y$ as follows, 
\beal{es05a16}
g(\overline{t}_1,\overline{t}_2,x,y,f;\mathcal{M}^{mes}_{\mathcal{C}_{+-}})=\sum_{n_2=0}^{\infty} \sum_{n_1=0}^{\infty} [2 n_1;n_1+n_2]f^{2n_1- 2 n_2} \overline{t}_1^{2n_1+2n_2} \overline{t}_2^{n_1+n_2}~,~
\eea 
where
$[n;m]=[n]_x [m]_y$ are the characters of irreducible representation of $SU(2)_x \times SU(2)_y$ 
with highest weight $(n),(m)$. 
The corresponding highest weight generating function \cite{Hanany:2014dia} is given by,
\beal{es05a17}
h(\overline{t}_1,\overline{t}_2,\mu,\nu,f;\mathcal{M}^{mes}_{\mathcal{C}_{+-}})=\frac{1}{(1-\nu f^{-2} \overline{t}_1^2 \overline{t}_2)(1-\mu^2 \nu f^2 \overline{t}_1^2 \overline{t}_2)} 
~,~
\eea
where $\mu^m \nu^n$ counts characters of the form $[m]_x [n]_y$.

The plethystic logarithm of the refined Hilbert series in \eref{es05a16} is as follows,
\beal{es05a18}
PL[g(\overline{t}_2,\overline{t}_2,x,y,f;\mathcal{M}^{mes}_{\mathcal{C}_{+-}})] &=& ([1]_y f^{-2} +[2]_x [1]_y f^2) \overline{t}_1^2 \overline{t}_2 -([2]_x+[2]_x f^4
\nn\\
&&
+[2]_y f^4)\overline{t}_1^4 \overline{t}_2^2+\dots ~,~
\eea
where we can see that the mesonic moduli space $\mathcal{M}^{mes}_{\mathcal{C}_{+-}}$
is not a complete intersection.
We can extract the generators for $\mathcal{M}^{mes}_{\mathcal{C}_{+-}}$ from the first positive terms in the plethystic logarithm. 
The generators are listed in \tref{tab_90}
with their corresponding mesonic flavor symmetry charges.
\\

\begin{table}[ht!]
\centering
\resizebox{0.95\textwidth}{!}{
\begin{tabular}{|c|c|c|c|c|c|c|c|}
\hline
PL term&generators  & GLSM fields & $SU(2)_{x}$ & $SU(2)_{y}$ & $U(1)_{f}$  & fugacity\\
\hline\hline
\multirow{2}{*}{$+[1]_y f^{-2} \overline{t}_1^2 \overline{t}_2$}&$Y_{12} Y_{21}=Y_{34} Y_{43}$ & $ p_3 p_4 o  $ & $0$ &$+1$&$-2$& $y f^{-2}\overline{t}_1^2 \overline{t}_2$\\
&$Y_{12} X_{21}=Y_{34} X_{43} $ & $ p_4 p_5 o $ & $0$ &$-1$&$-2$ & $y^{-1} f^{-2}\overline{t}_1^2 \overline{t}_2$\\
\hline
\multirow{6}{*}{$+[2]_x [1]_y f^2 \overline{t}_1^2 \overline{t}_2$}&$Q_{31} X_{14} Y_{43}=X_{14} Q_{42} Y_{21}$ & $ p_1^2 p_3 q o $ & $+2$ &$+1$&$2$ & $x^2 y f^2 \overline{t}_1^2 \overline{t}_2$\\
&$Q_{31} X_{14} X_{43}=X_{14} Q_{42} X_{21}$ & $ p_1^2 p_5 p_7 q o $ & $+2$ &$-1$&$2$ & $x^2 y^{-1} f^2 \overline{t}_1^2 \overline{t}_2$\\
&$X_{14} P_{42} Y_{21}=V_{14} Q_{42} Y_{21} =X_{14} Y_{43} P_{31}=V_{14} Y_{43} Q_{31} $
 & $p_1 p_2 p_3 q o  $ & $0$ &$+1$&$2$ & $y f^2 \overline{t}_1^2 \overline{t}_2$\\
&$ X_{14} P_{42} X_{21}=V_{14} Q_{42} X_{21} =X_{14} X_{43} P_{31}=V_{14} X_{43} Q_{31} $
 & $p_1 p_2 p_5 q o  $ & $0$ &$-1$&$2$ & $y^{-1} f^2 \overline{t}_1^2 \overline{t}_2$\\
&$P_{31} V_{14} Y_{43}=V_{14} P_{42} Y_{21}$ & $p_2^2 p_3 q o  $ & $-2$ &$+1$&$2$ & $x^{-2} y f^2 \overline{t}_1^2 \overline{t}_2$\\
&$P_{31} V_{14} X_{43}=V_{14} P_{42} X_{21}$ & $p_2^2 p_5 q o  $ & $-2$ &$-1$&$2$ & $x^{-2} y^{-1} f^2 \overline{t}_1^2 \overline{t}_2$\\
\hline
\end{tabular}}
\caption{
Generators of the mesonic moduli space $\mathcal{M}^{mes}_{\mathcal{C}_{+-}}$
of the $\mathcal{C}_{+-}$ brane brick model
in terms of chiral fields and GLSM fields with corresponding mesonic flavor symmetry charges. 
Here, $q$ and $o$ denote products of GLSM fields $q_1 q_2$ and $o_1 o_2$, respectively.
\label{tab_90}
}
\end{table}

\paragraph{Mass Deformation and the Brane Brick Model for $Q^{1,1,1}$.}
We can introduce the following mass terms to the $E$-terms of the $\mathcal{C}_{+-}$ brane brick model as first studied in \cite{Franco:2016fxm},
\beal{es05a20}
\resizebox{0.65\textwidth}{!}{$
\begin{array}{rrrclcrcl}
& & & J   & & & E & \textcolor{blue}{+}  & \textcolor{blue}{\Delta E} \\
 \Lambda_{12} : & \ \ \  & X_{21} \cdot Y_{12} \cdot Y_{21}  &-&   Y_{21} \cdot Y_{12} \cdot X_{21}  & \ \ \ \  & \textcolor{blue}{ - Y_{12}}+ V_{14} \cdot Q_{42}  &-&  X_{14} \cdot P_{42} \\ 
 \Lambda_{34} : & \ \ \  & X_{43} \cdot Y_{34} \cdot Y_{43}  &-&  Y_{43} \cdot Y_{34} \cdot X_{43}  & \ \ \ \  &  \textcolor{blue}{+Y_{34}}+P_{31} \cdot X_{14}  &-&  Q_{31} \cdot V_{14}  \\
 \end{array} 
 $}
 ~.~
\eea
This is the first example of a mass deformation connecting $2d$ $(0,2)$ theories associated to toric Calabi-Yau 4-folds as originally studied in \cite{Franco:2016fxm}. 
Here, we note that the original $E$-terms above consist of chiral fields
that form extremal brick matchings $p_1$ and $p_2$ as summarized in the $P$-matrix in \eref{es05a11}, 
whereas the mass terms introduced in \eref{es05a20}
consist of chiral fields that form the extremal brick matching $p_4$.
Based on the $U(1)_R$ charges of these extremal brick matchings summarized in \tref{tab_30}, 
we note that the mass terms introduced in \eref{es05a20} satisfy the overall $U(1)_R$ charge constraint set by the original $E$-terms as follows, 
\beal{es05a21}
r[p_4] = r[p_1] + r[p_2] = 2r_1~.~
\eea

By integrating out the mass terms in \eref{es05a20}, we obtain the following $J$- and $E$-terms 
that correspond to the brane brick model for $Q^{1,1,1}$ \cite{Franco:2015tna},
\beal{es05a22}
\resizebox{0.8\textwidth}{!}{$
\begin{array}{rrrclcrcl}
& & & J  & & &  & E   \\
 \Lambda_{41}^{1} : & \ & V_{14} \cdot Q_{42} \cdot Y_{21} \cdot X_{14} &-&  X_{14} \cdot Y_{43} \cdot Q_{31} \cdot V_{14}  & \ \ \ \ & P_{42} \cdot X_{21} &-& X_{43} \cdot P_{31}  \\
 \Lambda_{41}^{2} : & \ &  X_{14} \cdot X_{43} \cdot Q_{31} \cdot V_{14} &-& V_{14} \cdot Q_{42} \cdot X_{21} \cdot X_{14}  & \ \ \ \ &  P_{42} \cdot Y_{21} &-& Y_{43} \cdot P_{31}  \\
 \Lambda_{41}^{3} : & \ & V_{14} \cdot Y_{43} \cdot P_{31} \cdot X_{14}  &-& X_{14} \cdot P_{42} \cdot Y_{21} \cdot V_{14}  & \ \ \ \ &  Q_{42} \cdot X_{21} &-& X_{43} \cdot Q_{31}   \\ 
 \Lambda_{41}^{4} : & \ &X_{14} \cdot P_{42} \cdot X_{21} \cdot V_{14}   &-& V_{14} \cdot X_{43} \cdot P_{31} \cdot X_{14}  & \ \ \ \ &  Q_{42} \cdot Y_{21} &-& Y_{43} \cdot Q_{31} \\
 \Lambda_{32}^{1} : & \ & Y_{21} \cdot X_{14} \cdot X_{43} &-& X_{21} \cdot X_{14} \cdot Y_{43}   & \ \ \ \ &   P_{31} \cdot V_{14} \cdot Q_{42} &-& Q_{31} \cdot V_{14} \cdot P_{42} \\
 \Lambda_{32}^{2} : & \ &  Y_{21} \cdot V_{14} \cdot X_{43} &-& X_{21} \cdot V_{14} \cdot Y_{43}  & \ \ \ \ &  P_{31} \cdot X_{14} \cdot Q_{42} &-& Q_{31} \cdot X_{14} \cdot P_{42}  
\end{array}
$}
~,~
\eea
where the corresponding quiver diagram is shown in \fref{f_ref_ex3d}(b).

Using the forward algorithm \cite{Feng:2000mi, Franco:2015tna, Franco:2015tya}, we obtain the $P$-matrix for the $Q^{1,1,1}$ brane brick model, 
which takes the following form,
\beal{es05a23}
P =
\resizebox{0.25\textwidth}{!}{$
\left(
\ba{c|cccccc|cc}
 & p_2 & p_3 & p_4 & p_5 & p_6 & p_7 & o_1 & o_2 \\
\hline
 P_{31} & 1 & 0 & 0 & 0 & 0 & 0 & 0 & 1 \\ 
 P_{42} & 1 & 0 & 0 & 0 & 0 & 0 & 1 & 0 \\ 
 Q_{31} & 0 & 0 & 0 & 0 & 1 & 0 & 0 & 1 \\ 
 Q_{42} & 0 & 0 & 0 & 0 & 1 & 0 & 1 & 0 \\ 
 V_{14} & 0 & 0 & 1 & 0 & 0 & 0 & 0 & 0 \\ 
 X_{14} & 0 & 0 & 0 & 0 & 0 & 1 & 0 & 0 \\ 
 X_{21} & 0 & 0 & 0 & 1 & 0 & 0 & 0 & 1 \\ 
 X_{43} & 0 & 0 & 0 & 1 & 0 & 0 & 1 & 0 \\ 
 Y_{21} & 0 & 1 & 0 & 0 & 0 & 0 & 0 & 1 \\ 
 Y_{43} & 0 & 1 & 0 & 0 & 0 & 0 & 1 & 0
\ea
\right)
$}
~.~
\eea
The $U(1)$ charges under the $J$- and $E$-terms in \eref{es05a22}
as well as the $D$-terms are given by the following charge matrices, 
\beal{es05a24}
Q_{JE} =
\resizebox{0.22\textwidth}{!}{$
\left(
\ba{cccccc|cc}
 p_2 & p_3 & p_4 & p_5 & p_6 & p_7 & o_1 & o_2 \\
\hline
 1 & 1 & 0 & 1 & 1 & 0 & -1 & -1
\ea
\right)
$}
~,~
Q_{D} =
\resizebox{0.22\textwidth}{!}{$
\left(
\ba{cccccc|cc}
 p_2 & p_3 & p_4 & p_5 & p_6 & p_7 & o_1 & o_2 \\
\hline
 0 & 0 & 1 & 0 & 0 & 1 & 0 & -1 \\ 
 0 & 1 & 0 & 1 & 0 & 0 & -1 & 0 \\ 
 0 & -1 & 0 & -1 & 0 & 0 & 0 & 1
\ea
\right)
$}
~.~
\eea
The resulting toric diagram for the $Q^{1,1,1}$ model
is given by the following $G_t$-matrix, 
\beal{es05a25}
G_t =
\resizebox{0.22\textwidth}{!}{$
\left(
\ba{cccccc|cc}
 p_2 & p_3 & p_4 & p_5 & p_6 & p_7 & o_1 & o_2 \\
\hline
 -1 & 0 & 0 & 0 & 1 & 0 & 0 & 0 \\ 
 0 & 0 & 1 & 1 & 1 & 0 & 1 & 1 \\ 
 -1 & 0 & 1 & 0 & 1 & -1 & 0 & 0 \\ 
  \hline
 1 & 1 & 1 & 1 & 1 & 1 & 2 & 2
\ea
\right)
$}
~,~
\eea
where we can see that the $Q^{1,1,1}$ model has $6$ extremal GLSM fields corresponding to the $6$ extremal points in the toric diagram shown in \fref{f_ref_ex3c}(b).
Here, we also note that the $Q^{1,1,1}$ brane brick model has $2$ extra GLSM fields 
that over-parameterize the mesonic moduli space $\mathcal{M}^{mes}_{Q^{1,1,1}}$ and do not have corresponding points in the toric diagram for $Q^{1,1,1}$.

Based on the $Q_{JE}$ and $Q_D$ matrices, we can see that the $Q^{1,1,1}$ model has an enhanced
global symmetry of the form, 
\beal{es05a26}
SU(2)_x \times SU(2)_{y} \times SU(2)_{z} \times U(1)_R ~,~
\eea
where $SU(2)_x \times SU(2)_{y} \times SU(2)_{z}$ is the enhanced mesonic flavor symmetry.
\tref{tab_32} summarizes how the extremal GLSM fields $p_a$ are charged under the global symmetry of the $Q^{1,1,1}$ brane brick model.

\begin{table}[ht!]
\centering
\begin{tabular}{|c|c|c|c|c|l|}
\hline
\; & $SU(2)_x$ & $SU(2)_y$ & $SU(2)_z$ & $U(1)_R$ & fugacity \\
\hline
$p_2$ & $ 0 $ & $ +1 $ & $ 0 $ & $ r_1 $ & $t_2 = y \overline{t}_1 $ \\
$p_3$ & $ 0 $ & $ 0 $ & $ +1 $ & $ r_2 $ & $t_3 =z \overline{t}_2 $ \\
$p_4$ & $ +1 $ & $ 0 $ & $ 0 $ & $ r_1 $ & $t_4 =x \overline{t}_1$ \\
$p_5$ & $ 0 $ & $ 0 $ & $ -1 $ & $ r_2 $ & $t_5 =z^{-1} \overline{t}_2$ \\
$p_6$ & $ 0 $ & $ -1 $ & $ 0 $ & $ r_1 $ & $t_6=y^{-1} \overline{t}_1$ \\
$p_7$ & $ -1 $ & $ 0 $ & $ 0 $ & $ r_1 $ & $t_7 =x^{-1} \overline{t}_1$ \\
\hline
\end{tabular}
\caption{Charges under the global symmetry of the $Q^{1,1,1}$ model on the extremal GLSM fields $p_a$.
Here, $U(1)_R$ charges $r_1$ and $r_2$ are chosen such that 
the $J$- and $E$-terms coupled to Fermi fields have an overall $U(1)_R$ charge of $2$
with $4r_1+2r_2=2$. \label{tab_32}}
\end{table}

Using the Molien integral formula in \eref{es01a15}, we can calculate the Hilbert series of the mesonic moduli space $\mathcal{M}^{mes}_{Q^{1,1,1}}$, 
which takes the following form,
\beal{es05a27}
&&
g(t_a ,y_o;\mathcal{M}^{mes}_{Q^{1,1,1}})= \frac{P(t_a ,y_o;\mathcal{M}^{mes}_{Q^{1,1,1}})}{(1-y_o t_2 t_3 t_4) (1-y_o t_2 t_4 t_5) (1-y_o t_3 t_4 t_6) 
(1-y_o t_4 t_5 t_6)  }\nn\\&&\hspace{1cm}\times
\frac{1}{(1-y_o t_2 t_3 t_7) (1-y_o t_2 t_5 t_7) 
(1-y_o t_3 t_6 t_7) (1-y_o t_5 t_6 t_7)} ~,~
\eea
where $t_a$ are the fugacities for the extremal GLSM fields $p_a$ and $y_o$ counts the product of extra GLSM fields $o_1 o_2$ \cite{Franco:2015tna}. 
The numerator $P(t_a,y_o;\mathcal{M}^{mes}_{Q_{1,1,1}})$ for the Hilbert series in \eref{es05a27} is presented in appendix \sref{app_num_06}.

We can rewrite the Hilbert series in \eref{es05a27}
in terms of characters of irreducible representations of the enhanced mesonic flavor symmetry $SU(2)_x\times SU(2)_y\times SU(2)_z$.
This can be done using the following fugacity map, 
\beal{es05a28}
\overline{t}_1= (t_2 t_4 t_6 t_7)^{1/4}~,~ 
\overline{t}_2= (t_3 t_5)^{1/2}~,~ 
x= \frac{t_4^{1/2}}{t_7^{1/2}}~,~ 
y= \frac{t_2^{1/2}}{t_6^{1/2}}~,~  
z= \frac{t_3^{1/2}}{t_5^{1/2}}~,~
\eea
where the fugacities $x$,$y$ and $z$ correspond to the mesonic flavor symmetry factors $SU(2)_x$, $SU(2)_y$ and $SU(2)_z$, respectively.
By further setting $y_o=1$, we obtain the following character expansion of the Hilbert series in \eref{es05a27}, 
\beal{es05a29}
g(\overline{t}_1,\overline{t}_2,x,y,z;\mathcal{M}^{mes}_{Q^{1,1,1}})=\sum_{n=0}^{\infty} [n;n;n] \overline{t}_1^{2n} \overline{t}_2^n 
~,~
\eea
where $[n;m;k]=[n]_x [m]_y [k]_z$ are characters of irreducible representations of $SU(2)_x \times SU(2)_y \times SU(2)_z$ with highest weight given by $(m),(n),(k)$. 
The corresponding highest weight generating function for the Hilbert series in \eref{es05a29} takes the following form,
\beal{es05a30}
h(\overline{t}_1,\overline{t}_2,\mu,\nu,\kappa;\mathcal{M}^{mes}_{Q^{1,1,1}})=\frac{1}{(1-\mu \nu \kappa \overline{t}_1^2 \overline{t}_2)} 
~,~
\eea
where $\mu^m \nu^n \kappa ^k$ counts characters of irreducible representations of $SU(2)_x \times SU(2)_y \times SU(2)_z$ of the form $[m]_x [n]_y [k]_z$. 

The plethystic logarithm of the Hilbert series in \eref{es05a29} is given by 
\beal{es05a31}
&&
PL[g(\overline{t}_1,\overline{t}_2,x,y,z;\mathcal{M}^{mes}_{Q^{1,1,1}})]=[1]_x [1]_y [1]_z \overline{t}_1^2 \overline{t}_2 -([2]_x+[2]_y+[2]_z)\overline{t}_1^4 \overline{t}_2^2+\dots 
~,~
\nn\\
\eea
where we can see that the mesonic moduli space $\mathcal{M}^{mes}_{Q^{1,1,1}}$ is not a complete intersection.
We can extract the generators for $\mathcal{M}^{mes}_{Q^{1,1,1}}$ from the first positive terms in the plethystic logarithm. 
They are summarized with their mesonic flavor symmetry charges in \tref{tab_70}.
\\

\begin{table}[ht!]
\centering
\resizebox{0.95\textwidth}{!}{
\begin{tabular}{|c|c|c|c|c|c|c|c|}
\hline
PL term&generators  & GLSM fields & $SU(2)_{x}$ & $SU(2)_{y}$ & $SU(2)_{z}$  & fugacity\\
\hline\hline
\multirow{8}{*}{$+[1]_x [1]_y [1]_z \overline{t}_1^2 \overline{t}_2$}&$P_{31} V_{14} X_{43}=P_{42} X_{21} V_{14} $ & $ p_2 p_4 p_5 o $ & $+1$ &$+1$&$-1$ & $x y z^{-1} \overline{t}_1^2 \overline{t}_2$\\
&$P_{31} X_{14} X_{43}=P_{42} X_{21} X_{14} $ & $ p_2 p_5 p_7 o$ & $-1$ &$+1$&$-1$ & $x^{-1} y z^{-1} \overline{t}_1^2 \overline{t}_2$\\
&$V_{14} X_{43} Q_{31}=V_{14} Q_{42} X_{21} $ & $ p_4 p_5 p_6 o$ & $+1$ &$-1$&$-1$ & $x y^{-1} z^{-1} \overline{t}_1^2 \overline{t}_2$\\
&$X_{14} X_{43} Q_{31}=X_{14} Q_{42} X_{21} $ & $ p_5 p_6 p_7 o$ & $-1$ &$-1$&$-1$ & $x^{-1} y^{-1} z^{-1} \overline{t}_1^2 \overline{t}_2$\\
&$Y_{43} P_{31} V_{14}=P_{42} Y_{21} V_{14}$ & $p_2 p_3 p_4 o$ & $+1$ &$+1$&$+1$ & $x y z \overline{t}_1^2 \overline{t}_2$\\
&$Y_{43} P_{31} X_{14}=P_{42} Y_{21} X_{14}$ & $p_2 p_3 p_7 o$ & $-1$ &$+1$&$+1$ & $x^{-1} y z \overline{t}_1^2 \overline{t}_2$\\
&$Y_{43} Q_{31} V_{14}=V_{14} Q_{42} Y_{21} $ & $p_3 p_4 p_6 o$ & $+1$ &$-1$&$+1$ & $x y^{-1} z \overline{t}_1^2 \overline{t}_2$\\
&$Y_{43} Q_{31} X_{14}=X_{14} Q_{42} Y_{21}$ & $p_3 p_6 p_7 o$ & $-1$ &$-1$&$+1$  & $x^{-1} y^{-1} z \overline{t}_1^2 \overline{t}_2$\\
\hline
\end{tabular}}
\caption{
Generators of the mesonic moduli space $\mathcal{M}^{mes}_{Q^{1,1,1}}$ of the $Q^{1,1,1}$ brane brick model in terms of chiral fields and GLSM fields with the corresponding mesonic flavor symmetry charges. Here, $o$ denotes the product of extra GLSM fields $o_1 o_2$.
\label{tab_70}
}
\end{table}

\paragraph{Comparison between $\mathcal{C}_{+-}$ and $Q^{1,1,1}$.}
Based on the plethystic logarithm in \eref{es05a18}, we note that the generators of the mesonic moduli space $\mathcal{M}^{mes}_{\mathcal{C}_{+-}}$
transform under the following irreducible representations of the mesonic flavor symmetry
$SU(2)_x \times SU(2)_y \times U(1)_f$,
\beal{es05a40}
[1]_y f^{-2} ~,~
[2]_x [1]_y f^2 ~,~
\eea
giving a total of $8$ generators. 
In comparison, the generators of the mesonic moduli space $\mathcal{M}^{mes}_{Q^{1,1,1}}$
transform under the following irreducible representations of the mesonic flavor symmetry
$SU(2)_x \times SU(2)_y \times U(1)_f$ based on the plethystic logarithm in \eref{es05a31},
\beal{es05a41}
[1]_x [1]_y [1]_z ~,~
\eea
which gives in total $8$ generators.
We can here observe
that the total number of generators of the mesonic moduli space stays invariant between brane brick models 
$\mathcal{C}_{+-}$ and $Q^{1,1,1}$. 
This confirms the general observation that the total number of generators of the mesonic moduli spaces stays invariant if the corresponding brane brick models are associated to toric Calabi-Yau 4-folds, which are related by a birational transformation of the form in \eref{es01a52}. 

We can also take a closer look at the refined Hilbert series in \eref{es05a16} for
the mesonic moduli space $\mathcal{M}^{mes}_{\mathcal{C}_{+-}}$.
By setting the mesonic flavor fugacities to $x=y=f=1$, 
the Hilbert series in \eref{es05a16} for the mesonic moduli space $\mathcal{M}^{mes}_{\mathcal{C}_{+-}}$
can be unrefined such that it is only in terms of $U(1)_R$ fugacities $\overline{t}_1, \overline{t}_2$ corresponding to the $U(1)_R$ charges $r_1, r_2$ defined in \tref{tab_30}.
The resulting unrefined Hilbert series takes the following form, 
\beal{es05a42}
g(\overline{t}_1,\overline{t}_2;\mathcal{M}^{mes}_{\mathcal{C}_{+-}})= \frac{1-9 \overline{t}_1^4 \overline{t}_2^2+16 \overline{t}_1^6 \overline{t}_2^3-9 \overline{t}_1^8 \overline{t}_2^4+\overline{t}_1^{12} \overline{t}_2^6}{(1-\overline{t}_1^2 \overline{t}_2)^8}
~.~
\eea
We can also express the Hilbert series for the mesonic moduli space $\mathcal{M}^{mes}_{Q^{1,1,1}}$
in \eref{es05a29} in terms of only $U(1)_R$ fugacities $\overline{t}_1, \overline{t}_1$
corresponding to $U(1)_R$ charges $r_1,r_2$ in \tref{tab_32}
by setting the mesonic flavor symmetry fugacities to $x=y=f=1$.
The resulting unrefined Hilbert series takes the following form, 
\beal{es05a43} 
g(\overline{t}_1,\overline{t}_2;\mathcal{M}^{\text{mes}}_{Q^{1,1,1}}) = \frac{1-9 \overline{t}_1^4 \overline{t}_2^2+16 \overline{t}_1^6 \overline{t}_2^3-9 \overline{t}_1^8 \overline{t}_2^4+\overline{t}_1^{12} \overline{t}_2^6}{(1-\overline{t}_1^2 \overline{t}_2)^8}
~.~
\eea

From \eref{es05a42} and \eref{es05a43}, 
we observe that the unrefined Hilbert series in terms of only $U(1)_R$ fugacities
are identical for brane brick models $\mathcal{C}_{+-}$ and $Q^{1,1,1}$.
This further supports our observation that the Hilbert series for mesonic moduli spaces refined under only $U(1)_R$ fugacities
remain invariant between brane brick models that correspond to toric Calabi-Yau 4-folds, which are related by a birational transformation of the form in \eref{es01a52}.
In this example, the corresponding birational transformation is given in \eref{es05a03}.
Our observation also shows here that birational transformations connecting toric Calabi-Yau 4-folds with non-reflexive toric diagrams 
leave the number of generators and the unrefined Hilbert series invariant between corresponding brane brick models. 
This observation complements the original observations in \cite{Ghim:2024asj} in terms of toric Fano 3-folds
whose toric diagrams are reflexive polytopes.
\\

\subsection{Non-reflexive Case: $P_{+-}^{1}[\mathbb{C}^3/\mathbb{Z}_5~(1,1,3)]$ and $P_{+-}^{2}[\mathbb{C}^3/\mathbb{Z}_5~(1,1,3)]$ \label{sec:hpmc3z5_deform}} 

\begin{figure}[H]
\begin{center}
\resizebox{0.85\hsize}{!}{
\includegraphics[height=6cm]{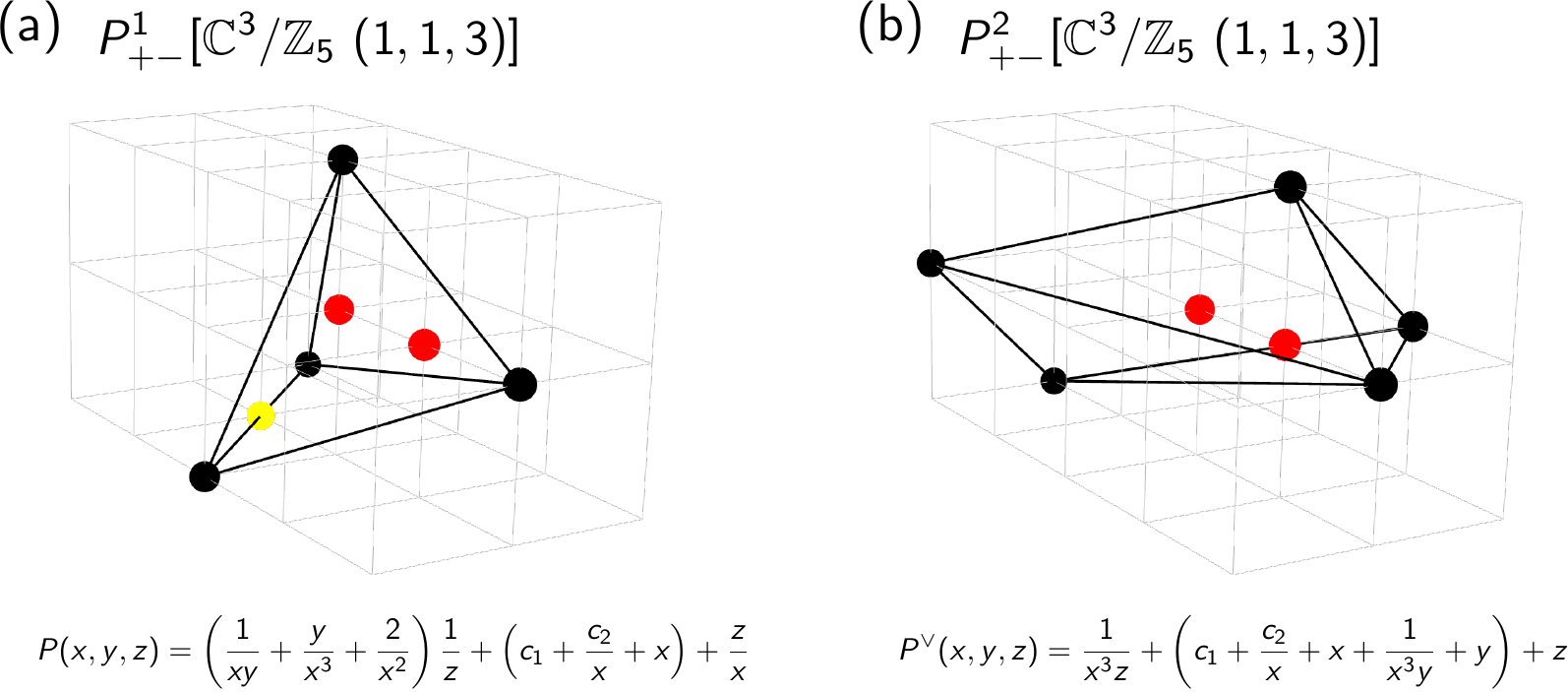} 
}
\caption{
The toric diagrams for the toric Calabi-Yau 4-folds known as (a) $P_{+-}^{1}[\mathbb{C}^3/\mathbb{Z}_5~(1,1,3)]$ and (b) $P_{+-}^{2}[\mathbb{C}^3/\mathbb{Z}_5~(1,1,3)]$,
and the corresponding Newton polynomials. 
\label{f_ref_ex4}}
 \end{center}
 \end{figure}

\fref{f_ref_ex4}(a) shows the toric diagram for the
$P_{+-}^{1}[\mathbb{C}^3/\mathbb{Z}_5~(1,1,3)]$ brane brick model.
The coordinates of the points of this toric diagram are given by,
\beal{es08a01} \label{H+-C3Z5_points}
\Delta = \{ (-1,-1,-1) , (1,0,0) , (-1,0,1),
(-3,1,-1) , (-2,0,-1) , (-1,0,0) , (0,0,0) \} ~.~
\nn\\
\eea
\\

\paragraph{Algebraic Mutation.}
The Newton polynomial corresponding to the toric diagram for $P_{+-}^{1}[\mathbb{C}^3/\mathbb{Z}_5~(1,1,3)]$
can be written using the point coordinates in \eref{es08a01} as follows, 
\beal{es08a02}
P ( x, y, z) = \left(
\frac{1}{xy}+\frac{y}{x^3}+\frac{2}{x^2} 
\right) \frac{1}{z} + \left(c_1 +\frac{c_2}{x} + x \right) + \frac{z}{x} ~,~
\eea
where we have chosen the coefficients for the extremal points to be $1$,
for the point on the edge to be $2$, 
and for the two internal points to be $c_1, c_2 \in \mathbb{C}^*$, respectively. 
Using these choices of coefficients in the Newton polynomial, 
we can introduce the following birational transformation, 
\beal{es08a03} \label{alg-hpm}
\varphi_A ~:~
(x,y,z) \mapsto 
\left(
x,y, \left(x+y \right)z
\right)
~,~ 
\text{where $A(x,y) = x+y$}
~,~
\eea
which maps the Newton polynomial in \eref{es08a02} to the following form, 
\beal{es08a04}
P^\prime (x,y,z) =
\frac{x+y}{x^3 y z}
+
\left(
c_1 + \frac{c_2 }{x} + x
\right)
+ 
\left(
1+ \frac{y}{x}
\right)
z
~.~
\eea 
Introducing a further $GL(3, \mathbb{Z})$ transformation on the coordinates $x,y,z \in \mathbb{C}^*$
in \eref{es08a04}
given by,
\beal{es08a03b}
N =
    \left(
    \begin{array}{ccc}
    1 & 0 & 0 \\
    1 & 1 & -1 \\
    0 & 0 & 1
    \end{array}
    \right) 
    ~,~
\eea
we obtain the following Newton polynomial,
\beal{es08a05} \label{newt-hpm2}
P^\vee (x, y, z) = 
\frac{1}{x^3 z}
+ 
\left(
c_1 + \frac{c_2}{x} + x + \frac{1}{x^3 y} + y
\right)
+ 
z
~.~
\eea
The Newton polynomial above corresponds to the toric diagram for 
the
$P_{+-}^{2}[\mathbb{C}^3/\mathbb{Z}_5~(1,1,3)]$ brane brick model.
The toric diagram is shown in \fref{f_ref_ex4}(b).
We note here that like in the original toric diagram for 
$P_{+-}^{1}[\mathbb{C}^3/\mathbb{Z}_5~(1,1,3)]$, 
the toric diagram for $P_{+-}^{2}[\mathbb{C}^3/\mathbb{Z}_5~(1,1,3)]$ after the algebraic mutation has also 2 internal points.
We can see here that the coefficients $c_1, c_2 \in \mathbb{C}^*$
assigned to the two internal points in the original Newton polynomial in \eref{es08a02}
remain unaffected under the algebraic mutation as seen in \eref{es08a05}.
In other words, the birational transformation and the corresponding algebraic mutation with \eref{es08a03b} remains 
unchanged for any choice of coefficients $c_1, c_2 \in \mathbb{C}^*$ assigned to the two internal points in the Newton polynomial for $P_{+-}^{1}[\mathbb{C}^3/\mathbb{Z}_5~(1,1,3)]$.
\\

\begin{figure}[H]
\begin{center}
\resizebox{0.85\hsize}{!}{
\includegraphics[height=6cm]{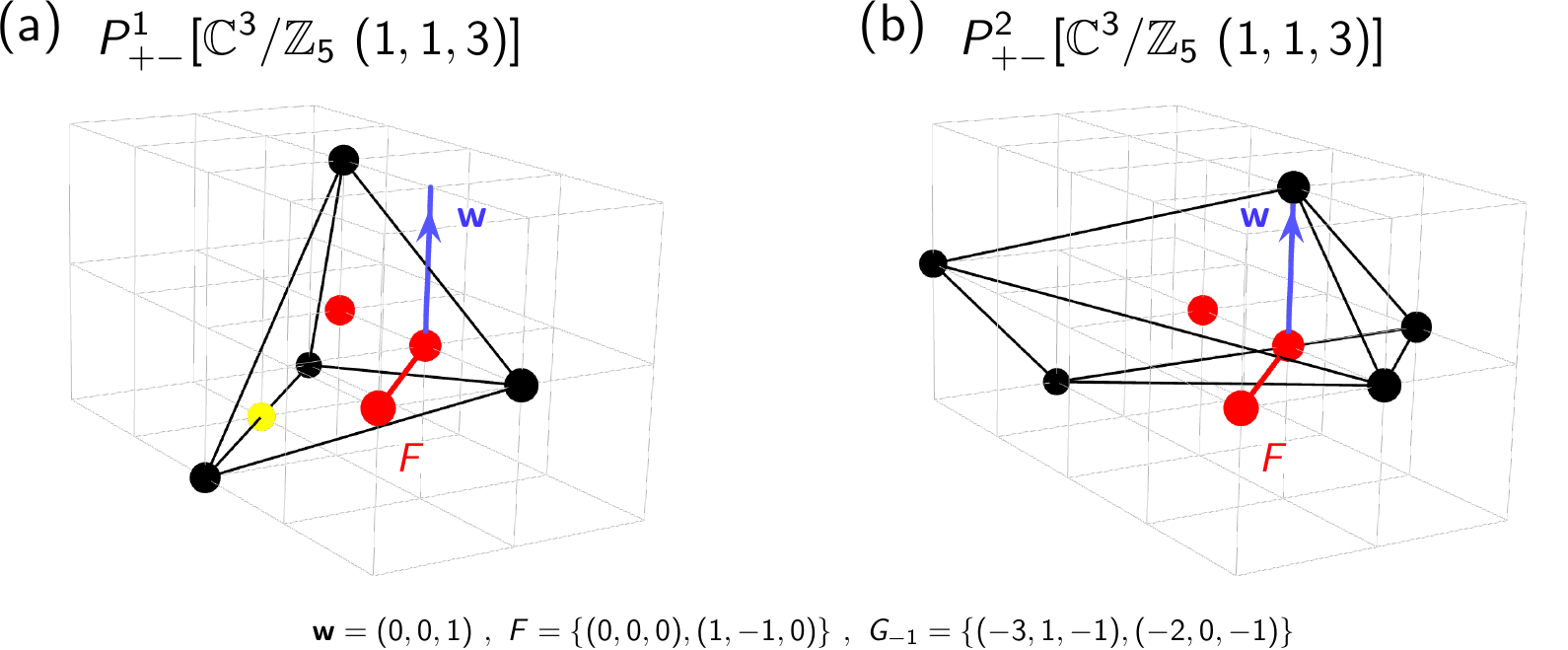} 
}
\caption{
The toric diagrams for the toric Calabi-Yau 4-folds known as (a) $P_{+-}^{1}[\mathbb{C}^3/\mathbb{Z}_5~(1,1,3)]$ and (b) $P_{+-}^{2}[\mathbb{C}^3/\mathbb{Z}_5~(1,1,3)]$,
with height measurement vector $\textbf{w}=(0,0,1)$ and factor $F$ used for combinatorial mutation.
\label{f_ref_ex4b}}
 \end{center}
 \end{figure}

\paragraph{Combinatorial Mutation.}
Let us choose
the height measurement vector $\textbf{w}=(0,0,1)$ in order to perform the combinatorial mutation on the toric diagram for the $P_{+-}^{1}[\mathbb{C}^3/\mathbb{Z}_5~(1,1,3)]$ brane brick model shown in \fref{f_ref_ex4b}(a).
We further choose the factor $F$ and consider the corresponding polytopes $G_h$ as follows.
\beal{es08a06}
F = \{ (0,0,0) ~,~ (1,-1,0) \} ~,~ G_{-1} = \{ (-3,1,-1) ~,~ (-2,0,-1) \} ~.~
\eea
The above choices define a combinatorial mutation of the toric diagram for $P_{+-}^{1}[\mathbb{C}^3/\mathbb{Z}_5~(1,1,3)]$,
leading to the following convex polytope in $\mathbb{Z}^3$, 
\beal{es08a07}
\begin{tabular}{rcl}
$\underline{ h=-1 : }$ & $(-3,1,-1) ~,~ (-2,0,-1)$ & $\in G_{-1}$~,~\\
$\underline{ h=  0  : }$  & $(-1,0,0) ~,~ (0,0,0) ~,~ (1,0,0)$ & $\in w_0 (\Delta)$ ~,~ \\
$\underline{ h=+1 : }$ & $(-1,0,1) ~,~ (0,-1,1)$& $\in w_1 (\Delta) + F$ ~,~ \\
\end{tabular}
\eea
which we identify as the toric diagram for the $P_{+-}^{2}[\mathbb{C}^3/\mathbb{Z}_5~(1,1,3)]$ brane brick model as shown in \fref{f_ref_ex4b}(b).
\\

\begin{figure}[H]
\begin{center}
\resizebox{0.94\hsize}{!}{
\includegraphics[height=6cm]{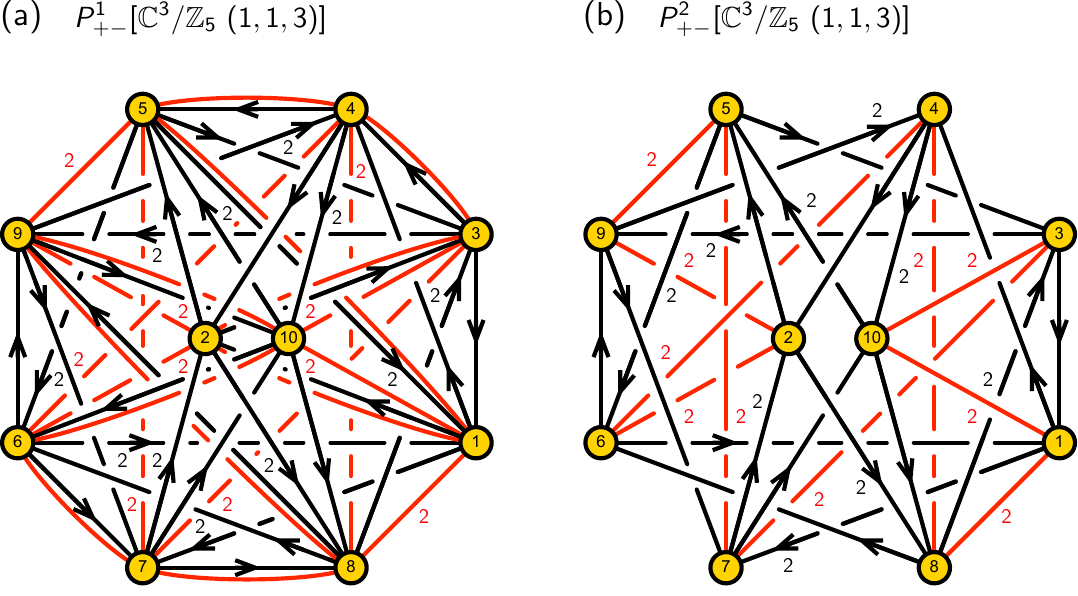} 
}
\caption{
The quiver diagrams for the brane brick models corresponding to (a) $P_{+-}^{1}[\mathbb{C}^3/\mathbb{Z}_5~(1,1,3)]$ and (b) $P_{+-}^{2}[\mathbb{C}^3/\mathbb{Z}_5~(1,1,3)]$.
\label{f_ref_ex4d}}
 \end{center}
 \end{figure}

\begin{figure}[ht!!]
\begin{center}
\resizebox{0.85\hsize}{!}{
\includegraphics[height=6cm]{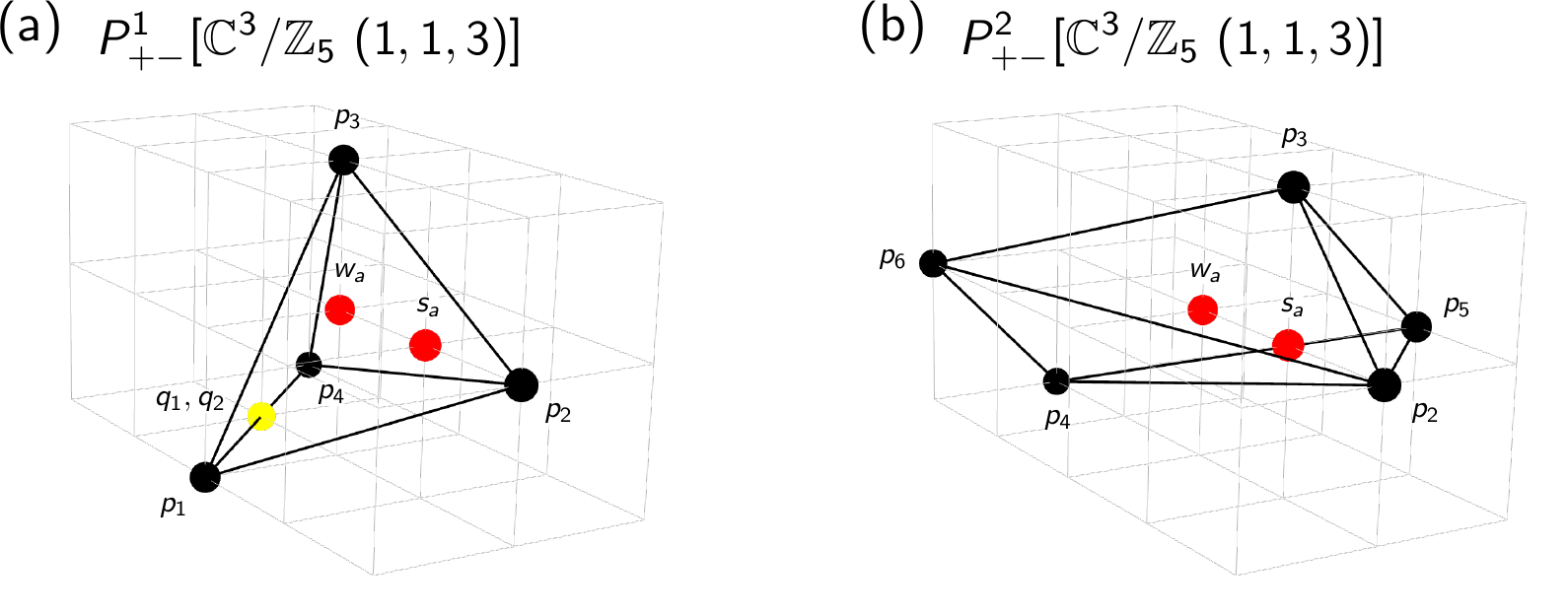} 
}
\caption{
The toric diagrams for the toric Calabi-Yau 4-folds known as (a) $P_{+-}^{1}[\mathbb{C}^3/\mathbb{Z}_5~(1,1,3)]$ and (b) $P_{+-}^{2}[\mathbb{C}^3/\mathbb{Z}_5~(1,1,3)]$,
with points labelled by the GLSM fields in the corresponding brane brick models.
\label{f_ref_ex4c}}
 \end{center}
 \end{figure}

\paragraph{Brane Brick Model for $P_{+-}^{1}[\mathbb{C}^3/\mathbb{Z}_5~(1,1,3)]$ .}
The quiver of the $P_{+-}^{1}[\mathbb{C}^3/\mathbb{Z}_5~(1,1,3)]$ brane brick model is shown in \fref{f_ref_ex4d}(a) and the $J$-and $E$-terms take the following form, 
\beal{es08a10}
\resizebox{0.45\textwidth}{!}{$
\begin{array}{rrrclcrcl}
& & & J  & & &  & E  \\
\Lambda_{12}: & \ \ \ \ & X_{23}\cdot  X_{31}&-& X_{25}\cdot  X_{51}& \ \ \ \ & Q_{17}\cdot  P_{72} &-& K_{17}\cdot  V_{72}\\ 
\Lambda_{23}: & \ \ \ \ &  X_{34}\cdot  X_{42}&-& X_{31}\cdot  X_{12}& \ \ \ \ & Q_{28}\cdot  P_{83} &-& K_{28}\cdot  V_{83}\\ 
\Lambda_{34}: & \ \ \ \ &  X_{42}\cdot  X_{23}&-& X_{45}\cdot  X_{53}& \ \ \ \ & K_{39}\cdot  V_{94}&-&Q_{39}\cdot  P_{94}\\ 
\Lambda_{45}: & \ \ \ \ &  X_{51}\cdot  X_{14}&-& X_{53}\cdot  X_{34}& \ \ \ \ & Q_{4,10}\cdot P_{10,5} &-& K_{4,10}\cdot  V_{10,5}\\ 
\Lambda_{51}: & \ \ \ \ &  X_{12}\cdot  X_{25}&-& X_{14}\cdot  X_{45}& \ \ \ \ & Q_{56}\cdot  P_{61} &-& K_{56}\cdot  V_{61}\\ 
\Lambda_{67}: & \ \ \ \ &  X_{7,10}\cdot  X_{10,6}&-& X_{78}\cdot  X_{86}& \ \ \ \ & P_{61}\cdot  Q_{17} &-& V_{61}\cdot  K_{17} \\ 
\Lambda_{78}: & \ \ \ \ &  X_{89}\cdot  X_{97}&-& X_{86}\cdot  X_{67}& \ \ \ \ & V_{72}\cdot  K_{28}&-& P_{72}\cdot  Q_{28}\\ 
\Lambda_{89}: & \ \ \ \ &  X_{97}\cdot  X_{78}&-& X_{9,10}\cdot  X_{10,8}& \ \ \ \ & P_{83}\cdot  Q_{39} &-& V_{83}\cdot  K_{39}\\ 
\Lambda_{9,10}: & \ \ \ \ &  X_{10,8}\cdot  X_{89}&-& X_{10,6}\cdot  X_{69}& \ \ \ \ & P_{94}\cdot  Q_{4,10} &-& V_{94}\cdot  K_{4,10}\\ 
\Lambda_{10,6}: & \ \ \ \ &  X_{69}\cdot  X_{9,10}&-& X_{67}\cdot  X_{7,10}& \ \ \ \ & P_{10,5}\cdot  Q_{56} &-& V_{10,5}\cdot  K_{56} \\ 
\Lambda_{62}^{1}: & \ \ \ \ &  K_{28}\cdot  X_{86}&-& X_{25}\cdot  K_{56}& \ \ \ \ & X_{67}\cdot  V_{72}&-& V_{61}\cdot  X_{12}\\ 
\Lambda_{73}^{1}: & \ \ \ \ &  X_{31}\cdot  K_{17}&-& K_{39}\cdot  X_{97}& \ \ \ \ & V_{72}\cdot  X_{23}&-& X_{78}\cdot  V_{83}\\ 
\Lambda_{84}^{1}: & \ \ \ \ &  X_{42}\cdot  K_{28} &-& K_{4,10}\cdot  X_{10,8} & \ \ \ \ & V_{83}\cdot  X_{34}&-& X_{89}\cdot  V_{94}\\ 
\Lambda_{95}^{1}: & \ \ \ \ &  K_{56}\cdot  X_{69} &-& X_{53}\cdot  K_{39} & \ \ \ \ & X_{9,10}\cdot  V_{10,5}&-& V_{94}\cdot  X_{45}\\ 
\Lambda_{10,1}^{1}: & \ \ \ \ & K_{17}\cdot  X_{7,10} &-& X_{14}\cdot  K_{4,10} & \ \ \ \ & X_{10,6}\cdot   V_{61}&-& V_{10,5}\cdot  X_{51}\\ 
\Lambda_{64}^{1}: & \ \ \ \ &  K_{4,10}\cdot  X_{10,6}&-& X_{45}\cdot  K_{56}& \ \ \ \ & V_{61}\cdot  X_{14} &-& X_{69}\cdot  V_{94} \\ 
\Lambda_{92}^{1}: & \ \ \ \ &  X_{23}\cdot  K_{39}&-& K_{28}\cdot  X_{89}& \ \ \ \ & X_{97}\cdot  V_{72}&-& V_{94}\cdot  X_{42}\\ 
\Lambda_{75}^{1}: & \ \ \ \ &  K_{56}\cdot  X_{67}&-& X_{51}K_{17}& \ \ \ \ & V_{72}\cdot  X_{25} &-& X_{7,10}\cdot  V_{10,5} \\ 
\Lambda_{10,3}^{1}: & \ \ \ \ & X_{34}\cdot  K_{4,10}&-& K_{39}\cdot  X_{9,10}& \ \ \ \ & X_{10,8}\cdot  V_{83}&-& V_{10,5}\cdot  X_{53}\\ 
\Lambda_{81}^{1}: & \ \ \ \ &  K_{17}\cdot  X_{78}&-& X_{12}\cdot  K_{28}& \ \ \ \ & V_{83}\cdot  X_{31} &-& X_{86}\cdot  V_{61} \\ 
\Lambda_{62}^{2}: & \ \ \ \ & X_{25}\cdot  Q_{56} &-& Q_{28}\cdot  X_{86} & \ \ \ \ & X_{67}\cdot  P_{72}&-& P_{61}\cdot  X_{12}\\ 
\Lambda_{73}^{2}: & \ \ \ \ & Q_{39}\cdot  X_{97} &-& X_{31}\cdot  Q_{17} & \ \ \ \ & P_{72}\cdot  X_{23}&-& X_{78}\cdot  P_{83}\\ 
\Lambda_{84}^{2}: & \ \ \ \ &  Q_{4,10}\cdot  X_{10,8}&-& X_{42}\cdot  Q_{28}& \ \ \ \ & P_{83}\cdot  X_{34}&-& X_{89}\cdot  P_{94}\\ 
\Lambda_{95}^{2}: & \ \ \ \ &  X_{53}\cdot  Q_{39}&-& Q_{56}\cdot  X_{69}& \ \ \ \ & X_{9,10}\cdot  P_{10,5}&-& P_{94}\cdot  X_{45}\\ 
\Lambda_{10,1}^{2}: & \ \ \ \ &  X_{14}\cdot  Q_{4,10}&-& Q_{17}\cdot  X_{7,10}& \ \ \ \ & X_{10,6}\cdot   P_{61}&-& P_{10,5}\cdot  X_{51}\\ 
\Lambda_{64}^{2}: & \ \ \ \ &  Q_{4,10}\cdot  X_{10,6}&-& X_{45}\cdot  Q_{56}& \ \ \ \ & X_{69}\cdot  P_{94}&-& P_{61}\cdot  X_{14}\\ 
\Lambda_{92}^{2}: & \ \ \ \ &  X_{23}\cdot  Q_{39}&-& Q_{28}\cdot  X_{89}& \ \ \ \ & P_{94}\cdot  X_{42} &-& X_{97}\cdot  P_{72} \\ 
\Lambda_{75}^{2}: & \ \ \ \ &  Q_{56}\cdot  X_{67}&-& X_{51}\cdot  Q_{17}& \ \ \ \ & X_{7,10}\cdot  P_{10,5}&-& P_{72}\cdot  X_{25}\\ 
\Lambda_{10,3}^{2}: & \ \ \ \ & X_{34}\cdot  Q_{4,10}&-& Q_{39}\cdot  X_{9,10}& \ \ \ \ & P_{10,5}\cdot  X_{53} &-& X_{10,8}\cdot  P_{83} \\ 
\Lambda_{81}^{2}: & \ \ \ \ &  Q_{17}\cdot  X_{78}&-& X_{12}\cdot  Q_{28}& \ \ \ \ & X_{86}\cdot  P_{61}&-& P_{83}\cdot  X_{31}\\
 \end{array}
$}
~.~
\eea
Using the forward algorithm \cite{Feng:2000mi, Franco:2015tna, Franco:2015tya},
we obtain from the $J$- and $E$-terms the 
corresponding $P$-matrix,
\beal{es08a11}
&&
P =
\nn\\
&&
\resizebox{0.95\textwidth}{!}{$
\left(
\begin{array}{c|cccc|cc|cccccccccc|ccccccccccccccc|ccccccccccccccc|cccccccccc|cccccccccc}
& p_1 & p_2 & p_3 & p_4 & q_1 & q_2 & s_1 & s_2 & s_3 & s_4 & s_5 & s_6 & s_7 & s_8 & s_9 & s_{10} & w_1 & w_2 & w_3 & w_4 & w_5 & w_6 & w_7 & w_8 & w_9 & w_{10} & w_{11} & w_{12} & w_{13} & w_{14} & w_{15} & o_1 & o_2 & o_3 & o_4 & o_5 & o_6 & o_7 & o_8 & o_9 & o_{10} & o_{11} & o_{12} & o_{13} & o_{14} & o_{15} & u_{1} & u_{2} & u_{3} & u_{4} & u_{5} & u_{6} & u_{7} & u_{8} & u_{9} & u_{10} & v_{1} & v_{2} & v_{3} & v_{4} & v_{5} & v_{6} & v_{7} & v_{8} & v_{9} & v_{10} \\
   \hline
K_{17}&             0 & 0 & 0 & 1 & 0 & 1 & 0 & 0 & 0 & 0 & 0 & 0 & 0 & 0 & 0 & 1 & 0 & 0 & 0 & 0 & 0 & 0 & 0 & 0 & 0 & 0 & 0 & 0 & 1 & 1 & 1 & 0 & 0 & 0 & 0 & 0 & 0 & 1 & 1 & 1 & 1 & 1 & 1 & 1 & 1 & 1 & 0 & 0 & 0 & 1 & 1 & 1 & 1 & 1 & 1 & 1 & 0 & 0 & 0 & 0 & 1 & 1 & 1 & 2 & 2 & 1 \\
K_{28}&              0 & 0 & 0 & 1 & 0 & 1 & 0 & 0 & 0 & 0 & 0 & 0 & 0 & 0 & 1 & 0 & 0 & 0 & 0 & 0 & 0 & 0 & 0 & 0 & 0 & 1 & 1 & 1 & 0 & 0 & 0 & 0 & 0 & 0 & 1 & 1 & 1 & 0 & 0 & 0 & 1 & 1 & 1 & 1 & 1 & 1 & 1 & 1 & 1 & 0 & 0 & 0 & 1 & 1 & 1 & 1 & 0 & 0 & 1 & 1 & 0 & 0 & 1 & 1 & 2 & 2 \\
K_{39}&              0 & 0 & 0 & 1 & 0 & 1 & 0 & 0 & 0 & 0 & 0 & 0 & 0 & 1 & 0 & 0 & 0 & 0 & 0 & 0 & 0 & 0 & 0 & 1 & 1 & 0 & 0 & 0 & 0 & 0 & 1 & 0 & 1 & 1 & 1 & 1 & 1 & 0 & 0 & 1 & 0 & 0 & 0 & 1 & 1 & 1 & 0 & 1 & 1 & 1 & 1 & 1 & 0 & 0 & 1 & 1 & 0 & 1 & 1 & 2 & 0 & 1 & 0 & 0 & 1 & 2 \\
K_{4,10}&           0 & 0 & 0 & 1 & 0 & 1 & 0 & 0 & 0 & 0 & 0 & 0 & 1 & 0 & 0 & 0 & 0 & 0 & 0 & 0 & 0 & 0 & 1 & 0 & 0 & 0 & 0 & 1 & 0 & 1 & 0 & 1 & 1 & 1 & 0 & 1 & 1 & 0 & 1 & 1 & 0 & 0 & 1 & 0 & 0 & 1 & 1 & 0 & 1 & 0 & 1 & 1 & 1 & 1 & 0 & 1 & 1 & 1 & 0 & 2 & 0 & 2 & 0 & 1 & 0 & 1 \\
K_{56}&             0 & 0 & 0 & 1 & 0 & 1 & 0 & 0 & 0 & 0 & 0 & 1 & 0 & 0 & 0 & 0 & 0 & 0 & 0 & 0 & 0 & 1 & 0 & 0 & 1 & 0 & 1 & 0 & 0 & 0 & 0 & 1 & 0 & 1 & 0 & 0 & 1 & 1 & 1 & 1 & 0 & 1 & 1 & 0 & 1 & 0 & 1 & 1 & 1 & 1 & 0 & 1 & 0 & 1 & 1 & 0 & 1 & 0 & 0 & 1 & 1 & 2 & 0 & 2 & 1 & 0 \\
Q_{17}&     1 & 0 & 0 & 0 & 0 & 1 & 0 & 0 & 0 & 0 & 0 & 0 & 0 & 0 & 0 & 1 & 0 & 0 & 0 & 0 & 0 & 0 & 0 & 0 & 0 & 0 & 0 & 0 & 1 & 1 & 1 & 0 & 0 & 0 & 0 & 0 & 0 & 1 & 1 & 1 & 1 & 1 & 1 & 1 & 1 & 1 & 0 & 0 & 0 & 1 & 1 & 1 & 1 & 1 & 1 & 1 & 0 & 0 & 0 & 0 & 1 & 1 & 1 & 2 & 2 & 1 \\
Q_{28}&     1 & 0 & 0 & 0 & 0 & 1 & 0 & 0 & 0 & 0 & 0 & 0 & 0 & 0 & 1 & 0 & 0 & 0 & 0 & 0 & 0 & 0 & 0 & 0 & 0 & 1 & 1 & 1 & 0 & 0 & 0 & 0 & 0 & 0 & 1 & 1 & 1 & 0 & 0 & 0 & 1 & 1 & 1 & 1 & 1 & 1 & 1 & 1 & 1 & 0 & 0 & 0 & 1 & 1 & 1 & 1 & 0 & 0 & 1 & 1 & 0 & 0 & 1 & 1 & 2 & 2 \\
Q_{39}&     1 & 0 & 0 & 0 & 0 & 1 & 0 & 0 & 0 & 0 & 0 & 0 & 0 & 1 & 0 & 0 & 0 & 0 & 0 & 0 & 0 & 0 & 0 & 1 & 1 & 0 & 0 & 0 & 0 & 0 & 1 & 0 & 1 & 1 & 1 & 1 & 1 & 0 & 0 & 1 & 0 & 0 & 0 & 1 & 1 & 1 & 0 & 1 & 1 & 1 & 1 & 1 & 0 & 0 & 1 & 1 & 0 & 1 & 1 & 2 & 0 & 1 & 0 & 0 & 1 & 2 \\
Q_{4,10}&  1 & 0 & 0 & 0 & 0 & 1 & 0 & 0 & 0 & 0 & 0 & 0 & 1 & 0 & 0 & 0 & 0 & 0 & 0 & 0 & 0 & 0 & 1 & 0 & 0 & 0 & 0 & 1 & 0 & 1 & 0 & 1 & 1 & 1 & 0 & 1 & 1 & 0 & 1 & 1 & 0 & 0 & 1 & 0 & 0 & 1 & 1 & 0 & 1 & 0 & 1 & 1 & 1 & 1 & 0 & 1 & 1 & 1 & 0 & 2 & 0 & 2 & 0 & 1 & 0 & 1 \\
Q_{56}&     1 & 0 & 0 & 0 & 0 & 1 & 0 & 0 & 0 & 0 & 0 & 1 & 0 & 0 & 0 & 0 & 0 & 0 & 0 & 0 & 0 & 1 & 0 & 0 & 1 & 0 & 1 & 0 & 0 & 0 & 0 & 1 & 0 & 1 & 0 & 0 & 1 & 1 & 1 & 1 & 0 & 1 & 1 & 0 & 1 & 0 & 1 & 1 & 1 & 1 & 0 & 1 & 0 & 1 & 1 & 0 & 1 & 0 & 0 & 1 & 1 & 2 & 0 & 2 & 1 & 0 \\
V_{61}&              1 & 0 & 0 & 0 & 1 & 0 & 0 & 0 & 0 & 0 & 1 & 0 & 0 & 0 & 0 & 0 & 0 & 0 & 0 & 1 & 1 & 0 & 0 & 1 & 0 & 0 & 0 & 0 & 0 & 0 & 0 & 1 & 1 & 1 & 1 & 1 & 1 & 0 & 0 & 0 & 1 & 0 & 0 & 1 & 0 & 1 & 1 & 1 & 1 & 0 & 1 & 0 & 1 & 1 & 0 & 1 & 1 & 2 & 2 & 1 & 0 & 0 & 1 & 0 & 0 & 1 \\
V_{72}&              1 & 0 & 0 & 0 & 1 & 0 & 0 & 0 & 0 & 1 & 0 & 0 & 0 & 0 & 0 & 0 & 0 & 1 & 1 & 0 & 0 & 0 & 1 & 0 & 0 & 0 & 0 & 0 & 0 & 0 & 0 & 1 & 1 & 1 & 1 & 1 & 1 & 1 & 1 & 1 & 0 & 0 & 0 & 0 & 0 & 0 & 1 & 1 & 1 & 1 & 1 & 1 & 0 & 0 & 1 & 0 & 2 & 2 & 1 & 1 & 1 & 1 & 0 & 0 & 0 & 0 \\
V_{83}&              1 & 0 & 0 & 0 & 1 & 0 & 0 & 0 & 1 & 0 & 0 & 0 & 0 & 0 & 0 & 0 & 1 & 0 & 0 & 0 & 1 & 1 & 0 & 0 & 0 & 0 & 0 & 0 & 0 & 0 & 0 & 1 & 1 & 1 & 0 & 0 & 0 & 1 & 1 & 1 & 1 & 1 & 1 & 0 & 0 & 0 & 1 & 0 & 0 & 1 & 1 & 1 & 1 & 1 & 0 & 1 & 2 & 1 & 0 & 0 & 2 & 1 & 1 & 1 & 0 & 0 \\
V_{94}&              1 & 0 & 0 & 0 & 1 & 0 & 0 & 1 & 0 & 0 & 0 & 0 & 0 & 0 & 0 & 0 & 0 & 0 & 1 & 1 & 0 & 0 & 0 & 0 & 0 & 0 & 0 & 0 & 1 & 0 & 0 & 1 & 0 & 0 & 1 & 0 & 0 & 1 & 1 & 0 & 1 & 1 & 1 & 1 & 1 & 0 & 1 & 1 & 1 & 1 & 0 & 0 & 1 & 1 & 1 & 0 & 1 & 0 & 1 & 0 & 2 & 0 & 2 & 1 & 1 & 0 \\
V_{10,5}&           1 & 0 & 0 & 0 & 1 & 0 & 1 & 0 & 0 & 0 & 0 & 0 & 0 & 0 & 0 & 0 & 1 & 1 & 0 & 0 & 0 & 0 & 0 & 0 & 0 & 1 & 0 & 0 & 0 & 0 & 0 & 0 & 1 & 0 & 1 & 1 & 0 & 1 & 0 & 0 & 1 & 1 & 0 & 1 & 1 & 1 & 0 & 1 & 0 & 1 & 1 & 1 & 1 & 0 & 1 & 1 & 0 & 1 & 2 & 0 & 1 & 0 & 2 & 0 & 1 & 1 \\
P_{61}&     0 & 0 & 0 & 1 & 1 & 0 & 0 & 0 & 0 & 0 & 1 & 0 & 0 & 0 & 0 & 0 & 0 & 0 & 0 & 1 & 1 & 0 & 0 & 1 & 0 & 0 & 0 & 0 & 0 & 0 & 0 & 1 & 1 & 1 & 1 & 1 & 1 & 0 & 0 & 0 & 1 & 0 & 0 & 1 & 0 & 1 & 1 & 1 & 1 & 0 & 1 & 0 & 1 & 1 & 0 & 1 & 1 & 2 & 2 & 1 & 0 & 0 & 1 & 0 & 0 & 1 \\
P_{72}&     0 & 0 & 0 & 1 & 1 & 0 & 0 & 0 & 0 & 1 & 0 & 0 & 0 & 0 & 0 & 0 & 0 & 1 & 1 & 0 & 0 & 0 & 1 & 0 & 0 & 0 & 0 & 0 & 0 & 0 & 0 & 1 & 1 & 1 & 1 & 1 & 1 & 1 & 1 & 1 & 0 & 0 & 0 & 0 & 0 & 0 & 1 & 1 & 1 & 1 & 1 & 1 & 0 & 0 & 1 & 0 & 2 & 2 & 1 & 1 & 1 & 1 & 0 & 0 & 0 & 0 \\
P_{83}&     0 & 0 & 0 & 1 & 1 & 0 & 0 & 0 & 1 & 0 & 0 & 0 & 0 & 0 & 0 & 0 & 1 & 0 & 0 & 0 & 1 & 1 & 0 & 0 & 0 & 0 & 0 & 0 & 0 & 0 & 0 & 1 & 1 & 1 & 0 & 0 & 0 & 1 & 1 & 1 & 1 & 1 & 1 & 0 & 0 & 0 & 1 & 0 & 0 & 1 & 1 & 1 & 1 & 1 & 0 & 1 & 2 & 1 & 0 & 0 & 2 & 1 & 1 & 1 & 0 & 0 \\
P_{94}&     0 & 0 & 0 & 1 & 1 & 0 & 0 & 1 & 0 & 0 & 0 & 0 & 0 & 0 & 0 & 0 & 0 & 0 & 1 & 1 & 0 & 0 & 0 & 0 & 0 & 0 & 0 & 0 & 1 & 0 & 0 & 1 & 0 & 0 & 1 & 0 & 0 & 1 & 1 & 0 & 1 & 1 & 1 & 1 & 1 & 0 & 1 & 1 & 1 & 1 & 0 & 0 & 1 & 1 & 1 & 0 & 1 & 0 & 1 & 0 & 2 & 0 & 2 & 1 & 1 & 0 \\
P_{10,5}&  0 & 0 & 0 & 1 & 1 & 0 & 1 & 0 & 0 & 0 & 0 & 0 & 0 & 0 & 0 & 0 & 1 & 1 & 0 & 0 & 0 & 0 & 0 & 0 & 0 & 1 & 0 & 0 & 0 & 0 & 0 & 0 & 1 & 0 & 1 & 1 & 0 & 1 & 0 & 0 & 1 & 1 & 0 & 1 & 1 & 1 & 0 & 1 & 0 & 1 & 1 & 1 & 1 & 0 & 1 & 1 & 0 & 1 & 2 & 0 & 1 & 0 & 2 & 0 & 1 & 1 \\
X_{12}&              0 & 0 & 1 & 0 & 0 & 0 & 0 & 0 & 0 & 1 & 0 & 0 & 0 & 0 & 0 & 1 & 0 & 1 & 1 & 0 & 0 & 0 & 1 & 0 & 0 & 0 & 0 & 0 & 1 & 1 & 1 & 0 & 0 & 0 & 0 & 0 & 0 & 1 & 1 & 1 & 0 & 0 & 0 & 0 & 0 & 0 & 0 & 0 & 0 & 1 & 1 & 1 & 0 & 0 & 1 & 0 & 1 & 1 & 0 & 0 & 1 & 1 & 0 & 1 & 1 & 0 \\
X_{14}&              0 & 1 & 0 & 0 & 0 & 0 & 0 & 1 & 1 & 1 & 0 & 0 & 0 & 1 & 1 & 1 & 0 & 0 & 1 & 0 & 0 & 0 & 0 & 0 & 0 & 0 & 0 & 0 & 1 & 0 & 1 & 0 & 0 & 0 & 1 & 0 & 0 & 1 & 1 & 1 & 1 & 1 & 1 & 1 & 1 & 0 & 0 & 0 & 0 & 1 & 0 & 0 & 0 & 0 & 1 & 0 & 1 & 0 & 0 & 0 & 2 & 0 & 1 & 1 & 2 & 1 \\
X_{23}&              0 & 0 & 1 & 0 & 0 & 0 & 0 & 0 & 1 & 0 & 0 & 0 & 0 & 0 & 1 & 0 & 1 & 0 & 0 & 0 & 1 & 1 & 0 & 0 & 0 & 1 & 1 & 1 & 0 & 0 & 0 & 0 & 0 & 0 & 0 & 0 & 0 & 0 & 0 & 0 & 1 & 1 & 1 & 0 & 0 & 0 & 1 & 0 & 0 & 0 & 0 & 0 & 1 & 1 & 0 & 1 & 1 & 0 & 0 & 0 & 1 & 0 & 1 & 1 & 1 & 1 \\
X_{25}&              0 & 1 & 0 & 0 & 0 & 0 & 1 & 1 & 1 & 0 & 0 & 0 & 1 & 1 & 1 & 0 & 1 & 0 & 0 & 0 & 0 & 0 & 0 & 0 & 0 & 1 & 0 & 1 & 0 & 0 & 0 & 0 & 1 & 0 & 1 & 1 & 0 & 0 & 0 & 0 & 1 & 1 & 1 & 1 & 1 & 1 & 0 & 0 & 0 & 0 & 0 & 0 & 1 & 0 & 0 & 1 & 0 & 0 & 1 & 1 & 1 & 0 & 2 & 0 & 1 & 2 \\
X_{31}&              0 & 1 & 0 & 0 & 0 & 0 & 1 & 1 & 0 & 0 & 1 & 1 & 1 & 1 & 0 & 0 & 0 & 0 & 0 & 1 & 0 & 0 & 0 & 1 & 1 & 0 & 0 & 0 & 0 & 0 & 0 & 1 & 1 & 1 & 1 & 1 & 1 & 0 & 0 & 0 & 0 & 0 & 0 & 1 & 1 & 1 & 0 & 1 & 1 & 0 & 0 & 0 & 0 & 0 & 0 & 0 & 0 & 1 & 2 & 2 & 0 & 1 & 1 & 0 & 0 & 1 \\
X_{34}&              0 & 0 & 1 & 0 & 0 & 0 & 0 & 1 & 0 & 0 & 0 & 0 & 0 & 1 & 0 & 0 & 0 & 0 & 1 & 1 & 0 & 0 & 0 & 1 & 1 & 0 & 0 & 0 & 1 & 0 & 1 & 0 & 0 & 0 & 1 & 0 & 0 & 0 & 0 & 0 & 0 & 0 & 0 & 1 & 1 & 0 & 0 & 1 & 1 & 1 & 0 & 0 & 0 & 0 & 1 & 0 & 0 & 0 & 1 & 1 & 1 & 0 & 1 & 0 & 1 & 1 \\
X_{42}&             0 & 1 & 0 & 0 & 0 & 0 & 1 & 0 & 0 & 1 & 1 & 1 & 1 & 0 & 0 & 1 & 0 & 1 & 0 & 0 & 0 & 0 & 1 & 0 & 0 & 0 & 0 & 0 & 0 & 1 & 0 & 1 & 1 & 1 & 0 & 1 & 1 & 1 & 1 & 1 & 0 & 0 & 0 & 0 & 0 & 1 & 0 & 0 & 0 & 0 & 1 & 1 & 0 & 0 & 0 & 0 & 1 & 2 & 1 & 1 & 0 & 2 & 0 & 1 & 0 & 0 \\
X_{45}&              0 & 0 & 1 & 0 & 0 & 0 & 1 & 0 & 0 & 0 & 0 & 0 & 1 & 0 & 0 & 0 & 1 & 1 & 0 & 0 & 0 & 0 & 1 & 0 & 0 & 1 & 0 & 1 & 0 & 1 & 0 & 0 & 1 & 0 & 0 & 1 & 0 & 0 & 0 & 0 & 0 & 0 & 0 & 0 & 0 & 1 & 0 & 0 & 0 & 0 & 1 & 1 & 1 & 0 & 0 & 1 & 0 & 1 & 1 & 1 & 0 & 1 & 1 & 0 & 0 & 1 \\
X_{51}&              0 & 0 & 1 & 0 & 0 & 0 & 0 & 0 & 0 & 0 & 1 & 1 & 0 & 0 & 0 & 0 & 0 & 0 & 0 & 1 & 1 & 1 & 0 & 1 & 1 & 0 & 1 & 0 & 0 & 0 & 0 & 1 & 0 & 1 & 0 & 0 & 1 & 0 & 0 & 0 & 0 & 0 & 0 & 0 & 0 & 0 & 1 & 1 & 1 & 0 & 0 & 0 & 0 & 1 & 0 & 0 & 1 & 1 & 1 & 1 & 0 & 1 & 0 & 1 & 0 & 0 \\
X_{53}&              0 & 1 & 0 & 0 & 0 & 0 & 0 & 0 & 1 & 1 & 1 & 1 & 0 & 0 & 1 & 1 & 0 & 0 & 0 & 0 & 1 & 1 & 0 & 0 & 0 & 0 & 1 & 0 & 0 & 0 & 0 & 1 & 0 & 1 & 0 & 0 & 1 & 1 & 1 & 1 & 1 & 1 & 1 & 0 & 0 & 0 & 1 & 0 & 0 & 0 & 0 & 0 & 0 & 1 & 0 & 0 & 2 & 1 & 0 & 0 & 1 & 1 & 0 & 2 & 1 & 0 \\
X_{67}&              0 & 0 & 1 & 0 & 0 & 0 & 0 & 0 & 0 & 0 & 1 & 0 & 0 & 0 & 0 & 1 & 0 & 0 & 0 & 1 & 1 & 0 & 0 & 1 & 0 & 0 & 0 & 0 & 1 & 1 & 1 & 0 & 0 & 0 & 0 & 0 & 0 & 0 & 0 & 0 & 1 & 0 & 0 & 1 & 0 & 1 & 0 & 0 & 0 & 0 & 1 & 0 & 1 & 1 & 0 & 1 & 0 & 1 & 1 & 0 & 0 & 0 & 1 & 1 & 1 & 1 \\
X_{69}&              0 & 1 & 0 & 0 & 0 & 0 & 0 & 0 & 1 & 1 & 1 & 0 & 0 & 1 & 1 & 1 & 0 & 0 & 0 & 0 & 1 & 0 & 0 & 1 & 0 & 0 & 0 & 0 & 0 & 0 & 1 & 0 & 1 & 1 & 1 & 1 & 1 & 0 & 0 & 1 & 1 & 0 & 0 & 1 & 0 & 1 & 0 & 0 & 0 & 0 & 1 & 0 & 0 & 0 & 0 & 1 & 1 & 2 & 1 & 1 & 0 & 0 & 0 & 0 & 1 & 2 \\
X_{78}&              0 & 0 & 1 & 0 & 0 & 0 & 0 & 0 & 0 & 1 & 0 & 0 & 0 & 0 & 1 & 0 & 0 & 1 & 1 & 0 & 0 & 0 & 1 & 0 & 0 & 1 & 1 & 1 & 0 & 0 & 0 & 0 & 0 & 0 & 1 & 1 & 1 & 0 & 0 & 0 & 0 & 0 & 0 & 0 & 0 & 0 & 1 & 1 & 1 & 0 & 0 & 0 & 0 & 0 & 1 & 0 & 1 & 1 & 1 & 1 & 0 & 0 & 0 & 0 & 1 & 1 \\
X_{7,10}&           0 & 1 & 0 & 0 & 0 & 0 & 0 & 1 & 1 & 1 & 0 & 0 & 1 & 1 & 1 & 0 & 0 & 0 & 1 & 0 & 0 & 0 & 1 & 0 & 0 & 0 & 0 & 1 & 0 & 0 & 0 & 1 & 1 & 1 & 1 & 1 & 1 & 0 & 1 & 1 & 0 & 0 & 1 & 0 & 0 & 0 & 1 & 0 & 1 & 0 & 0 & 0 & 0 & 0 & 0 & 0 & 2 & 1 & 0 & 2 & 1 & 1 & 0 & 0 & 0 & 1 \\
X_{86}&              0 & 1 & 0 & 0 & 0 & 0 & 1 & 1 & 1 & 0 & 0 & 1 & 1 & 1 & 0 & 0 & 1 & 0 & 0 & 0 & 0 & 1 & 0 & 0 & 1 & 0 & 0 & 0 & 0 & 0 & 0 & 1 & 1 & 1 & 0 & 0 & 0 & 1 & 1 & 1 & 0 & 1 & 1 & 0 & 1 & 0 & 0 & 0 & 0 & 1 & 0 & 1 & 0 & 0 & 0 & 0 & 1 & 0 & 0 & 1 & 2 & 2 & 1 & 1 & 0 & 0 \\
X_{89}&              0 & 0 & 1 & 0 & 0 & 0 & 0 & 0 & 1 & 0 & 0 & 0 & 0 & 1 & 0 & 0 & 1 & 0 & 0 & 0 & 1 & 1 & 0 & 1 & 1 & 0 & 0 & 0 & 0 & 0 & 1 & 0 & 1 & 1 & 0 & 0 & 0 & 0 & 0 & 1 & 0 & 0 & 0 & 0 & 0 & 0 & 0 & 0 & 0 & 1 & 1 & 1 & 0 & 0 & 0 & 1 & 1 & 1 & 0 & 1 & 1 & 1 & 0 & 0 & 0 & 1 \\
X_{97}&              0 & 1 & 0 & 0 & 0 & 0 & 1 & 1 & 0 & 0 & 1 & 1 & 1 & 0 & 0 & 1 & 0 & 0 & 0 & 1 & 0 & 0 & 0 & 0 & 0 & 0 & 0 & 0 & 1 & 1 & 0 & 1 & 0 & 0 & 0 & 0 & 0 & 1 & 1 & 0 & 1 & 1 & 1 & 1 & 1 & 1 & 0 & 0 & 0 & 0 & 0 & 0 & 1 & 1 & 0 & 0 & 0 & 0 & 1 & 0 & 1 & 1 & 2 & 2 & 1 & 0 \\
X_{9,10}&           0 & 0 & 1 & 0 & 0 & 0 & 0 & 1 & 0 & 0 & 0 & 0 & 1 & 0 & 0 & 0 & 0 & 0 & 1 & 1 & 0 & 0 & 1 & 0 & 0 & 0 & 0 & 1 & 1 & 1 & 0 & 1 & 0 & 0 & 0 & 0 & 0 & 0 & 1 & 0 & 0 & 0 & 1 & 0 & 0 & 0 & 1 & 0 & 1 & 0 & 0 & 0 & 1 & 1 & 0 & 0 & 1 & 0 & 0 & 1 & 1 & 1 & 1 & 1 & 0 & 0 \\
X_{10,6}&           0 & 0 & 1 & 0 & 0 & 0 & 1 & 0 & 0 & 0 & 0 & 1 & 0 & 0 & 0 & 0 & 1 & 1 & 0 & 0 & 0 & 1 & 0 & 0 & 1 & 1 & 1 & 0 & 0 & 0 & 0 & 0 & 0 & 0 & 0 & 0 & 0 & 1 & 0 & 0 & 0 & 1 & 0 & 0 & 1 & 0 & 0 & 1 & 0 & 1 & 0 & 1 & 0 & 0 & 1 & 0 & 0 & 0 & 1 & 0 & 1 & 1 & 1 & 1 & 1 & 0 \\
X_{10,8}&           0 & 1 & 0 & 0 & 0 & 0 & 1 & 0 & 0 & 1 & 1 & 1 & 0 & 0 & 1 & 1 & 0 & 1 & 0 & 0 & 0 & 0 & 0 & 0 & 0 & 1 & 1 & 0 & 0 & 0 & 0 & 0 & 0 & 0 & 1 & 1 & 1 & 1 & 0 & 0 & 1 & 1 & 0 & 1 & 1 & 1 & 0 & 1 & 0 & 0 & 0 & 0 & 0 & 0 & 1 & 0 & 0 & 1 & 2 & 0 & 0 & 0 & 1 & 1 & 2 & 1 \\
\end{array}
\right)
$}
.~\nn\\
\eea
where we have $4$ GLSM fields corresponding to the 4 extremal points of the toric diagram in \fref{f_ref_ex4c}(a). 
The $U(1)$ charges on GLSM fields corresponding to the $J$- and $E$-terms as well as the $D$-terms are given respectively as follows, 
\beal{es08a12}
&&
Q_{JE} =
\nn\\
&&
\resizebox{0.92\textwidth}{!}{$
\left(
\ba{cccc|cc|cccccccccc|ccccccccccccccc|ccccccccccccccc|cccccccccc|cccccccccc}
p_1 & p_2 & p_3 & p_4 & q_1 & q_2 & s_1 & s_2 & s_3 & s_4 & s_5 & s_6 & s_7 & s_8 & s_9 & s_{10} & w_1 & w_2 & w_3 & w_4 & w_5 & w_6 & w_7 & w_8 & w_9 & w_{10} & w_{11} & w_{12} & w_{13} & w_{14} & w_{15} & o_1 & o_2 & o_3 & o_4 & o_5 & o_6 & o_7 & o_8 & o_9 & o_{10} & o_{11} & o_{12} & o_{13} & o_{14} & o_{15} & u_{1} & u_{2} & u_{3} & u_{4} & u_{5} & u_{6} & u_{7} & u_{8} & u_{9} & u_{10} & v_{1} & v_{2} & v_{3} & v_{4} & v_{5} & v_{6} & v_{7} & v_{8} & v_{9} & v_{10} \\
\hline
 -1 & -2 & -1 & -1 & 0 & 0 & 0 & 1 & 1 & 1 & 0 & 1 & 0 & -1 & -1 & 0 & 0 & 0 & 0 & 0 & 0 & 0 & 0 & 0 & 0 & 0 & 0 & 0 & 0 & 0 & 0 & 0 & 0 & 0 & 0 & 0 & 0 & 0 & 0 & 0 & 0 & 0 & 0 & 0 & 0 & 0 & 0 & 0 & 0 & 0 & 0 & 0 & 0 & 0 & 0 & 0 & 0 & 0 & 0 & 0 & 0 & 0 & 0 & 0 & 0 & 1 \\
 -2 & -2 & -1 & -2 & 2 & 0 & -1 & -1 & 0 & 0 & 0 & 1 & 2 & 1 & 0 & 0 & 0 & 0 & 0 & 0 & 0 & 0 & 0 & 0 & 0 & 0 & 0 & 0 & 0 & 0 & 0 & 0 & 0 & 0 & 0 & 0 & 0 & 0 & 0 & 0 & 0 & 0 & 0 & 0 & 0 & 0 & 0 & 0 & 0 & 0 & 0 & 0 & 0 & 0 & 0 & 0 & 0 & 0 & 0 & 0 & 0 & 0 & 0 & 0 & 1 & 0 \\
 -2 & -2 & -1 & -2 & 2 & 0 & 0 & -1 & -1 & 0 & 0 & 0 & 1 & 2 & 1 & 0 & 0 & 0 & 0 & 0 & 0 & 0 & 0 & 0 & 0 & 0 & 0 & 0 & 0 & 0 & 0 & 0 & 0 & 0 & 0 & 0 & 0 & 0 & 0 & 0 & 0 & 0 & 0 & 0 & 0 & 0 & 0 & 0 & 0 & 0 & 0 & 0 & 0 & 0 & 0 & 0 & 0 & 0 & 0 & 0 & 0 & 0 & 0 & 1 & 0 & 0 \\
 -1 & -2 & -1 & -1 & 0 & 0 & -1 & -1 & 0 & 1 & 0 & 1 & 1 & 1 & 0 & 0 & 0 & 0 & 0 & 0 & 0 & 0 & 0 & 0 & 0 & 0 & 0 & 0 & 0 & 0 & 0 & 0 & 0 & 0 & 0 & 0 & 0 & 0 & 0 & 0 & 0 & 0 & 0 & 0 & 0 & 0 & 0 & 0 & 0 & 0 & 0 & 0 & 0 & 0 & 0 & 0 & 0 & 0 & 0 & 0 & 0 & 0 & 1 & 0 & 0 & 0 \\
 -1 & 1 & 0 & -1 & 1 & 0 & 0 & 0 & -1 & -1 & 0 & -1 & -1 & 0 & 1 & 0 & 0 & 0 & 0 & 0 & 0 & 0 & 0 & 0 & 0 & 0 & 0 & 0 & 0 & 0 & 0 & 0 & 0 & 0 & 0 & 0 & 0 & 0 & 0 & 0 & 0 & 0 & 0 & 0 & 0 & 0 & 0 & 0 & 0 & 0 & 0 & 0 & 0 & 0 & 0 & 0 & 0 & 0 & 0 & 0 & 0 & 1 & 0 & 0 & 0 & 0 \\
 -1 & 1 & 0 & -1 & 1 & 0 & -1 & -2 & -2 & -1 & 0 & 0 & 1 & 1 & 1 & 0 & 0 & 0 & 0 & 0 & 0 & 0 & 0 & 0 & 0 & 0 & 0 & 0 & 0 & 0 & 0 & 0 & 0 & 0 & 0 & 0 & 0 & 0 & 0 & 0 & 0 & 0 & 0 & 0 & 0 & 0 & 0 & 0 & 0 & 0 & 0 & 0 & 0 & 0 & 0 & 0 & 0 & 0 & 0 & 0 & 1 & 0 & 0 & 0 & 0 & 0 \\
 0 & 1 & 0 & 0 & -1 & 0 & 1 & 1 & 1 & 0 & 0 & -1 & -2 & -2 & -1 & 0 & 0 & 0 & 0 & 0 & 0 & 0 & 0 & 0 & 0 & 0 & 0 & 0 & 0 & 0 & 0 & 0 & 0 & 0 & 0 & 0 & 0 & 0 & 0 & 0 & 0 & 0 & 0 & 0 & 0 & 0 & 0 & 0 & 0 & 0 & 0 & 0 & 0 & 0 & 0 & 0 & 0 & 0 & 0 & 1 & 0 & 0 & 0 & 0 & 0 & 0 \\
 0 & -2 & -1 & 0 & -2 & 0 & 0 & 1 & 2 & 1 & 0 & 0 & 0 & -1 & -1 & 0 & 0 & 0 & 0 & 0 & 0 & 0 & 0 & 0 & 0 & 0 & 0 & 0 & 0 & 0 & 0 & 0 & 0 & 0 & 0 & 0 & 0 & 0 & 0 & 0 & 0 & 0 & 0 & 0 & 0 & 0 & 0 & 0 & 0 & 0 & 0 & 0 & 0 & 0 & 0 & 0 & 0 & 0 & 1 & 0 & 0 & 0 & 0 & 0 & 0 & 0 \\
 0 & -2 & -1 & 0 & -2 & 0 & 1 & 2 & 1 & 0 & 0 & 0 & -1 & -1 & 0 & 0 & 0 & 0 & 0 & 0 & 0 & 0 & 0 & 0 & 0 & 0 & 0 & 0 & 0 & 0 & 0 & 0 & 0 & 0 & 0 & 0 & 0 & 0 & 0 & 0 & 0 & 0 & 0 & 0 & 0 & 0 & 0 & 0 & 0 & 0 & 0 & 0 & 0 & 0 & 0 & 0 & 0 & 1 & 0 & 0 & 0 & 0 & 0 & 0 & 0 & 0 \\
 0 & 1 & 0 & 0 & -1 & 0 & 1 & 0 & -1 & -1 & 0 & -1 & -1 & 0 & 0 & 0 & 0 & 0 & 0 & 0 & 0 & 0 & 0 & 0 & 0 & 0 & 0 & 0 & 0 & 0 & 0 & 0 & 0 & 0 & 0 & 0 & 0 & 0 & 0 & 0 & 0 & 0 & 0 & 0 & 0 & 0 & 0 & 0 & 0 & 0 & 0 & 0 & 0 & 0 & 0 & 0 & 1 & 0 & 0 & 0 & 0 & 0 & 0 & 0 & 0 & 0 \\
 -1 & -2 & -1 & -1 & 0 & 0 & 0 & 1 & 0 & 1 & 0 & 1 & 0 & 0 & 0 & 0 & 0 & 0 & 0 & 0 & 0 & 0 & 0 & 0 & 0 & 0 & 0 & 0 & 0 & 0 & 0 & 0 & 0 & 0 & 0 & 0 & 0 & 0 & 0 & 0 & 0 & 0 & 0 & 0 & 0 & 0 & 0 & 0 & 0 & 0 & 0 & 0 & 0 & 0 & 0 & 1 & 0 & 0 & 0 & 0 & 0 & 0 & 0 & 0 & 0 & 0 \\
 -1 & 1 & 0 & -1 & 1 & 0 & -1 & -1 & 0 & -1 & 0 & 0 & 1 & 0 & 0 & 0 & 0 & 0 & 0 & 0 & 0 & 0 & 0 & 0 & 0 & 0 & 0 & 0 & 0 & 0 & 0 & 0 & 0 & 0 & 0 & 0 & 0 & 0 & 0 & 0 & 0 & 0 & 0 & 0 & 0 & 0 & 0 & 0 & 0 & 0 & 0 & 0 & 0 & 0 & 1 & 0 & 0 & 0 & 0 & 0 & 0 & 0 & 0 & 0 & 0 & 0 \\
 -1 & -2 & -1 & -1 & 0 & 0 & 1 & 0 & 0 & 1 & 0 & 0 & 0 & 1 & 0 & 0 & 0 & 0 & 0 & 0 & 0 & 0 & 0 & 0 & 0 & 0 & 0 & 0 & 0 & 0 & 0 & 0 & 0 & 0 & 0 & 0 & 0 & 0 & 0 & 0 & 0 & 0 & 0 & 0 & 0 & 0 & 0 & 0 & 0 & 0 & 0 & 0 & 0 & 1 & 0 & 0 & 0 & 0 & 0 & 0 & 0 & 0 & 0 & 0 & 0 & 0 \\
 -1 & -2 & -1 & -1 & 0 & 0 & 0 & 0 & 0 & 1 & 0 & 1 & 0 & 1 & 0 & 0 & 0 & 0 & 0 & 0 & 0 & 0 & 0 & 0 & 0 & 0 & 0 & 0 & 0 & 0 & 0 & 0 & 0 & 0 & 0 & 0 & 0 & 0 & 0 & 0 & 0 & 0 & 0 & 0 & 0 & 0 & 0 & 0 & 0 & 0 & 0 & 0 & 1 & 0 & 0 & 0 & 0 & 0 & 0 & 0 & 0 & 0 & 0 & 0 & 0 & 0 \\
 -1 & 1 & 0 & -1 & 1 & 0 & -1 & 0 & -1 & -1 & 0 & 0 & 0 & 0 & 1 & 0 & 0 & 0 & 0 & 0 & 0 & 0 & 0 & 0 & 0 & 0 & 0 & 0 & 0 & 0 & 0 & 0 & 0 & 0 & 0 & 0 & 0 & 0 & 0 & 0 & 0 & 0 & 0 & 0 & 0 & 0 & 0 & 0 & 0 & 0 & 0 & 1 & 0 & 0 & 0 & 0 & 0 & 0 & 0 & 0 & 0 & 0 & 0 & 0 & 0 & 0 \\
 -1 & -2 & -1 & -1 & 0 & 0 & 0 & 1 & 0 & 0 & 0 & 1 & 0 & 0 & 1 & 0 & 0 & 0 & 0 & 0 & 0 & 0 & 0 & 0 & 0 & 0 & 0 & 0 & 0 & 0 & 0 & 0 & 0 & 0 & 0 & 0 & 0 & 0 & 0 & 0 & 0 & 0 & 0 & 0 & 0 & 0 & 0 & 0 & 0 & 0 & 1 & 0 & 0 & 0 & 0 & 0 & 0 & 0 & 0 & 0 & 0 & 0 & 0 & 0 & 0 & 0 \\
 -1 & 1 & 0 & -1 & 1 & 0 & -1 & -1 & -1 & -1 & 0 & 0 & 1 & 0 & 1 & 0 & 0 & 0 & 0 & 0 & 0 & 0 & 0 & 0 & 0 & 0 & 0 & 0 & 0 & 0 & 0 & 0 & 0 & 0 & 0 & 0 & 0 & 0 & 0 & 0 & 0 & 0 & 0 & 0 & 0 & 0 & 0 & 0 & 0 & 1 & 0 & 0 & 0 & 0 & 0 & 0 & 0 & 0 & 0 & 0 & 0 & 0 & 0 & 0 & 0 & 0 \\
 0 & 1 & 0 & 0 & -1 & 0 & 1 & 0 & 1 & 0 & 0 & -1 & -1 & -1 & -1 & 0 & 0 & 0 & 0 & 0 & 0 & 0 & 0 & 0 & 0 & 0 & 0 & 0 & 0 & 0 & 0 & 0 & 0 & 0 & 0 & 0 & 0 & 0 & 0 & 0 & 0 & 0 & 0 & 0 & 0 & 0 & 0 & 0 & 1 & 0 & 0 & 0 & 0 & 0 & 0 & 0 & 0 & 0 & 0 & 0 & 0 & 0 & 0 & 0 & 0 & 0 \\
 0 & 1 & 0 & 0 & -1 & 0 & 0 & 0 & 1 & 0 & 0 & -1 & 0 & -1 & -1 & 0 & 0 & 0 & 0 & 0 & 0 & 0 & 0 & 0 & 0 & 0 & 0 & 0 & 0 & 0 & 0 & 0 & 0 & 0 & 0 & 0 & 0 & 0 & 0 & 0 & 0 & 0 & 0 & 0 & 0 & 0 & 0 & 1 & 0 & 0 & 0 & 0 & 0 & 0 & 0 & 0 & 0 & 0 & 0 & 0 & 0 & 0 & 0 & 0 & 0 & 0 \\
 0 & 1 & 0 & 0 & -1 & 0 & 1 & 0 & 0 & 0 & 0 & -1 & -1 & 0 & -1 & 0 & 0 & 0 & 0 & 0 & 0 & 0 & 0 & 0 & 0 & 0 & 0 & 0 & 0 & 0 & 0 & 0 & 0 & 0 & 0 & 0 & 0 & 0 & 0 & 0 & 0 & 0 & 0 & 0 & 0 & 0 & 1 & 0 & 0 & 0 & 0 & 0 & 0 & 0 & 0 & 0 & 0 & 0 & 0 & 0 & 0 & 0 & 0 & 0 & 0 & 0 \\
 -1 & -3 & -1 & -1 & 0 & 0 & 0 & 1 & 1 & 1 & 0 & 1 & 0 & 0 & 0 & 0 & 0 & 0 & 0 & 0 & 0 & 0 & 0 & 0 & 0 & 0 & 0 & 0 & 0 & 0 & 0 & 0 & 0 & 0 & 0 & 0 & 0 & 0 & 0 & 0 & 0 & 0 & 0 & 0 & 0 & 1 & 0 & 0 & 0 & 0 & 0 & 0 & 0 & 0 & 0 & 0 & 0 & 0 & 0 & 0 & 0 & 0 & 0 & 0 & 0 & 0 \\
 -1 & 0 & 0 & -1 & 1 & 0 & -1 & -1 & 0 & 0 & 0 & 0 & 1 & 0 & 0 & 0 & 0 & 0 & 0 & 0 & 0 & 0 & 0 & 0 & 0 & 0 & 0 & 0 & 0 & 0 & 0 & 0 & 0 & 0 & 0 & 0 & 0 & 0 & 0 & 0 & 0 & 0 & 0 & 0 & 1 & 0 & 0 & 0 & 0 & 0 & 0 & 0 & 0 & 0 & 0 & 0 & 0 & 0 & 0 & 0 & 0 & 0 & 0 & 0 & 0 & 0 \\
 -1 & -3 & -1 & -1 & 0 & 0 & 0 & 0 & 1 & 1 & 0 & 1 & 1 & 0 & 0 & 0 & 0 & 0 & 0 & 0 & 0 & 0 & 0 & 0 & 0 & 0 & 0 & 0 & 0 & 0 & 0 & 0 & 0 & 0 & 0 & 0 & 0 & 0 & 0 & 0 & 0 & 0 & 0 & 1 & 0 & 0 & 0 & 0 & 0 & 0 & 0 & 0 & 0 & 0 & 0 & 0 & 0 & 0 & 0 & 0 & 0 & 0 & 0 & 0 & 0 & 0 \\
 -1 & 0 & 0 & -1 & 1 & 0 & 0 & -1 & -1 & 0 & 0 & 0 & 0 & 1 & 0 & 0 & 0 & 0 & 0 & 0 & 0 & 0 & 0 & 0 & 0 & 0 & 0 & 0 & 0 & 0 & 0 & 0 & 0 & 0 & 0 & 0 & 0 & 0 & 0 & 0 & 0 & 0 & 1 & 0 & 0 & 0 & 0 & 0 & 0 & 0 & 0 & 0 & 0 & 0 & 0 & 0 & 0 & 0 & 0 & 0 & 0 & 0 & 0 & 0 & 0 & 0 \\
 -1 & 0 & 0 & -1 & 1 & 0 & -1 & -1 & -1 & 0 & 0 & 0 & 1 & 1 & 0 & 0 & 0 & 0 & 0 & 0 & 0 & 0 & 0 & 0 & 0 & 0 & 0 & 0 & 0 & 0 & 0 & 0 & 0 & 0 & 0 & 0 & 0 & 0 & 0 & 0 & 0 & 1 & 0 & 0 & 0 & 0 & 0 & 0 & 0 & 0 & 0 & 0 & 0 & 0 & 0 & 0 & 0 & 0 & 0 & 0 & 0 & 0 & 0 & 0 & 0 & 0 \\
 -1 & -3 & -1 & -1 & 0 & 0 & 0 & 0 & 0 & 1 & 0 & 1 & 1 & 1 & 0 & 0 & 0 & 0 & 0 & 0 & 0 & 0 & 0 & 0 & 0 & 0 & 0 & 0 & 0 & 0 & 0 & 0 & 0 & 0 & 0 & 0 & 0 & 0 & 0 & 0 & 1 & 0 & 0 & 0 & 0 & 0 & 0 & 0 & 0 & 0 & 0 & 0 & 0 & 0 & 0 & 0 & 0 & 0 & 0 & 0 & 0 & 0 & 0 & 0 & 0 & 0 \\
 -1 & 0 & 0 & -1 & 1 & 0 & 0 & 0 & -1 & -1 & 0 & 0 & 0 & 0 & 1 & 0 & 0 & 0 & 0 & 0 & 0 & 0 & 0 & 0 & 0 & 0 & 0 & 0 & 0 & 0 & 0 & 0 & 0 & 0 & 0 & 0 & 0 & 0 & 0 & 1 & 0 & 0 & 0 & 0 & 0 & 0 & 0 & 0 & 0 & 0 & 0 & 0 & 0 & 0 & 0 & 0 & 0 & 0 & 0 & 0 & 0 & 0 & 0 & 0 & 0 & 0 \\
 -1 & 0 & 0 & -1 & 1 & 0 & 0 & -1 & -1 & -1 & 0 & 0 & 0 & 1 & 1 & 0 & 0 & 0 & 0 & 0 & 0 & 0 & 0 & 0 & 0 & 0 & 0 & 0 & 0 & 0 & 0 & 0 & 0 & 0 & 0 & 0 & 0 & 0 & 1 & 0 & 0 & 0 & 0 & 0 & 0 & 0 & 0 & 0 & 0 & 0 & 0 & 0 & 0 & 0 & 0 & 0 & 0 & 0 & 0 & 0 & 0 & 0 & 0 & 0 & 0 & 0 \\
 -1 & 0 & 0 & -1 & 1 & 0 & -1 & -1 & -1 & -1 & 0 & 0 & 1 & 1 & 1 & 0 & 0 & 0 & 0 & 0 & 0 & 0 & 0 & 0 & 0 & 0 & 0 & 0 & 0 & 0 & 0 & 0 & 0 & 0 & 0 & 0 & 0 & 1 & 0 & 0 & 0 & 0 & 0 & 0 & 0 & 0 & 0 & 0 & 0 & 0 & 0 & 0 & 0 & 0 & 0 & 0 & 0 & 0 & 0 & 0 & 0 & 0 & 0 & 0 & 0 & 0 \\
 0 & 0 & 0 & 0 & -1 & 0 & 1 & 1 & 1 & 0 & 0 & -1 & -1 & -1 & -1 & 0 & 0 & 0 & 0 & 0 & 0 & 0 & 0 & 0 & 0 & 0 & 0 & 0 & 0 & 0 & 0 & 0 & 0 & 0 & 0 & 0 & 1 & 0 & 0 & 0 & 0 & 0 & 0 & 0 & 0 & 0 & 0 & 0 & 0 & 0 & 0 & 0 & 0 & 0 & 0 & 0 & 0 & 0 & 0 & 0 & 0 & 0 & 0 & 0 & 0 & 0 \\
 0 & 0 & 0 & 0 & -1 & 0 & 0 & 1 & 1 & 0 & 0 & 0 & -1 & -1 & -1 & 0 & 0 & 0 & 0 & 0 & 0 & 0 & 0 & 0 & 0 & 0 & 0 & 0 & 0 & 0 & 0 & 0 & 0 & 0 & 0 & 1 & 0 & 0 & 0 & 0 & 0 & 0 & 0 & 0 & 0 & 0 & 0 & 0 & 0 & 0 & 0 & 0 & 0 & 0 & 0 & 0 & 0 & 0 & 0 & 0 & 0 & 0 & 0 & 0 & 0 & 0 \\
 0 & 0 & 0 & 0 & -1 & 0 & 0 & 0 & 1 & 0 & 0 & 0 & 0 & -1 & -1 & 0 & 0 & 0 & 0 & 0 & 0 & 0 & 0 & 0 & 0 & 0 & 0 & 0 & 0 & 0 & 0 & 0 & 0 & 0 & 1 & 0 & 0 & 0 & 0 & 0 & 0 & 0 & 0 & 0 & 0 & 0 & 0 & 0 & 0 & 0 & 0 & 0 & 0 & 0 & 0 & 0 & 0 & 0 & 0 & 0 & 0 & 0 & 0 & 0 & 0 & 0 \\
 0 & 0 & 0 & 0 & -1 & 0 & 1 & 1 & 0 & 0 & 0 & -1 & -1 & -1 & 0 & 0 & 0 & 0 & 0 & 0 & 0 & 0 & 0 & 0 & 0 & 0 & 0 & 0 & 0 & 0 & 0 & 0 & 0 & 1 & 0 & 0 & 0 & 0 & 0 & 0 & 0 & 0 & 0 & 0 & 0 & 0 & 0 & 0 & 0 & 0 & 0 & 0 & 0 & 0 & 0 & 0 & 0 & 0 & 0 & 0 & 0 & 0 & 0 & 0 & 0 & 0 \\
 0 & 0 & 0 & 0 & -1 & 0 & 0 & 1 & 0 & 0 & 0 & 0 & -1 & -1 & 0 & 0 & 0 & 0 & 0 & 0 & 0 & 0 & 0 & 0 & 0 & 0 & 0 & 0 & 0 & 0 & 0 & 0 & 1 & 0 & 0 & 0 & 0 & 0 & 0 & 0 & 0 & 0 & 0 & 0 & 0 & 0 & 0 & 0 & 0 & 0 & 0 & 0 & 0 & 0 & 0 & 0 & 0 & 0 & 0 & 0 & 0 & 0 & 0 & 0 & 0 & 0 \\
 0 & 0 & 0 & 0 & -1 & 0 & 1 & 0 & 0 & 0 & 0 & -1 & -1 & 0 & 0 & 0 & 0 & 0 & 0 & 0 & 0 & 0 & 0 & 0 & 0 & 0 & 0 & 0 & 0 & 0 & 0 & 1 & 0 & 0 & 0 & 0 & 0 & 0 & 0 & 0 & 0 & 0 & 0 & 0 & 0 & 0 & 0 & 0 & 0 & 0 & 0 & 0 & 0 & 0 & 0 & 0 & 0 & 0 & 0 & 0 & 0 & 0 & 0 & 0 & 0 & 0 \\
 -1 & -2 & -1 & -1 & 1 & 0 & 0 & 0 & 0 & 0 & 0 & 1 & 1 & 0 & 1 & 0 & 0 & 0 & 0 & 0 & 0 & 0 & 0 & 0 & 0 & 0 & 0 & 0 & 0 & 0 & 1 & 0 & 0 & 0 & 0 & 0 & 0 & 0 & 0 & 0 & 0 & 0 & 0 & 0 & 0 & 0 & 0 & 0 & 0 & 0 & 0 & 0 & 0 & 0 & 0 & 0 & 0 & 0 & 0 & 0 & 0 & 0 & 0 & 0 & 0 & 0 \\
 -1 & -2 & -1 & -1 & 1 & 0 & 0 & 0 & 0 & 0 & 0 & 1 & 0 & 1 & 1 & 0 & 0 & 0 & 0 & 0 & 0 & 0 & 0 & 0 & 0 & 0 & 0 & 0 & 0 & 1 & 0 & 0 & 0 & 0 & 0 & 0 & 0 & 0 & 0 & 0 & 0 & 0 & 0 & 0 & 0 & 0 & 0 & 0 & 0 & 0 & 0 & 0 & 0 & 0 & 0 & 0 & 0 & 0 & 0 & 0 & 0 & 0 & 0 & 0 & 0 & 0 \\
 -1 & -2 & -1 & -1 & 1 & 0 & 0 & -1 & 0 & 0 & 0 & 1 & 1 & 1 & 1 & 0 & 0 & 0 & 0 & 0 & 0 & 0 & 0 & 0 & 0 & 0 & 0 & 0 & 1 & 0 & 0 & 0 & 0 & 0 & 0 & 0 & 0 & 0 & 0 & 0 & 0 & 0 & 0 & 0 & 0 & 0 & 0 & 0 & 0 & 0 & 0 & 0 & 0 & 0 & 0 & 0 & 0 & 0 & 0 & 0 & 0 & 0 & 0 & 0 & 0 & 0 \\
 0 & 1 & 0 & 0 & 0 & 0 & 0 & 0 & 0 & 0 & 0 & 0 & -1 & 0 & -1 & 0 & 0 & 0 & 0 & 0 & 0 & 0 & 0 & 0 & 0 & 0 & 0 & 1 & 0 & 0 & 0 & 0 & 0 & 0 & 0 & 0 & 0 & 0 & 0 & 0 & 0 & 0 & 0 & 0 & 0 & 0 & 0 & 0 & 0 & 0 & 0 & 0 & 0 & 0 & 0 & 0 & 0 & 0 & 0 & 0 & 0 & 0 & 0 & 0 & 0 & 0 \\
 0 & 1 & 0 & 0 & 0 & 0 & 0 & 0 & 0 & 0 & 0 & -1 & 0 & 0 & -1 & 0 & 0 & 0 & 0 & 0 & 0 & 0 & 0 & 0 & 0 & 0 & 1 & 0 & 0 & 0 & 0 & 0 & 0 & 0 & 0 & 0 & 0 & 0 & 0 & 0 & 0 & 0 & 0 & 0 & 0 & 0 & 0 & 0 & 0 & 0 & 0 & 0 & 0 & 0 & 0 & 0 & 0 & 0 & 0 & 0 & 0 & 0 & 0 & 0 & 0 & 0 \\
 0 & 1 & 0 & 0 & 0 & 0 & -1 & 0 & 0 & 0 & 0 & 0 & 0 & 0 & -1 & 0 & 0 & 0 & 0 & 0 & 0 & 0 & 0 & 0 & 0 & 1 & 0 & 0 & 0 & 0 & 0 & 0 & 0 & 0 & 0 & 0 & 0 & 0 & 0 & 0 & 0 & 0 & 0 & 0 & 0 & 0 & 0 & 0 & 0 & 0 & 0 & 0 & 0 & 0 & 0 & 0 & 0 & 0 & 0 & 0 & 0 & 0 & 0 & 0 & 0 & 0 \\
 0 & 1 & 0 & 0 & 0 & 0 & 0 & 0 & 0 & 0 & 0 & -1 & 0 & -1 & 0 & 0 & 0 & 0 & 0 & 0 & 0 & 0 & 0 & 0 & 1 & 0 & 0 & 0 & 0 & 0 & 0 & 0 & 0 & 0 & 0 & 0 & 0 & 0 & 0 & 0 & 0 & 0 & 0 & 0 & 0 & 0 & 0 & 0 & 0 & 0 & 0 & 0 & 0 & 0 & 0 & 0 & 0 & 0 & 0 & 0 & 0 & 0 & 0 & 0 & 0 & 0 \\
 0 & -2 & -1 & 0 & -1 & 0 & 1 & 1 & 1 & 1 & 0 & 0 & 0 & -1 & 0 & 0 & 0 & 0 & 0 & 0 & 0 & 0 & 0 & 1 & 0 & 0 & 0 & 0 & 0 & 0 & 0 & 0 & 0 & 0 & 0 & 0 & 0 & 0 & 0 & 0 & 0 & 0 & 0 & 0 & 0 & 0 & 0 & 0 & 0 & 0 & 0 & 0 & 0 & 0 & 0 & 0 & 0 & 0 & 0 & 0 & 0 & 0 & 0 & 0 & 0 & 0 \\
 0 & 1 & 0 & 0 & 0 & 0 & 0 & 0 & 0 & -1 & 0 & 0 & -1 & 0 & 0 & 0 & 0 & 0 & 0 & 0 & 0 & 0 & 1 & 0 & 0 & 0 & 0 & 0 & 0 & 0 & 0 & 0 & 0 & 0 & 0 & 0 & 0 & 0 & 0 & 0 & 0 & 0 & 0 & 0 & 0 & 0 & 0 & 0 & 0 & 0 & 0 & 0 & 0 & 0 & 0 & 0 & 0 & 0 & 0 & 0 & 0 & 0 & 0 & 0 & 0 & 0 \\
 0 & 1 & 0 & 0 & 0 & 0 & 0 & 0 & -1 & 0 & 0 & -1 & 0 & 0 & 0 & 0 & 0 & 0 & 0 & 0 & 0 & 1 & 0 & 0 & 0 & 0 & 0 & 0 & 0 & 0 & 0 & 0 & 0 & 0 & 0 & 0 & 0 & 0 & 0 & 0 & 0 & 0 & 0 & 0 & 0 & 0 & 0 & 0 & 0 & 0 & 0 & 0 & 0 & 0 & 0 & 0 & 0 & 0 & 0 & 0 & 0 & 0 & 0 & 0 & 0 & 0 \\
 0 & -2 & -1 & 0 & -1 & 0 & 1 & 1 & 0 & 1 & 0 & 0 & 0 & 0 & 0 & 0 & 0 & 0 & 0 & 0 & 1 & 0 & 0 & 0 & 0 & 0 & 0 & 0 & 0 & 0 & 0 & 0 & 0 & 0 & 0 & 0 & 0 & 0 & 0 & 0 & 0 & 0 & 0 & 0 & 0 & 0 & 0 & 0 & 0 & 0 & 0 & 0 & 0 & 0 & 0 & 0 & 0 & 0 & 0 & 0 & 0 & 0 & 0 & 0 & 0 & 0 \\
 0 & -2 & -1 & 0 & -1 & 0 & 1 & 0 & 1 & 1 & 0 & 0 & 0 & 0 & 0 & 0 & 0 & 0 & 0 & 1 & 0 & 0 & 0 & 0 & 0 & 0 & 0 & 0 & 0 & 0 & 0 & 0 & 0 & 0 & 0 & 0 & 0 & 0 & 0 & 0 & 0 & 0 & 0 & 0 & 0 & 0 & 0 & 0 & 0 & 0 & 0 & 0 & 0 & 0 & 0 & 0 & 0 & 0 & 0 & 0 & 0 & 0 & 0 & 0 & 0 & 0 \\
 0 & 1 & 0 & 0 & 0 & 0 & 0 & -1 & 0 & -1 & 0 & 0 & 0 & 0 & 0 & 0 & 0 & 0 & 1 & 0 & 0 & 0 & 0 & 0 & 0 & 0 & 0 & 0 & 0 & 0 & 0 & 0 & 0 & 0 & 0 & 0 & 0 & 0 & 0 & 0 & 0 & 0 & 0 & 0 & 0 & 0 & 0 & 0 & 0 & 0 & 0 & 0 & 0 & 0 & 0 & 0 & 0 & 0 & 0 & 0 & 0 & 0 & 0 & 0 & 0 & 0 \\
 0 & 1 & 0 & 0 & 0 & 0 & -1 & 0 & 0 & -1 & 0 & 0 & 0 & 0 & 0 & 0 & 0 & 1 & 0 & 0 & 0 & 0 & 0 & 0 & 0 & 0 & 0 & 0 & 0 & 0 & 0 & 0 & 0 & 0 & 0 & 0 & 0 & 0 & 0 & 0 & 0 & 0 & 0 & 0 & 0 & 0 & 0 & 0 & 0 & 0 & 0 & 0 & 0 & 0 & 0 & 0 & 0 & 0 & 0 & 0 & 0 & 0 & 0 & 0 & 0 & 0 \\
 0 & 1 & 0 & 0 & 0 & 0 & -1 & 0 & -1 & 0 & 0 & 0 & 0 & 0 & 0 & 0 & 1 & 0 & 0 & 0 & 0 & 0 & 0 & 0 & 0 & 0 & 0 & 0 & 0 & 0 & 0 & 0 & 0 & 0 & 0 & 0 & 0 & 0 & 0 & 0 & 0 & 0 & 0 & 0 & 0 & 0 & 0 & 0 & 0 & 0 & 0 & 0 & 0 & 0 & 0 & 0 & 0 & 0 & 0 & 0 & 0 & 0 & 0 & 0 & 0 & 0 \\
 -1 & -3 & -1 & -1 & 1 & 0 & 0 & 0 & 0 & 0 & 0 & 1 & 1 & 1 & 1 & 1 & 0 & 0 & 0 & 0 & 0 & 0 & 0 & 0 & 0 & 0 & 0 & 0 & 0 & 0 & 0 & 0 & 0 & 0 & 0 & 0 & 0 & 0 & 0 & 0 & 0 & 0 & 0 & 0 & 0 & 0 & 0 & 0 & 0 & 0 & 0 & 0 & 0 & 0 & 0 & 0 & 0 & 0 & 0 & 0 & 0 & 0 & 0 & 0 & 0 & 0 \\
 0 & -3 & -1 & 0 & -1 & 0 & 1 & 1 & 1 & 1 & 1 & 0 & 0 & 0 & 0 & 0 & 0 & 0 & 0 & 0 & 0 & 0 & 0 & 0 & 0 & 0 & 0 & 0 & 0 & 0 & 0 & 0 & 0 & 0 & 0 & 0 & 0 & 0 & 0 & 0 & 0 & 0 & 0 & 0 & 0 & 0 & 0 & 0 & 0 & 0 & 0 & 0 & 0 & 0 & 0 & 0 & 0 & 0 & 0 & 0 & 0 & 0 & 0 & 0 & 0 & 0 \\
 -1 & 0 & 0 & -1 & 1 & 1 & 0 & 0 & 0 & 0 & 0 & 0 & 0 & 0 & 0 & 0 & 0 & 0 & 0 & 0 & 0 & 0 & 0 & 0 & 0 & 0 & 0 & 0 & 0 & 0 & 0 & 0 & 0 & 0 & 0 & 0 & 0 & 0 & 0 & 0 & 0 & 0 & 0 & 0 & 0 & 0 & 0 & 0 & 0 & 0 & 0 & 0 & 0 & 0 & 0 & 0 & 0 & 0 & 0 & 0 & 0 & 0 & 0 & 0 & 0 & 0 \\
\ea
\right)
$}
~,~\nn\\
&&
Q_D =
\nn\\
&&
\resizebox{0.92\textwidth}{!}{$
\left(
\ba{cccc|cc|cccccccccc|ccccccccccccccc|ccccccccccccccc|cccccccccc|cccccccccc}
p_1 & p_2 & p_3 & p_4 & q_1 & q_2 & s_1 & s_2 & s_3 & s_4 & s_5 & s_6 & s_7 & s_8 & s_9 & s_{10} & w_1 & w_2 & w_3 & w_4 & w_5 & w_6 & w_7 & w_8 & w_9 & w_{10} & w_{11} & w_{12} & w_{13} & w_{14} & w_{15} & o_1 & o_2 & o_3 & o_4 & o_5 & o_6 & o_7 & o_8 & o_9 & o_{10} & o_{11} & o_{12} & o_{13} & o_{14} & o_{15} & u_{1} & u_{2} & u_{3} & u_{4} & u_{5} & u_{6} & u_{7} & u_{8} & u_{9} & u_{10} & v_{1} & v_{2} & v_{3} & v_{4} & v_{5} & v_{6} & v_{7} & v_{8} & v_{9} & v_{10} \\
\hline
 0 & 0 & 0 & 0 & 0 & 0 & 1 & 0 & 0 & 0 & 0 & 0 & 0 & 0 & 0 & -1 & 0 & 0 & 0 & 0 & 0 & 0 & 0 & 0 & 0 & 0 & 0 & 0 & 0 & 0 & 0 & 0 & 0 & 0 & 0 & 0 & 0 & 0 & 0 & 0 & 0 & 0 & 0 & 0 & 0 & 0 & 0 & 0 & 0 & 0 & 0 & 0 & 0 & 0 & 0 & 0 & 0 & 0 & 0 & 0 & 0 & 0 & 0 & 0 & 0 & 0 \\
 0 & 0 & 0 & 0 & 0 & 0 & 0 & 1 & 0 & 0 & 0 & 0 & 0 & 0 & 0 & -1 & 0 & 0 & 0 & 0 & 0 & 0 & 0 & 0 & 0 & 0 & 0 & 0 & 0 & 0 & 0 & 0 & 0 & 0 & 0 & 0 & 0 & 0 & 0 & 0 & 0 & 0 & 0 & 0 & 0 & 0 & 0 & 0 & 0 & 0 & 0 & 0 & 0 & 0 & 0 & 0 & 0 & 0 & 0 & 0 & 0 & 0 & 0 & 0 & 0 & 0 \\
 0 & 0 & 0 & 0 & 0 & 0 & 0 & 0 & 1 & 0 & 0 & 0 & 0 & 0 & 0 & -1 & 0 & 0 & 0 & 0 & 0 & 0 & 0 & 0 & 0 & 0 & 0 & 0 & 0 & 0 & 0 & 0 & 0 & 0 & 0 & 0 & 0 & 0 & 0 & 0 & 0 & 0 & 0 & 0 & 0 & 0 & 0 & 0 & 0 & 0 & 0 & 0 & 0 & 0 & 0 & 0 & 0 & 0 & 0 & 0 & 0 & 0 & 0 & 0 & 0 & 0 \\
 0 & 0 & 0 & 0 & 0 & 0 & 0 & 0 & 0 & 1 & 0 & 0 & 0 & 0 & 0 & -1 & 0 & 0 & 0 & 0 & 0 & 0 & 0 & 0 & 0 & 0 & 0 & 0 & 0 & 0 & 0 & 0 & 0 & 0 & 0 & 0 & 0 & 0 & 0 & 0 & 0 & 0 & 0 & 0 & 0 & 0 & 0 & 0 & 0 & 0 & 0 & 0 & 0 & 0 & 0 & 0 & 0 & 0 & 0 & 0 & 0 & 0 & 0 & 0 & 0 & 0 \\
 0 & 0 & 0 & 0 & 0 & 0 & 0 & 0 & 0 & 0 & 1 & 0 & 0 & 0 & 0 & -1 & 0 & 0 & 0 & 0 & 0 & 0 & 0 & 0 & 0 & 0 & 0 & 0 & 0 & 0 & 0 & 0 & 0 & 0 & 0 & 0 & 0 & 0 & 0 & 0 & 0 & 0 & 0 & 0 & 0 & 0 & 0 & 0 & 0 & 0 & 0 & 0 & 0 & 0 & 0 & 0 & 0 & 0 & 0 & 0 & 0 & 0 & 0 & 0 & 0 & 0 \\
 0 & 0 & 0 & 0 & 0 & 0 & 0 & 0 & 0 & 0 & 0 & 1 & 0 & 0 & 0 & -1 & 0 & 0 & 0 & 0 & 0 & 0 & 0 & 0 & 0 & 0 & 0 & 0 & 0 & 0 & 0 & 0 & 0 & 0 & 0 & 0 & 0 & 0 & 0 & 0 & 0 & 0 & 0 & 0 & 0 & 0 & 0 & 0 & 0 & 0 & 0 & 0 & 0 & 0 & 0 & 0 & 0 & 0 & 0 & 0 & 0 & 0 & 0 & 0 & 0 & 0 \\
 0 & 0 & 0 & 0 & 0 & 0 & 0 & 0 & 0 & 0 & 0 & 0 & 1 & 0 & 0 & -1 & 0 & 0 & 0 & 0 & 0 & 0 & 0 & 0 & 0 & 0 & 0 & 0 & 0 & 0 & 0 & 0 & 0 & 0 & 0 & 0 & 0 & 0 & 0 & 0 & 0 & 0 & 0 & 0 & 0 & 0 & 0 & 0 & 0 & 0 & 0 & 0 & 0 & 0 & 0 & 0 & 0 & 0 & 0 & 0 & 0 & 0 & 0 & 0 & 0 & 0 \\
 0 & 0 & 0 & 0 & 0 & 0 & 0 & 0 & 0 & 0 & 0 & 0 & 0 & 1 & 0 & -1 & 0 & 0 & 0 & 0 & 0 & 0 & 0 & 0 & 0 & 0 & 0 & 0 & 0 & 0 & 0 & 0 & 0 & 0 & 0 & 0 & 0 & 0 & 0 & 0 & 0 & 0 & 0 & 0 & 0 & 0 & 0 & 0 & 0 & 0 & 0 & 0 & 0 & 0 & 0 & 0 & 0 & 0 & 0 & 0 & 0 & 0 & 0 & 0 & 0 & 0 \\
 0 & 0 & 0 & 0 & 0 & 0 & 0 & 0 & 0 & 0 & 0 & 0 & 0 & 0 & 1 & -1 & 0 & 0 & 0 & 0 & 0 & 0 & 0 & 0 & 0 & 0 & 0 & 0 & 0 & 0 & 0 & 0 & 0 & 0 & 0 & 0 & 0 & 0 & 0 & 0 & 0 & 0 & 0 & 0 & 0 & 0 & 0 & 0 & 0 & 0 & 0 & 0 & 0 & 0 & 0 & 0 & 0 & 0 & 0 & 0 & 0 & 0 & 0 & 0 & 0 & 0 \\
\ea
\right)
$}
~.~\nn\\
\eea
The corresponding toric diagram in \fref{f_ref_ex4c}(a) is then given by, 
\beal{es08a13}
&&
G_t =
\nn\\
&&
\resizebox{0.92 \textwidth}{!}{$
\left(
\ba{cccc|cc|cccccccccc|ccccccccccccccc|ccccccccccccccc|cccccccccc|cccccccccc}
p_1 & p_2 & p_3 & p_4 & q_1 & q_2 & s_1 & s_2 & s_3 & s_4 & s_5 & s_6 & s_7 & s_8 & s_9 & s_{10} & w_1 & w_2 & w_3 & w_4 & w_5 & w_6 & w_7 & w_8 & w_9 & w_{10} & w_{11} & w_{12} & w_{13} & w_{14} & w_{15} & o_1 & o_2 & o_3 & o_4 & o_5 & o_6 & o_7 & o_8 & o_9 & o_{10} & o_{11} & o_{12} & o_{13} & o_{14} & o_{15} & u_{1} & u_{2} & u_{3} & u_{4} & u_{5} & u_{6} & u_{7} & u_{8} & u_{9} & u_{10} & v_{1} & v_{2} & v_{3} & v_{4} & v_{5} & v_{6} & v_{7} & v_{8} & v_{9} & v_{10} \\
\hline
 -1 & 1 & -1 & -3 & -2 & -2 & 0 & 0 & 0 & 0 & 0 & 0 & 0 & 0 & 0 & 0 & -1 & -1 & -1 & -1 & -1 & -1 & -1 & -1 & -1 & -1 & -1 & -1 & -1 & -1 & -1 & -2 & -2 & -2 & -2 & -2 & -2 & -2 & -2 & -2 & -2 & -2 & -2 & -2 & -2 & -2 & -3 & -3 & -3 & -3 & -3 & -3 & -3 & -3 & -3 & -3 & -3 & -3 & -3 & -3
   & -3 & -3 & -3 & -3 & -3 & -3 \\
 -1 & 0 & 0 & 1 & 0 & 0 & 0 & 0 & 0 & 0 & 0 & 0 & 0 & 0 & 0 & 0 & 0 & 0 & 0 & 0 & 0 & 0 & 0 & 0 & 0 & 0 & 0 & 0 & 0 & 0 & 0 & 0 & 0 & 0 & 0 & 0 & 0 & 0 & 0 & 0 & 0 & 0 & 0 & 0 & 0 & 0 & 0 & 0 & 0 & 0 & 0 & 0 & 0 & 0 & 0 & 0 & 0 & 0 & 0 & 0 & 0 & 0 & 0 & 0 & 0 & 0 \\
  -1 & 0 & 1 & -1 & -1 & -1 & 0 & 0 & 0 & 0 & 0 & 0 & 0 & 0 & 0 & 0 & 0 & 0 & 0 & 0 & 0 & 0 & 0 & 0 & 0 & 0 & 0 & 0 & 0 & 0 & 0 & -1 & -1 & -1 & -1 & -1 & -1 & -1 & -1 & -1 & -1 & -1 & -1 & -1 & -1 & -1 & -1 & -1 & -1 & -1 & -1 & -1 & -1 & -1 & -1 & -1 & -1 & -1 & -1 & -1 & -1 & -1 & -1 &
   -1 & -1 & -1 \\
 \hline
 1 & 1 & 1 & 1 & 1 & 1 & 1 & 1 & 1 & 1 & 1 & 1 & 1 & 1 & 1 & 1 & 1 & 1 & 1 & 1 & 1 & 1 & 1 & 1 & 1 & 1 & 1 & 1 & 1 & 1 & 1 & 2 & 2 & 2 & 2 & 2 & 2 & 2 & 2 & 2 & 2 & 2 & 2 & 2 & 2 & 2 & 2 & 2 & 2 & 2 & 2 & 2 & 2 & 2 & 2 & 2 & 3 & 3 & 3 & 3 & 3 & 3 & 3 & 3 & 3 & 3 \\
\end{array}
\right)
$}
~,~
\nn\\
\eea
where we note that we have in total 35 extra GLSM fields \cite{Franco:2015tna}, which do not correspond to any points in the toric diagram for $P_{+-}^{1}[\mathbb{C}^3/\mathbb{Z}_5~(1,1,3)]$.

From the $Q_{JE}$ and $Q_D$ charge matrices, we note that the global symmetry of the brane brick model for $P_{+-}^{1}[\mathbb{C}^3/\mathbb{Z}_5~(1,1,3)]$ is enhanced to the following form, 
\beal{es08a14}
SU(2)_x \times U(1)_{f_1} \times U(1)_{f_2} \times U(1)_R ~,~
\eea
where $SU(2)_x \times U(1)_{f_1} \times U(1)_{f_2}$ is the mesonic flavor symmetry. 
The charges on GLSM fields due to the global symmetry above are summarized in \tref{tab_40}.

\begin{table}[ht!]
\centering
\begin{tabular}{|c|c|c|c|c|c|}
\hline
\; & $SU(2)_x$ & $U(1)_{f_1}$ & $U(1)_{f_2}$ & $U(1)_R$ & fugacity \\
\hline
$p_1$ & $ +1 $ & $ 0 $ & $ +1 $ & $ r_1 $ & $t_1= x f_2 \overline{t}_1 $ \\
$p_2$ & $ 0 $ & $ +1 $ & $ -2 $ & $ r_2 $ & $t_2 = f_1 f_2^{-2} \overline{t}_2 $\\
$p_3$ & $ 0 $ & $ -1 $ & $ 0 $ & $ 2 r_1 $ & $t_3= f_1^{-1} \overline{t}_1^2$ \\
$p_4$ & $ -1 $ & $ 0 $ & $ +1 $ & $ r_1 $ & $t_4= x^{-1} f_2 \overline{t}_1 $ \\
\hline
\end{tabular}
\caption{Charges under the global symmetry of the $P_{+-}^{1}[\mathbb{C}^3/\mathbb{Z}_5~(1,1,3)]$ model on the extremal GLSM fields $p_a$.
Here, $U(1)_R$ charges $r_1$, $r_2$ and $r_3$ are chosen such that 
the $J$- and $E$-terms coupled to Fermi fields have an overall $U(1)_R$ charge of $2$
with $4r_1 + r_2 =2$.  \label{tab_40}}
\end{table}

The Hilbert series of the mesonic moduli space $\mathcal{M}^{mes}_{P_{+-}^{1}[\mathbb{C}^3/\mathbb{Z}_5~(1,1,3)]}$ can be obtained using the Molien integral formula in \eref{es01a15}. The Hilbert series takes the following form,
\beal{es08a15}
&&
g(t_a,y_q,y_s,y_w,y_o,y_u,y_v;\mathcal{M}^{mes}_{{P_{+-}^{1}[\mathbb{C}^3/\mathbb{Z}_5~(1,1,3)]}})
=
\nn\\
&& 
\hspace{0.2cm}
\frac{
P(t_a,y_q,y_s,y_w,y_o,y_u,y_v;\mathcal{M}^{mes}_{{P_{+-}^{1}[\mathbb{C}^3/\mathbb{Z}_5~(1,1,3)]} })
}{
(1-y_q^5 y_s y_w^2 y_o^6 y_u^7 y_v^8 t_1^{10}) (1-y_s^3 y_w y_o^3 y_u y_v^4 t_2^5)
(1-y_s y_w^2 y_o y_u^2 y_v^3 t_3^5) (1-y_q^5 y_s y_w^2 y_o^6 y_u^7 y_v^8 t_4^{10})
}
~,~
\nn\\
\eea
where $t_a$ are the fugacities associated to the extremal GLSM fields $p_a$ 
and $y_q$,$y_s$,$y_w$,$y_o$,$y_u$,$y_v$ correspond to products of GLSM fields $q_1q_2$ , $s_1\dots s_{10}$, $w_1 \dots w_{15}$ , $o_1 \dots o_{15}$, $u_1 \dots u_{10}$, $v_1 \dots v_{10}$, respectively. 
The numerator $P(t_a,y_q,y_s,y_w,y_o,y_u,y_v;\mathcal{M}^{mes}_{P_{+-}^{1}[\mathbb{C}^3/\mathbb{Z}_5~(1,1,3)] })$ in \eref{es08a15}
is presented in appendix \sref{app_num_07}.

Using the following fugacity map,
\beal{es08a15b}
\overline{t}_1=t_1^{1/4} t_4^{1/2}~,~
\overline{t}_2=\frac{t_2 t_3}{t_4}~,~
x=\frac{t_1^{1/2}}{t_4^{1/2}}~,~
f_1=\frac{t_1^{1/2} t_4}{t_3}~,~
f_2=t_1^{1/4}
~,~
\eea
where 
$\overline{t}_1, \overline{t}_2$ correspond respectively to $U(1)_R$ charges $r_1, r_2$ defined in \tref{tab_40},
and $x$ corresponds to $SU(2)_x$ and $f_1, f_2$ correspond to $U(1)_{f_1}$ and $U(1)_{f_2}$ of the mesonic flavor symmetry, respectively, 
we can rewrite the Hilbert series in \eref{es08a15}
as a character expansion in terms of irreducible representations of $SU(2)_x$.
By additionally setting
$y_q=y_s=y_w=y_o=y_u=y_v=1$, we have the following highest weight generating function \cite{Hanany:2014dia} given by, 
\beal{es08a16}
&&
h(\overline{t}_1,\overline{t}_2,\mu,f_1,f_2;\mathcal{M}^{mes}_{P_{+-}^{1}[\mathbb{C}^3/\mathbb{Z}_5~(1,1,3)]})
=
\frac{1}{(1 - f_1^{-5} \overline{t}_1^{10}) (1 - f_1^5 f_2^{-10} \overline{t}_2^5) (1 -\mu^{10}  f_2^{10}  \overline{t}_1^{10})} \nn \\ 
&& 
\hspace{1cm}
\times
\big (1 + f_1^{-4} f_2^2 \mu^2 \overline{t}_1^{10}  +f_1^{-3} f_2^4 \mu^4 \overline{t}_1^{10}  
 +f_1^{-2} f_2^6 \mu^6 \overline{t}_1^{10} + f_1^{-1} f_2^8 \mu^8 \overline{t}_1^{10} \nn \\ 
&& 
\hspace{1.5cm}
 +f_1 f_2^{-2} \overline{t}_1^4 \overline{t}_2 + \mu^2 \overline{t}_1^4 \overline{t}_2 + f_1 f_2^2 \mu^4 \overline{t}_1^4 \overline{t}_2 + 
 f_1^{-3} f_2^4 \mu^6 \overline{t}_1^{14} \overline{t}_2  +f_1^{-2}  f_2^6 \mu^8 \overline{t}_1^{14} \overline{t}_2  \nn \\ 
&&
\hspace{1.5cm}
+ f_1^2 f_2^{-4}  \overline{t}_1^8 \overline{t}_2^2 +  f_1 f_2^{-2} \mu^2 \overline{t}_1^8 \overline{t}_2^2+ 
 \mu^4 \overline{t}_1^8 \overline{t}_2^2 + f_1 f_2^2 \mu^6 \overline{t}_1^8 \overline{t}_2^2 + f_1^2 f_2^4 \mu^8 \overline{t}_1^8 \overline{t}_2^2  \nn \\ 
&& 
\hspace{1.5cm}
+ f_1^2 f_2^{-6} \overline{t}_1^2 \overline{t}_2^3 + f_1^3 f_2^{-4} \mu^2 \overline{t}_1^2 \overline{t}_2^3  + 
f_1 f_2^{-2}  \mu^4 \overline{t}_1^{12} \overline{t}_2^3 + \mu^6 \overline{t}_1^{12} \overline{t}_2^3 + 
 f_1 f_2^2 \mu^8 \overline{t}_1^{12} \overline{t}_2^3 \nn \\ 
&& 
\hspace{1.5cm}
+ f_1 f_2^{-8} \overline{t}_1^6 \overline{t}_2^4 + 
 f_1^2 f_2^{-6} \mu^2 \overline{t}_1^6 \overline{t}_2^4  + f_1^3 f_2^{-4} \mu^4 \overline{t}_1^6 \overline{t}_2^4 + 
 f_1^4 f_2^{-2} \mu^6 \overline{t}_1^6 \overline{t}_2^4  + \mu^8 \overline{t}_1^{16} \overline{t}_2^4
\big ) 
~,~
\nn\\
\eea
where $\mu^m$ counts characters of the form $[m]_x$ corresponding to irreducible representations of $SU(2)_x$ with highest weight $(m)$.

The plethystic logarithm of the refined Hilbert series corresponding to 
\eref{es08a16} is given by,
\beal{es08a17}
&&
PL[g(\overline{t}_1,\overline{t}_2,x,f_1,f_2;\mathcal{M}^{mes}_{P_{+-}^{1}[\mathbb{C}^3/\mathbb{Z}_5~(1,1,3)]})]
=
f_1^5 f_2^{-10} \overline{t}_2^5
+f_1^2 f_2^{-6} \overline{t}_1^2 \overline{t}_2^3
+ [2]_x   f_1^3 f_2^{-4} \overline{t}_1^2 \overline{t}_2^3 
\nn\\
&&
\hspace{1cm}
+f_1^{-1} f_2^{-2} \overline{t}_1^4 \overline{t}_2
+ [4]_x  f_1 f_2^2  \overline{t}_1^4 \overline{t}_2 
+ [2]_x  \overline{t}_1^4 \overline{t}_2 
+ [10]_x  f_2^{10}  \overline{t}_1^{10} 
+ [8]_x  f_1^{-1} f_2^8  \overline{t}_1^{10} 
 \nn\\
 &&
 \hspace{1cm}
+ [6]_x  f_1^{-2} f_2^6  \overline{t}_1^{10} 
+ [4]_x  f_1^{-3} f_2^4  \overline{t}_1^{10}  
+ [2]_x  f_1^{-4} f_2^2  \overline{t}_1^{10}  
+ f_1^{-5} \overline{t}_1^{10}
- (f_1^3 f_2^{-4} \overline{t}_1^6 \overline{t}_2^4
+\overline{t}_1^8 \overline{t}_2^2
 \nn\\
 &&
 \hspace{1cm}
 +f_1^2 f_2^4 \overline{t}_1^8 \overline{t}_2^2
+[2]_x  f_1 f_2^2 \overline{t}_1^8 \overline{t}_2^2 
+[4]_x  f_1^4 f_2^{-2} \overline{t}_1^6 \overline{t}_2^4 
+[4]_x  f_1^3 f_2^{-4}  \overline{t}_1^6 \overline{t}_2^4 
+[4]_x  f_1^2 f_2^4  \overline{t}_1^8 \overline{t}_2^2 
 \nn\\
 &&
 \hspace{1cm} 
+[4]_x   f_1 f_2^2  \overline{t}_1^8 \overline{t}_2^2 
+[4]_x  \overline{t}_1^8 \overline{t}_2^2 
+[2]_x  f_1^4 f_2^{-2}  \overline{t}_1^6 \overline{t}_2^4 
+[2]_x  f_1^3 f_2^{-4}  \overline{t}_1^6 \overline{t}_2^4 
+[2]_x  f_1^2 f_2^{-6}  \overline{t}_1^6 \overline{t}_2^4)
+
\dots
~,~
\nn\\
\eea
where we can see that the mesonic moduli space $\mathcal{M}^{mes}_{P_{+-}^{1}[\mathbb{C}^3/\mathbb{Z}_5~(1,1,3)]}$
is not a complete intersection. 
We can extract the generators of 
$\mathcal{M}^{mes}_{P_{+-}^{1}[\mathbb{C}^3/\mathbb{Z}_5~(1,1,3)]}$ from the first positive terms in the plethystic logarithm. 
The generators are listed in \tref{tab_100}
with their corresponding mesonic flavor symmetry charges.
\\

\begin{table}[H]
\centering
\resizebox{0.55\textwidth}{!}{
\begin{tabular}{|c|c|c|c|c|c|c|}
\hline
PL term  & GLSM fields & $SU(2)_{x}$ & $U(1)_{f_1}$ & $U(1)_{f_2}$  & fugacity\\
\hline\hline
\multirow{1}{*}{$+f_1^5 f_2^{-10} \overline{t}_2^5$}&$p_2^5 w s^3 o^3 u v^4$ & $0$ &$+5$&$-10$ & $f_1^5 f_2^{-10} \overline{t}_2^5$\\
\hline
\multirow{1}{*}{$+f_1^2 f_2^{-6} \overline{t}_1^2 \overline{t}_3^3$}&  $p_2^3 p_3 w s^2 o^2 u v^3$ & $0$ &$+2$&$-6$ & $f_1^2 f_2^{-6}\overline{t}_1^2 \overline{t}_3^3$\\
\hline
\multirow{3}{*}{$+[2]_x f_1^3 f_2^{-4} \overline{t}_1^2 \overline{t}_2^3$}& $p_1^2 p_2^3 q w s^2 o^3 u^2 v^4$ & $+2$ &$+3$&$-4$ & $x^2 f_1^3 f_2^{-4} \overline{t}_1^2 \overline{t}_2^3$\\
& $p_1 p_2^3 p_4 q w s^2 o^3 u^2 v^4$ & $0$ &$+3$&$-4$ & $f_1^3 f_2^{-4} \overline{t}_1^2 \overline{t}_2^3$\\
& $p_2^3 p_4^2 q w s^2 o^3 u^2 v^4$ & $-2$ &$+3$&$-4$ & $x^{-2} f_1^3 f_2^{-4} \overline{t}_1^2 \overline{t}_2^3$\\
\hline
\multirow{1}{*}{$+f_1^{-1} f_2^{-2} \overline{t}_1^4 \overline{t}_2$}&  $ p_2 p_3^2 w s o u v^2$ & $0$ &$-1$&$-2$ & $f_1^{-1} f_2^{-2} \overline{t}_1^4 \overline{t}_2$\\
\hline
\multirow{3}{*}{$+[2]_x \overline{t}_1^4 \overline{t}_2$}& $p_1^2 p_2 p_3 q w s o^2 u^2 v^3$ & $+2$ &$0$&$0$ & $x^2\overline{t}_1^4 \overline{t}_2$\\
& $p_1 p_2 p_3 p_4 q w s o^2 u^2 v^3$ & $0$ &$0$&$0$ & $\overline{t}_1^4 \overline{t}_2$\\
& $p_2 p_3 p_4^2 q w s o^2 u^2 v^3$ & $-2$ &$0$&$0$ & $x^{-2} \overline{t}_1^4 \overline{t}_2$\\
\hline
\multirow{5}{*}{$+[4]_x f_1 f_2^2 \overline{t}_1^4 \overline{t}_2$}& $p_1^4 p_2 q^2 w s o^3 u^3 v^4$ & $+4$ &$+1$&$+2$ & $x^4 f_1 f_2^2  \overline{t}_1^4 \overline{t}_2$\\
&$p_1^3 p_2 p_4 w s o^3 u^3 v^4$ & $+2$  &$+1$&$+2$ & $x^2 f_1 f_2^2  \overline{t}_1^4 \overline{t}_2$\\
&$p_1^2 p_2 p_4^2 w s o^3 u^3 v^4$ & $0$  &$+1$&$+2$ & $f_1 f_2^2  \overline{t}_1^4 \overline{t}_2$\\
&$p_1 p_2 p_4^3 w s o^3 u^3 v^4$ & $-2$  &$+1$&$+2$ & $x^{-2} f_1 f_2^2  \overline{t}_1^4 \overline{t}_2$\\
&$p_2 p_4^4 w s o^3 u^3 v^4$ & $-4$  &$+1$&$+2$ & $x^{-4} f_1 f_2^2  \overline{t}_1^4 \overline{t}_2$\\
\hline
\multirow{11}{*}{$+[10]_x f_2^{10} \overline{t}_1^{10}$}&$p_1^{10} q^5 w^2 s o^6 u^7 v^8$ & $+10$ &$0$&$+10$ & $x^{10} f_2^{10} \overline{t}_1^{10}$\\
&$p_1^9 p_4 q^5 w^2 s o^6 u^7 v^8$ & $+8$ &$0$&$+10$ & $x^8 f_2^{10}\overline{t}_1^{10}$\\
&$p_1^8 p_4^2 q^5 w^2 s o^6 u^7 v^8$ & $+6$ &$0$&$+10$  & $x^6 f_2^{10}\overline{t}_1^{10}$\\
&$p_1^7 p_4^3 q^5 w^2 s o^6 u^7 v^8$ & $+4$ &$0$&$+10$  & $x^4 f_2^{10}\overline{t}_1^{10}$\\
&$ p_1^6 p_4^4 q^5 w^2 s o^6 u^7 v^8$ & $+2$ &$0$&$+10$ & $x^2 f_2^{10}\overline{t}_1^{10}$\\
&$ p_1^5 p_4^5 q^5 w^2 s o^6 u^7 v^8$ & $0$ &$0$&$+10$ & $f_2^{10}\overline{t}_1^{10}$\\
&$ p_1^4 p_4^6 q^5 w^2 s o^6 u^7 v^8$ & $-2$ &$0$&$+10$  & $x^{-2} f_2^{10}\overline{t}_1^{10}$\\
&$p_1^3 p_4^7 q^5 w^2 s o^6 u^7 v^8$ & $-4$ &$0$&$+10$ & $x^{-4}f_2^{10}\overline{t}_1^{10}$\\
&$p_1^2 p_4^8 q^5 w^2 s o^6 u^7 v^8$ & $-6$ &$0$&$+10$  & $x^{-6}f_2^{10}\overline{t}_1^{10}$\\
&$p_1 p_4^9 q^5 w^2 s o^6 u^7 v^8$ & $-8$ &$0$&$+10$ & $x^{-8}f_2^{10}\overline{t}_1^{10}$\\
&$p_4^{10} q^5 w^2 s o^6 u^7 v^8$ & $-10$ &$0$&$+10$ & $x^{-10}f_2^{10}\overline{t}_1^{10}$\\
\hline
\multirow{9}{*}{$+[8]_x f_1^{-1} f_2^8 \overline{t}_1^{10}$}& $p_1^8 p_3 q^4 w^2 s o^5 u^6 v^7$ & $+8$ &$-1$&$+8$ & $x^8 f_1^{-1} f_2^8 \overline{t}_1^{10}$\\
&$p_1^7 p_3 p_4 q^4 w^2 s o^5 u^6 v^7$ & $+6$ &$-1$&$+8$ & $x^6 f_1^{-1} f_2^8\overline{t}_1^{10}$\\
&$p_1^6 p_3 p_4^2 q^4 w^2 s o^5 u^6 v^7$ & $+4$ &$-1$&$+8$  & $x^4 f_1^{-1} f_2^8\overline{t}_1^{10}$\\
&$p_1^5 p_3 p_4^3 q^4 w^2 s o^5 u^6 v^7$ & $+2$ &$-1$&$+8$ & $x^2 f_1^{-1} f_2^8\overline{t}_1^{10}$\\
&$p_1^4 p_3 p_4^4 q^4 w^2 s o^5 u^6 v^7$ & $0$ &$-1$&$+8$ & $f_1^{-1} f_2^8\overline{t}_1^{10}$\\
&$p_1^3 p_3 p_4^5 q^4 w^2 s o^5 u^6 v^7$ & $-2$ &$-1$&$+8$ & $x^{-2} f_1^{-1} f_2^8\overline{t}_1^{10}$\\
&$p_1^2 p_3 p_4^6 q^4 w^2 s o^5 u^6 v^7$ & $-4$ &$-1$&$+8$ & $x^{-4} f_1^{-1} f_2^8\overline{t}_1^{10}$\\
&$p_1 p_3 p_4^7 q^4 w^2 s o^5 u^6 v^7$ & $-6$ &$-1$&$+8$ & $x^{-6} f_1^{-1} f_2^8\overline{t}_1^{10}$\\
&$p_3 p_4^8 q^4 w^2 s o^5 u^6 v^7$ & $-8$ &$-1$&$+8$ & $x^{-8} f_1^{-1} f_2^8\overline{t}_1^{10}$\\
\hline
\multirow{7}{*}{$+[6]_x f_1^{-2} f_2^6 \overline{t}_1^{10}$}&$p_1^6 p_3^2 q^3 w^2 s o^4 u^5 v^6$ & $+6$ &$-2$&$+6$ & $x^6 f_1^{-2} f_2^6\overline{t}_1^{10}$\\
&$p_1^5 p_3^2 p_4 q^3 w^2 s o^4 u^5 v^6$ & $+4$ &$-2$&$+6$ & $x^4 f_1^{-2} f_2^6\overline{t}_1^{10}$\\
&$ p_1^4 p_3^2 p_4^2 q^3 w^2 s o^4 u^5 v^6$ & $+2$ &$-2$&$+6$  & $x^2 f_1^{-2} f_2^6\overline{t}_1^{10}$\\
&$p_1^3 p_3^2 p_4^3 q^3 w^2 s o^4 u^5 v^6$ & $0$ &$-2$&$+6$ & $f_1^{-2} f_2^6\overline{t}_1^{10}$\\
&$p_1^2 p_3^2 p_4^4 q^3 w^2 s o^4 u^5 v^6$ & $-2$ &$-2$&$+6$ & $x^{-2} f_1^{-2} f_2^6 \overline{t}_1^{10}$\\
&$p_1 p_3^2 p_4^5 q^3 w^2 s o^4 u^5 v^6$ & $-4$ &$-2$&$+6$ & $x^{-4} f_1^{-2} f_2^6 \overline{t}_1^{10}$\\
&$p_3^2 p_4^6 q^3 w^2 s o^4 u^5 v^6$ & $-6$ &$-2$&$+6$ & $x^{-6} f_1^{-2} f_2^6 \overline{t}_1^{10}$\\
\hline
\multirow{5}{*}{$+[4]_x f_1^{-3} f_2^4 \overline{t}_1^{10}$}&$p_1^4 p_3^3 q^2 w^2 s o^3 u^4 v^5$ & $+4$ &$-3$&$+4$ & $x^4f_1^{-3} f_2^4 \overline{t}_1^{10}$\\
&$p_1^3 p_3^3 p_4 q^2 w^2 s o^3 u^4 v^5$ & $+2$  &$-3$&$+4$ & $x^2f_1^{-3} f_2^4 \overline{t}_1^{10}$\\
&$p_1^2 p_3^3 p_4^2 q^2 w^2 s o^3 u^4 v^5$ & $0$  &$-3$&$+4$ & $f_1^{-3} f_2^4 \overline{t}_1^{10}$\\
&$p_1 p_3^3 p_4^3 q^2 w^2 s o^3 u^4 v^5$ & $-2$  &$-3$&$+4$ & $x^{-2} f_1^{-3} f_2^4 \overline{t}_1^{10}$\\
&$p_3^3 p_4^4 q^2 w^2 s o^3 u^4 v^5$ & $-4$  &$-3$&$+4$ & $x^{-4} f_1^{-3} f_2^4 \overline{t}_1^{10}$\\
\hline
\multirow{3}{*}{$+[2]_x f_1^{-4} f_2^2 \overline{t}_1^{10}$}&$p_1^2 p_3^4 q w^2 s o^2 u^3 v^4$ & $+2$ &$-4$&$+2$ & $x^2 f_1^{-4} f_2^2 \overline{t}_1^{10}$\\
&$p_1 p_3^4 p_4 q w^2 s o^2 u^3 v^4$ & $0$ &$-4$&$+2$ & $f_1^{-4} f_2^2 \overline{t}_1^{10}$\\
&$p_3^4 p_4^2 q w^2 s o^2 u^3 v^4$ & $-2$ &$-4$&$+2$ & $x^{-2} f_1^{-4} f_2^2 \overline{t}_1^{10}$\\
\hline
\multirow{1}{*}{$+f_1^{-5} \overline{t}_1^{10}$}&$p_3^5 w^2 s o u^2 v^3$ & $0$ &$-5$&$0$ & $f_1^{-5} \overline{t}_1^{10}$\\
\hline
\end{tabular}}
\caption{
Generators of the mesonic moduli space $\mathcal{M}^{mes}_{P_{+-}^{1}[\mathbb{C}^3/\mathbb{Z}_5~(1,1,3)]}$ of the $P_{+-}^{1}[\mathbb{C}^3/\mathbb{Z}_5~(1,1,3)]$ brane brick model in terms of chiral fields and GLSM fields with corresponding mesonic flavor symmetry charges. Here, $q$, $s$, $r$, $o$, $u$ and $v$ denote products of GLSM fields $q_1 q_2$, $s_1 \dots s_{10}$, $w_1 \dots w_{15}$, $o_1 \dots o_{15}$, $u_1 \dots u_{10}$ and $v_1 \dots v_{10}$, respectively.
\label{tab_100}
}
\end{table}

\paragraph{Mass Deformation and the Brane Brick Model for $P_{+-}^{2}[\mathbb{C}^3/\mathbb{Z}_5~(1,1,3)]$.} 
We can introduce mass terms to the $E$-terms of the $P_{+-}^{1}[\mathbb{C}^3/\mathbb{Z}_5~(1,1,3)]$ model in \eref{es08a10},
which take the following form,
\beal{es08a18}
\resizebox{0.63\textwidth}{!}{$
\begin{array}{rrrclcrcl}
& & & J   & &  & E & \textcolor{blue}{+}  & \textcolor{blue}{\Delta E} \\
 \Lambda_{12} : & \ \ \  & X_{23} \cdot X_{31} 
 &-&   X_{25} \cdot X_{51} & \ \ \ \  & \textcolor{blue}{- X_{12}}+ Q_{17} \cdot P_{72} &-& K_{17} \cdot V_{72} \\ 
 \Lambda_{23} : & \ \ \  & X_{34} \cdot X_{42} 
 &-&   X_{31} \cdot X_{12} & \ \ \ \  & \textcolor{blue}{ - X_{23}}+ Q_{28} \cdot P_{83} &-& K_{28} \cdot V_{83} \\
 \Lambda_{34} : & \ \ \  & X_{42} \cdot X_{23} 
 &-&   X_{45} \cdot X_{53} & \ \ \ \  & \textcolor{blue}{+ X_{34}}+ K_{39} \cdot V_{94}  &-&  Q_{39} \cdot P_{94} \\ 
 \Lambda_{45} : & \ \ \  & X_{51} \cdot X_{14} 
 &-&   X_{53} \cdot X_{34} & \ \ \ \  & \textcolor{blue}{- X_{45}}+ Q_{4,10} \cdot P_{10,5}  &-& K_{4,10} \cdot V_{10,5}  \\
 \Lambda_{51} : & \ \ \  & X_{12} \cdot X_{25} 
 &-&   X_{14} \cdot X_{45} & \ \ \ \  & \textcolor{blue}{- X_{51}}+ Q_{56} \cdot P_{61}  &-&   K_{56} \cdot V_{61}  \\
 \Lambda_{67} : & \ \ \  & X_{7,10} \cdot X_{10,6} 
 &-&   X_{78} \cdot X_{89} & \ \ \ \  & \textcolor{blue}{- X_{67}}+  P_{61} \cdot Q_{17} &-&  V_{61} \cdot K_{17}   \\
 \Lambda_{78} : & \ \ \  & X_{89} \cdot X_{97} 
 &-&   X_{86} \cdot X_{67} & \ \ \ \  & \textcolor{blue}{+ X_{78}}+ V_{72} \cdot K_{28}  &-&  P_{72} \cdot Q_{28} \\
 \Lambda_{89} : & \ \ \  & X_{97} \cdot X_{78} 
 &-&   X_{9,10} \cdot X_{10,8} & \ \ \ \  & \textcolor{blue}{-X_{89}}+ P_{83} \cdot Q_{39} &-&  V_{83} \cdot K_{39} \\
 \Lambda_{9,10} : & \ \ \  & X_{10,8} \cdot X_{89} 
 &-&   X_{10,6} \cdot X_{69} & \ \ \ \  & \textcolor{blue}{ - X_{9,10}}+ P_{94} \cdot Q_{4,10} &-& V_{94} \cdot K_{4,10}  \\
 \Lambda_{10,6} : & \ \ \  & X_{69} \cdot X_{9,10} 
 &-&   X_{67} \cdot X_{7,10} & \ \ \ \  & \textcolor{blue}{ - X_{10,6}}+ P_{10,5} \cdot Q_{56} &-&   V_{10,5} \cdot K_{56}  \\ 
\end{array} 
$}
 ~.~
\eea
Here, we note that the original $E$-terms above consist of chiral fields that form extremal brick matchings $p_{1}$ and $p_{4}$
as summarized in the $P$-matrix in \eref{es08a11}.
The mass terms introduced in \eref{es08a18} consist of chiral fields that form the extremal brick matching $p_{3}$.
Based on the $U(1)_R$ charges of these extremal brick matchings as summarized in \tref{tab_40}, 
we note that the mass terms introduced in \eref{es08a18} satisfy the following overall $U(1)_R$ charge constraint on the original $E$-terms, 
\beal{es08a19b}
r[p_{3}] = r[p_{1}] + r[p_{4}] = 2 r_1~.~
\eea

By integrating out the mass terms in \eref{es08a18}, we obtain the $J$- and $E$-terms corresponding to the $P_{+-}^{2}[\mathbb{C}^3/\mathbb{Z}_5~(1,1,3)]$ brane brick model given by,
\beal{es08a19}
\resizebox{0.72\textwidth}{!}{$
\begin{array}{rrrclcrcl}
& & & J  & & &  & E  \\
\Lambda^{1}_{62}:   & \ \ \ \ &  K_{28}\cdot X_{86}&-&   X_{25}\cdot K_{56}& \ \ \ \ & P_{61}\cdot Q_{17}\cdot V_{72}&-&   V_{61}\cdot Q_{17}\cdot P_{72} \\
 \Lambda^{1}_{73}:   & \ \ \ \ &  X_{31}\cdot K_{17}&-&   K_{39}\cdot X_{97}& \ \ \ \ &  V_{72}\cdot Q_{28}\cdot P_{83}&-&   P_{72}\cdot Q_{28}\cdot V_{83} \\
\Lambda^{1}_{84}:   & \ \ \ \ &  X_{42} \cdot K_{28} &-&   K_{4,10}\cdot X_{10,8} & \ \ \ \ &  V_{83}\cdot Q_{39}\cdot P_{94}&-&   P_{83}\cdot Q_{39}\cdot V_{94} \\
\Lambda^{1}_{95}:   & \ \ \ \ & K_{56}\cdot X_{69} &-&   X_{53}\cdot K_{39} & \ \ \ \ &  P_{94}\cdot Q_{4,10}\cdot V_{10,5}&-&   V_{94}\cdot Q_{4,10}\cdot P_{10,5} \\
\Lambda^{1}_{10,1}:   & \ \ \ \ & K_{17}\cdot X_{7,10} &-& X_{14}\cdot K_{4,10}  & \ \ \ \ &  P_{10,5}\cdot Q_{56}\cdot V_{61}&-&   V_{10,5}\cdot Q_{56}\cdot P_{61} \\
\Lambda^{1}_{64}:   & \ \ \ \ & K_{4,10}\cdot P_{10,5}\cdot Q_{56}&-&   Q_{4,10}\cdot P_{10,5}\cdot K_{56}& \ \ \ \ & V_{61}\cdot X_{14} &-&  X_{69}\cdot V_{94}  \\
\Lambda^{1}_{92}:   & \ \ \ \ & Q_{28}\cdot P_{83}\cdot K_{39}&-&   K_{28}\cdot P_{83}\cdot Q_{39}& \ \ \ \ &  X_{97}\cdot V_{72}&-&   V_{94}\cdot X_{42} \\
\Lambda^{1}_{75}:   & \ \ \ \ &  K_{56}\cdot P_{61}\cdot Q_{17}&-&   Q_{56}\cdot P_{61}\cdot K_{17}& \ \ \ \ &  V_{72}\cdot X_{25} &-& X_{7,10}\cdot V_{10,5} \\
\Lambda^{1}_{10,3}:   & \ \ \ \ & Q_{39}\cdot P_{94}\cdot K_{4,10}&-&   K_{39}\cdot P_{94}\cdot Q_{4,10}& \ \ \ \ &  X_{10,8}\cdot V_{83}&-&   V_{10,5}\cdot X_{53} \\
\Lambda^{1}_{81}:   & \ \ \ \ & K_{17}\cdot P_{72}\cdot Q_{28}&-&   Q_{17}\cdot P_{72}\cdot K_{28}& \ \ \ \ &  V_{83}\cdot X_{31} &-&  X_{86}\cdot V_{61} \\
\Lambda^{2}_{62}:   & \ \ \ \ &  X_{25}\cdot Q_{56} &-&  Q_{28}\cdot X_{86} & \ \ \ \ &  P_{61}\cdot K_{17}\cdot V_{72}&-&   V_{61}\cdot K_{17}\cdot P_{72} \\
\Lambda^{2}_{73}:   & \ \ \ \ & Q_{39}\cdot X_{97} &-& X_{31}\cdot Q_{17}  & \ \ \ \ &  V_{72}\cdot K_{28}\cdot P_{83}&-&   P_{72}\cdot K_{28}\cdot V_{83} \\
\Lambda^{2}_{84}:   & \ \ \ \ &  Q_{4,10}\cdot X_{10,8}&-&   X_{42}\cdot Q_{28}& \ \ \ \ &  V_{83}\cdot K_{39}\cdot P_{94}&-&   P_{83}\cdot K_{39}\cdot V_{94} \\
\Lambda^{2}_{95}:   & \ \ \ \ & X_{53}\cdot Q_{39}&-&   Q_{56}\cdot X_{69}& \ \ \ \ &  P_{94}\cdot K_{4,10}\cdot V_{10,5}&-&   V_{94}\cdot K_{4,10}\cdot P_{10,5} \\
\Lambda^{2}_{10,1}:   & \ \ \ \ &         X_{14}\cdot Q_{4,10}&-&   Q_{17}\cdot X_{7,10}& \ \ \ \ &  P_{10,5}\cdot K_{56}\cdot V_{61}&-& V_{10,5}\cdot K_{56}\cdot P_{61} \\
\Lambda^{2}_{64}:   & \ \ \ \ & K_{4,10}\cdot V_{10,5}\cdot Q_{56}&-&   Q_{4,10}\cdot V_{10,5}\cdot K_{56}& \ \ \ \ &  X_{69}\cdot P_{94}&-&   P_{61}\cdot X_{14} \\
\Lambda^{2}_{92}:   & \ \ \ \ &  Q_{28}\cdot V_{83}\cdot K_{39}&-&   K_{28}\cdot V_{83}\cdot Q_{39}& \ \ \ \ & P_{94}\cdot X_{42} &-& X_{97}\cdot P_{72}  \\
\Lambda^{2}_{75}:   & \ \ \ \ & K_{56}\cdot V_{61}\cdot Q_{17}&-&   Q_{56}\cdot V_{61}\cdot K_{17}& \ \ \ \ &  X_{7,10}\cdot P_{10,5}&-&   P_{72}\cdot X_{25} \\
\Lambda^{2}_{10,3}:   & \ \ \ \ & Q_{39}\cdot V_{94}\cdot K_{4,10}&-&   K_{39}\cdot V_{94}\cdot Q_{4,10}& \ \ \ \ & P_{10,5}\cdot X_{53} &-& X_{10,8}\cdot P_{83}   \\
\Lambda^{2}_{81}:   & \ \ \ \ &  K_{17}\cdot V_{72}\cdot Q_{28}&-&   Q_{17}\cdot V_{72}\cdot K_{28}& \ \ \ \ & X_{86}\cdot P_{61}&-&   P_{83}\cdot X_{31} \\
\end{array}
$}
~,~
\eea
where the associated quiver diagram in shown in \fref{f_ref_ex4d}(b).

Using the forward algorithm \cite{Feng:2000mi, Franco:2015tna, Franco:2015tya}, we obtain the $P$-matrix for the $P_{+-}^{2}[\mathbb{C}^3/\mathbb{Z}_5~(1,1,3)]$ brane brick model, 
\beal{es08a20}
P =
\resizebox{0.78\textwidth}{!}{$
\left(
\begin{array}{c|ccccc|cccccccccc|ccccccccccccccc|cccccccccc}
~  & p_2 & p_3 & p_4 & p_5 & p_6 & s_1 &
   s_2 & s_3 & s_4 & s_5 & s_6 & s_7 & s_8 &
   s_9 & s_{10} & w_1 & w_2 & w_3 & w_4 & w_5 &
   w_6 & w_7 & w_8 & w_9 & w_{10} & w_{11} & w_{12} &
   w_{13} & w_{14} & w_{15} &  u_{1} &  u_{2} &  u_{3} &  u_{4} &
    u_{5} &  u_{6} &  u_{7} &  u_{8} &  u_{9} &  u_{10} \\
   \hline
 K_{17}     &  0 & 0 & 0 & 0 & 1 & 0 & 0 & 0 & 0 & 0 & 0 & 0 & 0 & 0 & 1 & 0 & 0 & 0 & 0 & 0 & 0 & 0 & 0 & 0 & 0 & 0 & 0 & 1 & 1 & 1 & 0 & 0 & 0 & 1 & 1 & 1 & 1 & 1 & 1 & 1 \\
 K_{28}     &   0 & 0 & 0 & 0 & 1 & 0 & 0 & 0 & 0 & 0 & 0 & 0 & 0 & 1 & 0 & 0 & 0 & 0 & 0 & 0 & 0 & 0 & 0 & 0 & 1 & 1 & 1 & 0 & 0 & 0 & 1 & 1 & 1 & 0 & 0 & 0 & 1 & 1 & 1 & 1 \\
 K_{39}     &   0 & 0 & 0 & 0 & 1 & 0 & 0 & 0 & 0 & 0 & 0 & 0 & 1 & 0 & 0 & 0 & 0 & 0 & 0 & 0 & 0 & 0 & 1 & 1 & 0 & 0 & 0 & 0 & 0 & 1 & 0 & 1 & 1 & 1 & 1 & 1 & 0 & 0 & 1 & 1 \\
 K_{4,10}   &   0 & 0 & 0 & 0 & 1 & 0 & 0 & 0 & 0 & 0 & 0 & 1 & 0 & 0 & 0 & 0 & 0 & 0 & 0 & 0 & 0 & 1 & 0 & 0 & 0 & 0 & 1 & 0 & 1 & 0 & 1 & 0 & 1 & 0 & 1 & 1 & 1 & 1 & 0 & 1 \\
 K_{56}     &   0 & 0 & 0 & 0 & 1 & 0 & 0 & 0 & 0 & 0 & 1 & 0 & 0 & 0 & 0 & 0 & 0 & 0 & 0 & 0 & 1 & 0 & 0 & 1 & 0 & 1 & 0 & 0 & 0 & 0 & 1 & 1 & 1 & 1 & 0 & 1 & 0 & 1 & 1 & 0 \\
 P_{61}     &   0 & 0 & 1 & 0 & 0 & 0 & 0 & 0 & 0 & 1 & 0 & 0 & 0 & 0 & 0 & 0 & 0 & 0 & 1 & 1 & 0 & 0 & 1 & 0 & 0 & 0 & 0 & 0 & 0 & 0 & 1 & 1 & 1 & 0 & 1 & 0 & 1 & 1 & 0 & 1 \\
 P_{72}     &   0 & 0 & 1 & 0 & 0 & 0 & 0 & 0 & 1 & 0 & 0 & 0 & 0 & 0 & 0 & 0 & 1 & 1 & 0 & 0 & 0 & 1 & 0 & 0 & 0 & 0 & 0 & 0 & 0 & 0 & 1 & 1 & 1 & 1 & 1 & 1 & 0 & 0 & 1 & 0 \\
 P_{83}     &   0 & 0 & 1 & 0 & 0 & 0 & 0 & 1 & 0 & 0 & 0 & 0 & 0 & 0 & 0 & 1 & 0 & 0 & 0 & 1 & 1 & 0 & 0 & 0 & 0 & 0 & 0 & 0 & 0 & 0 & 1 & 0 & 0 & 1 & 1 & 1 & 1 & 1 & 0 & 1 \\
 P_{94}     &   0 & 0 & 1 & 0 & 0 & 0 & 1 & 0 & 0 & 0 & 0 & 0 & 0 & 0 & 0 & 0 & 0 & 1 & 1 & 0 & 0 & 0 & 0 & 0 & 0 & 0 & 0 & 1 & 0 & 0 & 1 & 1 & 1 & 1 & 0 & 0 & 1 & 1 & 1 & 0 \\
 P_{10,5}   &   0 & 0 & 1 & 0 & 0 & 1 & 0 & 0 & 0 & 0 & 0 & 0 & 0 & 0 & 0 & 1 & 1 & 0 & 0 & 0 & 0 & 0 & 0 & 0 & 1 & 0 & 0 & 0 & 0 & 0 & 0 & 1 & 0 & 1 & 1 & 1 & 1 & 0 & 1 & 1 \\
 Q_{17}     &   0 & 1 & 0 & 0 & 0 & 0 & 0 & 0 & 0 & 0 & 0 & 0 & 0 & 0 & 1 & 0 & 0 & 0 & 0 & 0 & 0 & 0 & 0 & 0 & 0 & 0 & 0 & 1 & 1 & 1 & 0 & 0 & 0 & 1 & 1 & 1 & 1 & 1 & 1 & 1 \\
 Q_{28}     &   0 & 1 & 0 & 0 & 0 & 0 & 0 & 0 & 0 & 0 & 0 & 0 & 0 & 1 & 0 & 0 & 0 & 0 & 0 & 0 & 0 & 0 & 0 & 0 & 1 & 1 & 1 & 0 & 0 & 0 & 1 & 1 & 1 & 0 & 0 & 0 & 1 & 1 & 1 & 1 \\
 Q_{39}     &   0 & 1 & 0 & 0 & 0 & 0 & 0 & 0 & 0 & 0 & 0 & 0 & 1 & 0 & 0 & 0 & 0 & 0 & 0 & 0 & 0 & 0 & 1 & 1 & 0 & 0 & 0 & 0 & 0 & 1 & 0 & 1 & 1 & 1 & 1 & 1 & 0 & 0 & 1 & 1 \\
 Q_{4,10}   &   0 & 1 & 0 & 0 & 0 & 0 & 0 & 0 & 0 & 0 & 0 & 1 & 0 & 0 & 0 & 0 & 0 & 0 & 0 & 0 & 0 & 1 & 0 & 0 & 0 & 0 & 1 & 0 & 1 & 0 & 1 & 0 & 1 & 0 & 1 & 1 & 1 & 1 & 0 & 1 \\
 Q_{56}     &   0 & 1 & 0 & 0 & 0 & 0 & 0 & 0 & 0 & 0 & 1 & 0 & 0 & 0 & 0 & 0 & 0 & 0 & 0 & 0 & 1 & 0 & 0 & 1 & 0 & 1 & 0 & 0 & 0 & 0 & 1 & 1 & 1 & 1 & 0 & 1 & 0 & 1 & 1 & 0 \\
 V_{61}     &   0 & 0 & 0 & 1 & 0 & 0 & 0 & 0 & 0 & 1 & 0 & 0 & 0 & 0 & 0 & 0 & 0 & 0 & 1 & 1 & 0 & 0 & 1 & 0 & 0 & 0 & 0 & 0 & 0 & 0 & 1 & 1 & 1 & 0 & 1 & 0 & 1 & 1 & 0 & 1 \\
 V_{72}     &   0 & 0 & 0 & 1 & 0 & 0 & 0 & 0 & 1 & 0 & 0 & 0 & 0 & 0 & 0 & 0 & 1 & 1 & 0 & 0 & 0 & 1 & 0 & 0 & 0 & 0 & 0 & 0 & 0 & 0 & 1 & 1 & 1 & 1 & 1 & 1 & 0 & 0 & 1 & 0 \\
 V_{83}     &   0 & 0 & 0 & 1 & 0 & 0 & 0 & 1 & 0 & 0 & 0 & 0 & 0 & 0 & 0 & 1 & 0 & 0 & 0 & 1 & 1 & 0 & 0 & 0 & 0 & 0 & 0 & 0 & 0 & 0 & 1 & 0 & 0 & 1 & 1 & 1 & 1 & 1 & 0 & 1 \\
 V_{94}     &   0 & 0 & 0 & 1 & 0 & 0 & 1 & 0 & 0 & 0 & 0 & 0 & 0 & 0 & 0 & 0 & 0 & 1 & 1 & 0 & 0 & 0 & 0 & 0 & 0 & 0 & 0 & 1 & 0 & 0 & 1 & 1 & 1 & 1 & 0 & 0 & 1 & 1 & 1 & 0 \\
 V_{10,5}   &   0 & 0 & 0 & 1 & 0 & 1 & 0 & 0 & 0 & 0 & 0 & 0 & 0 & 0 & 0 & 1 & 1 & 0 & 0 & 0 & 0 & 0 & 0 & 0 & 1 & 0 & 0 & 0 & 0 & 0 & 0 & 1 & 0 & 1 & 1 & 1 & 1 & 0 & 1 & 1 \\
 X_{14}     &   1 & 0 & 0 & 0 & 0 & 0 & 1 & 1 & 1 & 0 & 0 & 0 & 1 & 1 & 1 & 0 & 0 & 1 & 0 & 0 & 0 & 0 & 0 & 0 & 0 & 0 & 0 & 1 & 0 & 1 & 0 & 0 & 0 & 1 & 0 & 0 & 0 & 0 & 1 & 0 \\
 X_{25}     &   1 & 0 & 0 & 0 & 0 & 1 & 1 & 1 & 0 & 0 & 0 & 1 & 1 & 1 & 0 & 1 & 0 & 0 & 0 & 0 & 0 & 0 & 0 & 0 & 1 & 0 & 1 & 0 & 0 & 0 & 0 & 0 & 0 & 0 & 0 & 0 & 1 & 0 & 0 & 1 \\
 X_{31}     &   1 & 0 & 0 & 0 & 0 & 1 & 1 & 0 & 0 & 1 & 1 & 1 & 1 & 0 & 0 & 0 & 0 & 0 & 1 & 0 & 0 & 0 & 1 & 1 & 0 & 0 & 0 & 0 & 0 & 0 & 0 & 1 & 1 & 0 & 0 & 0 & 0 & 0 & 0 & 0 \\
 X_{42}     &   1 & 0 & 0 & 0 & 0 & 1 & 0 & 0 & 1 & 1 & 1 & 1 & 0 & 0 & 1 & 0 & 1 & 0 & 0 & 0 & 0 & 1 & 0 & 0 & 0 & 0 & 0 & 0 & 1 & 0 & 0 & 0 & 0 & 0 & 1 & 1 & 0 & 0 & 0 & 0 \\
 X_{53}     &   1 & 0 & 0 & 0 & 0 & 0 & 0 & 1 & 1 & 1 & 1 & 0 & 0 & 1 & 1 & 0 & 0 & 0 & 0 & 1 & 1 & 0 & 0 & 0 & 0 & 1 & 0 & 0 & 0 & 0 & 1 & 0 & 0 & 0 & 0 & 0 & 0 & 1 & 0 & 0 \\
 X_{69}     &   1 & 0 & 0 & 0 & 0 & 0 & 0 & 1 & 1 & 1 & 0 & 0 & 1 & 1 & 1 & 0 & 0 & 0 & 0 & 1 & 0 & 0 & 1 & 0 & 0 & 0 & 0 & 0 & 0 & 1 & 0 & 0 & 0 & 0 & 1 & 0 & 0 & 0 & 0 & 1 \\
 X_{7,10}   &   1 & 0 & 0 & 0 & 0 & 0 & 1 & 1 & 1 & 0 & 0 & 1 & 1 & 1 & 0 & 0 & 0 & 1 & 0 & 0 & 0 & 1 & 0 & 0 & 0 & 0 & 1 & 0 & 0 & 0 & 1 & 0 & 1 & 0 & 0 & 0 & 0 & 0 & 0 & 0 \\
 X_{86}     &   1 & 0 & 0 & 0 & 0 & 1 & 1 & 1 & 0 & 0 & 1 & 1 & 1 & 0 & 0 & 1 & 0 & 0 & 0 & 0 & 1 & 0 & 0 & 1 & 0 & 0 & 0 & 0 & 0 & 0 & 0 & 0 & 0 & 1 & 0 & 1 & 0 & 0 & 0 & 0 \\
 X_{97}     &   1 & 0 & 0 & 0 & 0 & 1 & 1 & 0 & 0 & 1 & 1 & 1 & 0 & 0 & 1 & 0 & 0 & 0 & 1 & 0 & 0 & 0 & 0 & 0 & 0 & 0 & 0 & 1 & 1 & 0 & 0 & 0 & 0 & 0 & 0 & 0 & 1 & 1 & 0 & 0 \\
 X_{10,8}   &   1 & 0 & 0 & 0 & 0 & 1 & 0 & 0 & 1 & 1 & 1 & 0 & 0 & 1 & 1 & 0 & 1 & 0 & 0 & 0 & 0 & 0 & 0 & 0 & 1 & 1 & 0 & 0 & 0 & 0 & 0 & 1 & 0 & 0 & 0 & 0 & 0 & 0 & 1 & 0 \\
\end{array}
\right)
$}
.~
\eea
The $U(1)$ charges under the $J$- and $E$-terms in \eref{es08a19}
as well as the $D$-terms are given by the following charge matrices, 
\beal{es08a21}
Q_{JE} =
\resizebox{0.77\textwidth}{!}{$
\left(
\ba{ccccc|cccccccccc|ccccccccccccccc|cccccccccc}
p_2 & p_3 & p_4 & p_5 & p_6 & s_1 &
   s_2 & s_3 & s_4 & s_5 & s_6 & s_7 & s_8 &
   s_9 & s_{10} & w_1 & w_2 & w_3 & w_4 & w_5 &
   w_6 & w_7 & w_8 & w_9 & w_{10} & w_{11} & w_{12} &
   w_{13} & w_{14} & w_{15} &  u_{1} &  u_{2} &  u_{3} &  u_{4} &
    u_{5} &  u_{6} &  u_{7} &  u_{8} &  u_{9} &  u_{10} \\
\hline
 1 & 0 & 0 & 0 & 0 & 0 & 0 & 0 & 0 & 0 & 0 & 0 & 0 & 0 & 0 & 0 & 0 & 0 & 0 & 0 & 0 & 0 & 0 & 0 & 0 & 0 & 0 & 0 & 0 & -7 & -4 & -4 & 3 & 3 & 3 & -4 & -4 & 3 & 3 & 3 \\
 0 & 1 & 0 & 0 & 1 & 0 & 0 & 0 & 0 & 0 & 0 & 0 & 0 & 0 & 0 & 0 & 0 & 0 & 0 & 0 & 0 & 0 & 0 & 0 & 0 & 0 & 0 & 0 & 0 & 1 & 1 & 1 & -1 & 0 & 0 & 0 & 1 & -1 & -1 & -1 \\
 0 & 0 & 1 & 1 & 0 & 0 & 0 & 0 & 0 & 0 & 0 & 0 & 0 & 0 & 0 & 0 & 0 & 0 & 0 & 0 & 0 & 0 & 0 & 0 & 0 & 0 & 0 & 0 & 0 & 1 & 0 & 0 & 0 & -1 & -1 & 1 & 0 & 0 & 0 & 0 \\
 0 & 0 & 0 & 0 & 0 & 1 & 0 & 0 & 0 & 0 & 0 & 0 & 0 & 0 & 0 & 0 & 0 & 0 & 0 & 0 & 0 & 0 & 0 & 0 & 0 & 0 & 0 & 0 & 0 & -4 & -2 & -3 & 2 & 2 & 2 & -3 & -3 & 2 & 2 & 2 \\
 0 & 0 & 0 & 0 & 0 & 0 & 1 & 0 & 0 & 0 & 0 & 0 & 0 & 0 & 0 & 0 & 0 & 0 & 0 & 0 & 0 & 0 & 0 & 0 & 0 & 0 & 0 & 0 & 0 & -4 & -2 & -2 & 1 & 1 & 2 & -2 & -3 & 2 & 2 & 2 \\
 0 & 0 & 0 & 0 & 0 & 0 & 0 & 1 & 0 & 0 & 0 & 0 & 0 & 0 & 0 & 0 & 0 & 0 & 0 & 0 & 0 & 0 & 0 & 0 & 0 & 0 & 0 & 0 & 0 & -4 & -3 & -2 & 2 & 1 & 2 & -2 & -2 & 2 & 2 & 1 \\
 0 & 0 & 0 & 0 & 0 & 0 & 0 & 0 & 1 & 0 & 0 & 0 & 0 & 0 & 0 & 0 & 0 & 0 & 0 & 0 & 0 & 0 & 0 & 0 & 0 & 0 & 0 & 0 & 0 & -4 & -3 & -2 & 2 & 2 & 1 & -2 & -2 & 2 & 1 & 2 \\
 0 & 0 & 0 & 0 & 0 & 0 & 0 & 0 & 0 & 1 & 0 & 0 & 0 & 0 & 0 & 0 & 0 & 0 & 0 & 0 & 0 & 0 & 0 & 0 & 0 & 0 & 0 & 0 & 0 & -4 & -2 & -3 & 2 & 2 & 1 & -2 & -2 & 1 & 2 & 2 \\
 0 & 0 & 0 & 0 & 0 & 0 & 0 & 0 & 0 & 0 & 1 & 0 & 0 & 0 & 0 & 0 & 0 & 0 & 0 & 0 & 0 & 0 & 0 & 0 & 0 & 0 & 0 & 0 & 0 & -4 & -2 & -3 & 2 & 2 & 2 & -3 & -2 & 1 & 2 & 2 \\
 0 & 0 & 0 & 0 & 0 & 0 & 0 & 0 & 0 & 0 & 0 & 1 & 0 & 0 & 0 & 0 & 0 & 0 & 0 & 0 & 0 & 0 & 0 & 0 & 0 & 0 & 0 & 0 & 0 & -4 & -2 & -2 & 1 & 2 & 2 & -3 & -3 & 2 & 2 & 2 \\
 0 & 0 & 0 & 0 & 0 & 0 & 0 & 0 & 0 & 0 & 0 & 0 & 1 & 0 & 0 & 0 & 0 & 0 & 0 & 0 & 0 & 0 & 0 & 0 & 0 & 0 & 0 & 0 & 0 & -4 & -2 & -2 & 1 & 1 & 2 & -2 & -2 & 2 & 2 & 1 \\
 0 & 0 & 0 & 0 & 0 & 0 & 0 & 0 & 0 & 0 & 0 & 0 & 0 & 1 & 0 & 0 & 0 & 0 & 0 & 0 & 0 & 0 & 0 & 0 & 0 & 0 & 0 & 0 & 0 & -4 & -3 & -2 & 2 & 2 & 2 & -2 & -2 & 2 & 1 & 1 \\
 0 & 0 & 0 & 0 & 0 & 0 & 0 & 0 & 0 & 0 & 0 & 0 & 0 & 0 & 1 & 0 & 0 & 0 & 0 & 0 & 0 & 0 & 0 & 0 & 0 & 0 & 0 & 0 & 0 & -4 & -2 & -2 & 2 & 2 & 1 & -2 & -2 & 1 & 1 & 2 \\
 0 & 0 & 0 & 0 & 0 & 0 & 0 & 0 & 0 & 0 & 0 & 0 & 0 & 0 & 0 & 1 & 0 & 0 & 0 & 0 & 0 & 0 & 0 & 0 & 0 & 0 & 0 & 0 & 0 & -1 & -1 & -1 & 1 & 0 & 1 & -1 & -1 & 1 & 1 & 0 \\
 0 & 0 & 0 & 0 & 0 & 0 & 0 & 0 & 0 & 0 & 0 & 0 & 0 & 0 & 0 & 0 & 1 & 0 & 0 & 0 & 0 & 0 & 0 & 0 & 0 & 0 & 0 & 0 & 0 & -1 & -1 & -1 & 1 & 1 & 0 & -1 & -1 & 1 & 0 & 1 \\
 0 & 0 & 0 & 0 & 0 & 0 & 0 & 0 & 0 & 0 & 0 & 0 & 0 & 0 & 0 & 0 & 0 & 1 & 0 & 0 & 0 & 0 & 0 & 0 & 0 & 0 & 0 & 0 & 0 & -1 & -1 & 0 & 0 & 0 & 0 & 0 & -1 & 1 & 0 & 1 \\
 0 & 0 & 0 & 0 & 0 & 0 & 0 & 0 & 0 & 0 & 0 & 0 & 0 & 0 & 0 & 0 & 0 & 0 & 1 & 0 & 0 & 0 & 0 & 0 & 0 & 0 & 0 & 0 & 0 & -1 & 0 & -1 & 0 & 0 & 0 & 0 & -1 & 0 & 1 & 1 \\
 0 & 0 & 0 & 0 & 0 & 0 & 0 & 0 & 0 & 0 & 0 & 0 & 0 & 0 & 0 & 0 & 0 & 0 & 0 & 1 & 0 & 0 & 0 & 0 & 0 & 0 & 0 & 0 & 0 & -1 & -1 & -1 & 1 & 0 & 0 & 0 & 0 & 0 & 1 & 0 \\
 0 & 0 & 0 & 0 & 0 & 0 & 0 & 0 & 0 & 0 & 0 & 0 & 0 & 0 & 0 & 0 & 0 & 0 & 0 & 0 & 1 & 0 & 0 & 0 & 0 & 0 & 0 & 0 & 0 & -1 & -1 & -1 & 1 & 0 & 1 & -1 & 0 & 0 & 1 & 0 \\
 0 & 0 & 0 & 0 & 0 & 0 & 0 & 0 & 0 & 0 & 0 & 0 & 0 & 0 & 0 & 0 & 0 & 0 & 0 & 0 & 0 & 1 & 0 & 0 & 0 & 0 & 0 & 0 & 0 & -1 & -1 & 0 & 0 & 1 & 0 & -1 & -1 & 1 & 0 & 1 \\
 0 & 0 & 0 & 0 & 0 & 0 & 0 & 0 & 0 & 0 & 0 & 0 & 0 & 0 & 0 & 0 & 0 & 0 & 0 & 0 & 0 & 0 & 1 & 0 & 0 & 0 & 0 & 0 & 0 & -1 & 0 & -1 & 0 & 0 & 0 & 0 & 0 & 0 & 1 & 0 \\
 0 & 0 & 0 & 0 & 0 & 0 & 0 & 0 & 0 & 0 & 0 & 0 & 0 & 0 & 0 & 0 & 0 & 0 & 0 & 0 & 0 & 0 & 0 & 1 & 0 & 0 & 0 & 0 & 0 & -1 & 0 & -1 & 0 & 0 & 1 & -1 & 0 & 0 & 1 & 0 \\
 0 & 0 & 0 & 0 & 0 & 0 & 0 & 0 & 0 & 0 & 0 & 0 & 0 & 0 & 0 & 0 & 0 & 0 & 0 & 0 & 0 & 0 & 0 & 0 & 1 & 0 & 0 & 0 & 0 & -1 & -1 & -1 & 1 & 1 & 1 & -1 & -1 & 1 & 0 & 0 \\
 0 & 0 & 0 & 0 & 0 & 0 & 0 & 0 & 0 & 0 & 0 & 0 & 0 & 0 & 0 & 0 & 0 & 0 & 0 & 0 & 0 & 0 & 0 & 0 & 0 & 1 & 0 & 0 & 0 & -1 & -1 & -1 & 1 & 1 & 1 & -1 & 0 & 0 & 0 & 0 \\
 0 & 0 & 0 & 0 & 0 & 0 & 0 & 0 & 0 & 0 & 0 & 0 & 0 & 0 & 0 & 0 & 0 & 0 & 0 & 0 & 0 & 0 & 0 & 0 & 0 & 0 & 1 & 0 & 0 & -1 & -1 & 0 & 0 & 1 & 1 & -1 & -1 & 1 & 0 & 0 \\
 0 & 0 & 0 & 0 & 0 & 0 & 0 & 0 & 0 & 0 & 0 & 0 & 0 & 0 & 0 & 0 & 0 & 0 & 0 & 0 & 0 & 0 & 0 & 0 & 0 & 0 & 0 & 1 & 0 & -1 & 0 & 0 & 0 & 0 & 0 & 0 & -1 & 0 & 0 & 1 \\
 0 & 0 & 0 & 0 & 0 & 0 & 0 & 0 & 0 & 0 & 0 & 0 & 0 & 0 & 0 & 0 & 0 & 0 & 0 & 0 & 0 & 0 & 0 & 0 & 0 & 0 & 0 & 0 & 1 & -1 & 0 & 0 & 0 & 1 & 0 & -1 & -1 & 0 & 0 & 1 \\
\ea
\right)
$}
~,~
\nn\\
\eea
\beal{es08a22}
Q_D =
\resizebox{0.77\textwidth}{!}{$
\left(
\ba{ccccc|cccccccccc|ccccccccccccccc|cccccccccc}
p_2 & p_3 & p_4 & p_5 & p_6 & s_1 &
   s_2 & s_3 & s_4 & s_5 & s_6 & s_7 & s_8 &
   s_9 & s_{10} & w_1 & w_2 & w_3 & w_4 & w_5 &
   w_6 & w_7 & w_8 & w_9 & w_{10} & w_{11} & w_{12} &
   w_{13} & w_{14} & w_{15} &  u_{1} &  u_{2} &  u_{3} &  u_{4} &
    u_{5} &  u_{6} &  u_{7} &  u_{8} &  u_{9} &  u_{10} \\
\hline
0 & 0 & 0 & 0 & 0 & 1 & 0 & 0 & 0 & 0 & 0 & 0 & 0 & 0 & -1 & 0 & 0 & 0 & 0 & 0 & 0 & 0 & 0 & 0 & 0 & 0 & 0 & 0 & 0 & 0 & 0 & 0 & 0 & 0 & 0 & 0 & 0 & 0 & 0 & 0 \\
 0 & 0 & 0 & 0 & 0 & 0 & 1 & 0 & 0 & 0 & 0 & 0 & 0 & 0 & -1 & 0 & 0 & 0 & 0 & 0 & 0 & 0 & 0 & 0 & 0 & 0 & 0 & 0 & 0 & 0 & 0 & 0 & 0 & 0 & 0 & 0 & 0 & 0 & 0 & 0 \\
 0 & 0 & 0 & 0 & 0 & 0 & 0 & 1 & 0 & 0 & 0 & 0 & 0 & 0 & -1 & 0 & 0 & 0 & 0 & 0 & 0 & 0 & 0 & 0 & 0 & 0 & 0 & 0 & 0 & 0 & 0 & 0 & 0 & 0 & 0 & 0 & 0 & 0 & 0 & 0 \\
 0 & 0 & 0 & 0 & 0 & 0 & 0 & 0 & 1 & 0 & 0 & 0 & 0 & 0 & -1 & 0 & 0 & 0 & 0 & 0 & 0 & 0 & 0 & 0 & 0 & 0 & 0 & 0 & 0 & 0 & 0 & 0 & 0 & 0 & 0 & 0 & 0 & 0 & 0 & 0 \\
 0 & 0 & 0 & 0 & 0 & 0 & 0 & 0 & 0 & 1 & 0 & 0 & 0 & 0 & -1 & 0 & 0 & 0 & 0 & 0 & 0 & 0 & 0 & 0 & 0 & 0 & 0 & 0 & 0 & 0 & 0 & 0 & 0 & 0 & 0 & 0 & 0 & 0 & 0 & 0 \\
 0 & 0 & 0 & 0 & 0 & 0 & 0 & 0 & 0 & 0 & 1 & 0 & 0 & 0 & -1 & 0 & 0 & 0 & 0 & 0 & 0 & 0 & 0 & 0 & 0 & 0 & 0 & 0 & 0 & 0 & 0 & 0 & 0 & 0 & 0 & 0 & 0 & 0 & 0 & 0 \\
 0 & 0 & 0 & 0 & 0 & 0 & 0 & 0 & 0 & 0 & 0 & 1 & 0 & 0 & -1 & 0 & 0 & 0 & 0 & 0 & 0 & 0 & 0 & 0 & 0 & 0 & 0 & 0 & 0 & 0 & 0 & 0 & 0 & 0 & 0 & 0 & 0 & 0 & 0 & 0 \\
 0 & 0 & 0 & 0 & 0 & 0 & 0 & 0 & 0 & 0 & 0 & 0 & 1 & 0 & -1 & 0 & 0 & 0 & 0 & 0 & 0 & 0 & 0 & 0 & 0 & 0 & 0 & 0 & 0 & 0 & 0 & 0 & 0 & 0 & 0 & 0 & 0 & 0 & 0 & 0 \\
 0 & 0 & 0 & 0 & 0 & 0 & 0 & 0 & 0 & 0 & 0 & 0 & 0 & 1 & -1 & 0 & 0 & 0 & 0 & 0 & 0 & 0 & 0 & 0 & 0 & 0 & 0 & 0 & 0 & 0 & 0 & 0 & 0 & 0 & 0 & 0 & 0 & 0 & 0 & 0 \\
\ea
\right)
$}
~.~ \nn\\
\eea
The resulting toric diagram for the $P_{+-}^{2}[\mathbb{C}^3/\mathbb{Z}_5~(1,1,3)]$ model
is given by the following $G_t$-matrix, 
\beal{es08a23}
G_t =
\resizebox{0.77\textwidth}{!}{$
\left(
\ba{ccccc|cccccccccc|ccccccccccccccc|cccccccccc}
p_2 & p_3 & p_4 & p_5 & p_6 & s_1 &
   s_2 & s_3 & s_4 & s_5 & s_6 & s_7 & s_8 &
   s_9 & s_{10} & w_1 & w_2 & w_3 & w_4 & w_5 &
   w_6 & w_7 & w_8 & w_9 & w_{10} & w_{11} & w_{12} &
   w_{13} & w_{14} & w_{15} &  u_{1} &  u_{2} &  u_{3} &  u_{4} &
    u_{5} &  u_{6} &  u_{7} &  u_{8} &  u_{9} &  u_{10} \\
\hline
 1 & 0 & -3 & 0 & -3 & 0 & 0 & 0 & 0 & 0 & 0 & 0 & 0 & 0 & 0 & -1 & -1 & -1 & -1 & -1 & -1 & -1 & -1 & -1 & -1 & -1 & -1 & -1 & -1 & -1 & -4
   & -4 & -4 & -4 & -4 & -4 & -4 & -4 & -4 & -4 \\
 0 & 0 & -1 & 1 & 0 & 0 & 0 & 0 & 0 & 0 & 0 & 0 & 0 & 0 & 0 & 0 & 0 & 0 & 0 & 0 & 0 & 0 & 0 & 0 & 0 & 0 & 0 & 0 & 0 & 0 & 0 & 0 & 0 & 0 & 0 &
   0 & 0 & 0 & 0 & 0 \\
 0 & 1 & 0 & 0 & -1 & 0 & 0 & 0 & 0 & 0 & 0 & 0 & 0 & 0 & 0 & 0 & 0 & 0 & 0 & 0 & 0 & 0 & 0 & 0 & 0 & 0 & 0 & 0 & 0 & 0 & 0 & 0 & 0 & 0 & 0 &
   0 & 0 & 0 & 0 & 0 \\
   \hline
 1 & 1 & 1 & 1 & 1 & 1 & 1 & 1 & 1 & 1 & 1 & 1 & 1 & 1 & 1 & 1 & 1 & 1 & 1 & 1 & 1 & 1 & 1 & 1 & 1 & 1 & 1 & 1 & 1 & 1 & 3 & 3 & 3 & 3 & 3 &
   3 & 3 & 3 & 3 & 3 \\
\end{array}
\right)
$}
~.~\nn\\
\eea
where the toric diagram is shown in \fref{f_ref_ex4c}(b).
Here, we note that there are $5$ extremal GLSM fields corresponding to the $5$ extremal points in the toric diagram of the $P_{+-}^{2}[\mathbb{C}^3/\mathbb{Z}_5~(1,1,3)]$ model.
Furthermore, we have $10$ extra GLSM fields that over-parameterize the mesonic moduli space $\mathcal{M}^{mes}_{P_{+-}^{2}[\mathbb{C}^3/\mathbb{Z}_5~(1,1,3)]}$
and do not correspond to any points in the toric diagram for $P_{+-}^{2}[\mathbb{C}^3/\mathbb{Z}_5~(1,1,3)]$. 

Based on the $Q_{JE}$ and $Q_D$ matrices, we can see that the $P_{+-}^{2}[\mathbb{C}^3/\mathbb{Z}_5~(1,1,3)]$ model has an enhanced
global symmetry of the form, 
\beal{es08a24}
SU(2)_x \times SU(2)_{y} \times U(1)_{f} \times U(1)_R ~,~
\eea
where $SU(2)_x \times SU(2)_{y} \times U(1)_{f}$ is the enhanced mesonic flavor symmetry.
\tref{tab_41} summarizes how the extremal GLSM fields $p_a$ are charged under the global symmetry of the $P_{+-}^{2}[\mathbb{C}^3/\mathbb{Z}_5~(1,1,3)]$ model.

\begin{table}[ht!]
\centering
\begin{tabular}{|c|c|c|c|c|l|}
\hline
\; & $SU(2)_x$ & $SU(2)_y$ & $U(1)_f$ & $U(1)_R$ &  fugacity \\
\hline
$p_2$ & $ 0 $ & $ 0 $ & $ +2 $ & $ r_2$ & $t_2 = f^{2} \overline{t}_2$ \\
$p_3$ & $ +1 $ & $ 0 $ & $ -1 $ & $ r_1 $ & $t_3 = x f^{-1} \overline{t}_1 $ \\
$p_4$ & $ 0 $ & $ +1  $ & $0  $ & $ r_1 $ & $t_4 = y \overline{t}_1 $ \\
$p_5$ & $ 0 $ & $ -1 $ & $ 0 $ & $ r_1 $ & $t_5= y^{-1} \overline{t}_1 $ \\
$p_6$ & $ -1 $ & $ 0 $ & $ -1 $ & $ r_1 $ & $t_6= x^{-1} f^{-1}\overline{t}_1$ \\
\hline
\end{tabular}
\caption{Charges under the global symmetry of the $P_{+-}^{2}[\mathbb{C}^3/\mathbb{Z}_5~(1,1,3)]$ model on the extremal GLSM fields $p_a$.
Here, $U(1)_R$ charges $r_1$ and $r_2$ are chosen such that 
the $J$- and $E$-terms coupled to Fermi fields have an overall $U(1)_R$ charge of $2$
with $4r_1+r_2=2$. \label{tab_41}}
\end{table}

The Hilbert series of the mesonic moduli space $\mathcal{M}^{mes}_{P_{+-}^{2}[\mathbb{C}^3/\mathbb{Z}_5~(1,1,3)]}$ takes the following form, 
\beal{es08a25}
&&
g(t_a,y_s,y_w,y_u,;\mathcal{M}^{mes}_{P_{+-}^{2}[\mathbb{C}^3/\mathbb{Z}_5~(1,1,3)]})
= 
\nn\\
&&
\hspace{0.5cm}
\frac{
P(t_a,y_s,y_w,y_u,;\mathcal{M}^{mes}_{P_{+-}^{2}[\mathbb{C}^3/\mathbb{Z}_5~(1,1,3)]})
}{
(1-y_s^3 y_w y_u t_2^5) (1-y_s y_w^2 y_u^7 t_3^5 t_4^5) 
(1-y_s y_w^2 y_u^7 t_3^5 t_5^5) (1-y_s y_w^2 y_u^7 t_4^5 t_6^5)
(1-y_s y_w^2 y_u^7 t_5^5 t_6^5)
} 
~,~
\nn\\
\eea
where $t_a$ are the fugacities corresponding to the extremal GLSM fields $p_a$,
and $y_s$,$y_w$,$y_u$ are fugacities that count the products of GLSM fields $s_1\dots s_{10}$, $w_1 \dots w_{15}$ , $u_1 \dots u_{10}$, respectively. 
The numerator $P(t_a,y_s,y_w,y_u,;\mathcal{M}^{mes}_{P_{+-}^{2}[\mathbb{C}^3/\mathbb{Z}_5~(1,1,3)]})$ in \eref{es08a25} is presented in appendix \sref{app_num_08}.

The Hilbert series for $\mathcal{M}^{mes}_{P_{+-}^{2}[\mathbb{C}^3/\mathbb{Z}_5~(1,1,3)]}$ in \eref{es08a25}
can be rewritten in terms of characters of irreducible representations of the enhanced mesonic flavor symmetry $SU(2)_x \times SU(2)_{y} \times U(1)_{f} \times U(1)_R$.
This can be done using the following fugacity map,
\beal{es08a26}
\overline{t}_1= t_4^{1/2} t_5^{1/2}~,~
\overline{t}_2= \frac{t_2 t_3 t_6}{t_4 t_5}~,~
x=\frac{t_3^{1/2}}{t_6^{1/2}}~,~ 
y=\frac{t_4^{1/2}}{t_5^{1/2}}~,~ 
f=\frac{t_4^{1/2} t_5^{1/2}}{t_3^{1/2} t_6^{1/2}}
~,~
\eea
where the fugacities $\overline{t}_1, \overline{t}_2$ correspond to the $U(1)_R$ charges $r_1,r_2$
defined in \tref{tab_41}, 
and the fugacities $x$, $y$ and $f$ correspond to the mesonic flavor symmetry factors $SU(2)_x$, $SU(2)_y$ and $U(1)_f$, respectively.
By setting $y_s=y_r=y_e=1$, we can rewrite the Hilbert series in \eref{es08a25}
in terms of the characters of irreducible representations for $SU(2)_x \times SU(2)_y$.
The corresponding highest weight generating function takes the following form,
\beal{es08a27}
&&
h(\overline{t}_1,\overline{t}_2,f,\mu,\nu;\mathcal{M}^{mes}_{P_{+-}^{2}[\mathbb{C}^3/\mathbb{Z}_5~(1,1,3)]})
=
\nn\\
&&
\hspace{2.5cm}
\frac{1 + f^2 \mu^2 \nu^2 \overline{t}_1^4 \overline{t}_2 + f^4 \mu^4 \nu^4 \overline{t}_1^8 \overline{t}_2^2 + 
 f^6 \mu \nu \overline{t}_1^2 \overline{t}_2^3 + f^8 \mu^3 \nu^3 \overline{t}_1^6 \overline{t}_2^4}{(1-f^{10} \overline{t}_2^5)(1-\mu^5 \nu^5 \overline{t}_1^{10})} 
 ~,~
\eea
where $\mu^m \nu^n$ counts characters of irreducible representations of $SU(2)_x \times SU(2)_y$ of the form $[m]_x [n]_y$.

The plethystic logarithm of the refined Hilbert series 
corresponding to \eref{es08a27}
takes the following form, 
\beal{es08a28}
&&
PL[g(\overline{t}_1,\overline{t}_2,f,x,y; \mathcal{M}^{mes}_{P_{+-}^{2}[\mathbb{C}^3/\mathbb{Z}_5~(1,1,3)]} )]  
=
f^{10} \overline{t}_2^5
+[1]_x [1]_y f^5 \overline{t}_1^2 \overline{t}_2^3
+[2]_x [2]_y \overline{t}_1^4 \overline{t}_2
\nn\\
&&
\hspace{1cm} 
+f^{-5} [5]_x [5]_y  \overline{t}_1^{10}
- ( f^{10} \overline{t}_1^4 \overline{t}_2^6
+ \overline{t}_1^8 \overline{t}_2^2
+  [1]_x [1]_y f^5 \overline{t}_1^6 \overline{t}_2^4 
+ [1]_x [3]_y f^5 \overline{t}_1^6 \overline{t}_2^4
+[2]_x [2]_y \overline{t}_1^8 \overline{t}_2^2 
\nn\\
&&
\hspace{1cm} 
+ [2]_x [2]_y f^{10} \overline{t}_1^4 \overline{t}_2^6
+[3]_x [1]_y f^5 \overline{t}_1^6 \overline{t}_2^4+[4]_x \overline{t}_1^8 \overline{t}_2^2 + [4]_y \overline{t}_1^8 \overline{t}_2^2) 
+\dots 
~,~
\eea
where we note that the mesonic moduli space $\mathcal{M}^{mes}_{P_{+-}^{2}[\mathbb{C}^3/\mathbb{Z}_5~(1,1,3)]}$ is not a complete intersection.
We can extract the generators for $\mathcal{M}^{mes}_{P_{+-}^{2}[\mathbb{C}^3/\mathbb{Z}_5~(1,1,3)]}$ from the first positive terms in the plethystic logarithm. 
All generators for the mesonic moduli space $\mathcal{M}^{mes}_{P_{+-}^{2}[\mathbb{C}^3/\mathbb{Z}_5~(1,1,3)]}$ are summarized
with their mesonic flavor symmetry charges in \tref{tab_120}.
\\

\begin{table}[ht!]
\centering
\resizebox{0.53\textwidth}{!}{
\begin{tabular}{|c|c|c|c|c|c|c|}
\hline
PL term  & GLSM fields & $SU(2)_{x}$ & $SU(2)_{y}$ & $U(1)_{f}$  & fugacity\\
\hline\hline
\multirow{1}{*}{$+f^{10} \overline{t}_2^5$}&  $p_2^5 s^3 w u$ & $0$ &$0$&$+10$ & $f^{10} \overline{t}_2^5$\\
\hline
\multirow{4}{*}{$+[1]_x [1]_y  f^5 \overline{t}_1^2 \overline{t}_2^3$}&  $p_2^3 p_3 p_4 s^2 w u^2$ & $+1$ &$+1$&$+5$ & $x y f^5 \overline{t}_1^2 \overline{t}_2^3$\\
&  $p_2^3 p_3 p_5 s^2 w u^2$ & $+1$ &$-1$&$+5$  & $x y^{-1} f^5 \overline{t}_1^2 \overline{t}_2^3$\\
&  $p_2^3 p_4 p_6 s^2 w u^2$ & $-1$ &$+1$&$+5$  & $x^{-1} y f^5\overline{t}_1^2 \overline{t}_2^3$\\
&  $p_2^3 p_5 p_6 s^2 w u^2$ & $-1$ &$-1$&$+5$ & $x^{-1} y^{-1} f^5  \overline{t}_1^2 \overline{t}_2^3$\\
\hline
\multirow{9}{*}{$+[2]_x [2]_y \overline{t}_1^4 \overline{t}_2$}&  $p_2 p_3^2 p_4^2 s w u^3 $ & $+2$ &$+2$&$0$  & $x^2 y^2 \overline{t}_1^4 \overline{t}_2$\\
&  $p_2 p_3^2 p_4 p_5 s w u^3 $ & $+2$ &$0$&$0$  & $x^2 \overline{t}_1^4 \overline{t}_2$\\
&  $p_2 p_3^2 p_5^2 s w u^3 $ & $+2$ &$-2$&$0$ & $x^2 y^{-2} \overline{t}_1^4 \overline{t}_2$\\
&  $p_2 p_3 p_4^2 p_6 s w u^3 $ & $0$ &$+2$&$0$ & $y^2 \overline{t}_1^4 \overline{t}_2$\\
&  $p_2 p_3 p_4 p_5 p_6 s w u^3 $ & $0$ &$0$&$0$ & $ \overline{t}_1^4 \overline{t}_2$\\
&  $p_2 p_3 p_5^2 p_6 s w u^3 $ & $0$ &$-2$&$0$ & $y^{-2} \overline{t}_1^4 \overline{t}_2$\\
&  $p_2 p_4^2 p_6^2 s w u^3 $ & $-2$ &$+2$&$0$ & $x^{-2} y^2 \overline{t}_1^4 \overline{t}_2$\\
&  $p_2 p_4 p_5 p_6^2 s w u^3 $ & $-2$ &$0$&$0$  & $x^{-2} \overline{t}_1^4 \overline{t}_2$\\
&  $p_2 p_5^2 p_6^2 s w u^3 $ & $-2$ &$-2$&$0$ & $x^{-2} y^{-2} \overline{t}_1^4 \overline{t}_2$\\
\hline
\multirow{36}{*}{$+[5]_x [5]_y f^{-5} \overline{t}_1^{10}$}&  $p_3^5 p_4^5 s w^2 u^7 $ & $+5$ &$+5$&$-5$ & $x^5 y^5 f^{-5} \overline{t}_1^{10}$\\
&  $p_3^5 p_4^4 p_5 s w^2 u^7$ & $+5$ &$+3$&$-5$ & $x^5 y^3 f^{-5}\overline{t}_1^{10}$\\
&  $p_3^5 p_4^3 p_5^2 s w^2 u^7$ & $+5$ &$+1$&$-5$ & $x^5 y f^{-5}\overline{t}_1^{10}$\\
&  $p_3^5 p_4^2 p_5^3 s w^2 u^7$ & $+5$ &$-1$&$-5$ & $x^5 y^{-1} f^{-5} \overline{t}_1^{10}$\\
&  $p_3^5 p_4 p_5^4 s w^2 u^7$ & $+5$ &$-3$&$-5$  & $x^5 y^{-3} f^{-5} \overline{t}_1^{10}$\\
&  $p_3^5 p_5^5 s w^2 u^7$ & $+5$ &$-5$&$-5$  & $x^5 y^{-5} f^{-5} \overline{t}_1^{10}$\\
&  $p_3^4 p_4^5 p_6  s w^2 u^7$ & $+3$ &$+5$&$-5$  & $x^3 y^5 f^{-5} \overline{t}_1^{10}$\\
&  $p_3^4 p_4^4 p_5 p_6 s w^2 u^7$ & $+3$ &$+3$&$-5$  & $x^3 y^3 f^{-5} \overline{t}_1^{10}$\\
&  $p_3^4 p_4^3 p_5^2 p_6 s w^2 u^7$ & $+3$ &$+1$&$-5$  & $x^3 y f^{-5} \overline{t}_1^{10}$\\
&  $p_3^4 p_4^2 p_5^3 p_6 s w^2 u^7$ & $+3$ &$-1$&$-5$  & $x^3 y^{-1} f^{-5} \overline{t}_1^{10}$\\
&  $p_3^4 p_4 p_5^4 p_6 s w^2 u^7$ & $+3$ &$-3$&$-5$  & $x^3 y^{-3} f^{-5} \overline{t}_1^{10}$\\
&  $p_3^4 p_5^5 p_6 s w^2 u^7$ & $+3$ &$-5$&$-5$ & $x^3 y^{-5} f^{-5} \overline{t}_1^{10}$\\
&  $p_3^3 p_4^5 p_6^2 s w^2 u^7$ & $+1$ &$+5$&$-5$ & $x y^5 f^{-5} \overline{t}_1^{10}$\\
&  $p_3^3 p_4^4 p_5 p_6^2 s w^2 u^7$ & $+1$ &$+3$&$-5$  & $x y^3 f^{-5} \overline{t}_1^{10}$\\
&  $p_3^3 p_4^3 p_5^2 p_6^2 s w^2 u^7$ & $+1$ &$+1$&$-5$ & $x y f^{-5} \overline{t}_1^{10}$\\
&  $p_3^3 p_4^2 p_5^3 p_6^2 s w^2 u^7$ & $+1$ &$-1$&$-5$ & $x y^{-1} f^{-5} \overline{t}_1^{10}$\\
&  $p_3^3 p_4 p_5^4 p_6^2 s w^2 u^7$ & $+1$ &$-3$&$-5$ & $x y^{-3} f^{-5} \overline{t}_1^{10}$\\
&  $p_3^3 p_5^5 p_6^2 s w^2 u^7$ & $+1$ &$-5$&$-5$ & $x y^{-5} f^{-5} \overline{t}_1^{10}$\\
&  $p_3^2 p_4^5 p_6^3 s w^2 u^7$ & $-1$ &$+5$&$-5$ & $x^{-1} y^5 f^{-5} \overline{t}_1^{10}$\\
&  $p_3^2 p_4^4 p_5 p_6^3 s w^2 u^7$ & $-1$ &$+3$&$-5$& $x^{-1} y^3 f^{-5} \overline{t}_1^{10}$\\
&  $p_3^2 p_4^3 p_5^2 p_6^3 s w^2 u^7$ & $-1$ &$+1$&$-5$  & $x^{-1} y f^{-5} \overline{t}_1^{10}$\\
&  $p_3^2 p_4^2 p_5^3 p_6^3 s w^2 u^7$ & $-1$ &$-1$&$-5$  & $x^{-1} y^{-1} f^{-5} \overline{t}_1^{10}$\\
&  $p_3^2 p_4 p_5^4 p_6^3 s w^2 u^7$ & $-1$ &$-3$&$-5$ & $x^{-1} y^{-3} f^{-5} \overline{t}_1^{10}$\\
&  $p_3^2 p_5^5 p_6^3 s w^2 u^7$ & $-1$ &$-5$&$-5$  & $x^{-1} y^{-5} f^{-5} \overline{t}_1^{10}$\\
&  $p_3 p_4^5 p_6^4 s w^2 u^7$ & $-3$ &$+5$&$-5$ & $x^{-3} y^5 f^{-5} \overline{t}_1^{10}$\\
&  $p_3 p_4^4 p_5 p_6^4 s w^2 u^7$ & $-3$ &$+3$&$-5$ & $x^{-3} y^3 f^{-5} \overline{t}_1^{10}$\\
&  $p_3 p_4^3 p_5^2 p_6^4 s w^2 u^7$ & $-3$ &$+1$&$-5$ & $x^{-3} y f^{-5} \overline{t}_1^{10}$\\
&  $p_3 p_4^2 p_5^3 p_6^4 s w^2 u^7$ & $-3$ &$-1$&$-5$ & $x^{-3} y^{-1} f^{-5} \overline{t}_1^{10}$\\
&  $p_3 p_4 p_5^4 p_6^4 s w^2 u^7$ & $-3$ &$-3$&$-5$  & $x^{-3} y^{-3} f^{-5} \overline{t}_1^{10}$\\
&  $p_3 p_5^5 p_6^4 s w^2 u^7$ & $-3$ &$-5$&$-5$  & $x^{-3} y^{-5} f^{-5} \overline{t}_1^{10}$\\
&  $p_4^5 p_6^5 s w^2 u^7$ & $-5$ &$+5$&$-5$ & $x^{-5} y^5 f^{-5} \overline{t}_1^{10}$\\
&  $p_4^4 p_5 p_6^5 s w^2 u^7$ & $-5$ &$+3$&$-5$ & $x^{-5} y^3 f^{-5} \overline{t}_1^{10}$\\
&  $p_4^3 p_5^2 p_6^5 s w^2 u^7$ & $-5$ &$+1$&$-5$  & $x^{-5} y f^{-5} \overline{t}_1^{10}$\\
&  $p_4^2 p_5^3 p_6^5 s w^2 u^7$ & $-5$ &$-1$&$-5$ & $x^{-5} y^{-1} f^{-5} \overline{t}_1^{10}$\\
&  $p_4 p_5^4 p_6^5 s w^2 u^7$ & $-5$ &$-3$&$-5$ & $x^{-5} y^{-3} f^{-5} \overline{t}_1^{10}$\\
&  $p_5^5 p_6^5s w^2 u^7$ & $-5$ &$-5$&$-5$& $x^{-5} y^{-5} f^{-5} \overline{t}_1^{10}$\\
\hline
\end{tabular}}
\caption{
Generators of the mesonic moduli space $\mathcal{M}^{mes}_{P_{+-}^{2}[\mathbb{C}^3/\mathbb{Z}_5~(1,1,3)]}$ 
of the $P_{+-}^{2}[\mathbb{C}^3/\mathbb{Z}_5~(1,1,3)]$ brane brick model in terms of chiral fields and GLSM fields with the corresponding mesonic flavor symmetry charges.
Here, $s$, $w$ and $u$ denote the products of GLSM fields $s_1 \dots s_{10}$, $w_1 \dots w_{15}$ and $u_1 \dots u_{10}$, respectively.
\label{tab_120}
}
\end{table}

\paragraph{Comparison between $P_{+-}^{1}[\mathbb{C}^3/\mathbb{Z}_5~(1,1,3)]$ and $P_{+-}^{2}[\mathbb{C}^3/\mathbb{Z}_5~(1,1,3)]$.}
From the plethystic logarithm in \eref{es08a17}, we see that the generators of the mesonic moduli space $\mathcal{M}^{mes}_{P_{+-}^{1}[\mathbb{C}^3/\mathbb{Z}_5~(1,1,3)]}$ of the $P_{+-}^{1}[\mathbb{C}^3/\mathbb{Z}_5~(1,1,3)]$ brane brick model
transform under the following irreducible representations of the mesonic flavor symmetry $SU(2)_x \times U(1)_{f_1} \times U(1)_{f_2}$, 
\beal{es08a30}
&
f_1^5 f_2^{-10} ~,~
f_1^2 f_2^{-6} ~,~
[2]_x f_1^3 f_2^{-4} ~,~
f_1^{-1} f_2^{-2} ~,~
[2]_x ~,~
[4]_x f_1 f_2^2
~,~
&
\nn\\
&
[10]_x f_2^{10} ~,~
[8]_x f_1^{-1} ~,~
[6]_x f_1^{-2} f_2^6 ~,~
[4]_x f_1^{-3} f_2^4 ~,~
[2]_x f_1^{-4} f_2^2 ~,~
f_1^{-5}
~,~
&
\eea
giving a total of $50$ generators.
In comparison, the generators of the mesonic moduli space $\mathcal{M}^{mes}_{P_{+-}^{2}[\mathbb{C}^3/\mathbb{Z}_5~(1,1,3)]}$ of the $P_{+-}^{2}[\mathbb{C}^3/\mathbb{Z}_5~(1,1,3)]$ brane brick model transform under the following irreducible representations of the mesonic flavor symmetry
$SU(2)_x \times SU(2)_y \times U(1)_f$
based on the plethystic logarithm in \eref{es08a28},
\beal{es08a31}
&
f^{10} ~,~
[1]_x [1]_y f^5 ~,~
[2]_x [2]_y ~,~
[5]_x [5]_y ~,~
&
\eea
which gives in total $50$ generators. 
We can clearly see here that the total number of generators of the mesonic moduli space stays invariant 
when one goes from the brane brick model $P_{+-}^{1}[\mathbb{C}^3/\mathbb{Z}_5~(1,1,3)]$
to the brane brick model $P_{+-}^{2}[\mathbb{C}^3/\mathbb{Z}_5~(1,1,3)]$ under the mass deformation introduced in \eref{es08a18}.
This confirms the general result in this work that the total number of generators of the mesonic moduli spaces
stays invariant if the corresponding brane brick models are associated to toric Calabi-Yau 4-folds, which are related by a birational transformation of the general form in \eref{es01a52}.

We can also take a closer look at the refined Hilbert series in \eref{es08a16}
for the mesonic moduli space $\mathcal{M}^{mes}_{P_{+-}^{1}[\mathbb{C}^3/\mathbb{Z}_5~(1,1,3)]}$.
By setting all mesonic flavor fugacities to $x=f_1 = f_2 = 1$, the Hilbert series corresponding to \eref{es08a16} for $\mathcal{M}^{mes}_{P_{+-}^{1}[\mathbb{C}^3/\mathbb{Z}_5~(1,1,3)]}$
can be unrefined such that it is only in terms of $U(1)_R$ fugacities $\overline{t}_1, \overline{t}_2$ associated to the $U(1)_R$ charges $r_1, r_2$ defined in \tref{tab_40}.
The resulting unrefined Hilbert series for $\mathcal{M}^{mes}_{P_{+-}^{1}[\mathbb{C}^3/\mathbb{Z}_5~(1,1,3)]}$ takes the following form, 
\beal{es08a32}
&&
g(\overline{t}_1,\overline{t}_2; \mathcal{M}^{mes}_{P_{+-}^{1}[\mathbb{C}^3/\mathbb{Z}_5~(1,1,3)]})
= 
\frac{1}{
(1- \overline{t}_1^{10})^3 (1-\overline{t}_2^{5})
} 
\times
\Big[ 
1
+33 \overline{t}_1^{10}
+16 \overline{t}_1^{20}
+9 \overline{t}_1^4 \overline{t}_2
\nn\\
&& 
\hspace{1cm}
+37 \overline{t}_1^{14} \overline{t}_2
+4 \overline{t}_1^{24} \overline{t}_2
+25 \overline{t}_1^8 \overline{t}_2^2
+25 \overline{t}_1^{18} \overline{t}_2^2
+4 \overline{t}_1^2 \overline{t}_2^3
+37 \overline{t}_1^{12} \overline{t}_2^3
+9 \overline{t}_1^{22} \overline{t}_2^3
+16 \overline{t}_1^6 \overline{t}_2^4
\nn\\
&& 
\hspace{1cm}
+33 \overline{t}_1^{16} \overline{t}_2^4
+\overline{t}_1^{26} \overline{t}_2^4
\Big] 
~.~ 
\eea
We can also express the Hilbert series for the mesonic moduli space $\mathcal{M}^{mes}_{P_{+-}^{2}[\mathbb{C}^3/\mathbb{Z}_5~(1,1,3)]}$
corresponding to \eref{es08a27} in terms of only $U(1)_R$ fugacities $\overline{t}_1, \overline{t}_2$
corresponding to $U(1)_R$ charges $r_1, r_2$ defined in \tref{tab_41}.
This can be done by setting the mesonic flavor symmetry fugacities to $x=y=f=1$.
The resulting unrefined Hilbert series for $\mathcal{M}^{mes}_{P_{+-}^{2}[\mathbb{C}^3/\mathbb{Z}_5~(1,1,3)]}$ takes the following form,
\beal{es08a33}
&&
g(\overline{t}_1,\overline{t}_2; \mathcal{M}^{mes}_{P_{+-}^{2}[\mathbb{C}^3/\mathbb{Z}_5~(1,1,3)]})
= 
\frac{1}{
(1- \overline{t}_1^{10})^3 (1-\overline{t}_2^{5})
} 
\times
\Big[ 
1
+33 \overline{t}_1^{10}
+16 \overline{t}_1^{20}
+9 \overline{t}_1^4 \overline{t}_2
\nn\\
&& 
\hspace{1cm}
+37 \overline{t}_1^{14} \overline{t}_2
+4 \overline{t}_1^{24} \overline{t}_2
+25 \overline{t}_1^8 \overline{t}_2^2
+25 \overline{t}_1^{18} \overline{t}_2^2
+4 \overline{t}_1^2 \overline{t}_2^3
+37 \overline{t}_1^{12} \overline{t}_2^3
+9 \overline{t}_1^{22} \overline{t}_2^3
+16 \overline{t}_1^6 \overline{t}_2^4
\nn\\
&& 
\hspace{1cm}
+33 \overline{t}_1^{16} \overline{t}_2^4
+\overline{t}_1^{26} \overline{t}_2^4 
\Big]
~.~ 
\eea

We can see here that the unrefined Hilbert series in \eref{es08a32} and \eref{es08a33} in terms of only $U(1)_R$ fugacities are identical for brane brick models 
$P_{+-}^{1}[\mathbb{C}^3/\mathbb{Z}_5~(1,1,3)]$ and $P_{+-}^{2}[\mathbb{C}^3/\mathbb{Z}_5~(1,1,3)]$.
This further supports our observations that 
the Hilbert series for mesonic moduli spaces refined under only $U(1)_R$ fugacities remain invariant between brane brick models
that correspond to toric Calabi-Yau 4-folds, which are related by a birational transformation of the general form in \eref{es01a52}.
In this example, 
the birational transformation that relates
$P_{+-}^{1}[\mathbb{C}^3/\mathbb{Z}_5~(1,1,3)]$ with $P_{+-}^{2}[\mathbb{C}^3/\mathbb{Z}_5~(1,1,3)]$ is given in \eref{es08a03}.
Our observations also show here that birational transformations of the form in \eref{es01a52}
that connect
toric Calabi-Yau 4-folds with toric diagrams, which are not reflexive and have more than one internal point, 
do exist.
We also observe here that such birational transformations for toric Calabi-Yau 4-folds beyond toric Fano 3-folds
can be related to mass deformations between the corresponding brane brick models.
\\

\section{Discussions and Conclusions \label{sec:conc} }

In this work, we provide further evidence that 
birational transformations relating toric Fano 3-folds
can be identified with mass deformation of corresponding $2d$ $(0,2)$ supersymmetric gauge theories realized by brane brick models.
As a consequence of these birational transformations,
we show that 
the corresponding abelian $2d$ $(0,2)$ supersymmetric gauge theories
have mesonic moduli spaces with the same number of generators,
and that their Hilbert series coincide when refined only under the $U(1)_R$ symmetry. 

As first outlined in \cite{Ghim:2024asj}, 
we expect 
that, in general, 
whenever two brane brick models and their corresponding $2d$ $(0,2)$ supersymmetric gauge theories
are related by a mass deformation, 
the associated toric Calabi-Yau 4-folds
are related by a birational transformation
of the general form discussed in section \sref{sec:03}.
Consequently, we anticipate that,
among the 4319 reflexive polytopes in $\mathbb{Z}^3$, the 3025 
exhibiting non-trivial birational transformations 
also correspond to abelian brane brick models that
share the same number of generators for their mesonic moduli spaces
and the same Hilbert series when refined under the $U(1)_R$ symmetry. 

We further show in this work that the family of birational transformations
acting on toric Fano 3-folds
can be extended to more general toric Calabi-Yau 4-folds, whose toric diagrams are not necessary reflexive polytopes in $\mathbb{Z}^3$.
Our work identifies multiple examples
of such birational transformations on
non-reflexive toric diagrams -- both with and without internal lattice points -- 
and their associated toric Calabi-Yau 4-folds.
In several examples, we demonstrate that when two
$2d$ $(0,2)$ supersymmetric gauge theories realized by brane brick models
are related by a mass deformation, 
the corresponding toric Calabi-Yau 4-folds are likewise related by a birational transformation,
even if their toric diagrams are non-reflexive.
Moreover, we show that the birational transformations on non-reflexive toric diagrams
preserve the number of generators of the mesonic moduli space for the abelian $2d$ $(0,2)$ theories,
as well as their Hilbert series when refined under the $U(1)_R$ symmetry. 

Beyond toric Fano 3-folds,
we expect the birational transformations investigated here to organize toric Calabi-Yau 4-folds
and their corresponding brane brick models into large equivalence classes -- referred to as `buckets'
for toric Fano 3-folds in \cite{akhtar2012minkowski}.
Drawing on the insights from the Minimal Model Program in algebraic geometry, 
we anticipate that this work will set the stage for a systematic search for such equivalence classes among 
toric Calabi–Yau 4-folds and their corresponding
$2d$ $(0,2)$ supersymmetric gauge theories realized by brane brick models.

Moreover, 
our study is timely in light of the extensive work 
already carried out 
on the family of $4d$ $\mathcal{N}=1$ supersymmetric gauge theories arising 
as worldvolume theories of D3-branes probing toric Calabi-Yau 3-folds \cite{Douglas:1997de}.
These theories can be described by bipartite graphs on a 2-torus known as brane tilings \cite{Hanany:2005ve, Franco:2005rj, Franco:2005sm}
that encode Type IIB brane configurations 
T-dual to the probe D3-branes at the Calabi-Yau singularity. 
The work of \cite{Hanany:2012hi}
classified all $4d$ $\mathcal{N}=1$ supersymmetric gauge theories realized by brane tilings
corresponding to toric Calabi-Yau 3-folds
with reflexive toric diagrams given by the 16 reflexive polygons in $\mathbb{Z}^2$.
Many of these $4d$ $\mathcal{N}=1$ theories have been shown to be related to each other by mass deformations \cite{Bianchi:2014qma, Cremonesi:2023psg}, 
which can be identified with birational transformations of the corresponding reflexive toric diagrams and toric varieties.
These transformations constitute simpler analogues of the family of birational transformations that we study in this work 
and have been linked to local deformations of the brane tiling along its zig-zag paths \cite{higashitani2022deformations}.
More recently, these mutations on the toric diagrams for brane tilings
have been recognized as a pathway towards describing generalized toric polygons (GTPs)
and corresponding dual webs of 5-branes suspended from 7-branes \cite{Benini:2009gi, vanBeest:2020kou, Franco:2023flw, Arias-Tamargo:2024fjt, CarrenoBolla:2024fxy}.

It is promising to explore the parallels
between phenomena arising in brane tilings
and brane brick models due to birational transformations 
on their corresponding toric varieties, 
especially in light of the established connections via dimensional reduction, 
orbifold reduction \cite{Franco:2016fxm} and the more general framework of $3d$ printing \cite{Franco:2018qsc}.
Inspired by recent developments in brane tilings
and the results presented here for brane brick models,
we look forward to reporting further findings in the near future. 
\\

\section*{Acknowledgements}

D.G. is supported by JST PRESTO Grant Number JPMJPR2117.
D.G. would like to thank UNIST where this project was initiated.
R.-K. S. is supported by an Outstanding Young Scientist Grant (RS-2025-00516583) of the National Research Foundation of Korea (NRF).
He is also partly supported by the BK21 Program
(“Next Generation Education Program for Mathematical Sciences”, 4299990414089)
funded by the Ministry of Education in Korea and the National Research Foundation
of Korea.

\addtocontents{toc}{\protect\setcounter{tocdepth}{1}}
\appendix

\section{Numerators for the fully refined Hilbert Series \label{sec_app}}

We present here the palindromic 
numerators in the rational form of the
Hilbert series for mesonic moduli spaces of brane brick models studied in this work.

\subsection{$F_{0,+-}$ \label{app_num_01}}

\begingroup\makeatletter\def\f@size{7}\check@mathfonts
\begin{quote}\raggedright \label{es90a01}
$
P(t_a,y_q,y_s,y_o;\mathcal{M}^{mes}_{F_{0,+-}})=1+y_q^2 y_s y_o^3 t_1^4 t_2 t_3+y_q y_s y_o^2 t_1^2 t_2^2 t_4+y_q y_s 
y_o^2 t_1^2 t_2 t_3 t_4+y_q y_s y_o^2 t_1^2 t_3^2 t_4-y_q^3 y_s^2 
y_o^5 t_1^6 t_2^2 t_3^2 t_4+y_s y_o t_2 t_3 t_4^2-y_q^2 y_s^2 y_o^4 
t_1^4 t_2^3 t_3 t_4^2-y_q^2 y_s^2 y_o^4 t_1^4 t_2^2 t_3^2 t_4^2-y_q^2 
y_s^2 y_o^4 t_1^4 t_2 t_3^3 t_4^2-y_q y_s^2 y_o^3 t_1^2 t_2^2 t_3^2 
t_4^3-y_q^3 y_s^3 y_o^6 t_1^6 t_2^3 t_3^3 t_4^3+y_q^2 y_s y_o^3 t_1^3 
t_2^2 t_5+y_q^2 y_s y_o^3 t_1^3 t_2 t_3 t_5+y_q^2 y_s y_o^3 t_1^3 
t_3^2 t_5-y_q^4 y_s^2 y_o^6 t_1^7 t_2^2 t_3^2 t_5+y_q y_s y_o^2 t_1 
t_2^2 t_4 t_5+y_q y_s y_o^2 t_1 t_2 t_3 t_4 t_5+y_q y_s y_o^2 t_1 
t_3^2 t_4 t_5-y_q^3 y_s^2 y_o^5 t_1^5 t_2^2 t_3^2 t_4 t_5-y_q^2 y_s^2 
y_o^4 t_1^3 t_2^2 t_3^2 t_4^2 t_5-y_q^4 y_s^3 y_o^7 t_1^7 t_2^3 t_3^3 
t_4^2 t_5-y_q y_s^2 y_o^3 t_1 t_2^2 t_3^2 t_4^3 t_5-y_q^3 y_s^3 y_o^6 
t_1^5 t_2^3 t_3^3 t_4^3 t_5+y_q^2 y_s y_o^3 t_1^2 t_2^2 t_5^2+y_q^2 
y_s y_o^3 t_1^2 t_2 t_3 t_5^2+y_q^2 y_s y_o^3 t_1^2 t_3^2 t_5^2-y_q^4 
y_s^2 y_o^6 t_1^6 t_2^2 t_3^2 t_5^2+y_q y_s y_o^2 t_2^2 t_4 t_5^2+y_q 
y_s y_o^2 t_2 t_3 t_4 t_5^2+y_q y_s y_o^2 t_3^2 t_4 t_5^2-y_q^3 y_s^2 
y_o^5 t_1^4 t_2^2 t_3^2 t_4 t_5^2-y_q^2 y_s^2 y_o^4 t_1^2 t_2^2 t_3^2 
t_4^2 t_5^2-y_q^4 y_s^3 y_o^7 t_1^6 t_2^3 t_3^3 t_4^2 t_5^2-y_q y_s^2 
y_o^3 t_2^2 t_3^2 t_4^3 t_5^2-y_q^3 y_s^3 y_o^6 t_1^4 t_2^3 t_3^3 
t_4^3 t_5^2+y_q^2 y_s y_o^3 t_1 t_2^2 t_5^3+y_q^2 y_s y_o^3 t_1 t_2 
t_3 t_5^3+y_q^2 y_s y_o^3 t_1 t_3^2 t_5^3-y_q^4 y_s^2 y_o^6 t_1^5 
t_2^2 t_3^2 t_5^3+y_q^3 y_s^2 y_o^5 t_1^3 t_2^4 t_4 t_5^3+y_q^3 y_s^2 
y_o^5 t_1^3 t_2^3 t_3 t_4 t_5^3+y_q^3 y_s^2 y_o^5 t_1^3 t_2^2 t_3^2 
t_4 t_5^3-y_q^5 y_s^3 y_o^8 t_1^7 t_2^4 t_3^2 t_4 t_5^3+y_q^3 y_s^2 
y_o^5 t_1^3 t_2 t_3^3 t_4 t_5^3-y_q^5 y_s^3 y_o^8 t_1^7 t_2^3 t_3^3 
t_4 t_5^3+y_q^3 y_s^2 y_o^5 t_1^3 t_3^4 t_4 t_5^3-y_q^5 y_s^3 y_o^8 
t_1^7 t_2^2 t_3^4 t_4 t_5^3-y_q^2 y_s^2 y_o^4 t_1 t_2^2 t_3^2 t_4^2 
t_5^3-y_q^4 y_s^3 y_o^7 t_1^5 t_2^3 t_3^3 t_4^2 t_5^3-y_q^3 y_s^3 
y_o^6 t_1^3 t_2^4 t_3^2 t_4^3 t_5^3-y_q^3 y_s^3 y_o^6 t_1^3 t_2^3 
t_3^3 t_4^3 t_5^3-y_q^3 y_s^3 y_o^6 t_1^3 t_2^2 t_3^4 t_4^3 
t_5^3+y_q^5 y_s^4 y_o^9 t_1^7 t_2^4 t_3^4 t_4^3 t_5^3+y_q^2 y_s y_o^3 
t_2 t_3 t_5^4-y_q^4 y_s^2 y_o^6 t_1^4 t_2^3 t_3 t_5^4-y_q^4 y_s^2 
y_o^6 t_1^4 t_2^2 t_3^2 t_5^4-y_q^4 y_s^2 y_o^6 t_1^4 t_2 t_3^3 
t_5^4-y_q^3 y_s^2 y_o^5 t_1^2 t_2^2 t_3^2 t_4 t_5^4-y_q^5 y_s^3 y_o^8 
t_1^6 t_2^3 t_3^3 t_4 t_5^4-y_q^2 y_s^2 y_o^4 t_2^3 t_3 t_4^2 
t_5^4+y_q^4 y_s^3 y_o^7 t_1^4 t_2^5 t_3 t_4^2 t_5^4-y_q^2 y_s^2 y_o^4 
t_2^2 t_3^2 t_4^2 t_5^4+y_q^4 y_s^3 y_o^7 t_1^4 t_2^4 t_3^2 t_4^2 
t_5^4-y_q^2 y_s^2 y_o^4 t_2 t_3^3 t_4^2 t_5^4+y_q^4 y_s^3 y_o^7 t_1^4 
t_2^3 t_3^3 t_4^2 t_5^4+y_q^4 y_s^3 y_o^7 t_1^4 t_2^2 t_3^4 t_4^2 
t_5^4+y_q^4 y_s^3 y_o^7 t_1^4 t_2 t_3^5 t_4^2 t_5^4-y_q^3 y_s^3 y_o^6 
t_1^2 t_2^3 t_3^3 t_4^3 t_5^4+y_q^5 y_s^4 y_o^9 t_1^6 t_2^5 t_3^3 
t_4^3 t_5^4+y_q^5 y_s^4 y_o^9 t_1^6 t_2^4 t_3^4 t_4^3 t_5^4+y_q^5 
y_s^4 y_o^9 t_1^6 t_2^3 t_3^5 t_4^3 t_5^4-y_q^4 y_s^2 y_o^6 t_1^3 
t_2^2 t_3^2 t_5^5-y_q^6 y_s^3 y_o^9 t_1^7 t_2^3 t_3^3 t_5^5-y_q^3 
y_s^2 y_o^5 t_1 t_2^2 t_3^2 t_4 t_5^5-y_q^5 y_s^3 y_o^8 t_1^5 t_2^3 
t_3^3 t_4 t_5^5-y_q^4 y_s^3 y_o^7 t_1^3 t_2^3 t_3^3 t_4^2 t_5^5+y_q^6 
y_s^4 y_o^{10} t_1^7 t_2^5 t_3^3 t_4^2 t_5^5+y_q^6 y_s^4 y_o^{10} 
t_1^7 t_2^4 t_3^4 t_4^2 t_5^5+y_q^6 y_s^4 y_o^{10} t_1^7 t_2^3 t_3^5 
t_4^2 t_5^5-y_q^3 y_s^3 y_o^6 t_1 t_2^3 t_3^3 t_4^3 t_5^5+y_q^5 y_s^4 
y_o^9 t_1^5 t_2^5 t_3^3 t_4^3 t_5^5+y_q^5 y_s^4 y_o^9 t_1^5 t_2^4 
t_3^4 t_4^3 t_5^5+y_q^5 y_s^4 y_o^9 t_1^5 t_2^3 t_3^5 t_4^3 
t_5^5-y_q^4 y_s^2 y_o^6 t_1^2 t_2^2 t_3^2 t_5^6-y_q^6 y_s^3 y_o^9 
t_1^6 t_2^3 t_3^3 t_5^6-y_q^3 y_s^2 y_o^5 t_2^2 t_3^2 t_4 t_5^6-y_q^5 
y_s^3 y_o^8 t_1^4 t_2^3 t_3^3 t_4 t_5^6-y_q^4 y_s^3 y_o^7 t_1^2 t_2^3 
t_3^3 t_4^2 t_5^6+y_q^6 y_s^4 y_o^{10} t_1^6 t_2^5 t_3^3 t_4^2 
t_5^6+y_q^6 y_s^4 y_o^{10} t_1^6 t_2^4 t_3^4 t_4^2 t_5^6+y_q^6 y_s^4 
y_o^{10} t_1^6 t_2^3 t_3^5 t_4^2 t_5^6-y_q^3 y_s^3 y_o^6 t_2^3 t_3^3 
t_4^3 t_5^6+y_q^5 y_s^4 y_o^9 t_1^4 t_2^5 t_3^3 t_4^3 t_5^6+y_q^5 
y_s^4 y_o^9 t_1^4 t_2^4 t_3^4 t_4^3 t_5^6+y_q^5 y_s^4 y_o^9 t_1^4 
t_2^3 t_3^5 t_4^3 t_5^6-y_q^4 y_s^2 y_o^6 t_1 t_2^2 t_3^2 t_5^7-y_q^6 
y_s^3 y_o^9 t_1^5 t_2^3 t_3^3 t_5^7-y_q^5 y_s^3 y_o^8 t_1^3 t_2^4 
t_3^2 t_4 t_5^7-y_q^5 y_s^3 y_o^8 t_1^3 t_2^3 t_3^3 t_4 t_5^7-y_q^5 
y_s^3 y_o^8 t_1^3 t_2^2 t_3^4 t_4 t_5^7+y_q^7 y_s^4 y_o^{11} t_1^7 
t_2^4 t_3^4 t_4 t_5^7-y_q^4 y_s^3 y_o^7 t_1 t_2^3 t_3^3 t_4^2 
t_5^7+y_q^6 y_s^4 y_o^{10} t_1^5 t_2^5 t_3^3 t_4^2 t_5^7+y_q^6 y_s^4 
y_o^{10} t_1^5 t_2^4 t_3^4 t_4^2 t_5^7+y_q^6 y_s^4 y_o^{10} t_1^5 
t_2^3 t_3^5 t_4^2 t_5^7+y_q^5 y_s^4 y_o^9 t_1^3 t_2^4 t_3^4 t_4^3 
t_5^7+y_q^7 y_s^5 y_o^{12} t_1^7 t_2^5 t_3^5 t_4^3 t_5^7
~.~
$
\end{quote}
\endgroup

\subsection{$Q^{1,1,1}/\mathbb{Z}_2$ \label{app_num_02}}
\begingroup\makeatletter\def\f@size{7}\check@mathfonts
\begin{quote}\raggedright
$
P(t_a,y_s;\mathcal{M}^{mes}_{Q^{1,1,1}/ \mathbb{Z}_2})=
1+y_s t_2 t_3 t_4^2 t_5^2+y_s t_2^2 t_4^2 t_5 t_6+y_s t_2 t_3 t_4^2 
t_5 t_6+y_s t_3^2 t_4^2 t_5 t_6-y_s^2 t_2^2 t_3^2 t_4^4 t_5^3 t_6+y_s 
t_2 t_3 t_4^2 t_6^2-y_s^2 t_2^3 t_3 t_4^4 t_5^2 t_6^2-y_s^2 t_2^2 
t_3^2 t_4^4 t_5^2 t_6^2-y_s^2 t_2 t_3^3 t_4^4 t_5^2 t_6^2-y_s^2 t_2^2 
t_3^2 t_4^4 t_5 t_6^3-y_s^3 t_2^3 t_3^3 t_4^6 t_5^3 t_6^3+y_s t_2^2 
t_4 t_5^2 t_7+y_s t_2 t_3 t_4 t_5^2 t_7+y_s t_3^2 t_4 t_5^2 t_7-y_s^2 
t_2^2 t_3^2 t_4^3 t_5^4 t_7+y_s t_2^2 t_4 t_5 t_6 t_7+y_s t_2 t_3 t_4 
t_5 t_6 t_7+y_s t_3^2 t_4 t_5 t_6 t_7-y_s^2 t_2^2 t_3^2 t_4^3 t_5^3 
t_6 t_7+y_s t_2^2 t_4 t_6^2 t_7+y_s t_2 t_3 t_4 t_6^2 t_7+y_s t_3^2 
t_4 t_6^2 t_7-y_s^2 t_2^4 t_4^3 t_5^2 t_6^2 t_7-y_s^2 t_2^3 t_3 t_4^3 
t_5^2 t_6^2 t_7-3 y_s^2 t_2^2 t_3^2 t_4^3 t_5^2 t_6^2 t_7-y_s^2 t_2 
t_3^3 t_4^3 t_5^2 t_6^2 t_7-y_s^2 t_3^4 t_4^3 t_5^2 t_6^2 t_7+y_s^3 
t_2^4 t_3^2 t_4^5 t_5^4 t_6^2 t_7+y_s^3 t_2^2 t_3^4 t_4^5 t_5^4 t_6^2 
t_7-y_s^2 t_2^2 t_3^2 t_4^3 t_5 t_6^3 t_7-y_s^3 t_2^3 t_3^3 t_4^5 
t_5^3 t_6^3 t_7-y_s^2 t_2^2 t_3^2 t_4^3 t_6^4 t_7+y_s^3 t_2^4 t_3^2 
t_4^5 t_5^2 t_6^4 t_7+y_s^3 t_2^2 t_3^4 t_4^5 t_5^2 t_6^4 t_7-y_s^4 
t_2^4 t_3^4 t_4^7 t_5^4 t_6^4 t_7+y_s t_2 t_3 t_5^2 t_7^2-y_s^2 t_2^3 
t_3 t_4^2 t_5^4 t_7^2-y_s^2 t_2^2 t_3^2 t_4^2 t_5^4 t_7^2-y_s^2 t_2 
t_3^3 t_4^2 t_5^4 t_7^2+y_s t_2^2 t_5 t_6 t_7^2+y_s t_2 t_3 t_5 t_6 
t_7^2+y_s t_3^2 t_5 t_6 t_7^2-y_s^2 t_2^4 t_4^2 t_5^3 t_6 t_7^2-y_s^2 
t_2^3 t_3 t_4^2 t_5^3 t_6 t_7^2-3 y_s^2 t_2^2 t_3^2 t_4^2 t_5^3 t_6 
t_7^2-y_s^2 t_2 t_3^3 t_4^2 t_5^3 t_6 t_7^2-y_s^2 t_3^4 t_4^2 t_5^3 
t_6 t_7^2+y_s^3 t_2^4 t_3^2 t_4^4 t_5^5 t_6 t_7^2+y_s^3 t_2^2 t_3^4 
t_4^4 t_5^5 t_6 t_7^2+y_s t_2 t_3 t_6^2 t_7^2-y_s^2 t_2^4 t_4^2 t_5^2 
t_6^2 t_7^2-3 y_s^2 t_2^3 t_3 t_4^2 t_5^2 t_6^2 t_7^2-3 y_s^2 t_2^2 
t_3^2 t_4^2 t_5^2 t_6^2 t_7^2-3 y_s^2 t_2 t_3^3 t_4^2 t_5^2 t_6^2 
t_7^2-y_s^2 t_3^4 t_4^2 t_5^2 t_6^2 t_7^2+y_s^3 t_2^5 t_3 t_4^4 t_5^4 
t_6^2 t_7^2+2 y_s^3 t_2^4 t_3^2 t_4^4 t_5^4 t_6^2 t_7^2+2 y_s^3 t_2^3 
t_3^3 t_4^4 t_5^4 t_6^2 t_7^2+2 y_s^3 t_2^2 t_3^4 t_4^4 t_5^4 t_6^2 
t_7^2+y_s^3 t_2 t_3^5 t_4^4 t_5^4 t_6^2 t_7^2-y_s^2 t_2^4 t_4^2 t_5 
t_6^3 t_7^2-y_s^2 t_2^3 t_3 t_4^2 t_5 t_6^3 t_7^2-3 y_s^2 t_2^2 t_3^2 
t_4^2 t_5 t_6^3 t_7^2-y_s^2 t_2 t_3^3 t_4^2 t_5 t_6^3 t_7^2-y_s^2 
t_3^4 t_4^2 t_5 t_6^3 t_7^2+2 y_s^3 t_2^4 t_3^2 t_4^4 t_5^3 t_6^3 
t_7^2-y_s^3 t_2^3 t_3^3 t_4^4 t_5^3 t_6^3 t_7^2+2 y_s^3 t_2^2 t_3^4 
t_4^4 t_5^3 t_6^3 t_7^2+y_s^4 t_2^5 t_3^3 t_4^6 t_5^5 t_6^3 
t_7^2+y_s^4 t_2^3 t_3^5 t_4^6 t_5^5 t_6^3 t_7^2-y_s^2 t_2^3 t_3 t_4^2 
t_6^4 t_7^2-y_s^2 t_2^2 t_3^2 t_4^2 t_6^4 t_7^2-y_s^2 t_2 t_3^3 t_4^2 
t_6^4 t_7^2+y_s^3 t_2^5 t_3 t_4^4 t_5^2 t_6^4 t_7^2+2 y_s^3 t_2^4 
t_3^2 t_4^4 t_5^2 t_6^4 t_7^2+2 y_s^3 t_2^3 t_3^3 t_4^4 t_5^2 t_6^4 
t_7^2+2 y_s^3 t_2^2 t_3^4 t_4^4 t_5^2 t_6^4 t_7^2+y_s^3 t_2 t_3^5 
t_4^4 t_5^2 t_6^4 t_7^2-y_s^4 t_2^4 t_3^4 t_4^6 t_5^4 t_6^4 
t_7^2+y_s^3 t_2^4 t_3^2 t_4^4 t_5 t_6^5 t_7^2+y_s^3 t_2^2 t_3^4 t_4^4 
t_5 t_6^5 t_7^2+y_s^4 t_2^5 t_3^3 t_4^6 t_5^3 t_6^5 t_7^2+y_s^4 t_2^3 
t_3^5 t_4^6 t_5^3 t_6^5 t_7^2-y_s^2 t_2^2 t_3^2 t_4 t_5^4 t_7^3-y_s^3 
t_2^3 t_3^3 t_4^3 t_5^6 t_7^3-y_s^2 t_2^2 t_3^2 t_4 t_5^3 t_6 
t_7^3-y_s^3 t_2^3 t_3^3 t_4^3 t_5^5 t_6 t_7^3-y_s^2 t_2^4 t_4 t_5^2 
t_6^2 t_7^3-y_s^2 t_2^3 t_3 t_4 t_5^2 t_6^2 t_7^3-3 y_s^2 t_2^2 t_3^2 
t_4 t_5^2 t_6^2 t_7^3-y_s^2 t_2 t_3^3 t_4 t_5^2 t_6^2 t_7^3-y_s^2 
t_3^4 t_4 t_5^2 t_6^2 t_7^3+2 y_s^3 t_2^4 t_3^2 t_4^3 t_5^4 t_6^2 
t_7^3-y_s^3 t_2^3 t_3^3 t_4^3 t_5^4 t_6^2 t_7^3+2 y_s^3 t_2^2 t_3^4 
t_4^3 t_5^4 t_6^2 t_7^3+y_s^4 t_2^5 t_3^3 t_4^5 t_5^6 t_6^2 
t_7^3+y_s^4 t_2^3 t_3^5 t_4^5 t_5^6 t_6^2 t_7^3-y_s^2 t_2^2 t_3^2 t_4 
t_5 t_6^3 t_7^3-y_s^3 t_2^6 t_4^3 t_5^3 t_6^3 t_7^3-y_s^3 t_2^5 t_3 
t_4^3 t_5^3 t_6^3 t_7^3-y_s^3 t_2^4 t_3^2 t_4^3 t_5^3 t_6^3 t_7^3-3 
y_s^3 t_2^3 t_3^3 t_4^3 t_5^3 t_6^3 t_7^3-y_s^3 t_2^2 t_3^4 t_4^3 
t_5^3 t_6^3 t_7^3-y_s^3 t_2 t_3^5 t_4^3 t_5^3 t_6^3 t_7^3-y_s^3 t_3^6 
t_4^3 t_5^3 t_6^3 t_7^3+y_s^4 t_2^6 t_3^2 t_4^5 t_5^5 t_6^3 t_7^3+2 
y_s^4 t_2^5 t_3^3 t_4^5 t_5^5 t_6^3 t_7^3+2 y_s^4 t_2^4 t_3^4 t_4^5 
t_5^5 t_6^3 t_7^3+2 y_s^4 t_2^3 t_3^5 t_4^5 t_5^5 t_6^3 t_7^3+y_s^4 
t_2^2 t_3^6 t_4^5 t_5^5 t_6^3 t_7^3-y_s^2 t_2^2 t_3^2 t_4 t_6^4 
t_7^3+2 y_s^3 t_2^4 t_3^2 t_4^3 t_5^2 t_6^4 t_7^3-y_s^3 t_2^3 t_3^3 
t_4^3 t_5^2 t_6^4 t_7^3+2 y_s^3 t_2^2 t_3^4 t_4^3 t_5^2 t_6^4 t_7^3+2 
y_s^4 t_2^5 t_3^3 t_4^5 t_5^4 t_6^4 t_7^3-y_s^4 t_2^4 t_3^4 t_4^5 
t_5^4 t_6^4 t_7^3+2 y_s^4 t_2^3 t_3^5 t_4^5 t_5^4 t_6^4 t_7^3-y_s^5 
t_2^5 t_3^5 t_4^7 t_5^6 t_6^4 t_7^3-y_s^3 t_2^3 t_3^3 t_4^3 t_5 t_6^5 
t_7^3+y_s^4 t_2^6 t_3^2 t_4^5 t_5^3 t_6^5 t_7^3+2 y_s^4 t_2^5 t_3^3 
t_4^5 t_5^3 t_6^5 t_7^3+2 y_s^4 t_2^4 t_3^4 t_4^5 t_5^3 t_6^5 t_7^3+2 
y_s^4 t_2^3 t_3^5 t_4^5 t_5^3 t_6^5 t_7^3+y_s^4 t_2^2 t_3^6 t_4^5 
t_5^3 t_6^5 t_7^3-y_s^5 t_2^6 t_3^4 t_4^7 t_5^5 t_6^5 t_7^3-y_s^5 
t_2^5 t_3^5 t_4^7 t_5^5 t_6^5 t_7^3-y_s^5 t_2^4 t_3^6 t_4^7 t_5^5 
t_6^5 t_7^3-y_s^3 t_2^3 t_3^3 t_4^3 t_6^6 t_7^3+y_s^4 t_2^5 t_3^3 
t_4^5 t_5^2 t_6^6 t_7^3+y_s^4 t_2^3 t_3^5 t_4^5 t_5^2 t_6^6 
t_7^3-y_s^5 t_2^5 t_3^5 t_4^7 t_5^4 t_6^6 t_7^3-y_s^2 t_2^2 t_3^2 
t_5^3 t_6 t_7^4+y_s^3 t_2^4 t_3^2 t_4^2 t_5^5 t_6 t_7^4+y_s^3 t_2^2 
t_3^4 t_4^2 t_5^5 t_6 t_7^4-y_s^4 t_2^4 t_3^4 t_4^4 t_5^7 t_6 
t_7^4-y_s^2 t_2^3 t_3 t_5^2 t_6^2 t_7^4-y_s^2 t_2^2 t_3^2 t_5^2 t_6^2 
t_7^4-y_s^2 t_2 t_3^3 t_5^2 t_6^2 t_7^4+y_s^3 t_2^5 t_3 t_4^2 t_5^4 
t_6^2 t_7^4+2 y_s^3 t_2^4 t_3^2 t_4^2 t_5^4 t_6^2 t_7^4+2 y_s^3 t_2^3 
t_3^3 t_4^2 t_5^4 t_6^2 t_7^4+2 y_s^3 t_2^2 t_3^4 t_4^2 t_5^4 t_6^2 
t_7^4+y_s^3 t_2 t_3^5 t_4^2 t_5^4 t_6^2 t_7^4-y_s^4 t_2^4 t_3^4 t_4^4 
t_5^6 t_6^2 t_7^4-y_s^2 t_2^2 t_3^2 t_5 t_6^3 t_7^4+2 y_s^3 t_2^4 
t_3^2 t_4^2 t_5^3 t_6^3 t_7^4-y_s^3 t_2^3 t_3^3 t_4^2 t_5^3 t_6^3 
t_7^4+2 y_s^3 t_2^2 t_3^4 t_4^2 t_5^3 t_6^3 t_7^4+2 y_s^4 t_2^5 t_3^3 
t_4^4 t_5^5 t_6^3 t_7^4-y_s^4 t_2^4 t_3^4 t_4^4 t_5^5 t_6^3 t_7^4+2 
y_s^4 t_2^3 t_3^5 t_4^4 t_5^5 t_6^3 t_7^4-y_s^5 t_2^5 t_3^5 t_4^6 
t_5^7 t_6^3 t_7^4+y_s^3 t_2^5 t_3 t_4^2 t_5^2 t_6^4 t_7^4+2 y_s^3 
t_2^4 t_3^2 t_4^2 t_5^2 t_6^4 t_7^4+2 y_s^3 t_2^3 t_3^3 t_4^2 t_5^2 
t_6^4 t_7^4+2 y_s^3 t_2^2 t_3^4 t_4^2 t_5^2 t_6^4 t_7^4+y_s^3 t_2 
t_3^5 t_4^2 t_5^2 t_6^4 t_7^4-y_s^4 t_2^7 t_3 t_4^4 t_5^4 t_6^4 
t_7^4-y_s^4 t_2^6 t_3^2 t_4^4 t_5^4 t_6^4 t_7^4-y_s^4 t_2^5 t_3^3 
t_4^4 t_5^4 t_6^4 t_7^4-3 y_s^4 t_2^4 t_3^4 t_4^4 t_5^4 t_6^4 
t_7^4-y_s^4 t_2^3 t_3^5 t_4^4 t_5^4 t_6^4 t_7^4-y_s^4 t_2^2 t_3^6 
t_4^4 t_5^4 t_6^4 t_7^4-y_s^4 t_2 t_3^7 t_4^4 t_5^4 t_6^4 t_7^4-y_s^5 
t_2^5 t_3^5 t_4^6 t_5^6 t_6^4 t_7^4+y_s^3 t_2^4 t_3^2 t_4^2 t_5 t_6^5 
t_7^4+y_s^3 t_2^2 t_3^4 t_4^2 t_5 t_6^5 t_7^4+2 y_s^4 t_2^5 t_3^3 
t_4^4 t_5^3 t_6^5 t_7^4-y_s^4 t_2^4 t_3^4 t_4^4 t_5^3 t_6^5 t_7^4+2 
y_s^4 t_2^3 t_3^5 t_4^4 t_5^3 t_6^5 t_7^4-y_s^5 t_2^7 t_3^3 t_4^6 
t_5^5 t_6^5 t_7^4-y_s^5 t_2^6 t_3^4 t_4^6 t_5^5 t_6^5 t_7^4-3 y_s^5 
t_2^5 t_3^5 t_4^6 t_5^5 t_6^5 t_7^4-y_s^5 t_2^4 t_3^6 t_4^6 t_5^5 
t_6^5 t_7^4-y_s^5 t_2^3 t_3^7 t_4^6 t_5^5 t_6^5 t_7^4-y_s^4 t_2^4 
t_3^4 t_4^4 t_5^2 t_6^6 t_7^4-y_s^5 t_2^5 t_3^5 t_4^6 t_5^4 t_6^6 
t_7^4-y_s^4 t_2^4 t_3^4 t_4^4 t_5 t_6^7 t_7^4-y_s^5 t_2^5 t_3^5 t_4^6 
t_5^3 t_6^7 t_7^4+y_s^3 t_2^4 t_3^2 t_4 t_5^4 t_6^2 t_7^5+y_s^3 t_2^2 
t_3^4 t_4 t_5^4 t_6^2 t_7^5+y_s^4 t_2^5 t_3^3 t_4^3 t_5^6 t_6^2 
t_7^5+y_s^4 t_2^3 t_3^5 t_4^3 t_5^6 t_6^2 t_7^5-y_s^3 t_2^3 t_3^3 t_4 
t_5^3 t_6^3 t_7^5+y_s^4 t_2^6 t_3^2 t_4^3 t_5^5 t_6^3 t_7^5+2 y_s^4 
t_2^5 t_3^3 t_4^3 t_5^5 t_6^3 t_7^5+2 y_s^4 t_2^4 t_3^4 t_4^3 t_5^5 
t_6^3 t_7^5+2 y_s^4 t_2^3 t_3^5 t_4^3 t_5^5 t_6^3 t_7^5+y_s^4 t_2^2 
t_3^6 t_4^3 t_5^5 t_6^3 t_7^5-y_s^5 t_2^6 t_3^4 t_4^5 t_5^7 t_6^3 
t_7^5-y_s^5 t_2^5 t_3^5 t_4^5 t_5^7 t_6^3 t_7^5-y_s^5 t_2^4 t_3^6 
t_4^5 t_5^7 t_6^3 t_7^5+y_s^3 t_2^4 t_3^2 t_4 t_5^2 t_6^4 t_7^5+y_s^3 
t_2^2 t_3^4 t_4 t_5^2 t_6^4 t_7^5+2 y_s^4 t_2^5 t_3^3 t_4^3 t_5^4 
t_6^4 t_7^5-y_s^4 t_2^4 t_3^4 t_4^3 t_5^4 t_6^4 t_7^5+2 y_s^4 t_2^3 
t_3^5 t_4^3 t_5^4 t_6^4 t_7^5-y_s^5 t_2^7 t_3^3 t_4^5 t_5^6 t_6^4 
t_7^5-y_s^5 t_2^6 t_3^4 t_4^5 t_5^6 t_6^4 t_7^5-3 y_s^5 t_2^5 t_3^5 
t_4^5 t_5^6 t_6^4 t_7^5-y_s^5 t_2^4 t_3^6 t_4^5 t_5^6 t_6^4 
t_7^5-y_s^5 t_2^3 t_3^7 t_4^5 t_5^6 t_6^4 t_7^5+y_s^4 t_2^6 t_3^2 
t_4^3 t_5^3 t_6^5 t_7^5+2 y_s^4 t_2^5 t_3^3 t_4^3 t_5^3 t_6^5 t_7^5+2 
y_s^4 t_2^4 t_3^4 t_4^3 t_5^3 t_6^5 t_7^5+2 y_s^4 t_2^3 t_3^5 t_4^3 
t_5^3 t_6^5 t_7^5+y_s^4 t_2^2 t_3^6 t_4^3 t_5^3 t_6^5 t_7^5-y_s^5 
t_2^7 t_3^3 t_4^5 t_5^5 t_6^5 t_7^5-3 y_s^5 t_2^6 t_3^4 t_4^5 t_5^5 
t_6^5 t_7^5-3 y_s^5 t_2^5 t_3^5 t_4^5 t_5^5 t_6^5 t_7^5-3 y_s^5 t_2^4 
t_3^6 t_4^5 t_5^5 t_6^5 t_7^5-y_s^5 t_2^3 t_3^7 t_4^5 t_5^5 t_6^5 
t_7^5+y_s^6 t_2^6 t_3^6 t_4^7 t_5^7 t_6^5 t_7^5+y_s^4 t_2^5 t_3^3 
t_4^3 t_5^2 t_6^6 t_7^5+y_s^4 t_2^3 t_3^5 t_4^3 t_5^2 t_6^6 
t_7^5-y_s^5 t_2^7 t_3^3 t_4^5 t_5^4 t_6^6 t_7^5-y_s^5 t_2^6 t_3^4 
t_4^5 t_5^4 t_6^6 t_7^5-3 y_s^5 t_2^5 t_3^5 t_4^5 t_5^4 t_6^6 
t_7^5-y_s^5 t_2^4 t_3^6 t_4^5 t_5^4 t_6^6 t_7^5-y_s^5 t_2^3 t_3^7 
t_4^5 t_5^4 t_6^6 t_7^5+y_s^6 t_2^7 t_3^5 t_4^7 t_5^6 t_6^6 
t_7^5+y_s^6 t_2^6 t_3^6 t_4^7 t_5^6 t_6^6 t_7^5+y_s^6 t_2^5 t_3^7 
t_4^7 t_5^6 t_6^6 t_7^5-y_s^5 t_2^6 t_3^4 t_4^5 t_5^3 t_6^7 
t_7^5-y_s^5 t_2^5 t_3^5 t_4^5 t_5^3 t_6^7 t_7^5-y_s^5 t_2^4 t_3^6 
t_4^5 t_5^3 t_6^7 t_7^5+y_s^6 t_2^6 t_3^6 t_4^7 t_5^5 t_6^7 
t_7^5-y_s^3 t_2^3 t_3^3 t_5^3 t_6^3 t_7^6+y_s^4 t_2^5 t_3^3 t_4^2 
t_5^5 t_6^3 t_7^6+y_s^4 t_2^3 t_3^5 t_4^2 t_5^5 t_6^3 t_7^6-y_s^5 
t_2^5 t_3^5 t_4^4 t_5^7 t_6^3 t_7^6-y_s^4 t_2^4 t_3^4 t_4^2 t_5^4 
t_6^4 t_7^6-y_s^5 t_2^5 t_3^5 t_4^4 t_5^6 t_6^4 t_7^6+y_s^4 t_2^5 
t_3^3 t_4^2 t_5^3 t_6^5 t_7^6+y_s^4 t_2^3 t_3^5 t_4^2 t_5^3 t_6^5 
t_7^6-y_s^5 t_2^7 t_3^3 t_4^4 t_5^5 t_6^5 t_7^6-y_s^5 t_2^6 t_3^4 
t_4^4 t_5^5 t_6^5 t_7^6-3 y_s^5 t_2^5 t_3^5 t_4^4 t_5^5 t_6^5 
t_7^6-y_s^5 t_2^4 t_3^6 t_4^4 t_5^5 t_6^5 t_7^6-y_s^5 t_2^3 t_3^7 
t_4^4 t_5^5 t_6^5 t_7^6+y_s^6 t_2^7 t_3^5 t_4^6 t_5^7 t_6^5 
t_7^6+y_s^6 t_2^6 t_3^6 t_4^6 t_5^7 t_6^5 t_7^6+y_s^6 t_2^5 t_3^7 
t_4^6 t_5^7 t_6^5 t_7^6-y_s^5 t_2^5 t_3^5 t_4^4 t_5^4 t_6^6 
t_7^6+y_s^6 t_2^7 t_3^5 t_4^6 t_5^6 t_6^6 t_7^6+y_s^6 t_2^6 t_3^6 
t_4^6 t_5^6 t_6^6 t_7^6+y_s^6 t_2^5 t_3^7 t_4^6 t_5^6 t_6^6 
t_7^6-y_s^5 t_2^5 t_3^5 t_4^4 t_5^3 t_6^7 t_7^6+y_s^6 t_2^7 t_3^5 
t_4^6 t_5^5 t_6^7 t_7^6+y_s^6 t_2^6 t_3^6 t_4^6 t_5^5 t_6^7 
t_7^6+y_s^6 t_2^5 t_3^7 t_4^6 t_5^5 t_6^7 t_7^6-y_s^4 t_2^4 t_3^4 t_4 
t_5^4 t_6^4 t_7^7-y_s^5 t_2^5 t_3^5 t_4^3 t_5^6 t_6^4 t_7^7-y_s^5 
t_2^6 t_3^4 t_4^3 t_5^5 t_6^5 t_7^7-y_s^5 t_2^5 t_3^5 t_4^3 t_5^5 
t_6^5 t_7^7-y_s^5 t_2^4 t_3^6 t_4^3 t_5^5 t_6^5 t_7^7+y_s^6 t_2^6 
t_3^6 t_4^5 t_5^7 t_6^5 t_7^7-y_s^5 t_2^5 t_3^5 t_4^3 t_5^4 t_6^6 
t_7^7+y_s^6 t_2^7 t_3^5 t_4^5 t_5^6 t_6^6 t_7^7+y_s^6 t_2^6 t_3^6 
t_4^5 t_5^6 t_6^6 t_7^7+y_s^6 t_2^5 t_3^7 t_4^5 t_5^6 t_6^6 
t_7^7+y_s^6 t_2^6 t_3^6 t_4^5 t_5^5 t_6^7 t_7^7+y_s^7 t_2^7 t_3^7 
t_4^7 t_5^7 t_6^7 t_7^7
~.~
$
\end{quote}
\endgroup
\subsection{$\mathcal{C}_{++}$ \label{app_num_03}}
\begingroup\makeatletter\def\f@size{7}\check@mathfonts
\begin{quote}\raggedright
$
P(t_a,y_q,y_o;\mathcal{M}^{mes}_{\mathcal{C}_{++}})=
1+y_q y_o t_1 t_2 t_3+y_q y_o t_1 t_2 t_5+y_q y_o t_2^2 t_3 t_5-y_q 
y_o^2 t_1 t_2 t_3 t_4 t_5-y_q y_o^2 t_2^2 t_3^2 t_4 t_5-y_q y_o^2 
t_2^2 t_3 t_4 t_5^2-y_q^2 y_o^3 t_1 t_2^3 t_3^2 t_4 t_5^2
~.~
$
\end{quote}
\endgroup
\subsection{$H_4$ \label{app_num_04}}
\begingroup\makeatletter\def\f@size{7}\check@mathfonts
\begin{quote}\raggedright
$
P(t_a,y_o;\mathcal{M}^{mes}_{H_4})=1+y_o t_2 t_3 t_4 t_5-y_o^2 t_2 t_3^2 t_4^2 t_5 t_6-y_o^2 t_2 t_3 
t_4^2 t_5^2 t_6-y_o^2 t_2^2 t_3^2 t_4 t_5 t_7-y_o^2 t_2^2 t_3 t_4 
t_5^2 t_7-y_o^2 t_2 t_3^2 t_4 t_6 t_7-2 y_o^2 t_2 t_3 t_4 t_5 t_6 
t_7+y_o^3 t_2^2 t_3^3 t_4^2 t_5 t_6 t_7-y_o^2 t_2 t_4 t_5^2 t_6 t_7+2 
y_o^3 t_2^2 t_3^2 t_4^2 t_5^2 t_6 t_7+y_o^3 t_2^2 t_3 t_4^2 t_5^3 t_6 
t_7+y_o^3 t_2 t_3^2 t_4^2 t_5 t_6^2 t_7+y_o^3 t_2 t_3 t_4^2 t_5^2 
t_6^2 t_7+y_o^3 t_2^2 t_3^2 t_4 t_5 t_6 t_7^2+y_o^3 t_2^2 t_3 t_4 
t_5^2 t_6 t_7^2-y_o^4 t_2^2 t_3^2 t_4^2 t_5^2 t_6^2 t_7^2-y_o^5 t_2^3 
t_3^3 t_4^3 t_5^3 t_6^2 t_7^2
~.~
$
\end{quote}
\endgroup
\subsection{$ \mathcal{C}_{+-} $ \label{app_num_05}}
\begingroup\makeatletter\def\f@size{7}\check@mathfonts
\begin{quote}\raggedright
$
P(t_i,y_q,y_o;\mathcal{M}^{mes}_{\mathcal{C}_{+-} })=1+y_q y_o t_1 t_2 t_3+y_q y_o t_1 t_2 t_5-y_q^2 y_o^2 t_1^3 t_2 t_3 
t_5-y_q^2 y_o^2 t_1^2 t_2^2 t_3 t_5-y_q^2 y_o^2 t_1 t_2^3 t_3 t_5-y_q 
y_o^2 t_1^2 t_3 t_4 t_5-y_q y_o^2 t_1 t_2 t_3 t_4 t_5-y_q y_o^2 t_2^2 
t_3 t_4 t_5+y_q^2 y_o^3 t_1^2 t_2^2 t_3^2 t_4 t_5+y_q^2 y_o^3 t_1^2 
t_2^2 t_3 t_4 t_5^2+y_q^3 y_o^4 t_1^3 t_2^3 t_3^2 t_4 t_5^2
~.~
$
\end{quote}
\endgroup
\subsection{$ Q^{1,1,1} $ \label{app_num_06}}
\begingroup\makeatletter\def\f@size{7}\check@mathfonts
\begin{quote}\raggedright
$
P(t_a,y_o;\mathcal{M}^{mes}_{Q^{1,1,1}})=1-y_o^2 t_2 t_3 t_4^2 t_5 t_6-y_o^2 t_2^2 t_3 t_4 t_5 t_7-y_o^2 t_2 
t_3^2 t_4 t_6 t_7-3 y_o^2 t_2 t_3 t_4 t_5 t_6 t_7+2 y_o^3 t_2^2 t_3^2 
t_4^2 t_5 t_6 t_7-y_o^2 t_2 t_4 t_5^2 t_6 t_7+2 y_o^3 t_2^2 t_3 t_4^2 
t_5^2 t_6 t_7-y_o^2 t_3 t_4 t_5 t_6^2 t_7+2 y_o^3 t_2 t_3^2 t_4^2 t_5 
t_6^2 t_7+2 y_o^3 t_2 t_3 t_4^2 t_5^2 t_6^2 t_7-y_o^4 t_2^2 t_3^2 
t_4^3 t_5^2 t_6^2 t_7-y_o^2 t_2 t_3 t_5 t_6 t_7^2+2 y_o^3 t_2^2 t_3^2 
t_4 t_5 t_6 t_7^2+2 y_o^3 t_2^2 t_3 t_4 t_5^2 t_6 t_7^2-y_o^4 t_2^3 
t_3^2 t_4^2 t_5^2 t_6 t_7^2+2 y_o^3 t_2 t_3^2 t_4 t_5 t_6^2 
t_7^2-y_o^4 t_2^2 t_3^3 t_4^2 t_5 t_6^2 t_7^2+2 y_o^3 t_2 t_3 t_4 
t_5^2 t_6^2 t_7^2-3 y_o^4 t_2^2 t_3^2 t_4^2 t_5^2 t_6^2 t_7^2-y_o^4 
t_2^2 t_3 t_4^2 t_5^3 t_6^2 t_7^2-y_o^4 t_2 t_3^2 t_4^2 t_5^2 t_6^3 
t_7^2-y_o^4 t_2^2 t_3^2 t_4 t_5^2 t_6^2 t_7^3+y_o^6 t_2^3 t_3^3 t_4^3 
t_5^3 t_6^3 t_7^3
~.~
$
\end{quote}
\endgroup
\subsection{$P_{+-}^{1}[\mathbb{C}^3/\mathbb{Z}_5~(1,1,3)]$ \label{app_num_07}}
\begingroup\makeatletter\def\f@size{7}\check@mathfonts
\begin{quote}\raggedright
$
P(t_a,y_q,y_s,y_w,y_o,y_u,y_v;\mathcal{M}^{mes}_{P_{+-}^{1}[\mathbb{C}^3/\mathbb{Z}_5~(1,1,3)]})=1+y_q^2 y_s y_w y_o^3 y_u^3 y_v^4 t_1^4 t_2+y_q^4 y_s^2 y_w^2 y_o^6 
y_u^6 y_v^8 t_1^8 t_2^2+y_q y_s^2 y_w y_o^3 y_u^2 y_v^4 t_1^2 
t_2^3+y_q^3 y_s^3 y_w^2 y_o^6 y_u^5 y_v^8 t_1^6 t_2^4+y_q^4 y_s y_w^2 
y_o^5 y_u^6 y_v^7 t_1^8 t_3+y_q y_s y_w y_o^2 y_u^2 y_v^3 t_1^2 t_2 
t_3+y_q^3 y_s^2 y_w^2 y_o^5 y_u^5 y_v^7 t_1^6 t_2^2 t_3+y_s^2 y_w 
y_o^2 y_u y_v^3 t_2^3 t_3+y_q^2 y_s^3 y_w^2 y_o^5 y_u^4 y_v^7 t_1^4 
t_2^4 t_3+y_q^3 y_s y_w^2 y_o^4 y_u^5 y_v^6 t_1^6 t_3^2+y_s y_w y_o 
y_u y_v^2 t_2 t_3^2+y_q^2 y_s^2 y_w^2 y_o^4 y_u^4 y_v^6 t_1^4 t_2^2 
t_3^2+y_q^4 y_s^3 y_w^3 y_o^7 y_u^7 y_v^{10} t_1^8 t_2^3 t_3^2+y_q 
y_s^3 y_w^2 y_o^4 y_u^3 y_v^6 t_1^2 t_2^4 t_3^2+y_q^2 y_s y_w^2 y_o^3 
y_u^4 y_v^5 t_1^4 t_3^3+y_q^4 y_s^2 y_w^3 y_o^6 y_u^7 y_v^9 t_1^8 t_2 
t_3^3+y_q y_s^2 y_w^2 y_o^3 y_u^3 y_v^5 t_1^2 t_2^2 t_3^3+y_q^3 y_s^3 
y_w^3 y_o^6 y_u^6 y_v^9 t_1^6 t_2^3 t_3^3+y_s^3 y_w^2 y_o^3 y_u^2 
y_v^5 t_2^4 t_3^3+y_q y_s y_w^2 y_o^2 y_u^3 y_v^4 t_1^2 t_3^4+y_q^3 
y_s^2 y_w^3 y_o^5 y_u^6 y_v^8 t_1^6 t_2 t_3^4+y_s^2 y_w^2 y_o^2 y_u^2 
y_v^4 t_2^2 t_3^4+y_q^2 y_s^3 y_w^3 y_o^5 y_u^5 y_v^8 t_1^4 t_2^3 
t_3^4+y_q^4 y_s^4 y_w^4 y_o^8 y_u^8 y_v^{12} t_1^8 t_2^4 t_3^4+y_q^5 
y_s y_w^2 y_o^6 y_u^7 y_v^8 t_1^9 t_4+y_q^2 y_s y_w y_o^3 y_u^3 y_v^4 
t_1^3 t_2 t_4+y_q^4 y_s^2 y_w^2 y_o^6 y_u^6 y_v^8 t_1^7 t_2^2 t_4+y_q 
y_s^2 y_w y_o^3 y_u^2 y_v^4 t_1 t_2^3 t_4+y_q^3 y_s^3 y_w^2 y_o^6 
y_u^5 y_v^8 t_1^5 t_2^4 t_4+y_q^4 y_s y_w^2 y_o^5 y_u^6 y_v^7 t_1^7 
t_3 t_4+y_q y_s y_w y_o^2 y_u^2 y_v^3 t_1 t_2 t_3 t_4+y_q^3 y_s^2 
y_w^2 y_o^5 y_u^5 y_v^7 t_1^5 t_2^2 t_3 t_4+y_q^5 y_s^3 y_w^3 y_o^8 
y_u^8 y_v^{11} t_1^9 t_2^3 t_3 t_4+y_q^2 y_s^3 y_w^2 y_o^5 y_u^4 
y_v^7 t_1^3 t_2^4 t_3 t_4+y_q^3 y_s y_w^2 y_o^4 y_u^5 y_v^6 t_1^5 
t_3^2 t_4+y_q^5 y_s^2 y_w^3 y_o^7 y_u^8 y_v^{10} t_1^9 t_2 t_3^2 
t_4+y_q^2 y_s^2 y_w^2 y_o^4 y_u^4 y_v^6 t_1^3 t_2^2 t_3^2 t_4+y_q^4 
y_s^3 y_w^3 y_o^7 y_u^7 y_v^{10} t_1^7 t_2^3 t_3^2 t_4+y_q y_s^3 
y_w^2 y_o^4 y_u^3 y_v^6 t_1 t_2^4 t_3^2 t_4+y_q^2 y_s y_w^2 y_o^3 
y_u^4 y_v^5 t_1^3 t_3^3 t_4+y_q^4 y_s^2 y_w^3 y_o^6 y_u^7 y_v^9 t_1^7 
t_2 t_3^3 t_4+y_q y_s^2 y_w^2 y_o^3 y_u^3 y_v^5 t_1 t_2^2 t_3^3 
t_4+y_q^3 y_s^3 y_w^3 y_o^6 y_u^6 y_v^9 t_1^5 t_2^3 t_3^3 t_4+y_q^5 
y_s^4 y_w^4 y_o^9 y_u^9 y_v^{13} t_1^9 t_2^4 t_3^3 t_4+y_q y_s y_w^2 
y_o^2 y_u^3 y_v^4 t_1 t_3^4 t_4+y_q^3 y_s^2 y_w^3 y_o^5 y_u^6 y_v^8 
t_1^5 t_2 t_3^4 t_4+y_q^5 y_s^3 y_w^4 y_o^8 y_u^9 y_v^{12} t_1^9 
t_2^2 t_3^4 t_4+y_q^2 y_s^3 y_w^3 y_o^5 y_u^5 y_v^8 t_1^3 t_2^3 t_3^4 
t_4+y_q^4 y_s^4 y_w^4 y_o^8 y_u^8 y_v^{12} t_1^7 t_2^4 t_3^4 
t_4+y_q^5 y_s y_w^2 y_o^6 y_u^7 y_v^8 t_1^8 t_4^2+y_q^2 y_s y_w y_o^3 
y_u^3 y_v^4 t_1^2 t_2 t_4^2+y_q^4 y_s^2 y_w^2 y_o^6 y_u^6 y_v^8 t_1^6 
t_2^2 t_4^2+y_q y_s^2 y_w y_o^3 y_u^2 y_v^4 t_2^3 t_4^2+y_q^3 y_s^3 
y_w^2 y_o^6 y_u^5 y_v^8 t_1^4 t_2^4 t_4^2+y_q^4 y_s y_w^2 y_o^5 y_u^6 
y_v^7 t_1^6 t_3 t_4^2+y_q y_s y_w y_o^2 y_u^2 y_v^3 t_2 t_3 
t_4^2+y_q^3 y_s^2 y_w^2 y_o^5 y_u^5 y_v^7 t_1^4 t_2^2 t_3 t_4^2+y_q^5 
y_s^3 y_w^3 y_o^8 y_u^8 y_v^{11} t_1^8 t_2^3 t_3 t_4^2+y_q^2 y_s^3 
y_w^2 y_o^5 y_u^4 y_v^7 t_1^2 t_2^4 t_3 t_4^2+y_q^3 y_s y_w^2 y_o^4 
y_u^5 y_v^6 t_1^4 t_3^2 t_4^2+y_q^5 y_s^2 y_w^3 y_o^7 y_u^8 y_v^{10} 
t_1^8 t_2 t_3^2 t_4^2+y_q^2 y_s^2 y_w^2 y_o^4 y_u^4 y_v^6 t_1^2 t_2^2 
t_3^2 t_4^2+y_q^4 y_s^3 y_w^3 y_o^7 y_u^7 y_v^{10} t_1^6 t_2^3 t_3^2 
t_4^2+y_q y_s^3 y_w^2 y_o^4 y_u^3 y_v^6 t_2^4 t_3^2 t_4^2+y_q^2 y_s 
y_w^2 y_o^3 y_u^4 y_v^5 t_1^2 t_3^3 t_4^2+y_q^4 y_s^2 y_w^3 y_o^6 
y_u^7 y_v^9 t_1^6 t_2 t_3^3 t_4^2+y_q y_s^2 y_w^2 y_o^3 y_u^3 y_v^5 
t_2^2 t_3^3 t_4^2+y_q^3 y_s^3 y_w^3 y_o^6 y_u^6 y_v^9 t_1^4 t_2^3 
t_3^3 t_4^2+y_q^5 y_s^4 y_w^4 y_o^9 y_u^9 y_v^{13} t_1^8 t_2^4 t_3^3 
t_4^2+y_q y_s y_w^2 y_o^2 y_u^3 y_v^4 t_3^4 t_4^2+y_q^3 y_s^2 y_w^3 
y_o^5 y_u^6 y_v^8 t_1^4 t_2 t_3^4 t_4^2+y_q^5 y_s^3 y_w^4 y_o^8 y_u^9 
y_v^{12} t_1^8 t_2^2 t_3^4 t_4^2+y_q^2 y_s^3 y_w^3 y_o^5 y_u^5 y_v^8 
t_1^2 t_2^3 t_3^4 t_4^2+y_q^4 y_s^4 y_w^4 y_o^8 y_u^8 y_v^{12} t_1^6 
t_2^4 t_3^4 t_4^2+y_q^5 y_s y_w^2 y_o^6 y_u^7 y_v^8 t_1^7 t_4^3+y_q^2 
y_s y_w y_o^3 y_u^3 y_v^4 t_1 t_2 t_4^3+y_q^4 y_s^2 y_w^2 y_o^6 y_u^6 
y_v^8 t_1^5 t_2^2 t_4^3+y_q^6 y_s^3 y_w^3 y_o^9 y_u^9 y_v^{12} t_1^9 
t_2^3 t_4^3+y_q^3 y_s^3 y_w^2 y_o^6 y_u^5 y_v^8 t_1^3 t_2^4 
t_4^3+y_q^4 y_s y_w^2 y_o^5 y_u^6 y_v^7 t_1^5 t_3 t_4^3+y_q^6 y_s^2 
y_w^3 y_o^8 y_u^9 y_v^{11} t_1^9 t_2 t_3 t_4^3+y_q^3 y_s^2 y_w^2 
y_o^5 y_u^5 y_v^7 t_1^3 t_2^2 t_3 t_4^3+y_q^5 y_s^3 y_w^3 y_o^8 y_u^8 
y_v^{11} t_1^7 t_2^3 t_3 t_4^3+y_q^2 y_s^3 y_w^2 y_o^5 y_u^4 y_v^7 
t_1 t_2^4 t_3 t_4^3+y_q^3 y_s y_w^2 y_o^4 y_u^5 y_v^6 t_1^3 t_3^2 
t_4^3+y_q^5 y_s^2 y_w^3 y_o^7 y_u^8 y_v^{10} t_1^7 t_2 t_3^2 
t_4^3+y_q^2 y_s^2 y_w^2 y_o^4 y_u^4 y_v^6 t_1 t_2^2 t_3^2 t_4^3+y_q^4 
y_s^3 y_w^3 y_o^7 y_u^7 y_v^{10} t_1^5 t_2^3 t_3^2 t_4^3+y_q^6 y_s^4 
y_w^4 y_o^{10} y_u^{10} y_v^{14} t_1^9 t_2^4 t_3^2 t_4^3+y_q^2 y_s 
y_w^2 y_o^3 y_u^4 y_v^5 t_1 t_3^3 t_4^3+y_q^4 y_s^2 y_w^3 y_o^6 y_u^7 
y_v^9 t_1^5 t_2 t_3^3 t_4^3+y_q^6 y_s^3 y_w^4 y_o^9 y_u^{10} y_v^{13} 
t_1^9 t_2^2 t_3^3 t_4^3+y_q^3 y_s^3 y_w^3 y_o^6 y_u^6 y_v^9 t_1^3 
t_2^3 t_3^3 t_4^3+y_q^5 y_s^4 y_w^4 y_o^9 y_u^9 y_v^{13} t_1^7 t_2^4 
t_3^3 t_4^3+y_q^6 y_s^2 y_w^4 y_o^8 y_u^{10} y_v^{12} t_1^9 t_3^4 
t_4^3+y_q^3 y_s^2 y_w^3 y_o^5 y_u^6 y_v^8 t_1^3 t_2 t_3^4 t_4^3+y_q^5 
y_s^3 y_w^4 y_o^8 y_u^9 y_v^{12} t_1^7 t_2^2 t_3^4 t_4^3+y_q^2 y_s^3 
y_w^3 y_o^5 y_u^5 y_v^8 t_1 t_2^3 t_3^4 t_4^3+y_q^4 y_s^4 y_w^4 y_o^8 
y_u^8 y_v^{12} t_1^5 t_2^4 t_3^4 t_4^3+y_q^5 y_s y_w^2 y_o^6 y_u^7 
y_v^8 t_1^6 t_4^4+y_q^2 y_s y_w y_o^3 y_u^3 y_v^4 t_2 t_4^4+y_q^4 
y_s^2 y_w^2 y_o^6 y_u^6 y_v^8 t_1^4 t_2^2 t_4^4+y_q^6 y_s^3 y_w^3 
y_o^9 y_u^9 y_v^{12} t_1^8 t_2^3 t_4^4+y_q^3 y_s^3 y_w^2 y_o^6 y_u^5 
y_v^8 t_1^2 t_2^4 t_4^4+y_q^4 y_s y_w^2 y_o^5 y_u^6 y_v^7 t_1^4 t_3 
t_4^4+y_q^6 y_s^2 y_w^3 y_o^8 y_u^9 y_v^{11} t_1^8 t_2 t_3 
t_4^4+y_q^3 y_s^2 y_w^2 y_o^5 y_u^5 y_v^7 t_1^2 t_2^2 t_3 t_4^4+y_q^5 
y_s^3 y_w^3 y_o^8 y_u^8 y_v^{11} t_1^6 t_2^3 t_3 t_4^4+y_q^2 y_s^3 
y_w^2 y_o^5 y_u^4 y_v^7 t_2^4 t_3 t_4^4+y_q^3 y_s y_w^2 y_o^4 y_u^5 
y_v^6 t_1^2 t_3^2 t_4^4+y_q^5 y_s^2 y_w^3 y_o^7 y_u^8 y_v^{10} t_1^6 
t_2 t_3^2 t_4^4+y_q^2 y_s^2 y_w^2 y_o^4 y_u^4 y_v^6 t_2^2 t_3^2 
t_4^4+y_q^4 y_s^3 y_w^3 y_o^7 y_u^7 y_v^{10} t_1^4 t_2^3 t_3^2 
t_4^4+y_q^6 y_s^4 y_w^4 y_o^{10} y_u^{10} y_v^{14} t_1^8 t_2^4 t_3^2 
t_4^4+y_q^2 y_s y_w^2 y_o^3 y_u^4 y_v^5 t_3^3 t_4^4+y_q^4 y_s^2 y_w^3 
y_o^6 y_u^7 y_v^9 t_1^4 t_2 t_3^3 t_4^4+y_q^6 y_s^3 y_w^4 y_o^9 
y_u^{10} y_v^{13} t_1^8 t_2^2 t_3^3 t_4^4+y_q^3 y_s^3 y_w^3 y_o^6 
y_u^6 y_v^9 t_1^2 t_2^3 t_3^3 t_4^4+y_q^5 y_s^4 y_w^4 y_o^9 y_u^9 
y_v^{13} t_1^6 t_2^4 t_3^3 t_4^4+y_q^6 y_s^2 y_w^4 y_o^8 y_u^{10} 
y_v^{12} t_1^8 t_3^4 t_4^4+y_q^3 y_s^2 y_w^3 y_o^5 y_u^6 y_v^8 t_1^2 
t_2 t_3^4 t_4^4+y_q^5 y_s^3 y_w^4 y_o^8 y_u^9 y_v^{12} t_1^6 t_2^2 
t_3^4 t_4^4+y_q^2 y_s^3 y_w^3 y_o^5 y_u^5 y_v^8 t_2^3 t_3^4 
t_4^4+y_q^4 y_s^4 y_w^4 y_o^8 y_u^8 y_v^{12} t_1^4 t_2^4 t_3^4 
t_4^4+y_q^5 y_s y_w^2 y_o^6 y_u^7 y_v^8 t_1^5 t_4^5+y_q^7 y_s^2 y_w^3 
y_o^9 y_u^{10} y_v^{12} t_1^9 t_2 t_4^5+y_q^4 y_s^2 y_w^2 y_o^6 y_u^6 
y_v^8 t_1^3 t_2^2 t_4^5+y_q^6 y_s^3 y_w^3 y_o^9 y_u^9 y_v^{12} t_1^7 
t_2^3 t_4^5+y_q^3 y_s^3 y_w^2 y_o^6 y_u^5 y_v^8 t_1 t_2^4 t_4^5+y_q^4 
y_s y_w^2 y_o^5 y_u^6 y_v^7 t_1^3 t_3 t_4^5+y_q^6 y_s^2 y_w^3 y_o^8 
y_u^9 y_v^{11} t_1^7 t_2 t_3 t_4^5+y_q^3 y_s^2 y_w^2 y_o^5 y_u^5 
y_v^7 t_1 t_2^2 t_3 t_4^5+y_q^5 y_s^3 y_w^3 y_o^8 y_u^8 y_v^{11} 
t_1^5 t_2^3 t_3 t_4^5+y_q^7 y_s^4 y_w^4 y_o^{11} y_u^{11} y_v^{15} 
t_1^9 t_2^4 t_3 t_4^5+y_q^3 y_s y_w^2 y_o^4 y_u^5 y_v^6 t_1 t_3^2 
t_4^5+y_q^5 y_s^2 y_w^3 y_o^7 y_u^8 y_v^{10} t_1^5 t_2 t_3^2 
t_4^5+y_q^7 y_s^3 y_w^4 y_o^{10} y_u^{11} y_v^{14} t_1^9 t_2^2 t_3^2 
t_4^5+y_q^4 y_s^3 y_w^3 y_o^7 y_u^7 y_v^{10} t_1^3 t_2^3 t_3^2 
t_4^5+y_q^6 y_s^4 y_w^4 y_o^{10} y_u^{10} y_v^{14} t_1^7 t_2^4 t_3^2 
t_4^5+y_q^7 y_s^2 y_w^4 y_o^9 y_u^{11} y_v^{13} t_1^9 t_3^3 
t_4^5+y_q^4 y_s^2 y_w^3 y_o^6 y_u^7 y_v^9 t_1^3 t_2 t_3^3 t_4^5+y_q^6 
y_s^3 y_w^4 y_o^9 y_u^{10} y_v^{13} t_1^7 t_2^2 t_3^3 t_4^5+y_q^3 
y_s^3 y_w^3 y_o^6 y_u^6 y_v^9 t_1 t_2^3 t_3^3 t_4^5+y_q^5 y_s^4 y_w^4 
y_o^9 y_u^9 y_v^{13} t_1^5 t_2^4 t_3^3 t_4^5+y_q^6 y_s^2 y_w^4 y_o^8 
y_u^{10} y_v^{12} t_1^7 t_3^4 t_4^5+y_q^3 y_s^2 y_w^3 y_o^5 y_u^6 
y_v^8 t_1 t_2 t_3^4 t_4^5+y_q^5 y_s^3 y_w^4 y_o^8 y_u^9 y_v^{12} 
t_1^5 t_2^2 t_3^4 t_4^5+y_q^7 y_s^4 y_w^5 y_o^{11} y_u^{12} y_v^{16} 
t_1^9 t_2^3 t_3^4 t_4^5+y_q^4 y_s^4 y_w^4 y_o^8 y_u^8 y_v^{12} t_1^3 
t_2^4 t_3^4 t_4^5+y_q^5 y_s y_w^2 y_o^6 y_u^7 y_v^8 t_1^4 t_4^6+y_q^7 
y_s^2 y_w^3 y_o^9 y_u^{10} y_v^{12} t_1^8 t_2 t_4^6+y_q^4 y_s^2 y_w^2 
y_o^6 y_u^6 y_v^8 t_1^2 t_2^2 t_4^6+y_q^6 y_s^3 y_w^3 y_o^9 y_u^9 
y_v^{12} t_1^6 t_2^3 t_4^6+y_q^3 y_s^3 y_w^2 y_o^6 y_u^5 y_v^8 t_2^4 
t_4^6+y_q^4 y_s y_w^2 y_o^5 y_u^6 y_v^7 t_1^2 t_3 t_4^6+y_q^6 y_s^2 
y_w^3 y_o^8 y_u^9 y_v^{11} t_1^6 t_2 t_3 t_4^6+y_q^3 y_s^2 y_w^2 
y_o^5 y_u^5 y_v^7 t_2^2 t_3 t_4^6+y_q^5 y_s^3 y_w^3 y_o^8 y_u^8 
y_v^{11} t_1^4 t_2^3 t_3 t_4^6+y_q^7 y_s^4 y_w^4 y_o^{11} y_u^{11} 
y_v^{15} t_1^8 t_2^4 t_3 t_4^6+y_q^3 y_s y_w^2 y_o^4 y_u^5 y_v^6 
t_3^2 t_4^6+y_q^5 y_s^2 y_w^3 y_o^7 y_u^8 y_v^{10} t_1^4 t_2 t_3^2 
t_4^6+y_q^7 y_s^3 y_w^4 y_o^{10} y_u^{11} y_v^{14} t_1^8 t_2^2 t_3^2 
t_4^6+y_q^4 y_s^3 y_w^3 y_o^7 y_u^7 y_v^{10} t_1^2 t_2^3 t_3^2 
t_4^6+y_q^6 y_s^4 y_w^4 y_o^{10} y_u^{10} y_v^{14} t_1^6 t_2^4 t_3^2 
t_4^6+y_q^7 y_s^2 y_w^4 y_o^9 y_u^{11} y_v^{13} t_1^8 t_3^3 
t_4^6+y_q^4 y_s^2 y_w^3 y_o^6 y_u^7 y_v^9 t_1^2 t_2 t_3^3 t_4^6+y_q^6 
y_s^3 y_w^4 y_o^9 y_u^{10} y_v^{13} t_1^6 t_2^2 t_3^3 t_4^6+y_q^3 
y_s^3 y_w^3 y_o^6 y_u^6 y_v^9 t_2^3 t_3^3 t_4^6+y_q^5 y_s^4 y_w^4 
y_o^9 y_u^9 y_v^{13} t_1^4 t_2^4 t_3^3 t_4^6+y_q^6 y_s^2 y_w^4 y_o^8 
y_u^{10} y_v^{12} t_1^6 t_3^4 t_4^6+y_q^3 y_s^2 y_w^3 y_o^5 y_u^6 
y_v^8 t_2 t_3^4 t_4^6+y_q^5 y_s^3 y_w^4 y_o^8 y_u^9 y_v^{12} t_1^4 
t_2^2 t_3^4 t_4^6+y_q^7 y_s^4 y_w^5 y_o^{11} y_u^{12} y_v^{16} t_1^8 
t_2^3 t_3^4 t_4^6+y_q^4 y_s^4 y_w^4 y_o^8 y_u^8 y_v^{12} t_1^2 t_2^4 
t_3^4 t_4^6+y_q^5 y_s y_w^2 y_o^6 y_u^7 y_v^8 t_1^3 t_4^7+y_q^7 y_s^2 
y_w^3 y_o^9 y_u^{10} y_v^{12} t_1^7 t_2 t_4^7+y_q^4 y_s^2 y_w^2 y_o^6 
y_u^6 y_v^8 t_1 t_2^2 t_4^7+y_q^6 y_s^3 y_w^3 y_o^9 y_u^9 y_v^{12} 
t_1^5 t_2^3 t_4^7+y_q^8 y_s^4 y_w^4 y_o^{12} y_u^{12} y_v^{16} t_1^9 
t_2^4 t_4^7+y_q^4 y_s y_w^2 y_o^5 y_u^6 y_v^7 t_1 t_3 t_4^7+y_q^6 
y_s^2 y_w^3 y_o^8 y_u^9 y_v^{11} t_1^5 t_2 t_3 t_4^7+y_q^8 y_s^3 
y_w^4 y_o^{11} y_u^{12} y_v^{15} t_1^9 t_2^2 t_3 t_4^7+y_q^5 y_s^3 
y_w^3 y_o^8 y_u^8 y_v^{11} t_1^3 t_2^3 t_3 t_4^7+y_q^7 y_s^4 y_w^4 
y_o^{11} y_u^{11} y_v^{15} t_1^7 t_2^4 t_3 t_4^7+y_q^8 y_s^2 y_w^4 
y_o^{10} y_u^{12} y_v^{14} t_1^9 t_3^2 t_4^7+y_q^5 y_s^2 y_w^3 y_o^7 
y_u^8 y_v^{10} t_1^3 t_2 t_3^2 t_4^7+y_q^7 y_s^3 y_w^4 y_o^{10} 
y_u^{11} y_v^{14} t_1^7 t_2^2 t_3^2 t_4^7+y_q^4 y_s^3 y_w^3 y_o^7 
y_u^7 y_v^{10} t_1 t_2^3 t_3^2 t_4^7+y_q^6 y_s^4 y_w^4 y_o^{10} 
y_u^{10} y_v^{14} t_1^5 t_2^4 t_3^2 t_4^7+y_q^7 y_s^2 y_w^4 y_o^9 
y_u^{11} y_v^{13} t_1^7 t_3^3 t_4^7+y_q^4 y_s^2 y_w^3 y_o^6 y_u^7 
y_v^9 t_1 t_2 t_3^3 t_4^7+y_q^6 y_s^3 y_w^4 y_o^9 y_u^{10} y_v^{13} 
t_1^5 t_2^2 t_3^3 t_4^7+y_q^8 y_s^4 y_w^5 y_o^{12} y_u^{13} y_v^{17} 
t_1^9 t_2^3 t_3^3 t_4^7+y_q^5 y_s^4 y_w^4 y_o^9 y_u^9 y_v^{13} t_1^3 
t_2^4 t_3^3 t_4^7+y_q^6 y_s^2 y_w^4 y_o^8 y_u^{10} y_v^{12} t_1^5 
t_3^4 t_4^7+y_q^8 y_s^3 y_w^5 y_o^{11} y_u^{13} y_v^{16} t_1^9 t_2 
t_3^4 t_4^7+y_q^5 y_s^3 y_w^4 y_o^8 y_u^9 y_v^{12} t_1^3 t_2^2 t_3^4 
t_4^7+y_q^7 y_s^4 y_w^5 y_o^{11} y_u^{12} y_v^{16} t_1^7 t_2^3 t_3^4 
t_4^7+y_q^4 y_s^4 y_w^4 y_o^8 y_u^8 y_v^{12} t_1 t_2^4 t_3^4 
t_4^7+y_q^5 y_s y_w^2 y_o^6 y_u^7 y_v^8 t_1^2 t_4^8+y_q^7 y_s^2 y_w^3 
y_o^9 y_u^{10} y_v^{12} t_1^6 t_2 t_4^8+y_q^4 y_s^2 y_w^2 y_o^6 y_u^6 
y_v^8 t_2^2 t_4^8+y_q^6 y_s^3 y_w^3 y_o^9 y_u^9 y_v^{12} t_1^4 t_2^3 
t_4^8+y_q^8 y_s^4 y_w^4 y_o^{12} y_u^{12} y_v^{16} t_1^8 t_2^4 
t_4^8+y_q^4 y_s y_w^2 y_o^5 y_u^6 y_v^7 t_3 t_4^8+y_q^6 y_s^2 y_w^3 
y_o^8 y_u^9 y_v^{11} t_1^4 t_2 t_3 t_4^8+y_q^8 y_s^3 y_w^4 y_o^{11} 
y_u^{12} y_v^{15} t_1^8 t_2^2 t_3 t_4^8+y_q^5 y_s^3 y_w^3 y_o^8 y_u^8 
y_v^{11} t_1^2 t_2^3 t_3 t_4^8+y_q^7 y_s^4 y_w^4 y_o^{11} y_u^{11} 
y_v^{15} t_1^6 t_2^4 t_3 t_4^8+y_q^8 y_s^2 y_w^4 y_o^{10} y_u^{12} 
y_v^{14} t_1^8 t_3^2 t_4^8+y_q^5 y_s^2 y_w^3 y_o^7 y_u^8 y_v^{10} 
t_1^2 t_2 t_3^2 t_4^8+y_q^7 y_s^3 y_w^4 y_o^{10} y_u^{11} y_v^{14} 
t_1^6 t_2^2 t_3^2 t_4^8+y_q^4 y_s^3 y_w^3 y_o^7 y_u^7 y_v^{10} t_2^3 
t_3^2 t_4^8+y_q^6 y_s^4 y_w^4 y_o^{10} y_u^{10} y_v^{14} t_1^4 t_2^4 
t_3^2 t_4^8+y_q^7 y_s^2 y_w^4 y_o^9 y_u^{11} y_v^{13} t_1^6 t_3^3 
t_4^8+y_q^4 y_s^2 y_w^3 y_o^6 y_u^7 y_v^9 t_2 t_3^3 t_4^8+y_q^6 y_s^3 
y_w^4 y_o^9 y_u^{10} y_v^{13} t_1^4 t_2^2 t_3^3 t_4^8+y_q^8 y_s^4 
y_w^5 y_o^{12} y_u^{13} y_v^{17} t_1^8 t_2^3 t_3^3 t_4^8+y_q^5 y_s^4 
y_w^4 y_o^9 y_u^9 y_v^{13} t_1^2 t_2^4 t_3^3 t_4^8+y_q^6 y_s^2 y_w^4 
y_o^8 y_u^{10} y_v^{12} t_1^4 t_3^4 t_4^8+y_q^8 y_s^3 y_w^5 y_o^{11} 
y_u^{13} y_v^{16} t_1^8 t_2 t_3^4 t_4^8+y_q^5 y_s^3 y_w^4 y_o^8 y_u^9 
y_v^{12} t_1^2 t_2^2 t_3^4 t_4^8+y_q^7 y_s^4 y_w^5 y_o^{11} y_u^{12} 
y_v^{16} t_1^6 t_2^3 t_3^4 t_4^8+y_q^4 y_s^4 y_w^4 y_o^8 y_u^8 
y_v^{12} t_2^4 t_3^4 t_4^8+y_q^5 y_s y_w^2 y_o^6 y_u^7 y_v^8 t_1 
t_4^9+y_q^7 y_s^2 y_w^3 y_o^9 y_u^{10} y_v^{12} t_1^5 t_2 t_4^9+y_q^9 
y_s^3 y_w^4 y_o^{12} y_u^{13} y_v^{16} t_1^9 t_2^2 t_4^9+y_q^6 y_s^3 
y_w^3 y_o^9 y_u^9 y_v^{12} t_1^3 t_2^3 t_4^9+y_q^8 y_s^4 y_w^4 
y_o^{12} y_u^{12} y_v^{16} t_1^7 t_2^4 t_4^9+y_q^9 y_s^2 y_w^4 
y_o^{11} y_u^{13} y_v^{15} t_1^9 t_3 t_4^9+y_q^6 y_s^2 y_w^3 y_o^8 
y_u^9 y_v^{11} t_1^3 t_2 t_3 t_4^9+y_q^8 y_s^3 y_w^4 y_o^{11} 
y_u^{12} y_v^{15} t_1^7 t_2^2 t_3 t_4^9+y_q^5 y_s^3 y_w^3 y_o^8 y_u^8 
y_v^{11} t_1 t_2^3 t_3 t_4^9+y_q^7 y_s^4 y_w^4 y_o^{11} y_u^{11} 
y_v^{15} t_1^5 t_2^4 t_3 t_4^9+y_q^8 y_s^2 y_w^4 y_o^{10} y_u^{12} 
y_v^{14} t_1^7 t_3^2 t_4^9+y_q^5 y_s^2 y_w^3 y_o^7 y_u^8 y_v^{10} t_1 
t_2 t_3^2 t_4^9+y_q^7 y_s^3 y_w^4 y_o^{10} y_u^{11} y_v^{14} t_1^5 
t_2^2 t_3^2 t_4^9+y_q^9 y_s^4 y_w^5 y_o^{13} y_u^{14} y_v^{18} t_1^9 
t_2^3 t_3^2 t_4^9+y_q^6 y_s^4 y_w^4 y_o^{10} y_u^{10} y_v^{14} t_1^3 
t_2^4 t_3^2 t_4^9+y_q^7 y_s^2 y_w^4 y_o^9 y_u^{11} y_v^{13} t_1^5 
t_3^3 t_4^9+y_q^9 y_s^3 y_w^5 y_o^{12} y_u^{14} y_v^{17} t_1^9 t_2 
t_3^3 t_4^9+y_q^6 y_s^3 y_w^4 y_o^9 y_u^{10} y_v^{13} t_1^3 t_2^2 
t_3^3 t_4^9+y_q^8 y_s^4 y_w^5 y_o^{12} y_u^{13} y_v^{17} t_1^7 t_2^3 
t_3^3 t_4^9+y_q^5 y_s^4 y_w^4 y_o^9 y_u^9 y_v^{13} t_1 t_2^4 t_3^3 
t_4^9+y_q^6 y_s^2 y_w^4 y_o^8 y_u^{10} y_v^{12} t_1^3 t_3^4 
t_4^9+y_q^8 y_s^3 y_w^5 y_o^{11} y_u^{13} y_v^{16} t_1^7 t_2 t_3^4 
t_4^9+y_q^5 y_s^3 y_w^4 y_o^8 y_u^9 y_v^{12} t_1 t_2^2 t_3^4 
t_4^9+y_q^7 y_s^4 y_w^5 y_o^{11} y_u^{12} y_v^{16} t_1^5 t_2^3 t_3^4 
t_4^9+y_q^9 y_s^5 y_w^6 y_o^{14} y_u^{15} y_v^{20} t_1^9 t_2^4 t_3^4 
t_4^9
~.~
$
\end{quote}
\endgroup
\subsection{$P_{+-}^{2}[\mathbb{C}^3/\mathbb{Z}_5~(1,1,3)]$ \label{app_num_08}}
\begingroup\makeatletter\def\f@size{7}\check@mathfonts
\begin{quote}\raggedright
$
P(t_a,y_s,y_w,y_u;\mathcal{M}^{mes}_{P_{+-}^{2}[\mathbb{C}^3/\mathbb{Z}_5~(1,1,3)]})=1+y_s^2 y_w y_u^2 t_2^3 t_3 t_4+y_s y_w y_u^3 t_2 t_3^2 t_4^2+y_s^3 
y_w^2 y_u^5 t_2^4 t_3^3 t_4^3+y_s^2 y_w^2 y_u^6 t_2^2 t_3^4 
t_4^4+y_s^2 y_w y_u^2 t_2^3 t_3 t_5+y_s y_w y_u^3 t_2 t_3^2 t_4 
t_5+y_s^3 y_w^2 y_u^5 t_2^4 t_3^3 t_4^2 t_5+y_s^2 y_w^2 y_u^6 t_2^2 
t_3^4 t_4^3 t_5+y_s y_w^2 y_u^7 t_3^5 t_4^4 t_5+y_s y_w y_u^3 t_2 
t_3^2 t_5^2+y_s^3 y_w^2 y_u^5 t_2^4 t_3^3 t_4 t_5^2+y_s^2 y_w^2 y_u^6 
t_2^2 t_3^4 t_4^2 t_5^2+y_s y_w^2 y_u^7 t_3^5 t_4^3 t_5^2+y_s^3 y_w^3 
y_u^9 t_2^3 t_3^6 t_4^4 t_5^2+y_s^3 y_w^2 y_u^5 t_2^4 t_3^3 
t_5^3+y_s^2 y_w^2 y_u^6 t_2^2 t_3^4 t_4 t_5^3+y_s y_w^2 y_u^7 t_3^5 
t_4^2 t_5^3+y_s^3 y_w^3 y_u^9 t_2^3 t_3^6 t_4^3 t_5^3+y_s^2 y_w^3 
y_u^{10} t_2 t_3^7 t_4^4 t_5^3+y_s^2 y_w^2 y_u^6 t_2^2 t_3^4 
t_5^4+y_s y_w^2 y_u^7 t_3^5 t_4 t_5^4+y_s^3 y_w^3 y_u^9 t_2^3 t_3^6 
t_4^2 t_5^4+y_s^2 y_w^3 y_u^{10} t_2 t_3^7 t_4^3 t_5^4+y_s^4 y_w^4 
y_u^{12} t_2^4 t_3^8 t_4^4 t_5^4+y_s^2 y_w y_u^2 t_2^3 t_4 t_6+y_s 
y_w y_u^3 t_2 t_3 t_4^2 t_6+y_s^3 y_w^2 y_u^5 t_2^4 t_3^2 t_4^3 
t_6+y_s^2 y_w^2 y_u^6 t_2^2 t_3^3 t_4^4 t_6+y_s y_w^2 y_u^7 t_3^4 
t_4^5 t_6+y_s^2 y_w y_u^2 t_2^3 t_5 t_6+y_s y_w y_u^3 t_2 t_3 t_4 t_5 
t_6+y_s^3 y_w^2 y_u^5 t_2^4 t_3^2 t_4^2 t_5 t_6+y_s^2 y_w^2 y_u^6 
t_2^2 t_3^3 t_4^3 t_5 t_6+y_s y_w^2 y_u^7 t_3^4 t_4^4 t_5 t_6+y_s y_w 
y_u^3 t_2 t_3 t_5^2 t_6+y_s^3 y_w^2 y_u^5 t_2^4 t_3^2 t_4 t_5^2 
t_6+y_s^2 y_w^2 y_u^6 t_2^2 t_3^3 t_4^2 t_5^2 t_6+y_s y_w^2 y_u^7 
t_3^4 t_4^3 t_5^2 t_6+y_s^3 y_w^3 y_u^9 t_2^3 t_3^5 t_4^4 t_5^2 
t_6+y_s^3 y_w^2 y_u^5 t_2^4 t_3^2 t_5^3 t_6+y_s^2 y_w^2 y_u^6 t_2^2 
t_3^3 t_4 t_5^3 t_6+y_s y_w^2 y_u^7 t_3^4 t_4^2 t_5^3 t_6+y_s^3 y_w^3 
y_u^9 t_2^3 t_3^5 t_4^3 t_5^3 t_6+y_s^2 y_w^3 y_u^{10} t_2 t_3^6 
t_4^4 t_5^3 t_6+y_s^2 y_w^2 y_u^6 t_2^2 t_3^3 t_5^4 t_6+y_s y_w^2 
y_u^7 t_3^4 t_4 t_5^4 t_6+y_s^3 y_w^3 y_u^9 t_2^3 t_3^5 t_4^2 t_5^4 
t_6+y_s^2 y_w^3 y_u^{10} t_2 t_3^6 t_4^3 t_5^4 t_6+y_s^4 y_w^4 
y_u^{12} t_2^4 t_3^7 t_4^4 t_5^4 t_6+y_s y_w^2 y_u^7 t_3^4 t_5^5 
t_6-y_s^2 y_w^4 y_u^{14} t_3^9 t_4^5 t_5^5 t_6+y_s y_w y_u^3 t_2 
t_4^2 t_6^2+y_s^3 y_w^2 y_u^5 t_2^4 t_3 t_4^3 t_6^2+y_s^2 y_w^2 y_u^6 
t_2^2 t_3^2 t_4^4 t_6^2+y_s y_w^2 y_u^7 t_3^3 t_4^5 t_6^2+y_s^3 y_w^3 
y_u^9 t_2^3 t_3^4 t_4^6 t_6^2+y_s y_w y_u^3 t_2 t_4 t_5 t_6^2+y_s^3 
y_w^2 y_u^5 t_2^4 t_3 t_4^2 t_5 t_6^2+y_s^2 y_w^2 y_u^6 t_2^2 t_3^2 
t_4^3 t_5 t_6^2+y_s y_w^2 y_u^7 t_3^3 t_4^4 t_5 t_6^2+y_s^3 y_w^3 
y_u^9 t_2^3 t_3^4 t_4^5 t_5 t_6^2+y_s y_w y_u^3 t_2 t_5^2 t_6^2+y_s^3 
y_w^2 y_u^5 t_2^4 t_3 t_4 t_5^2 t_6^2+y_s^2 y_w^2 y_u^6 t_2^2 t_3^2 
t_4^2 t_5^2 t_6^2+y_s y_w^2 y_u^7 t_3^3 t_4^3 t_5^2 t_6^2+y_s^3 y_w^3 
y_u^9 t_2^3 t_3^4 t_4^4 t_5^2 t_6^2+y_s^3 y_w^2 y_u^5 t_2^4 t_3 t_5^3 
t_6^2+y_s^2 y_w^2 y_u^6 t_2^2 t_3^2 t_4 t_5^3 t_6^2+y_s y_w^2 y_u^7 
t_3^3 t_4^2 t_5^3 t_6^2+y_s^3 y_w^3 y_u^9 t_2^3 t_3^4 t_4^3 t_5^3 
t_6^2+y_s^2 y_w^3 y_u^{10} t_2 t_3^5 t_4^4 t_5^3 t_6^2+y_s^2 y_w^2 
y_u^6 t_2^2 t_3^2 t_5^4 t_6^2+y_s y_w^2 y_u^7 t_3^3 t_4 t_5^4 
t_6^2+y_s^3 y_w^3 y_u^9 t_2^3 t_3^4 t_4^2 t_5^4 t_6^2+y_s^2 y_w^3 
y_u^{10} t_2 t_3^5 t_4^3 t_5^4 t_6^2+y_s^4 y_w^4 y_u^{12} t_2^4 t_3^6 
t_4^4 t_5^4 t_6^2+y_s y_w^2 y_u^7 t_3^3 t_5^5 t_6^2+y_s^3 y_w^3 y_u^9 
t_2^3 t_3^4 t_4 t_5^5 t_6^2-y_s^2 y_w^4 y_u^{14} t_3^8 t_4^5 t_5^5 
t_6^2-y_s^4 y_w^5 y_u^{16} t_2^3 t_3^9 t_4^6 t_5^5 t_6^2+y_s^3 y_w^3 
y_u^9 t_2^3 t_3^4 t_5^6 t_6^2-y_s^4 y_w^5 y_u^{16} t_2^3 t_3^9 t_4^5 
t_5^6 t_6^2+y_s^3 y_w^2 y_u^5 t_2^4 t_4^3 t_6^3+y_s^2 y_w^2 y_u^6 
t_2^2 t_3 t_4^4 t_6^3+y_s y_w^2 y_u^7 t_3^2 t_4^5 t_6^3+y_s^3 y_w^3 
y_u^9 t_2^3 t_3^3 t_4^6 t_6^3+y_s^2 y_w^3 y_u^{10} t_2 t_3^4 t_4^7 
t_6^3+y_s^3 y_w^2 y_u^5 t_2^4 t_4^2 t_5 t_6^3+y_s^2 y_w^2 y_u^6 t_2^2 
t_3 t_4^3 t_5 t_6^3+y_s y_w^2 y_u^7 t_3^2 t_4^4 t_5 t_6^3+y_s^3 y_w^3 
y_u^9 t_2^3 t_3^3 t_4^5 t_5 t_6^3+y_s^2 y_w^3 y_u^{10} t_2 t_3^4 
t_4^6 t_5 t_6^3+y_s^3 y_w^2 y_u^5 t_2^4 t_4 t_5^2 t_6^3+y_s^2 y_w^2 
y_u^6 t_2^2 t_3 t_4^2 t_5^2 t_6^3+y_s y_w^2 y_u^7 t_3^2 t_4^3 t_5^2 
t_6^3+y_s^3 y_w^3 y_u^9 t_2^3 t_3^3 t_4^4 t_5^2 t_6^3+y_s^2 y_w^3 
y_u^{10} t_2 t_3^4 t_4^5 t_5^2 t_6^3+y_s^3 y_w^2 y_u^5 t_2^4 t_5^3 
t_6^3+y_s^2 y_w^2 y_u^6 t_2^2 t_3 t_4 t_5^3 t_6^3+y_s y_w^2 y_u^7 
t_3^2 t_4^2 t_5^3 t_6^3+y_s^3 y_w^3 y_u^9 t_2^3 t_3^3 t_4^3 t_5^3 
t_6^3+y_s^2 y_w^3 y_u^{10} t_2 t_3^4 t_4^4 t_5^3 t_6^3+y_s^2 y_w^2 
y_u^6 t_2^2 t_3 t_5^4 t_6^3+y_s y_w^2 y_u^7 t_3^2 t_4 t_5^4 
t_6^3+y_s^3 y_w^3 y_u^9 t_2^3 t_3^3 t_4^2 t_5^4 t_6^3+y_s^2 y_w^3 
y_u^{10} t_2 t_3^4 t_4^3 t_5^4 t_6^3+y_s^4 y_w^4 y_u^{12} t_2^4 t_3^5 
t_4^4 t_5^4 t_6^3+y_s y_w^2 y_u^7 t_3^2 t_5^5 t_6^3+y_s^3 y_w^3 y_u^9 
t_2^3 t_3^3 t_4 t_5^5 t_6^3+y_s^2 y_w^3 y_u^{10} t_2 t_3^4 t_4^2 
t_5^5 t_6^3-y_s^2 y_w^4 y_u^{14} t_3^7 t_4^5 t_5^5 t_6^3-y_s^4 y_w^5 
y_u^{16} t_2^3 t_3^8 t_4^6 t_5^5 t_6^3-y_s^3 y_w^5 y_u^{17} t_2 t_3^9 
t_4^7 t_5^5 t_6^3+y_s^3 y_w^3 y_u^9 t_2^3 t_3^3 t_5^6 t_6^3+y_s^2 
y_w^3 y_u^{10} t_2 t_3^4 t_4 t_5^6 t_6^3-y_s^4 y_w^5 y_u^{16} t_2^3 
t_3^8 t_4^5 t_5^6 t_6^3-y_s^3 y_w^5 y_u^{17} t_2 t_3^9 t_4^6 t_5^6 
t_6^3+y_s^2 y_w^3 y_u^{10} t_2 t_3^4 t_5^7 t_6^3-y_s^3 y_w^5 y_u^{17} 
t_2 t_3^9 t_4^5 t_5^7 t_6^3+y_s^2 y_w^2 y_u^6 t_2^2 t_4^4 t_6^4+y_s 
y_w^2 y_u^7 t_3 t_4^5 t_6^4+y_s^3 y_w^3 y_u^9 t_2^3 t_3^2 t_4^6 
t_6^4+y_s^2 y_w^3 y_u^{10} t_2 t_3^3 t_4^7 t_6^4+y_s^4 y_w^4 y_u^{12} 
t_2^4 t_3^4 t_4^8 t_6^4+y_s^2 y_w^2 y_u^6 t_2^2 t_4^3 t_5 t_6^4+y_s 
y_w^2 y_u^7 t_3 t_4^4 t_5 t_6^4+y_s^3 y_w^3 y_u^9 t_2^3 t_3^2 t_4^5 
t_5 t_6^4+y_s^2 y_w^3 y_u^{10} t_2 t_3^3 t_4^6 t_5 t_6^4+y_s^4 y_w^4 
y_u^{12} t_2^4 t_3^4 t_4^7 t_5 t_6^4+y_s^2 y_w^2 y_u^6 t_2^2 t_4^2 
t_5^2 t_6^4+y_s y_w^2 y_u^7 t_3 t_4^3 t_5^2 t_6^4+y_s^3 y_w^3 y_u^9 
t_2^3 t_3^2 t_4^4 t_5^2 t_6^4+y_s^2 y_w^3 y_u^{10} t_2 t_3^3 t_4^5 
t_5^2 t_6^4+y_s^4 y_w^4 y_u^{12} t_2^4 t_3^4 t_4^6 t_5^2 t_6^4+y_s^2 
y_w^2 y_u^6 t_2^2 t_4 t_5^3 t_6^4+y_s y_w^2 y_u^7 t_3 t_4^2 t_5^3 
t_6^4+y_s^3 y_w^3 y_u^9 t_2^3 t_3^2 t_4^3 t_5^3 t_6^4+y_s^2 y_w^3 
y_u^{10} t_2 t_3^3 t_4^4 t_5^3 t_6^4+y_s^4 y_w^4 y_u^{12} t_2^4 t_3^4 
t_4^5 t_5^3 t_6^4+y_s^2 y_w^2 y_u^6 t_2^2 t_5^4 t_6^4+y_s y_w^2 y_u^7 
t_3 t_4 t_5^4 t_6^4+y_s^3 y_w^3 y_u^9 t_2^3 t_3^2 t_4^2 t_5^4 
t_6^4+y_s^2 y_w^3 y_u^{10} t_2 t_3^3 t_4^3 t_5^4 t_6^4+y_s^4 y_w^4 
y_u^{12} t_2^4 t_3^4 t_4^4 t_5^4 t_6^4+y_s y_w^2 y_u^7 t_3 t_5^5 
t_6^4+y_s^3 y_w^3 y_u^9 t_2^3 t_3^2 t_4 t_5^5 t_6^4+y_s^2 y_w^3 
y_u^{10} t_2 t_3^3 t_4^2 t_5^5 t_6^4+y_s^4 y_w^4 y_u^{12} t_2^4 t_3^4 
t_4^3 t_5^5 t_6^4-y_s^2 y_w^4 y_u^{14} t_3^6 t_4^5 t_5^5 t_6^4-y_s^4 
y_w^5 y_u^{16} t_2^3 t_3^7 t_4^6 t_5^5 t_6^4-y_s^3 y_w^5 y_u^{17} t_2 
t_3^8 t_4^7 t_5^5 t_6^4-y_s^5 y_w^6 y_u^{19} t_2^4 t_3^9 t_4^8 t_5^5 
t_6^4+y_s^3 y_w^3 y_u^9 t_2^3 t_3^2 t_5^6 t_6^4+y_s^2 y_w^3 y_u^{10} 
t_2 t_3^3 t_4 t_5^6 t_6^4+y_s^4 y_w^4 y_u^{12} t_2^4 t_3^4 t_4^2 
t_5^6 t_6^4-y_s^4 y_w^5 y_u^{16} t_2^3 t_3^7 t_4^5 t_5^6 t_6^4-y_s^3 
y_w^5 y_u^{17} t_2 t_3^8 t_4^6 t_5^6 t_6^4-y_s^5 y_w^6 y_u^{19} t_2^4 
t_3^9 t_4^7 t_5^6 t_6^4+y_s^2 y_w^3 y_u^{10} t_2 t_3^3 t_5^7 
t_6^4+y_s^4 y_w^4 y_u^{12} t_2^4 t_3^4 t_4 t_5^7 t_6^4-y_s^3 y_w^5 
y_u^{17} t_2 t_3^8 t_4^5 t_5^7 t_6^4-y_s^5 y_w^6 y_u^{19} t_2^4 t_3^9 
t_4^6 t_5^7 t_6^4+y_s^4 y_w^4 y_u^{12} t_2^4 t_3^4 t_5^8 t_6^4-y_s^5 
y_w^6 y_u^{19} t_2^4 t_3^9 t_4^5 t_5^8 t_6^4+y_s y_w^2 y_u^7 t_4^4 
t_5 t_6^5-y_s^2 y_w^4 y_u^{14} t_3^5 t_4^9 t_5 t_6^5+y_s y_w^2 y_u^7 
t_4^3 t_5^2 t_6^5+y_s^3 y_w^3 y_u^9 t_2^3 t_3 t_4^4 t_5^2 t_6^5-y_s^2 
y_w^4 y_u^{14} t_3^5 t_4^8 t_5^2 t_6^5-y_s^4 y_w^5 y_u^{16} t_2^3 
t_3^6 t_4^9 t_5^2 t_6^5+y_s y_w^2 y_u^7 t_4^2 t_5^3 t_6^5+y_s^3 y_w^3 
y_u^9 t_2^3 t_3 t_4^3 t_5^3 t_6^5+y_s^2 y_w^3 y_u^{10} t_2 t_3^2 
t_4^4 t_5^3 t_6^5-y_s^2 y_w^4 y_u^{14} t_3^5 t_4^7 t_5^3 t_6^5-y_s^4 
y_w^5 y_u^{16} t_2^3 t_3^6 t_4^8 t_5^3 t_6^5-y_s^3 y_w^5 y_u^{17} t_2 
t_3^7 t_4^9 t_5^3 t_6^5+y_s y_w^2 y_u^7 t_4 t_5^4 t_6^5+y_s^3 y_w^3 
y_u^9 t_2^3 t_3 t_4^2 t_5^4 t_6^5+y_s^2 y_w^3 y_u^{10} t_2 t_3^2 
t_4^3 t_5^4 t_6^5+y_s^4 y_w^4 y_u^{12} t_2^4 t_3^3 t_4^4 t_5^4 
t_6^5-y_s^2 y_w^4 y_u^{14} t_3^5 t_4^6 t_5^4 t_6^5-y_s^4 y_w^5 
y_u^{16} t_2^3 t_3^6 t_4^7 t_5^4 t_6^5-y_s^3 y_w^5 y_u^{17} t_2 t_3^7 
t_4^8 t_5^4 t_6^5-y_s^5 y_w^6 y_u^{19} t_2^4 t_3^8 t_4^9 t_5^4 
t_6^5-y_s^2 y_w^4 y_u^{14} t_3^5 t_4^5 t_5^5 t_6^5-y_s^4 y_w^5 
y_u^{16} t_2^3 t_3^6 t_4^6 t_5^5 t_6^5-y_s^3 y_w^5 y_u^{17} t_2 t_3^7 
t_4^7 t_5^5 t_6^5-y_s^5 y_w^6 y_u^{19} t_2^4 t_3^8 t_4^8 t_5^5 
t_6^5-y_s^4 y_w^6 y_u^{20} t_2^2 t_3^9 t_4^9 t_5^5 t_6^5-y_s^2 y_w^4 
y_u^{14} t_3^5 t_4^4 t_5^6 t_6^5-y_s^4 y_w^5 y_u^{16} t_2^3 t_3^6 
t_4^5 t_5^6 t_6^5-y_s^3 y_w^5 y_u^{17} t_2 t_3^7 t_4^6 t_5^6 
t_6^5-y_s^5 y_w^6 y_u^{19} t_2^4 t_3^8 t_4^7 t_5^6 t_6^5-y_s^4 y_w^6 
y_u^{20} t_2^2 t_3^9 t_4^8 t_5^6 t_6^5-y_s^2 y_w^4 y_u^{14} t_3^5 
t_4^3 t_5^7 t_6^5-y_s^4 y_w^5 y_u^{16} t_2^3 t_3^6 t_4^4 t_5^7 
t_6^5-y_s^3 y_w^5 y_u^{17} t_2 t_3^7 t_4^5 t_5^7 t_6^5-y_s^5 y_w^6 
y_u^{19} t_2^4 t_3^8 t_4^6 t_5^7 t_6^5-y_s^4 y_w^6 y_u^{20} t_2^2 
t_3^9 t_4^7 t_5^7 t_6^5-y_s^2 y_w^4 y_u^{14} t_3^5 t_4^2 t_5^8 
t_6^5-y_s^4 y_w^5 y_u^{16} t_2^3 t_3^6 t_4^3 t_5^8 t_6^5-y_s^3 y_w^5 
y_u^{17} t_2 t_3^7 t_4^4 t_5^8 t_6^5-y_s^5 y_w^6 y_u^{19} t_2^4 t_3^8 
t_4^5 t_5^8 t_6^5-y_s^4 y_w^6 y_u^{20} t_2^2 t_3^9 t_4^6 t_5^8 
t_6^5-y_s^2 y_w^4 y_u^{14} t_3^5 t_4 t_5^9 t_6^5-y_s^4 y_w^5 y_u^{16} 
t_2^3 t_3^6 t_4^2 t_5^9 t_6^5-y_s^3 y_w^5 y_u^{17} t_2 t_3^7 t_4^3 
t_5^9 t_6^5-y_s^5 y_w^6 y_u^{19} t_2^4 t_3^8 t_4^4 t_5^9 t_6^5-y_s^4 
y_w^6 y_u^{20} t_2^2 t_3^9 t_4^5 t_5^9 t_6^5+y_s^3 y_w^3 y_u^9 t_2^3 
t_4^4 t_5^2 t_6^6-y_s^4 y_w^5 y_u^{16} t_2^3 t_3^5 t_4^9 t_5^2 
t_6^6+y_s^3 y_w^3 y_u^9 t_2^3 t_4^3 t_5^3 t_6^6+y_s^2 y_w^3 y_u^{10} 
t_2 t_3 t_4^4 t_5^3 t_6^6-y_s^4 y_w^5 y_u^{16} t_2^3 t_3^5 t_4^8 
t_5^3 t_6^6-y_s^3 y_w^5 y_u^{17} t_2 t_3^6 t_4^9 t_5^3 t_6^6+y_s^3 
y_w^3 y_u^9 t_2^3 t_4^2 t_5^4 t_6^6+y_s^2 y_w^3 y_u^{10} t_2 t_3 
t_4^3 t_5^4 t_6^6+y_s^4 y_w^4 y_u^{12} t_2^4 t_3^2 t_4^4 t_5^4 
t_6^6-y_s^4 y_w^5 y_u^{16} t_2^3 t_3^5 t_4^7 t_5^4 t_6^6-y_s^3 y_w^5 
y_u^{17} t_2 t_3^6 t_4^8 t_5^4 t_6^6-y_s^5 y_w^6 y_u^{19} t_2^4 t_3^7 
t_4^9 t_5^4 t_6^6-y_s^2 y_w^4 y_u^{14} t_3^4 t_4^5 t_5^5 t_6^6-y_s^4 
y_w^5 y_u^{16} t_2^3 t_3^5 t_4^6 t_5^5 t_6^6-y_s^3 y_w^5 y_u^{17} t_2 
t_3^6 t_4^7 t_5^5 t_6^6-y_s^5 y_w^6 y_u^{19} t_2^4 t_3^7 t_4^8 t_5^5 
t_6^6-y_s^4 y_w^6 y_u^{20} t_2^2 t_3^8 t_4^9 t_5^5 t_6^6-y_s^4 y_w^5 
y_u^{16} t_2^3 t_3^5 t_4^5 t_5^6 t_6^6-y_s^3 y_w^5 y_u^{17} t_2 t_3^6 
t_4^6 t_5^6 t_6^6-y_s^5 y_w^6 y_u^{19} t_2^4 t_3^7 t_4^7 t_5^6 
t_6^6-y_s^4 y_w^6 y_u^{20} t_2^2 t_3^8 t_4^8 t_5^6 t_6^6-y_s^3 y_w^6 
y_u^{21} t_3^9 t_4^9 t_5^6 t_6^6-y_s^4 y_w^5 y_u^{16} t_2^3 t_3^5 
t_4^4 t_5^7 t_6^6-y_s^3 y_w^5 y_u^{17} t_2 t_3^6 t_4^5 t_5^7 
t_6^6-y_s^5 y_w^6 y_u^{19} t_2^4 t_3^7 t_4^6 t_5^7 t_6^6-y_s^4 y_w^6 
y_u^{20} t_2^2 t_3^8 t_4^7 t_5^7 t_6^6-y_s^3 y_w^6 y_u^{21} t_3^9 
t_4^8 t_5^7 t_6^6-y_s^4 y_w^5 y_u^{16} t_2^3 t_3^5 t_4^3 t_5^8 
t_6^6-y_s^3 y_w^5 y_u^{17} t_2 t_3^6 t_4^4 t_5^8 t_6^6-y_s^5 y_w^6 
y_u^{19} t_2^4 t_3^7 t_4^5 t_5^8 t_6^6-y_s^4 y_w^6 y_u^{20} t_2^2 
t_3^8 t_4^6 t_5^8 t_6^6-y_s^3 y_w^6 y_u^{21} t_3^9 t_4^7 t_5^8 
t_6^6-y_s^4 y_w^5 y_u^{16} t_2^3 t_3^5 t_4^2 t_5^9 t_6^6-y_s^3 y_w^5 
y_u^{17} t_2 t_3^6 t_4^3 t_5^9 t_6^6-y_s^5 y_w^6 y_u^{19} t_2^4 t_3^7 
t_4^4 t_5^9 t_6^6-y_s^4 y_w^6 y_u^{20} t_2^2 t_3^8 t_4^5 t_5^9 
t_6^6-y_s^3 y_w^6 y_u^{21} t_3^9 t_4^6 t_5^9 t_6^6+y_s^2 y_w^3 
y_u^{10} t_2 t_4^4 t_5^3 t_6^7-y_s^3 y_w^5 y_u^{17} t_2 t_3^5 t_4^9 
t_5^3 t_6^7+y_s^2 y_w^3 y_u^{10} t_2 t_4^3 t_5^4 t_6^7+y_s^4 y_w^4 
y_u^{12} t_2^4 t_3 t_4^4 t_5^4 t_6^7-y_s^3 y_w^5 y_u^{17} t_2 t_3^5 
t_4^8 t_5^4 t_6^7-y_s^5 y_w^6 y_u^{19} t_2^4 t_3^6 t_4^9 t_5^4 
t_6^7-y_s^2 y_w^4 y_u^{14} t_3^3 t_4^5 t_5^5 t_6^7-y_s^4 y_w^5 
y_u^{16} t_2^3 t_3^4 t_4^6 t_5^5 t_6^7-y_s^3 y_w^5 y_u^{17} t_2 t_3^5 
t_4^7 t_5^5 t_6^7-y_s^5 y_w^6 y_u^{19} t_2^4 t_3^6 t_4^8 t_5^5 
t_6^7-y_s^4 y_w^6 y_u^{20} t_2^2 t_3^7 t_4^9 t_5^5 t_6^7-y_s^4 y_w^5 
y_u^{16} t_2^3 t_3^4 t_4^5 t_5^6 t_6^7-y_s^3 y_w^5 y_u^{17} t_2 t_3^5 
t_4^6 t_5^6 t_6^7-y_s^5 y_w^6 y_u^{19} t_2^4 t_3^6 t_4^7 t_5^6 
t_6^7-y_s^4 y_w^6 y_u^{20} t_2^2 t_3^7 t_4^8 t_5^6 t_6^7-y_s^3 y_w^6 
y_u^{21} t_3^8 t_4^9 t_5^6 t_6^7-y_s^3 y_w^5 y_u^{17} t_2 t_3^5 t_4^5 
t_5^7 t_6^7-y_s^5 y_w^6 y_u^{19} t_2^4 t_3^6 t_4^6 t_5^7 t_6^7-y_s^4 
y_w^6 y_u^{20} t_2^2 t_3^7 t_4^7 t_5^7 t_6^7-y_s^3 y_w^6 y_u^{21} 
t_3^8 t_4^8 t_5^7 t_6^7-y_s^5 y_w^7 y_u^{23} t_2^3 t_3^9 t_4^9 t_5^7 
t_6^7-y_s^3 y_w^5 y_u^{17} t_2 t_3^5 t_4^4 t_5^8 t_6^7-y_s^5 y_w^6 
y_u^{19} t_2^4 t_3^6 t_4^5 t_5^8 t_6^7-y_s^4 y_w^6 y_u^{20} t_2^2 
t_3^7 t_4^6 t_5^8 t_6^7-y_s^3 y_w^6 y_u^{21} t_3^8 t_4^7 t_5^8 
t_6^7-y_s^5 y_w^7 y_u^{23} t_2^3 t_3^9 t_4^8 t_5^8 t_6^7-y_s^3 y_w^5 
y_u^{17} t_2 t_3^5 t_4^3 t_5^9 t_6^7-y_s^5 y_w^6 y_u^{19} t_2^4 t_3^6 
t_4^4 t_5^9 t_6^7-y_s^4 y_w^6 y_u^{20} t_2^2 t_3^7 t_4^5 t_5^9 
t_6^7-y_s^3 y_w^6 y_u^{21} t_3^8 t_4^6 t_5^9 t_6^7-y_s^5 y_w^7 
y_u^{23} t_2^3 t_3^9 t_4^7 t_5^9 t_6^7+y_s^4 y_w^4 y_u^{12} t_2^4 
t_4^4 t_5^4 t_6^8-y_s^5 y_w^6 y_u^{19} t_2^4 t_3^5 t_4^9 t_5^4 
t_6^8-y_s^2 y_w^4 y_u^{14} t_3^2 t_4^5 t_5^5 t_6^8-y_s^4 y_w^5 
y_u^{16} t_2^3 t_3^3 t_4^6 t_5^5 t_6^8-y_s^3 y_w^5 y_u^{17} t_2 t_3^4 
t_4^7 t_5^5 t_6^8-y_s^5 y_w^6 y_u^{19} t_2^4 t_3^5 t_4^8 t_5^5 
t_6^8-y_s^4 y_w^6 y_u^{20} t_2^2 t_3^6 t_4^9 t_5^5 t_6^8-y_s^4 y_w^5 
y_u^{16} t_2^3 t_3^3 t_4^5 t_5^6 t_6^8-y_s^3 y_w^5 y_u^{17} t_2 t_3^4 
t_4^6 t_5^6 t_6^8-y_s^5 y_w^6 y_u^{19} t_2^4 t_3^5 t_4^7 t_5^6 
t_6^8-y_s^4 y_w^6 y_u^{20} t_2^2 t_3^6 t_4^8 t_5^6 t_6^8-y_s^3 y_w^6 
y_u^{21} t_3^7 t_4^9 t_5^6 t_6^8-y_s^3 y_w^5 y_u^{17} t_2 t_3^4 t_4^5 
t_5^7 t_6^8-y_s^5 y_w^6 y_u^{19} t_2^4 t_3^5 t_4^6 t_5^7 t_6^8-y_s^4 
y_w^6 y_u^{20} t_2^2 t_3^6 t_4^7 t_5^7 t_6^8-y_s^3 y_w^6 y_u^{21} 
t_3^7 t_4^8 t_5^7 t_6^8-y_s^5 y_w^7 y_u^{23} t_2^3 t_3^8 t_4^9 t_5^7 
t_6^8-y_s^5 y_w^6 y_u^{19} t_2^4 t_3^5 t_4^5 t_5^8 t_6^8-y_s^4 y_w^6 
y_u^{20} t_2^2 t_3^6 t_4^6 t_5^8 t_6^8-y_s^3 y_w^6 y_u^{21} t_3^7 
t_4^7 t_5^8 t_6^8-y_s^5 y_w^7 y_u^{23} t_2^3 t_3^8 t_4^8 t_5^8 
t_6^8-y_s^4 y_w^7 y_u^{24} t_2 t_3^9 t_4^9 t_5^8 t_6^8-y_s^5 y_w^6 
y_u^{19} t_2^4 t_3^5 t_4^4 t_5^9 t_6^8-y_s^4 y_w^6 y_u^{20} t_2^2 
t_3^6 t_4^5 t_5^9 t_6^8-y_s^3 y_w^6 y_u^{21} t_3^7 t_4^6 t_5^9 
t_6^8-y_s^5 y_w^7 y_u^{23} t_2^3 t_3^8 t_4^7 t_5^9 t_6^8-y_s^4 y_w^7 
y_u^{24} t_2 t_3^9 t_4^8 t_5^9 t_6^8-y_s^2 y_w^4 y_u^{14} t_3 t_4^5 
t_5^5 t_6^9-y_s^4 y_w^5 y_u^{16} t_2^3 t_3^2 t_4^6 t_5^5 t_6^9-y_s^3 
y_w^5 y_u^{17} t_2 t_3^3 t_4^7 t_5^5 t_6^9-y_s^5 y_w^6 y_u^{19} t_2^4 
t_3^4 t_4^8 t_5^5 t_6^9-y_s^4 y_w^6 y_u^{20} t_2^2 t_3^5 t_4^9 t_5^5 
t_6^9-y_s^4 y_w^5 y_u^{16} t_2^3 t_3^2 t_4^5 t_5^6 t_6^9-y_s^3 y_w^5 
y_u^{17} t_2 t_3^3 t_4^6 t_5^6 t_6^9-y_s^5 y_w^6 y_u^{19} t_2^4 t_3^4 
t_4^7 t_5^6 t_6^9-y_s^4 y_w^6 y_u^{20} t_2^2 t_3^5 t_4^8 t_5^6 
t_6^9-y_s^3 y_w^6 y_u^{21} t_3^6 t_4^9 t_5^6 t_6^9-y_s^3 y_w^5 
y_u^{17} t_2 t_3^3 t_4^5 t_5^7 t_6^9-y_s^5 y_w^6 y_u^{19} t_2^4 t_3^4 
t_4^6 t_5^7 t_6^9-y_s^4 y_w^6 y_u^{20} t_2^2 t_3^5 t_4^7 t_5^7 
t_6^9-y_s^3 y_w^6 y_u^{21} t_3^6 t_4^8 t_5^7 t_6^9-y_s^5 y_w^7 
y_u^{23} t_2^3 t_3^7 t_4^9 t_5^7 t_6^9-y_s^5 y_w^6 y_u^{19} t_2^4 
t_3^4 t_4^5 t_5^8 t_6^9-y_s^4 y_w^6 y_u^{20} t_2^2 t_3^5 t_4^6 t_5^8 
t_6^9-y_s^3 y_w^6 y_u^{21} t_3^6 t_4^7 t_5^8 t_6^9-y_s^5 y_w^7 
y_u^{23} t_2^3 t_3^7 t_4^8 t_5^8 t_6^9-y_s^4 y_w^7 y_u^{24} t_2 t_3^8 
t_4^9 t_5^8 t_6^9-y_s^4 y_w^6 y_u^{20} t_2^2 t_3^5 t_4^5 t_5^9 
t_6^9-y_s^3 y_w^6 y_u^{21} t_3^6 t_4^6 t_5^9 t_6^9-y_s^5 y_w^7 
y_u^{23} t_2^3 t_3^7 t_4^7 t_5^9 t_6^9-y_s^4 y_w^7 y_u^{24} t_2 t_3^8 
t_4^8 t_5^9 t_6^9-y_s^6 y_w^8 y_u^{26} t_2^4 t_3^9 t_4^9 t_5^9 t_6^9
~.~
$
\end{quote}
\endgroup
\bibliographystyle{JHEP}
\bibliography{mybib}

\providecommand{\href}[2]{#2}\begingroup\raggedright\begin{thebibliography}{10}

\bibitem{mori1979projective}
S.~Mori, \emph{Projective manifolds with ample tangent bundles}, {\emph{Annals
  of Mathematics} {\bf 110} (1979) 593--606}.

\bibitem{fa4d6420-e404-3d65-bad9-fee06a1f00ea}
S.~Mori, \emph{Threefolds whose canonical bundles are not numerically
  effective}, {\emph{Annals of Mathematics} {\bf 116} (1982) 133--176}.

\bibitem{ab3c5836-bdbd-3fe7-9721-d43dbbc02ff0}
S.~Mori, \emph{Flip theorem and the existence of minimal models for 3-folds},
  {\emph{Journal of the American Mathematical Society} {\bf 1} (1988)
  117--253}.

\bibitem{kawamata1985pluricanonical}
Y.~Kawamata, \emph{Pluricanonical systems on minimal algebraic varieties},
  {\emph{Inventiones mathematicae} {\bf 79} (1985) 567--588}.

\bibitem{MR1158625}
J.~Koll\'ar, Y.~Miyaoka and S.~Mori, \emph{Rationally connected varieties},
  {\emph{J. Algebraic Geom.} {\bf 1} (1992) 429--448}.

\bibitem{Kollar_Mori_1998}
J.~Kollár and S.~Mori, \emph{Birational Geometry of Algebraic Varieties}.
\newblock Cambridge Tracts in Mathematics. Cambridge University Press, 1998.

\bibitem{birkar2010existence}
C.~Birkar, P.~Cascini, C.~Hacon and J.~McKernan, \emph{Existence of minimal
  models for varieties of log general type}, {\emph{Journal of the American
  Mathematical Society} {\bf 23} (2010) 405--468}.

\bibitem{akhtar2012minkowski}
M.~Akhtar, T.~Coates, S.~Galkin, A.~M. Kasprzyk et~al., \emph{Minkowski
  polynomials and mutations}, {\emph{SIGMA. Symmetry, Integrability and
  Geometry: Methods and Applications} {\bf 8} (2012) 094}.

\bibitem{gross2013birational}
M.~Gross, P.~Hacking and S.~Keel, \emph{Birational geometry of cluster
  algebras},  \href{http://arxiv.org/abs/1309.2573}{{\tt 1309.2573}}.

\bibitem{coates2016quantum}
T.~Coates, A.~Corti, S.~Galkin and A.~Kasprzyk, \emph{Quantum periods for
  3--dimensional fano manifolds}, {\emph{Geometry \& Topology} {\bf 20} (2016)
  103--256}.

\bibitem{kasprzyk2017minimality}
A.~Kasprzyk, B.~Nill and T.~Prince, \emph{Minimality and mutation-equivalence
  of polygons},  in \emph{Forum of mathematics, Sigma}, vol.~5, p.~e18,
  Cambridge University Press, 2017.

\bibitem{coates2021maximally}
T.~Coates, A.~M. Kasprzyk, G.~Pitton and K.~Tveiten, \emph{Maximally mutable
  laurent polynomials}, {\emph{Proceedings of the Royal Society A} {\bf 477}
  (2021) 20210584}.

\bibitem{Coates2022MirrorSL}
T.~Coates, L.~Heuberger and A.~M. Kasprzyk, \emph{Mirror symmetry, laurent
  inversion and the classification of $\mathbb{Q}$-fano threefolds},
  \href{http://arxiv.org/abs/2210.07328}{{\tt 2210.07328}}.

\bibitem{corti2023cluster}
A.~Corti, \emph{Cluster varieties and toric specializations of fano varieties},
   \href{http://arxiv.org/abs/2304.04141}{{\tt 2304.04141}}.

\bibitem{Franco:2015tna}
S.~Franco, D.~Ghim, S.~Lee, R.-K. Seong and D.~Yokoyama, \emph{{2d (0,2) Quiver
  Gauge Theories and D-Branes}},
  \href{http://dx.doi.org/10.1007/JHEP09(2015)072}{\emph{JHEP} {\bf 09} (2015)
  072}, [\href{http://arxiv.org/abs/1506.03818}{{\tt 1506.03818}}].

\bibitem{Franco:2015tya}
S.~Franco, S.~Lee and R.-K. Seong, \emph{{Brane Brick Models, Toric Calabi-Yau
  4-Folds and 2d (0,2) Quivers}},
  \href{http://dx.doi.org/10.1007/JHEP02(2016)047}{\emph{JHEP} {\bf 02} (2016)
  047}, [\href{http://arxiv.org/abs/1510.01744}{{\tt 1510.01744}}].

\bibitem{Franco:2016nwv}
S.~Franco, S.~Lee and R.-K. Seong, \emph{{Brane brick models and 2d (0, 2)
  triality}}, \href{http://dx.doi.org/10.1007/JHEP05(2016)020}{\emph{JHEP} {\bf
  05} (2016) 020}, [\href{http://arxiv.org/abs/1602.01834}{{\tt 1602.01834}}].

\bibitem{Franco:2016qxh}
S.~Franco, S.~Lee, R.-K. Seong and C.~Vafa, \emph{{Brane Brick Models in the
  Mirror}}, \href{http://dx.doi.org/10.1007/JHEP02(2017)106}{\emph{JHEP} {\bf
  02} (2017) 106}, [\href{http://arxiv.org/abs/1609.01723}{{\tt 1609.01723}}].

\bibitem{Kreuzer:1995cd}
M.~Kreuzer and H.~Skarke, \emph{{On the classification of reflexive
  polyhedra}}, \href{http://dx.doi.org/10.1007/s002200050100}{\emph{Commun.
  Math. Phys.} {\bf 185} (1997) 495--508},
  [\href{http://arxiv.org/abs/hep-th/9512204}{{\tt hep-th/9512204}}].

\bibitem{Kreuzer:1998vb}
M.~Kreuzer and H.~Skarke, \emph{{Classification of reflexive polyhedra in
  three-dimensions}},
  \href{http://dx.doi.org/10.4310/ATMP.1998.v2.n4.a5}{\emph{Adv. Theor. Math.
  Phys.} {\bf 2} (1998) 853--871},
  [\href{http://arxiv.org/abs/hep-th/9805190}{{\tt hep-th/9805190}}].

\bibitem{Kreuzer:2000xy}
M.~Kreuzer and H.~Skarke, \emph{{Complete classification of reflexive polyhedra
  in four-dimensions}},
  \href{http://dx.doi.org/10.4310/ATMP.2000.v4.n6.a2}{\emph{Adv. Theor. Math.
  Phys.} {\bf 4} (2000) 1209--1230},
  [\href{http://arxiv.org/abs/hep-th/0002240}{{\tt hep-th/0002240}}].

\bibitem{Ghim:2024asj}
D.~Ghim, M.~Kho and R.-K. Seong, \emph{{Combinatorial and algebraic mutations
  of toric Fano 3-folds and mass deformations of 2d(0,2) quiver gauge
  theories}}, \href{http://dx.doi.org/10.1103/PhysRevD.110.086001}{\emph{Phys.
  Rev. D} {\bf 110} (2024) 086001},
  [\href{http://arxiv.org/abs/2407.19924}{{\tt 2407.19924}}].

\bibitem{Franco:2023tyf}
S.~Franco, D.~Ghim, G.~P. Goulas and R.-K. Seong, \emph{{Mass deformations of
  brane brick models}},
  \href{http://dx.doi.org/10.1007/JHEP09(2023)176}{\emph{JHEP} {\bf 09} (2023)
  176}, [\href{http://arxiv.org/abs/2307.03220}{{\tt 2307.03220}}].

\bibitem{Davey:2010px}
J.~Davey, A.~Hanany and R.-K. Seong, \emph{{Counting Orbifolds}},
  \href{http://dx.doi.org/10.1007/JHEP06(2010)010}{\emph{JHEP} {\bf 06} (2010)
  010}, [\href{http://arxiv.org/abs/1002.3609}{{\tt 1002.3609}}].

\bibitem{Hanany:2010ne}
A.~Hanany and R.-K. Seong, \emph{{Symmetries of Abelian Orbifolds}},
  \href{http://dx.doi.org/10.1007/JHEP01(2011)027}{\emph{JHEP} {\bf 01} (2011)
  027}, [\href{http://arxiv.org/abs/1009.3017}{{\tt 1009.3017}}].

\bibitem{Franco:2022gvl}
S.~Franco and R.-K. Seong, \emph{{Fano 3-folds, reflexive polytopes and brane
  brick models}}, \href{http://dx.doi.org/10.1007/JHEP08(2022)008}{\emph{JHEP}
  {\bf 08} (2022) 008}, [\href{http://arxiv.org/abs/2203.15816}{{\tt
  2203.15816}}].

\bibitem{Hori:2000kt}
K.~Hori and C.~Vafa, \emph{{Mirror symmetry}},
  \href{http://arxiv.org/abs/hep-th/0002222}{{\tt hep-th/0002222}}.

\bibitem{Hori:2000ck}
K.~Hori, A.~Iqbal and C.~Vafa, \emph{{D-branes and mirror symmetry}},
  \href{http://arxiv.org/abs/hep-th/0005247}{{\tt hep-th/0005247}}.

\bibitem{Cachazo:2001sg}
F.~Cachazo, B.~Fiol, K.~A. Intriligator, S.~Katz and C.~Vafa, \emph{{A
  Geometric unification of dualities}},
  \href{http://dx.doi.org/10.1016/S0550-3213(02)00078-0}{\emph{Nucl. Phys. B}
  {\bf 628} (2002) 3--78}, [\href{http://arxiv.org/abs/hep-th/0110028}{{\tt
  hep-th/0110028}}].

\bibitem{Feng:2005gw}
B.~Feng, Y.-H. He, K.~D. Kennaway and C.~Vafa, \emph{{Dimer models from mirror
  symmetry and quivering amoebae}},
  \href{http://dx.doi.org/10.4310/ATMP.2008.v12.n3.a2}{\emph{Adv. Theor. Math.
  Phys.} {\bf 12} (2008) 489--545},
  [\href{http://arxiv.org/abs/hep-th/0511287}{{\tt hep-th/0511287}}].

\bibitem{Franco:2016tcm}
S.~Franco, S.~Lee, R.-K. Seong and C.~Vafa, \emph{{Quadrality for
  Supersymmetric Matrix Models}},
  \href{http://dx.doi.org/10.1007/JHEP07(2017)053}{\emph{JHEP} {\bf 07} (2017)
  053}, [\href{http://arxiv.org/abs/1612.06859}{{\tt 1612.06859}}].

\bibitem{Seong:2023njx}
R.-K. Seong, \emph{{Unsupervised machine learning techniques for exploring
  tropical coamoeba, brane tilings and Seiberg duality}},
  \href{http://dx.doi.org/10.1103/PhysRevD.108.106009}{\emph{Phys. Rev. D} {\bf
  108} (2023) 106009}, [\href{http://arxiv.org/abs/2309.05702}{{\tt
  2309.05702}}].

\bibitem{Seong:2024wkt}
R.-K. Seong, \emph{{Generative AI for Brane Configurations, Tropical Coamoeba
  and 4d N=1 Quiver Gauge Theories}},
  \href{http://arxiv.org/abs/2411.16033}{{\tt 2411.16033}}.

\bibitem{fulton1993introduction}
W.~Fulton, \emph{Introduction to Toric Varieties}.
\newblock Annals of mathematics studies. Princeton University Press, 1993.

\bibitem{Cox:1993fz}
D.~A. Cox, \emph{{The Homogeneous coordinate ring of a toric variety, revised
  version}},  \href{http://arxiv.org/abs/alg-geom/9210008}{{\tt
  alg-geom/9210008}}.

\bibitem{Feng:2000mi}
B.~Feng, A.~Hanany and Y.-H. He, \emph{{D-brane gauge theories from toric
  singularities and toric duality}},
  \href{http://dx.doi.org/10.1016/S0550-3213(00)00699-4}{\emph{Nucl. Phys. B}
  {\bf 595} (2001) 165--200}, [\href{http://arxiv.org/abs/hep-th/0003085}{{\tt
  hep-th/0003085}}].

\bibitem{Benvenuti:2006qr}
S.~Benvenuti, B.~Feng, A.~Hanany and Y.-H. He, \emph{{Counting BPS Operators in
  Gauge Theories: Quivers, Syzygies and Plethystics}},
  \href{http://dx.doi.org/10.1088/1126-6708/2007/11/050}{\emph{JHEP} {\bf 11}
  (2007) 050}, [\href{http://arxiv.org/abs/hep-th/0608050}{{\tt
  hep-th/0608050}}].

\bibitem{Hanany:2006uc}
A.~Hanany and C.~Romelsberger, \emph{{Counting BPS operators in the chiral ring
  of N=2 supersymmetric gauge theories or N=2 braine surgery}},
  \href{http://dx.doi.org/10.4310/ATMP.2007.v11.n6.a4}{\emph{Adv. Theor. Math.
  Phys.} {\bf 11} (2007) 1091--1112},
  [\href{http://arxiv.org/abs/hep-th/0611346}{{\tt hep-th/0611346}}].

\bibitem{Butti:2007jv}
A.~Butti, D.~Forcella, A.~Hanany, D.~Vegh and A.~Zaffaroni, \emph{{Counting
  Chiral Operators in Quiver Gauge Theories}},
  \href{http://dx.doi.org/10.1088/1126-6708/2007/11/092}{\emph{JHEP} {\bf 11}
  (2007) 092}, [\href{http://arxiv.org/abs/0705.2771}{{\tt 0705.2771}}].

\bibitem{Feng:2007ur}
B.~Feng, A.~Hanany and Y.-H. He, \emph{{Counting gauge invariants: The
  Plethystic program}},
  \href{http://dx.doi.org/10.1088/1126-6708/2007/03/090}{\emph{JHEP} {\bf 03}
  (2007) 090}, [\href{http://arxiv.org/abs/hep-th/0701063}{{\tt
  hep-th/0701063}}].

\bibitem{Hanany:2007zz}
A.~Hanany, \emph{{Counting BPS operators in the chiral ring: The plethystic
  story}}, \href{http://dx.doi.org/10.1063/1.2803801}{\emph{AIP Conf. Proc.}
  {\bf 939} (2007) 165--175}.

\bibitem{Forcella:2008bb}
D.~Forcella, A.~Hanany, Y.-H. He and A.~Zaffaroni, \emph{{The Master Space of
  N=1 Gauge Theories}},
  \href{http://dx.doi.org/10.1088/1126-6708/2008/08/012}{\emph{JHEP} {\bf 08}
  (2008) 012}, [\href{http://arxiv.org/abs/0801.1585}{{\tt 0801.1585}}].

\bibitem{Forcella:2008eh}
D.~Forcella, A.~Hanany, Y.-H. He and A.~Zaffaroni, \emph{{Mastering the Master
  Space}}, \href{http://dx.doi.org/10.1007/s11005-008-0255-6}{\emph{Lett. Math.
  Phys.} {\bf 85} (2008) 163--171}, [\href{http://arxiv.org/abs/0801.3477}{{\tt
  0801.3477}}].

\bibitem{Kho:2023dcm}
M.~Kho and R.-K. Seong, \emph{{On the master space for brane brick models}},
  \href{http://dx.doi.org/10.1007/JHEP09(2023)150}{\emph{JHEP} {\bf 09} (2023)
  150}, [\href{http://arxiv.org/abs/2306.16616}{{\tt 2306.16616}}].

\bibitem{Gray:2006jb}
J.~Gray, Y.-H. He, V.~Jejjala and B.~D. Nelson, \emph{{Exploring the vacuum
  geometry of N=1 gauge theories}},
  \href{http://dx.doi.org/10.1016/j.nuclphysb.2006.06.001}{\emph{Nucl. Phys. B}
  {\bf 750} (2006) 1--27}, [\href{http://arxiv.org/abs/hep-th/0604208}{{\tt
  hep-th/0604208}}].

\bibitem{Batyrev_1982}
V.~V. Batyrev, \emph{Toroidal fano 3-folds},
  \href{http://dx.doi.org/10.1070/IM1982v019n01ABEH001404}{\emph{Mathematics of
  the USSR-Izvestiya} {\bf 19} (feb, 1982) 13}.

\bibitem{batyrev1993dual}
V.~V. Batyrev, \emph{Dual polyhedra and mirror symmetry for calabi-yau
  hypersurfaces in toric varieties},
  \href{http://arxiv.org/abs/alg-geom/9310003}{{\tt alg-geom/9310003}}.

\bibitem{borisov1993towards}
L.~Borisov, \emph{Towards the mirror symmetry for calabi-yau complete
  intersections in gorenstein toric fano varieties},
  \href{http://arxiv.org/abs/alg-geom/9310001}{{\tt alg-geom/9310001}}.

\bibitem{batyrev1994dual}
V.~V. Batyrev and L.~A. Borisov, \emph{Dual cones and mirror symmetry for
  generalized calabi-yau manifolds},
  \href{http://arxiv.org/abs/alg-geom/9402002}{{\tt alg-geom/9402002}}.

\bibitem{Batyrev:1994pg}
V.~V. Batyrev and L.~A. Borisov, \emph{{On Calabi-Yau complete intersections in
  toric varieties}},  \href{http://arxiv.org/abs/alg-geom/9412017}{{\tt
  alg-geom/9412017}}.

\bibitem{batyrev1999classification}
V.~V. Batyrev, \emph{On the classification of toric fano 4-folds},
  {\emph{Journal of Mathematical Sciences} {\bf 94} (1999) 1021--1050}.

\bibitem{He:2017gam}
Y.-H. He, R.-K. Seong and S.-T. Yau, \emph{{Calabi\textendash{}Yau Volumes and
  Reflexive Polytopes}},
  \href{http://dx.doi.org/10.1007/s00220-018-3128-6}{\emph{Commun. Math. Phys.}
  {\bf 361} (2018) 155--204}, [\href{http://arxiv.org/abs/1704.03462}{{\tt
  1704.03462}}].

\bibitem{Bao:2024nyu}
J.~Bao, E.~Choi, Y.-H. He, R.-K. Seong and S.-T. Yau, \emph{{Futaki Invariants
  and Reflexive Polygons}},  \href{http://arxiv.org/abs/2410.18476}{{\tt
  2410.18476}}.

\bibitem{cox2024toric}
D.~A. Cox, J.~B. Little and H.~K. Schenck, \emph{Toric varieties}, vol.~124.
\newblock American Mathematical Society, 2024.

\bibitem{ewald1988classification}
G.~Ewald, \emph{On the classification of toric fano varieties}, {\emph{Discrete
  \& computational geometry} {\bf 3} (1988) 49--54}.

\bibitem{watanabe1982classification}
K.~Watanabe and M.~Watanabe, \emph{The classification of fano 3-folds with
  torus embeddings}, {\emph{Tokyo Journal of Mathematics} {\bf 5} (1982)
  37--48}.

\bibitem{nill2005gorenstein}
B.~Nill, \emph{Gorenstein toric fano varieties}, {\emph{manuscripta
  mathematica} {\bf 116} (2005) 183--210}.

\bibitem{givental1996equivariant}
A.~B. Givental, \emph{Equivariant gromov-witten invariants},
  \href{http://arxiv.org/abs/alg-geom/9603021}{{\tt alg-geom/9603021}}.

\bibitem{beck2008maximal}
M.~Beck, S.~V. Sam and K.~M. Woods, \emph{Maximal periods of (ehrhart)
  quasi-polynomials}, {\emph{Journal of Combinatorial Theory, Series A} {\bf
  115} (2008) 517--525}.

\bibitem{ehrhart1977polynomes}
E.~Ehrhart, \emph{Polynomes arithmetiques et methode de polyedres en
  combinatoire}.
\newblock International Series of Numerical Mathematics. Birkh{\"a}user Basel,
  1977.

\bibitem{stanley1980decompositions}
R.~P. Stanley, \emph{Decompositions of rational convex polytopes}, {\emph{Ann.
  Discrete Math} {\bf 6} (1980) 333--342}.

\bibitem{Hanany:2014dia}
A.~Hanany and R.~Kalveks, \emph{{Highest Weight Generating Functions for
  Hilbert Series}},
  \href{http://dx.doi.org/10.1007/JHEP10(2014)152}{\emph{JHEP} {\bf 10} (2014)
  152}, [\href{http://arxiv.org/abs/1408.4690}{{\tt 1408.4690}}].

\bibitem{Franco:2017cjj}
S.~Franco, D.~Ghim, S.~Lee and R.-K. Seong, \emph{{Elliptic Genera of 2d (0,2)
  Gauge Theories from Brane Brick Models}},
  \href{http://dx.doi.org/10.1007/JHEP06(2017)068}{\emph{JHEP} {\bf 06} (2017)
  068}, [\href{http://arxiv.org/abs/1702.02948}{{\tt 1702.02948}}].

\bibitem{Franco:2016fxm}
S.~Franco, S.~Lee and R.-K. Seong, \emph{{Orbifold Reduction and 2d (0,2) Gauge
  Theories}}, \href{http://dx.doi.org/10.1007/JHEP03(2017)016}{\emph{JHEP} {\bf
  03} (2017) 016}, [\href{http://arxiv.org/abs/1609.07144}{{\tt 1609.07144}}].

\bibitem{Douglas:1997de}
M.~R. Douglas, B.~R. Greene and D.~R. Morrison, \emph{{Orbifold resolution by
  D-branes}},
  \href{http://dx.doi.org/10.1016/S0550-3213(97)00517-8}{\emph{Nucl. Phys. B}
  {\bf 506} (1997) 84--106}, [\href{http://arxiv.org/abs/hep-th/9704151}{{\tt
  hep-th/9704151}}].

\bibitem{Hanany:2005ve}
A.~Hanany and K.~D. Kennaway, \emph{{Dimer models and toric diagrams}},
  \href{http://arxiv.org/abs/hep-th/0503149}{{\tt hep-th/0503149}}.

\bibitem{Franco:2005rj}
S.~Franco, A.~Hanany, K.~D. Kennaway, D.~Vegh and B.~Wecht, \emph{{Brane dimers
  and quiver gauge theories}},
  \href{http://dx.doi.org/10.1088/1126-6708/2006/01/096}{\emph{JHEP} {\bf 01}
  (2006) 096}, [\href{http://arxiv.org/abs/hep-th/0504110}{{\tt
  hep-th/0504110}}].

\bibitem{Franco:2005sm}
S.~Franco, A.~Hanany, D.~Martelli, J.~Sparks, D.~Vegh and B.~Wecht,
  \emph{{Gauge theories from toric geometry and brane tilings}},
  \href{http://dx.doi.org/10.1088/1126-6708/2006/01/128}{\emph{JHEP} {\bf 01}
  (2006) 128}, [\href{http://arxiv.org/abs/hep-th/0505211}{{\tt
  hep-th/0505211}}].

\bibitem{Hanany:2012hi}
A.~Hanany and R.-K. Seong, \emph{{Brane Tilings and Reflexive Polygons}},
  \href{http://dx.doi.org/10.1002/prop.201200008}{\emph{Fortsch. Phys.} {\bf
  60} (2012) 695--803}, [\href{http://arxiv.org/abs/1201.2614}{{\tt
  1201.2614}}].

\bibitem{Bianchi:2014qma}
M.~Bianchi, S.~Cremonesi, A.~Hanany, J.~F. Morales, D.~Ricci~Pacifici and R.-K.
  Seong, \emph{{Mass-deformed Brane Tilings}},
  \href{http://dx.doi.org/10.1007/JHEP10(2014)027}{\emph{JHEP} {\bf 10} (2014)
  027}, [\href{http://arxiv.org/abs/1408.1957}{{\tt 1408.1957}}].

\bibitem{Cremonesi:2023psg}
S.~Cremonesi and J.~S\'a, \emph{{Zig-zag deformations of toric quiver gauge
  theories. Part I. Reflexive polytopes}},
  \href{http://dx.doi.org/10.1007/JHEP05(2024)114}{\emph{JHEP} {\bf 05} (2024)
  114}, [\href{http://arxiv.org/abs/2312.13909}{{\tt 2312.13909}}].

\bibitem{higashitani2022deformations}
A.~Higashitani, Y.~Nakajima et~al., \emph{Deformations of dimer models},
  {\emph{SIGMA. Symmetry, Integrability and Geometry: Methods and Applications}
  {\bf 18} (2022) 030}.

\bibitem{Benini:2009gi}
F.~Benini, S.~Benvenuti and Y.~Tachikawa, \emph{{Webs of five-branes and N=2
  superconformal field theories}},
  \href{http://dx.doi.org/10.1088/1126-6708/2009/09/052}{\emph{JHEP} {\bf 09}
  (2009) 052}, [\href{http://arxiv.org/abs/0906.0359}{{\tt 0906.0359}}].

\bibitem{vanBeest:2020kou}
M.~van Beest, A.~Bourget, J.~Eckhard and S.~Schafer-Nameki, \emph{{(Symplectic)
  Leaves and (5d Higgs) Branches in the Poly(go)nesian Tropical Rain Forest}},
  \href{http://dx.doi.org/10.1007/JHEP11(2020)124}{\emph{JHEP} {\bf 11} (2020)
  124}, [\href{http://arxiv.org/abs/2008.05577}{{\tt 2008.05577}}].

\bibitem{Franco:2023flw}
S.~Franco and R.-K. Seong, \emph{{Twin theories, polytope mutations and quivers
  for GTPs}}, \href{http://dx.doi.org/10.1007/JHEP07(2023)034}{\emph{JHEP} {\bf
  07} (2023) 034}, [\href{http://arxiv.org/abs/2302.10951}{{\tt 2302.10951}}].

\bibitem{Arias-Tamargo:2024fjt}
G.~Arias-Tamargo, S.~Franco and D.~Rodr\'\i{}guez-G\'omez, \emph{{The geometry
  of GTPs and 5d SCFTs}},
  \href{http://dx.doi.org/10.1007/JHEP07(2024)159}{\emph{JHEP} {\bf 07} (2024)
  159}, [\href{http://arxiv.org/abs/2403.09776}{{\tt 2403.09776}}].

\bibitem{CarrenoBolla:2024fxy}
I.~Carre\~no Bolla, S.~Franco and D.~Rodr\'\i{}guez-G\'omez, \emph{{The 5d
  Tangram: Brane Webs, 7-Branes and Primitive T-cones}},
  \href{http://arxiv.org/abs/2411.01510}{{\tt 2411.01510}}.

\bibitem{Franco:2018qsc}
S.~Franco and A.~Hasan, \emph{{$3d$ printing of $2d$ $
  \mathcal{N}=\left(0,2\right) $ gauge theories}},
  \href{http://dx.doi.org/10.1007/JHEP05(2018)082}{\emph{JHEP} {\bf 05} (2018)
  082}, [\href{http://arxiv.org/abs/1801.00799}{{\tt 1801.00799}}].

\end{thebibliography}\endgroup

\end{document}